\documentclass[twocolumn,11pt]{article}
\usepackage[displaymath, mathlines]{lineno}
\usepackage{authblk,natbib,amsmath,latexsym,pgf,vmargin,datetime,float,cancel,relsize,dblfloatfix,type1cm,placeins}
\usepackage{abstract}
\newcommand{\introduction}{\section{Introduction}}
\newenvironment{acknowledgements}{\section*{Acknowledgements}}{}
\setpapersize{A4}
\setmarginsrb{1.45cm}{1.45cm}{1.45cm}{1.45cm}{0mm}{0mm}{0mm}{11mm}
\date{}

\newcommand{\tsup}[1]{\textsuperscript{#1}}

\usepackage{color,enumitem}
\usepackage[colorlinks,allcolors=blue]{hyperref}
\newcommand{\prog}[1]{{\rm\bf#1}}

\usepackage{tikz-qtree}

\usepackage{fancyvrb}
\fvset{frame=single,framerule=.2mm,gobble=4,fontsize=\relscale{.86}}
\input{pygments}

\usepackage{float}
\newfloat{Listing}{hbtp}{lop}[section]
\newcommand*\FancyVerbStartString{}
\newcommand*\FancyVerbStopString{}

\usepackage{tikz,nicefrac,latexsym}
\usetikzlibrary{arrows,shadows}
\usepackage{pgf-umlsd}

\begin{document}
\title{
  libcloudph++ 0.2: single-moment bulk, double-moment bulk, and particle-based warm-rain microphysics library in C++
}

\renewcommand\Authands{, }

\author[1]{Sylwester~Arabas}
\author[1]{Anna~Jaruga}
\author[1]{Hanna~Pawlowska}
\author[2]{Wojciech~W.~Grabowski\footnote{Affiliate Professor at the University of Warsaw}}

\affil[1]{\small Institute of Geophysics, Faculty of Physics, University of Warsaw, Warsaw, Poland}
\affil[2]{\small National Center for Atmospheric Research (NCAR), Boulder, Colorado, USA}

\maketitle 

\begin{abstract}
This paper introduces a \underline{lib}rary of algorithms 
  for representing \underline{cloud} micro\underline{ph}ysics
  in numerical models.
The library is written in C\underline{++}, 
  hence the name {\it libcloudph++}.
In the current release, the library covers three warm-rain
  schemes: the single- and double-moment bulk schemes,
  and the particle-based scheme with Monte-Carlo coalescence.
The three schemes are intended for modelling frameworks of different 
  dimensionality and complexity ranging from parcel models
  to multi-dimensional cloud-resolving (e.g. large-eddy) simulations.
A two-dimensional prescribed-flow framework is used in example simulations
  presented in the paper with the aim of highlighting the library features.
The {\it libcloudph++} and all its mandatory dependencies are free and 
  open-source software.
The Boost.units library is used for zero-overhead dimensional analysis
  of the code at compile time.
The particle-based scheme is implemented using the Thrust library that
  allows to leverage the power of graphics processing units (GPU),
  retaining the possibility to compile the unchanged code for execution
  on single or multiple standard processors (CPUs).
The paper includes complete description of the programming interface (API) of
  the library and a performance analysis 
  including comparison of GPU and CPU setups.
\end{abstract}
\tableofcontents

\introduction\label{sec:intro}  

Representation of cloud processes in numerical models is crucial for 
  weather and climate prediction.
Taking climate modelling as an example, one may learn that numerous distinct 
  modelling systems are designed in similar ways, sharing not only the concepts
  but also the implementations of some of their components \citep{Pennel_and_Reichler_2010}.
This creates a perfect opportunity for code reuse which is one of the key ''best practices''
  for scientific computing \citep[][sec. 6]{Wilson_et_al_2014}.
The reality, however, is that the code to be shared is often ''transplanted'' from one model
  to another \citep[][sec. 4.6]{Easterbrook_and_Johns_2009} rather than reused in a way 
  enabling the users to benefit from ongoing development and updates of the shared code.
From the authors' experience, this practise is not uncommon in development
  of limited-area models as well (yet, such software-engineering issues are rarely
  the subject of discussion in literature).
As a consequence, there exist multiple implementations of the same algorithms 
  but it is difficult to dissect and attribute the differences among them.
Avoiding ''transplants'' in the code is not easy, as numerous software
  projects in atmospheric modelling feature monolithic design that hampers code reuse.

This brings us to the conclusion that there is a potential demand for a library-type
  cloud-microphysics software package that could be readily reused and that would enable
  its users to easily benefit from developments of other researchers (by gaining access
  to enhancements, corrections, or entirely new schemes).
Library approach would not only facilitate collaboration, but also reduce development time
  and maintenance effort by imposing separation of cloud microphysics logic from
  other source code components such as model dynamical core or parallelisation logic.
Such strict separation is also a prerequisite for genuine software testing.

Popularity of several geoscientific-modelling software packages 
  that offer shared-library functionality
  suggests soundness of such approach -- e.g., libRadtran 
  \citep{Mayer_and_Kylling_2005} and CLUBB \citep{Golaz_et_al_2002},
  cited nearly 350 and 100 times, respectively.

The motivation behind the development of the \mbox{libclouph++} library introduced herein is twofold.
First, we intend to exemplify the possibilities of library-based code reuse in the context of cloud modelling.
Second, in the long run, we intend to offer the community a range of tools applicable for research 
  on some of the key topics in atmospheric science such as the interactions between
  aerosol, clouds and precipitation -- phenomena that still pose significant challenges for 
  the existing tools and methodologies \citep{Stevens_and_Feingold_2009}.

The library is being designed with the aim of creating a collection of algorithms
  to be used within models of different dimensionality, 
  different dynamical cores, different parallelisation strategies, 
  and in principle models written in different programming languages.
Presented library is written in C++, a choice motivated by the availability of high-performance object-oriented
  libraries and the built-in ''template'' mechanism.
C++ templates allow the implemented algorithms not to be bound to a single data type,
  single array dimensionality or single hardware type (e.g. CPU/GPU choice).
The library code and documentation are released as free (meaning both gratis \& libre) 
  and open-source software -- a prerequisite for use in auditable and reproducible 
  research \citep{Morin_et_al_2012,Ince_et_al_2012}.

Openness, together with code brevity and documentation, 
  are also crucial for enabling the users not to treat the library
  as a ''black box''.
While the aim of creating a self-contained package with well-defined
  interface is black-box approach compatible,
  the authors encourage users to inspect and test the code.

Modelling of atmospheric clouds and precipitation implies
  employment of computational techniques for particle-laden flows.
These are divided into Eulerian and Lagrangian approaches
  \citep[see e.g.][Chapter 8]{Crowe_et_al_2012}.
In the Eulerian approach, the cloud and precipitation properties are assumed 
  to be continuous in space, like those of a fluid.
In the Lagrangian approach, the so-called computational particles are tracked
  through the model domain.
Information associated with those particles travels
  along their trajectories.
The local properties of a given volume can be diagnosed taking into 
  account the properties of particles contained within it.
The Eulerian approach is well suited for modelling transport
  of gaseous species in the atmosphere and hence is the most common 
  choice for modelling atmospheric flows in general.
This is why most cloud microphysics models are build using the Eulerian concept 
  \citep[][e.g~chapter~9.1]{Straka_2009}.
However, it is the Lagrangian approach that is particularly well suited for dilute flows
  such as those of cloud droplets and rain drops in the atmosphere.

In the current release, {\it libcloudph++} is equipped with implementations
  of three distinct models of cloud microphysics.
All three belong to the so-called warm-rain class of schemes, meaning they
  cover representation of processes leading to formation of rain but they
  do not cover representation of the ice phase (snow, hail, graupel, etc.).
The so-called single-moment bulk and double-moment bulk schemes described 
  in sections \ref{sec:bulk} and \ref{sec:mm} belong to the Eulerian class of methods.
In section \ref{sec:sdm}, a coupled Eulerian-Lagrangian particle-based scheme is presented.
In the particle-based scheme, Lagrangian tracking is used to represent the dispersed phase
  (atmospheric aerosol, cloud droplets, rain drops), while 
  the continuous phase (moisture, heat) is represented with the Eulerian approach.
Description of each of the three schemes is aimed at providing a complete set of
  information needed to use it, and includes:
\begin{itemize}
  \item{discussion of key assumptions,}
  \item{formulation of the scheme,}
  \item{definition of the Application Programming Interface (API)}
  \item{overview of the implementation,}
  \item{example results.}
\end{itemize}
The particle-based scheme, being a novel approach to modelling clouds and precipitation, 
  is~discussed in more detail than the bulk schemes.

Sections covering descriptions of the APIs include C++ code listings
  of all data structure definitions and function signatures needed to use the library.
In those sections, C++ nomenclature is used without introduction
  \citep[for reference, see][that includes C++11 used in the presented code]{Brokken_2013}.

Sections covering scheme formulation feature cloud-modelling nomenclature
  (see appendix~\ref{sec:common} for a brief introduction and further reading).
In general, it is our approach not to repeat in the text the referenced formul\ae~readily available in recent papers,
  but only to include equations that are specific to the presented formulation and its implementation.

Before introducing the three implemented schemes in sections \ref{sec:bulk}-\ref{sec:sdm},
  formulation of an example modelling context
  is presented in section \ref{sec:framework}.
Section \ref{sec:perf} presents a performance evaluation of all three schemes.
Section \ref{sec:concl} provides a summary of the key features of {\it libcloudph++}
  and outlines the development plans for the next releases.

Appendix~\ref{sec:common} contains an outline of governing equations
  for moist atmospheric flow.
Appendix~\ref{sec:symbols} contains a list of symbols 
  used throughout the~text.
Appendix~\ref{sec:icicle} covers description of a program called {\em icicle} 
  that depends on {\it libcloudph++} and is used to perform the example 
  2D simulations presented throughout the~text.

\section{Modelling context example}\label{sec:context}

Being a library, {\it libcloudph++} does not constitute a complete modelling system.
It is a set of reusable software components that need to be
  coupled at least with a dynamical core responsible for representing air motion.
In this section we describe an example context in which the library may be used.
The three following subsections cover description of a modelling framework,
  a set-up including initial conditions, and a conceptual numerical solver.
Example results obtained with these simulation components are presented
  alongside the microphysics schemes in sections~\ref{sec:bulk},~\ref{sec:mm}~and~\ref{sec:sdm}.

\subsection{2D kinematic framework}\label{sec:framework}

The formulation is inspired by the 2D kinematic framework described in
   \citet{Szumowski_et_al_1998,Morrison_and_Grabowski_2007,Rasinski_et_al_2011}.
A simple 2D kinematic framework mimicking 
  air motion in a cloud allows (and limits) one to study cloud 
  microphysical processes decoupled from cloud dynamics. 
In fact, the differences between simulations 
  when feedback on the dynamics is taken out can lead to better understanding 
  of the role of flow dynamics \citep[e.g.][]{Slawinska_et_al_2009}.
Such approach results in a computationally cheap yet still insightful set-up
  of potential use in: (i) development and testing of cloud-processes 
  parameterisations for larger scale models; (ii) studying such 
  processes as cloud processing of aerosols; and (iii) developing
  remote-sensing retrieval procedures involving detailed treatment 
  of cloud microphysics.

The primary constituting assumption is the stationarity of the
  dry-air density (here, a vertical profile $\rho_d(z)$ is used) 
  which allows to prescribe the 2D velocity field using a streamfunction:
\begin{eqnarray}\label{eq:uwstream}
    \begin{cases}
      \rho_d \cdot u =- \partial_z \psi\\
      \rho_d \cdot w = \partial_x \psi
    \end{cases}
\end{eqnarray}
where $\psi=\psi(x,z;t)$ is the streamfunction and 
$u$ and $w$ denote horizontal and vertical components of the velocity field $\vec{u}$.

As a side note, one may notice that the stationarity of the dry-air density field 
  together with phase-change-related variations in time of temperature and water vapour mixing ratio
  imply time variations of the pressure profile.
The deviations from the initial (hydrostatic) profile are insignificant.

\subsection{8\tsup{th} ICMW VOCALS set-up}\label{sec:setup}

\begin{figure}
  \center
  \includegraphics[width=.45\textwidth]{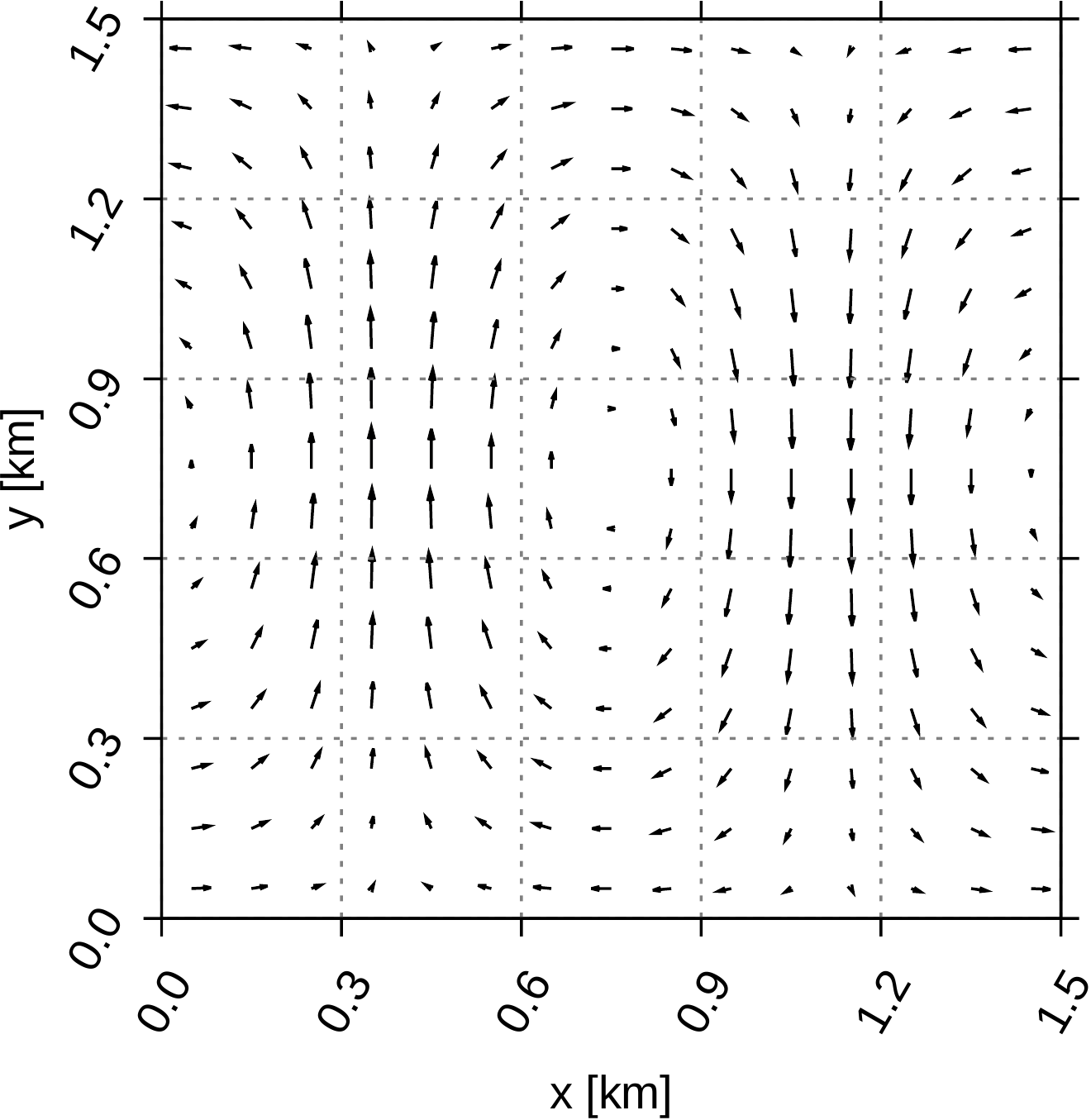}
  \caption{
    The constant-in-time velocity field used in the presented 2D simulations.
    See discussion of equations~\ref{eq:uwstream} and \ref{eq:rasinski}.
  }\label{fig:vel}
\end{figure}

Sample simulations presented in the following sections are based on a modelling set-up
  designed for the 8\tsup{th} International Cloud Modelling Workshop \citep[ICMW][case~1]{Muhlbauer_et_al_2013}.
It was designed as a simplest scenario applicable for benchmarking model capabilities for
  research on aerosol processing by clouds.
The cloud depth and aerosol characteristics are chosen to allow precipitation to develop over time and
  to mimic a drizzling stratocumulus cloud.

The set-up uses a kinematic framework of the type defined in the preceding subsection.
The definition of $\psi(x,z)$ is the same as in \citet[][Eq.~2]{Rasinski_et_al_2011}:
\begin{eqnarray}\label{eq:rasinski}
  \psi(x,z) = - w_{max} \frac{X}{\pi} {\rm sin}\!\left(\pi \frac{z}{Z}\right) {\rm cos}\!\left(2\pi \frac{x}{X} \right)
\end{eqnarray}
with $w_{max} = 0.6$\,m\,s\tsup{-1}, domain width $X=1.5\,\,{\rm km}$ and domain height $Z=1.5\,\,{\rm km}$. 
The resulting velocity field (depicted in Figure \ref{fig:vel}) 
  mimics an eddy spanning the whole domain, 
  and thus covering an updraught and a downdraught region.
The domain is periodic in horizontal.
To maintain flow incompressibility up to round-off error,
  velocity components (cf. eq.~\ref{eq:uwstream}) are derived from
  (\ref{eq:rasinski}) using numerical differentiation formul\ae~for
  a given grid type (Arakawa-C grid is used in the examples presented in the paper).

The initial profiles of liquid-water potential temperature $\theta_l$
  and the total water mixing ratio $r_t$ are defined as constant with altitude 
  ($\theta_l=289\,\,{\rm K}$; $r_t=7.5$\,g\,kg\tsup{-1}). 
The initial air-density profile corresponds to hydrostatic equilibrium with a pressure
  of $1015\,\,{\rm hPa}$ at the bottom of the domain.
This results in supersaturation in the upper part of the domain,
  where a cloud deck is formed in the simulations.

The domain is assumed to contain aerosol particles.
Their dry size spectrum is a bi-modal log-normal distribution:
\begin{eqnarray}\label{eq:lognormal}
  N(r_d) \!\!=\!\! \sum_m \! \frac{N_m}{\sqrt{2\,\pi}\,{\rm ln}(\sigma_m)} \frac{1}{r_d} {\rm exp}\!\!\left[\!-\!\left(\!\frac{{\rm ln}(\frac{r_d}{r_m})}{\sqrt{2}{\rm ln}(\sigma_m)}\!\right)^{\!\!\!2}\right]
\end{eqnarray}
  with the following parameters
  \citep[values close to those measured in the VOCALS campaign][Table~4]{Allen_et_al_2011}:
\begin{itemize}
  \item[~]{$\sigma_1\!=\!1.4$;~~ $d_1\!=\!0.04\,\,{\rm\mu m}$;~~ $N_{1}\!=\!60\,\,{\rm cm^{-3}}$}
  \item[~]{$\sigma_2\!=\!1.6$;~~ $d_2\!=\!0.15\,\,{\rm\mu m}$;~~ $N_{2}\!=\!40\,\,{\rm cm^{-3}}$}
\end{itemize} 
where $\sigma_{1,2}$ is the geometric standard deviation, $d_{1,2}=2\cdot r_{1,2}$ is the mode diameter
  and $N_{1,2}$ is the particle concentration at standard 
  conditions (T=20\tsup{$\circ$}C and p=1013.25~hPa).
This corresponds to a vertical gradient of concentration
  in the actual conditions of the model set-up due to air density changing with height, 
  and a gradual shift towards larger sizes of wet particle spectrum (due to relative humidity changing with height).
Both modes of the distribution are assumed to be composed of ammonium sulphate.

For models that include a description of the cloud droplet size spectrum,
  the initial data for the droplet concentration and size are to be  
  obtained by initialising the simulation with a two-hour-long spin-up period.
During the spin-up, precipitation formation and cloud drop sedimentation
  are switched off.
The spin-up period is intended to adjust an initial cloud droplet size spectrum
  (not specified by the setup) to an equilibrium state matching the 
  formulation of cloud microphysics with the prescribed flow.

One may chose to initialise the model with $\theta=\theta_l$ and $r_v=r_t$,
  and no condensed water (as it was done in the examples presented in this paper).
This simplifies initialisation, but results in an unrealistic initial supersaturation
  that may be an issue for a given microphysics scheme.
One may chose to impose a limit on the supersaturation, say $5\%$ (RH=1.05), when
  activating cloud drops during the spin-up.

To maintain steady mean temperature and moisture profiles (i.e.~to~compensate 
  for gradual water loss due to precipitation and warming of the boundary layer due to
  latent heating), mean temperature and moisture profiles are relaxed to the initial profile.
The temperature and moisture equations include an additional source term in the form 
  $-(\phi_0 - <\!\!\phi\!\!>)/\tau$, where $\phi_0$, $<\!\!\phi\!\!>$ and $\tau$ are 
  the initial profile, the horizontal mean of $\phi$ at a given height and 
  the relaxation time scale, respectively.
The relaxation time scale $\tau$ is height-dependant (mimicking effects of surface heat fluxes)
  and is prescribed as $\tau = \tau_{\rm rlx} \cdot {\rm exp}(z/z_{\rm rlx})$
  with $\tau_{\rm rlx} = 300$\,s and $z_{\rm rlx} = 200$\,m.
Note that such formulation does not dump small-scale perturbations of $\phi$, but simply shifts the horizontal
  mean toward $\phi_0$. 

The grid is composed of 75$\times$75 cells of equal size
  (hence the grid steps are 20\,m in both directions).
The advection-component timestep is one second.
Shorter sub-timesteps may be used within a microphysics component.

\begin{figure}[th!]
  \center
  \input{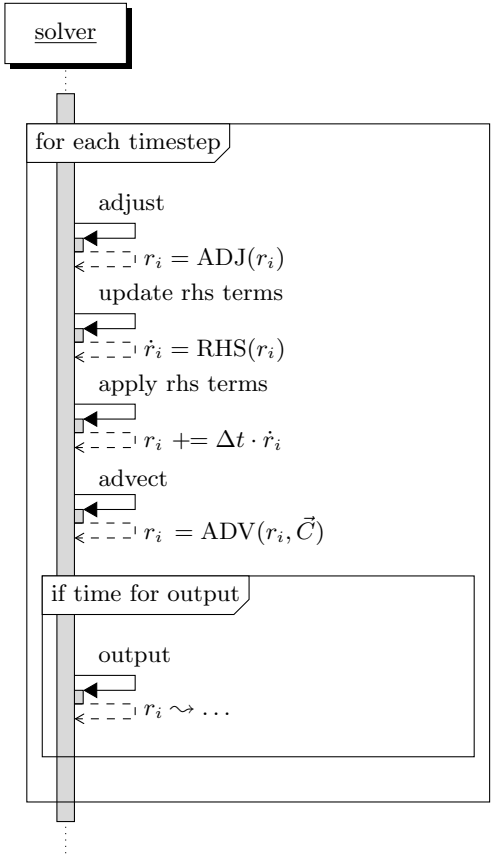}
  \caption{\label{fig:uml_proto}
    A sequence diagram depicting control flow in a conceptual solver
      described in section~\ref{sec:proto}.
    This solver design is extended with {\it libcloudph++} API calls 
      in diagrams presented in Figures~\ref{fig:uml_blk_1m}, 
      \ref{fig:uml_blk_2m} and \ref{fig:uml_lgrngn}.
    The diagram structure is modelled after the 
      Unified Modeling Language (UML) sequence diagrams.
    Arrows with solid lines depict calls, while the dashed arrows
      depict returns from the called code.
    Individual solver steps are annotated with labels 
      expressed in semi-mathematical notation and
      depicting key data dependencies.
    Model state variables are named $r_i$, their corresponding rhs
      terms are named $\dot{r}_i$.
    If a symbol appears on both sides of the equation, a programming-like
      assignment notation is meant, in which the old value of the symbol is used 
      prior to assignment, i.e. $r_i={\rm ADV}(r_i, \vec{C})$.
    ADV, ADJ and RHS depict all operations the solver
      does during the advection, adjustment and right-hand-side update
      steps, respectively.
  }
\end{figure}
\subsection{A conceptual solver}\label{sec:proto}

Example calling sequences for {\it libcloudph++}'s API are described in the following sections
  using a conceptual solver depicted in the diagram in Figure~\ref{fig:uml_proto}.
The conceptual solver is meant to perform numerical integration
  of a system of heterogeneous transport equations, each equation of the form:
\begin{eqnarray}
  \partial_t r_i + \frac{1}{\rho_d} \nabla \cdot (\vec{u} \rho_d r_i) = \dot{r}_i 
\end{eqnarray} 
where $r_i$ is the mixing ratio of the advected constituent, $\rho_d$ is the dry-air 
  ``carrier flow'' density, $\vec{u}$ is the velocity
  field, and the dotted right-hand-side term $\dot{r}_i$ depicts sources 
  (see also appendix~\ref{sec:common}).
The solver logic consists of five steps executed in a loop,
  with each loop repetition advancing the solution by one timestep.
Each of the first four integration steps is annotated in Figure~\ref{fig:uml_proto} 
  and described in the following paragraph.
The final step does data output and is performed conditionally every few timesteps.

The proposed solver design features uncentered-in-time integration of
  the right-hand-side terms.
Besides the right-hand-side terms, the integration procedure provides for representation 
  of sources using what is hereinafter referred to as adjustments.
Adjustments are basically all modifications of the model state that
  are not representable as right-hand-side terms
  (e.g. due to not being formulated as time derivatives).
Adjustments are done after advecting but before updating right-hand-side terms.

The library code itself is not bound to this particular solver logic --
  it is just a simple example intended to present the library API.
We refer the reader to \citet{Grabowski_and_Smolarkiewicz_2002} for discussion 
  of higher-order integration techniques for moist atmospheric flows.


\section{Single-moment bulk scheme}\label{sec:bulk}

\begin{table*}
  \caption{\label{tab:vars}
    State variables for the three implemented schemes.
    Number of state variables times the number of Eulerian grid cells plus
    number of particle attributes times the number of Lagrangian computational particles 
    gives an estimation of the memory requirement of a given scheme.
    See~appendix~\ref{sec:symbols} for symbol definitions.
  }
  \center
  \begin{tabular}{c|l|l}
                     &   Eulerian (PDE)                              & Lagrangian (ODE)                  \\
                     &    state variables                            & particle attributes               \\\hline
  1-moment bulk      &   $\theta$, $r_v$, $r_c$, $r_r$               & ---                               \\
  2-moment bulk      &   $\theta$, $r_v$, $r_c$, $r_r$, $n_c$, $n_r$ & ---                               \\
  particle-based     &   $\theta$, $r_v$                             & $r_d^3$, $r_w^2$, $N$, $\kappa$   \\
  \end{tabular}
\end{table*}

A common approach to represent cloud water and precipitation in
  a numerical simulation is the so-called single-moment bulk approach.
The concepts behind it date back to the seminal works of
  \citet[][section~3, and earlier works cited therein]{Kessler_1995}.
The constituting assumption of the scheme is the division of water
  condensate into two categories: cloud water and rain water.
The term single-moment refers to the fact that only the total mass 
  (proportional to the third moment of the particle size distribution)
  of water per category (cloud or rain) is considered in the model formulation.

In an Eulerian framework, two transport equations for the
  cloud water mixing ratio $r_c$ and the rain water mixing ratio $r_r$ are solved
  (in addition to the state variables $\theta$ and $r_v$ representing 
  heat and moisture content, respectively, 
  see Table~\ref{tab:vars} for a list of model-state variables
  in all schemes discussed in the paper).

Single-moment bulk microphysics is a simplistic approach.
Without information about the shape of droplet size distribution,
  the model is hardly capable of being coupled with a description
  of aerosol- or radiative-transfer processes.

\subsection{Formulation}
\subsubsection{Key assumptions}

The basic idea is to maintain saturation in the presence of cloud water.
Condensation/evaporation of cloud water triggered by supersaturation/subsaturation happens
  instantaneously.
Rain water forms through autoconversion
  of cloud water into rain (the negligible condensation 
  of rain water is not represented).
Autoconversion happens only after a prescribed threshold of the
  cloud water density is reached.
Subsequent increase in rain water is also possible through the
  accretion of cloud water by rain.

Cloud water is assumed to follow the airflow, whereas rain water falls
  relative to the air with sedimentation velocity.
Rain water evaporates only after all available cloud water
  has been evaporated and saturation is still not reached.
In contrast to cloud water, rain water evaporation 
  does not happen instantly.
Rain evaporation rate is a function of relative humidity, and
  is parameterised with an assumed shape of the raindrop size distribution.

\subsubsection{Phase changes}\label{sec:satadj}

Phase changes of water are represented with the so-called saturation adjustment
  procedure.
Unlike in several other formulations of the saturation adjustment procedure \citep[cf.][chapt.~4.2]{Straka_2009},
  the one implemented in {\em libcloudph++} covers not only cloud water 
  condensation and evaporation, but also rain water evaporation.

Any excess of water vapour with respect to saturation
  is instantly converted into cloud water, bringing the relative
  humidity to 100\%.
Similarly, any deficit with respect to saturation causes
  instantaneous evaporation of liquid water.
The formulation of the saturation adjustment procedure is given
  here making the latent heat release equation a starting point.
The heat source depicted with $\Delta \theta$ is defined through two integrals, 
  the first representing condensation or evaporation of cloud water,
  and the second one representing rain evaporation:
\begin{eqnarray} 
  \Delta \theta \,\, = \!&\! \mathlarger{\int\limits_{r_v }^{r_{v}' } \frac{d\theta}{dr_v} dr_v} &
                         \!+ \int\limits_{r_v'}^{r_{v}''} \frac{d\theta}{dr_v} dr_v \label{eq:satadj} \\
  \Delta r_v    \,\, = \!&\! \,\,\,\underbrace{(r_v' - r_v)}_{-\Delta r_c} &+\,\,\, \underbrace{(r_v'' - r_v')}_{-\Delta r_r}      
\end{eqnarray}
where $\nicefrac{d\theta}{dr_v}=\nicefrac{-\theta l_v}{c_{pd} T}$ 
  (cf. eq.~\ref{eq:rhod_th} in appendix~\ref{sec:common}) and the integration limit $r_v'$ 
  for cloud water condensation/evaporation is:
\begin{eqnarray} 
  r_v' = \begin{cases}
    r_{vs}'      & r_v > r_{vs} \\
    r_{vs}'      & r_v \le r_{vs} \,\,\,_\land\,\,\, r_c \ge r_{vs}' - r_v \\
    r_v + r_c & r_v \le r_{vs} \,\,\,_\land\,\,\, r_c < r_{vs}' - r_v \\
  \end{cases}
\end{eqnarray}
  where $r_{vs}' = r_{vs}(\rho_d, \theta', r_v')$ is the saturation vapour density
  evaluated after the adjustment.
The first case above corresponds to supersaturation.
The second and third cases correspond to subsaturation with either enough or insufficient amount of 
  cloud water to bring the air back to saturation.

When saturation is reached through condensation or evaporation of the cloud water,
  the second integral in~(\ref{eq:satadj}) representing evaporation of rain vanishes.
If not enough cloud water is available to reach saturation through evaporation, 
  the integration continues with the limit $\rho_v''$ defined as follows:
\begin{eqnarray}\label{eq:rho_v_bis}
  r_v'' = \begin{cases}
    r_v'                   & r_v' = r_{vs}' \\
    r_{vs}''               & r_v' < r_{vs}' \,\,\,_\land\,\,\, \delta r_r \ge r_{vs}'' - r_v' \\
    r_v' + \delta r_r      & r_v' < r_{vs}' \,\,\,_\land\,\,\, \delta r_r  <  r_{vs}'' - r_v'
  \end{cases}
\end{eqnarray}
where $\delta r_r$ depicts the limit of evaporation of rain within one timestep.
Here, it is parameterised as $\delta r_r = {\rm min}(r_r, \Delta t \cdot E_r)$ with $E_r$ being
  the evaporation rate of rain estimated
  following \citet[][eq.~5c]{Grabowski_and_Smolarkiewicz_1996} 
  using the formula of \citet[][eq.~25]{Ogura_and_Takahashi_1971}.
As with $r_{vs}'$, here $r_{vs}'' = r_{vs}(\rho_d, \theta'', r_v'')$.
  
Noteworthy, the name {\em adjustment} corresponds well with the 
  adjustments solver step introduced in section~\ref{sec:proto}
  as the procedure defined above is formulated through integration over vapour density
  rather than time \citep[see also discussion of eq.~3a in][]{Grabowski_and_Smolarkiewicz_1990}.
  
\subsubsection{Coalescence}

Collisions and coalescence between droplets are modelled with two separate processes:
  autoconversion and accretion.
Autoconversion represents collisions between cloud droplets only,
  while accretion refers to collisions between rain drops and cloud droplets.
Both are formulated (parameterised) in a phenomenological manner as right-hand-side (rhs) terms following 
  \citet[][eq.~5c]{Grabowski_and_Smolarkiewicz_1996} using the Kessler's formul\ae.
See \citet{Wood_2005} for a recent review of how these formulations compare with other bulk warm-rain schemes.

In the Kessler's formulation, autoconversion source term is proportional to ${\rm max}(r_c - r_{c0}, 0)$,
  where the value of the mixing-ratio threshold $r_{c0}$ effectively controls the onset of precipitation
  in the simulation.
Values of $r_{c0}$ found in literature vary from $10^{-4}$ to $10^{-3}$\,kg/kg
  \citep{Grabowski_and_Smolarkiewicz_1996}.

\subsubsection{Sedimentation}\label{sec:blk_1m_sedi}

Representation of sedimentation of rain water is formulated as a rhs term.
Another commonly used approach is to alter the vertical component of the 
  Courant number when calculating advection.
Here, the rhs term is formulated employing the upstream advective scheme:
\begin{eqnarray}\label{eq:flux_first}
  \!\!\!\!\!\!\!\!\!\!\!\!\!\!&\dot{r}_r^{\rm new} \!&\!   \,=\,  \dot{r}_r^{\rm old} \,\, - (F_{in} - F_{out}) / \rho_d \\
  \!\!\!\!\!\!\!\!\!\!\!\!\!\!&            F_{in}  \!&\!   \,=\,  \left. F_{out} \right|_{\text{above}} \\
  \!\!\!\!\!\!\!\!\!\!\!\!\!\!&            F_{out} \!&\!   \,=\,  -\frac{r_r}{\Delta z} \frac{\left[ \left.\rho_d\right|_{\text{below}} v_t(\left.r_r\right|_{\text{below}}) + \rho_d v_t(r_r) \right]}{2} \label{eq:flux_last}
\end{eqnarray}
  where $^{\rm old}$ and $^{\rm new}$ superscripts are introduced 
  to indicate that $\dot{r}_r$ is~a~sum of multiple terms, the one representing
  sedimentation being only one of them.
The $\left.\right|_{\text{above}}$ and $\left.\right|_{\text{below}}$ symbols refer to grid cell sequence in a column,
  $v_t$~is the rain terminal velocity parameterised as a function of rain water mixing ratio
  \citep[eq.~5d in][]{Grabowski_and_Smolarkiewicz_1996},
  and $F_{in}$ and $F_{out}$ symbolise fluxes of $r_r$ through the grid cell edges.

Employment of the upstream scheme brings several consequences.
First, unlike the cellwise phase-change and coalescence formulation,
  the sedimentation scheme is defined over a grid column.
Second, the combination of terminal velocity, vertical grid cell spacing $\Delta z$
  and the timestep $\Delta t$ must adhere to the Courant condition
  \citep[cf. discussion in][]{Grabowski_and_Smolarkiewicz_2002}.
Last but not least, the upstream algorithm introduces numerical diffusion,
  that can be alleviated by application of a higher-order advection scheme
  \citet[e.g., MPDATA, cf.][and references therein]{Smolarkiewicz_2006}.

\subsection{Programming interface}\label{sec:blk_1m_API}

\subsubsection{API elements}

The single-moment bulk scheme's API consists of one structure (composite data type)
  and three functions, all defined within the \prog{libcloudph::blk\_1m} namespace.
The separation of the scheme's logic into the three functions is done
  first according to the conceptual solver design (i.e.~separation of rhs terms
  and adjustments), and second according to a data-dependency criterion
  (i.e.~cellwise or columnwise calculations).
In case of the single-moment bulk scheme, the three functions actually
  correspond to the three represented processes, namely phase changes (cellwise adjustments),
  coalescence (cellwise rhs terms) and sedimentation (columnwise rhs term).
Sedimentation is the only process involving columnwise traversal of the domain
  (note the $\left.\right|_{\text{above}}$ and $\left.\right|_{\text{below}}$ symbols in~eq.~\ref{eq:flux_first}-\ref{eq:flux_last}).

\begin{Listing}
  \renewcommand*\FancyVerbStartString{\PY{c+c1}{//\PYZlt{}listing\PYZgt{}}}%
  \renewcommand*\FancyVerbStopString{\PY{c+c1}{//\PYZlt{}/listing\PYZgt{}}}%
  \begin{Verbatim}[commandchars=\\\{\}]
\PY{c+cm}{/** @file}
\PY{c+cm}{  * @copyright University of Warsaw}
\PY{c+cm}{  * @brief Definition of a structure holding options for single\PYZhy{}moment bulk microphysics}
\PY{c+cm}{  * @section LICENSE}
\PY{c+cm}{  * GPLv3+ (see the COPYING file or http://www.gnu.org/licenses/)}
\PY{c+cm}{  */}

\PY{c+cp}{\PYZsh{}}\PY{c+cp}{pragma once}

\PY{k}{namespace} \PY{n}{libcloudphxx}
\PY{p}{\PYZob{}}
  \PY{k}{namespace} \PY{n}{blk\PYZus{}1m} 
  \PY{p}{\PYZob{}} 
\PY{c+c1}{//\PYZlt{}listing\PYZgt{}}
    \PY{k}{template}\PY{o}{\PYZlt{}}\PY{k}{typename} \PY{k+kt}{real\PYZus{}t}\PY{o}{\PYZgt{}}
    \PY{k}{struct} \PY{k+kt}{opts\PYZus{}t} \PY{p}{\PYZob{}}
      \PY{k+kt}{bool} 
        \PY{n}{cond} \PY{o}{=} \PY{n+nb}{true}\PY{p}{,}    \PY{c+c1}{// condensation}
        \PY{n}{cevp} \PY{o}{=} \PY{n+nb}{true}\PY{p}{,}    \PY{c+c1}{// evaporation of cloud}
        \PY{n}{revp} \PY{o}{=} \PY{n+nb}{true}\PY{p}{,}    \PY{c+c1}{// evaporation of rain }
        \PY{n}{conv} \PY{o}{=} \PY{n+nb}{true}\PY{p}{,}    \PY{c+c1}{// autoconversion}
        \PY{n}{accr} \PY{o}{=} \PY{n+nb}{true}\PY{p}{,}    \PY{c+c1}{// accretion}
        \PY{n}{sedi} \PY{o}{=} \PY{n+nb}{true}\PY{p}{;}    \PY{c+c1}{// sedimentation}
      \PY{k+kt}{real\PYZus{}t} 
        \PY{n}{r\PYZus{}c0}  \PY{o}{=} \PY{l+m+mf}{5e\PYZhy{}4}\PY{p}{,}   \PY{c+c1}{// autoconv. threshold}
        \PY{n}{r\PYZus{}eps} \PY{o}{=} \PY{l+m+mf}{2e\PYZhy{}5}\PY{p}{;}   \PY{c+c1}{// absolute tolerance}
    \PY{p}{\PYZcb{}}\PY{p}{;}
\PY{c+c1}{//\PYZlt{}/listing\PYZgt{}}
  \PY{p}{\PYZcb{}}\PY{p}{;} 
\PY{p}{\PYZcb{}}\PY{p}{;}
\end{Verbatim}
  \vspace{-1.4em}%

  \caption{\label{lst:blk_1m_opt}
    \prog{blk\_1m::opts\_t} definition
  }
\end{Listing} 
The \prog{blk\_1m::opts\_t} structure (Listing~\ref{lst:blk_1m_opt})
  is intended for storing options of the scheme for a given simulation.
The template parameter \prog{real\_t} controls floating point format (e.g., float, double, long double, \ldots).
The structure fields include flags for toggling individual processes,
  a value of autoconversion threshold $r_{c0}$, and an absolute tolerance 
  used in numerically integrating the integrals in equation~\ref{eq:satadj}.
By default all processes are enabled, $r_{c0}=5\times10^{-4}$~kg/kg and the 
  tolerance is set to $2\times10^{-5}$~kg/kg.
All three functions from the single-moment bulk scheme's API expect 
  an instance of \prog{opts\_t} as their first parameter (see Listings \ref{lst:blk_1m_adj}-\ref{lst:blk_1m_clw}).

The saturation adjustment of state variables (cf. section~\ref{sec:satadj}) 
  is obtained through a call to the \prog{blk\_1m::adj\_cellwise()} 
  function (signature in Listing~\ref{lst:blk_1m_adj}).
The additional template parameter \prog{cont\_t} specifies the type of data
  container used for passing model state variables.
The function expects \prog{cont\_t} to be equipped with STL-style\footnote{C++ Standard Template Library} 
  iterator interface (e.g.,~the standard \prog{std::vector} class or a Blitz++ array slice as it is 
  used in the example code described in appendix~\ref{sec:icicle}).
\begin{Listing}
  \renewcommand*\FancyVerbStartString{\PY{c+c1}{//\PYZlt{}listing\PYZgt{}}}%
  \renewcommand*\FancyVerbStopString{\PY{c+c1}{//\PYZlt{}/listing\PYZgt{}}}%
  \begin{Verbatim}[commandchars=\\\{\}]
\PY{c+cm}{/** @file}
\PY{c+cm}{  * @copyright University of Warsaw}
\PY{c+cm}{  * @brief saturation adjustment routine using Boost.odeint for}
\PY{c+cm}{  *        solving latent\PYZhy{}heat release equation}
\PY{c+cm}{  * @section LICENSE}
\PY{c+cm}{  * GPLv3+ (see the COPYING file or http://www.gnu.org/licenses/)}
\PY{c+cm}{  */}

\PY{c+cp}{\PYZsh{}}\PY{c+cp}{pragma once}

\PY{c+cp}{\PYZsh{}}\PY{c+cp}{include \PYZlt{}libcloudph++}\PY{c+cp}{/}\PY{c+cp}{blk\PYZus{}1m}\PY{c+cp}{/}\PY{c+cp}{extincl.hpp\PYZgt{}}

\PY{k}{namespace} \PY{n}{libcloudphxx}
\PY{p}{\PYZob{}}
  \PY{k}{namespace} \PY{n}{blk\PYZus{}1m}
  \PY{p}{\PYZob{}}
    \PY{k}{namespace} \PY{n}{detail}
    \PY{p}{\PYZob{}}
      \PY{c+c1}{// ODE rhs describing latent\PYZhy{}heat release}
      \PY{k}{template} \PY{o}{\PYZlt{}}\PY{k}{typename} \PY{k+kt}{real\PYZus{}t}\PY{o}{\PYZgt{}}
      \PY{k}{class} \PY{n+nc}{rhs}
      \PY{p}{\PYZob{}}
	\PY{n+nl}{private:} 
         
        \PY{n}{quantity}\PY{o}{\PYZlt{}}\PY{n}{si}\PY{o}{:}\PY{o}{:}\PY{n}{mass\PYZus{}density}\PY{p}{,} \PY{k+kt}{real\PYZus{}t}\PY{o}{\PYZgt{}} \PY{n}{rhod}\PY{p}{;}
	
        \PY{n+nl}{public:} 
        
        \PY{k+kt}{void} \PY{n+nf}{init}\PY{p}{(}
	  \PY{k}{const} \PY{n}{quantity}\PY{o}{\PYZlt{}}\PY{n}{si}\PY{o}{:}\PY{o}{:}\PY{n}{mass\PYZus{}density}\PY{p}{,} \PY{k+kt}{real\PYZus{}t}\PY{o}{\PYZgt{}} \PY{o}{\PYZam{}}\PY{n}{\PYZus{}rhod}\PY{p}{,}
	  \PY{k}{const} \PY{n}{quantity}\PY{o}{\PYZlt{}}\PY{n}{si}\PY{o}{:}\PY{o}{:}\PY{n}{temperature}\PY{p}{,} \PY{k+kt}{real\PYZus{}t}\PY{o}{\PYZgt{}} \PY{o}{\PYZam{}}\PY{n}{th}\PY{p}{,}
	  \PY{k}{const} \PY{n}{quantity}\PY{o}{\PYZlt{}}\PY{n}{si}\PY{o}{:}\PY{o}{:}\PY{n}{dimensionless}\PY{p}{,} \PY{k+kt}{real\PYZus{}t}\PY{o}{\PYZgt{}} \PY{o}{\PYZam{}}\PY{n}{rv}
	\PY{p}{)}
	\PY{p}{\PYZob{}}
	  \PY{n}{rhod} \PY{o}{=} \PY{n}{\PYZus{}rhod}\PY{p}{;}
	  \PY{n}{update}\PY{p}{(}\PY{n}{th}\PY{p}{,} \PY{n}{rv}\PY{p}{)}\PY{p}{;}
	\PY{p}{\PYZcb{}}

	\PY{n}{quantity}\PY{o}{\PYZlt{}}\PY{n}{si}\PY{o}{:}\PY{o}{:}\PY{n}{dimensionless}\PY{p}{,} \PY{k+kt}{real\PYZus{}t}\PY{o}{\PYZgt{}} \PY{n}{r}\PY{p}{,} \PY{n}{rs}\PY{p}{;}
	\PY{n}{quantity}\PY{o}{\PYZlt{}}\PY{n}{si}\PY{o}{:}\PY{o}{:}\PY{n}{pressure}\PY{p}{,}      \PY{k+kt}{real\PYZus{}t}\PY{o}{\PYZgt{}} \PY{n}{p}\PY{p}{;}
	\PY{n}{quantity}\PY{o}{\PYZlt{}}\PY{n}{si}\PY{o}{:}\PY{o}{:}\PY{n}{temperature}\PY{p}{,}   \PY{k+kt}{real\PYZus{}t}\PY{o}{\PYZgt{}} \PY{n}{T}\PY{p}{;}

	\PY{n+nl}{private:} 

        \PY{k+kt}{void} \PY{n+nf}{update}\PY{p}{(}
	  \PY{k}{const} \PY{n}{quantity}\PY{o}{\PYZlt{}}\PY{n}{si}\PY{o}{:}\PY{o}{:}\PY{n}{temperature}\PY{p}{,} \PY{k+kt}{real\PYZus{}t}\PY{o}{\PYZgt{}} \PY{o}{\PYZam{}}\PY{n}{th}\PY{p}{,}
	  \PY{k}{const} \PY{n}{quantity}\PY{o}{\PYZlt{}}\PY{n}{si}\PY{o}{:}\PY{o}{:}\PY{n}{dimensionless}\PY{p}{,} \PY{k+kt}{real\PYZus{}t}\PY{o}{\PYZgt{}} \PY{o}{\PYZam{}}\PY{n}{rv}
	\PY{p}{)}
	\PY{p}{\PYZob{}}
          \PY{n}{r}  \PY{o}{=} \PY{n}{rv}\PY{p}{;}
	  \PY{n}{T}  \PY{o}{=} \PY{n}{common}\PY{o}{:}\PY{o}{:}\PY{n}{theta\PYZus{}dry}\PY{o}{:}\PY{o}{:}\PY{n}{T}\PY{o}{\PYZlt{}}\PY{k+kt}{real\PYZus{}t}\PY{o}{\PYZgt{}}\PY{p}{(}\PY{n}{th}\PY{p}{,} \PY{n}{rhod}\PY{p}{)}\PY{p}{;}
	  \PY{n}{p}  \PY{o}{=} \PY{n}{common}\PY{o}{:}\PY{o}{:}\PY{n}{theta\PYZus{}dry}\PY{o}{:}\PY{o}{:}\PY{n}{p}\PY{o}{\PYZlt{}}\PY{k+kt}{real\PYZus{}t}\PY{o}{\PYZgt{}}\PY{p}{(}\PY{n}{rhod}\PY{p}{,} \PY{n}{rv}\PY{p}{,} \PY{n}{T}\PY{p}{)}\PY{p}{;}
	  \PY{n}{rs} \PY{o}{=} \PY{n}{common}\PY{o}{:}\PY{o}{:}\PY{n}{const\PYZus{}cp}\PY{o}{:}\PY{o}{:}\PY{n}{r\PYZus{}vs}\PY{o}{\PYZlt{}}\PY{k+kt}{real\PYZus{}t}\PY{o}{\PYZgt{}}\PY{p}{(}\PY{n}{T}\PY{p}{,} \PY{n}{p}\PY{p}{)}\PY{p}{;}
	\PY{p}{\PYZcb{}}

	\PY{n+nl}{public:} 

	\PY{c+c1}{// F = d th / d rv }
        \PY{k+kt}{void} \PY{n+nf}{operator}\PY{p}{(}\PY{p}{)}\PY{p}{(}
	  \PY{k}{const} \PY{n}{quantity}\PY{o}{\PYZlt{}}\PY{n}{si}\PY{o}{:}\PY{o}{:}\PY{n}{temperature}\PY{p}{,} \PY{k+kt}{real\PYZus{}t}\PY{o}{\PYZgt{}} \PY{o}{\PYZam{}}\PY{n}{th}\PY{p}{,}
	  \PY{n}{quantity}\PY{o}{\PYZlt{}}\PY{n}{si}\PY{o}{:}\PY{o}{:}\PY{n}{temperature}\PY{p}{,} \PY{k+kt}{real\PYZus{}t}\PY{o}{\PYZgt{}} \PY{o}{\PYZam{}}\PY{n}{F}\PY{p}{,}
	  \PY{k}{const} \PY{n}{quantity}\PY{o}{\PYZlt{}}\PY{n}{si}\PY{o}{:}\PY{o}{:}\PY{n}{dimensionless}\PY{p}{,} \PY{k+kt}{real\PYZus{}t}\PY{o}{\PYZgt{}} \PY{o}{\PYZam{}}\PY{n}{rv}
	\PY{p}{)}
	\PY{p}{\PYZob{}}
	  \PY{n}{update}\PY{p}{(}\PY{n}{th}\PY{p}{,} \PY{n}{rv}\PY{p}{)}\PY{p}{;}
	  \PY{n}{F} \PY{o}{=} \PY{n}{common}\PY{o}{:}\PY{o}{:}\PY{n}{theta\PYZus{}dry}\PY{o}{:}\PY{o}{:}\PY{n}{d\PYZus{}th\PYZus{}d\PYZus{}rv}\PY{o}{\PYZlt{}}\PY{k+kt}{real\PYZus{}t}\PY{o}{\PYZgt{}}\PY{p}{(}\PY{n}{T}\PY{p}{,} \PY{n}{th}\PY{p}{)}\PY{p}{;} 
	\PY{p}{\PYZcb{}}
      \PY{p}{\PYZcb{}}\PY{p}{;}
    \PY{p}{\PYZcb{}}    

\PY{c+c1}{//\PYZlt{}listing\PYZgt{}}
    \PY{k}{template} \PY{o}{\PYZlt{}}\PY{k}{typename} \PY{k+kt}{real\PYZus{}t}\PY{p}{,} \PY{k}{class} \PY{n+nc}{cont\PYZus{}t}\PY{o}{\PYZgt{}}
    \PY{k+kt}{void} \PY{n}{adj\PYZus{}cellwise}\PY{p}{(}
      \PY{k}{const} \PY{k+kt}{opts\PYZus{}t}\PY{o}{\PYZlt{}}\PY{k+kt}{real\PYZus{}t}\PY{o}{\PYZgt{}} \PY{o}{\PYZam{}}\PY{n}{opts}\PY{p}{,}
      \PY{k}{const} \PY{k+kt}{cont\PYZus{}t} \PY{o}{\PYZam{}}\PY{n}{rhod\PYZus{}cont}\PY{p}{,} 
      \PY{k+kt}{cont\PYZus{}t} \PY{o}{\PYZam{}}\PY{n}{th\PYZus{}cont}\PY{p}{,} 
      \PY{k+kt}{cont\PYZus{}t} \PY{o}{\PYZam{}}\PY{n}{rv\PYZus{}cont}\PY{p}{,}
      \PY{k+kt}{cont\PYZus{}t} \PY{o}{\PYZam{}}\PY{n}{rc\PYZus{}cont}\PY{p}{,}
      \PY{k+kt}{cont\PYZus{}t} \PY{o}{\PYZam{}}\PY{n}{rr\PYZus{}cont}\PY{p}{,}
      \PY{k}{const} \PY{k+kt}{real\PYZus{}t} \PY{o}{\PYZam{}}\PY{n}{dt}
    \PY{p}{)}
\PY{c+c1}{//\PYZlt{}/listing\PYZgt{}}
    \PY{p}{\PYZob{}}
      \PY{k}{if} \PY{p}{(}\PY{o}{!}\PY{n}{opts}\PY{p}{.}\PY{n}{cond}\PY{p}{)} \PY{k}{return}\PY{p}{;} \PY{c+c1}{// ignoring values of opts.cevp and opts.revp}

      \PY{k}{namespace} \PY{n}{odeint} \PY{o}{=} \PY{n}{boost}\PY{o}{:}\PY{o}{:}\PY{n}{numeric}\PY{o}{:}\PY{o}{:}\PY{n}{odeint}\PY{p}{;}

      \PY{c+c1}{// odeint::euler\PYZlt{} // TODO: opcja?}
      \PY{n}{odeint}\PY{o}{:}\PY{o}{:}\PY{n}{runge\PYZus{}kutta4}\PY{o}{\PYZlt{}}
	\PY{n}{quantity}\PY{o}{\PYZlt{}}\PY{n}{si}\PY{o}{:}\PY{o}{:}\PY{n}{temperature}\PY{p}{,} \PY{k+kt}{real\PYZus{}t}\PY{o}{\PYZgt{}}\PY{p}{,}   \PY{c+c1}{// state\PYZus{}type}
	\PY{k+kt}{real\PYZus{}t}\PY{p}{,}                              \PY{c+c1}{// value\PYZus{}type}
	\PY{n}{quantity}\PY{o}{\PYZlt{}}\PY{n}{si}\PY{o}{:}\PY{o}{:}\PY{n}{temperature}\PY{p}{,} \PY{k+kt}{real\PYZus{}t}\PY{o}{\PYZgt{}}\PY{p}{,}   \PY{c+c1}{// deriv\PYZus{}type}
	\PY{n}{quantity}\PY{o}{\PYZlt{}}\PY{n}{si}\PY{o}{:}\PY{o}{:}\PY{n}{dimensionless}\PY{p}{,} \PY{k+kt}{real\PYZus{}t}\PY{o}{\PYZgt{}}\PY{p}{,} \PY{c+c1}{// time\PYZus{}type}
	\PY{n}{odeint}\PY{o}{:}\PY{o}{:}\PY{n}{vector\PYZus{}space\PYZus{}algebra}\PY{p}{,}
	\PY{n}{odeint}\PY{o}{:}\PY{o}{:}\PY{n}{default\PYZus{}operations}\PY{p}{,}
	\PY{n}{odeint}\PY{o}{:}\PY{o}{:}\PY{n}{never\PYZus{}resizer}
      \PY{o}{\PYZgt{}} \PY{n}{S}\PY{p}{;} \PY{c+c1}{// TODO: would be better to instantiate in the ctor (but what about thread safety! :()}
      \PY{k}{typename} \PY{n}{detail}\PY{o}{:}\PY{o}{:}\PY{n}{rhs}\PY{o}{\PYZlt{}}\PY{k+kt}{real\PYZus{}t}\PY{o}{\PYZgt{}} \PY{n}{F}\PY{p}{;}

      \PY{k}{for} \PY{p}{(}\PY{k}{auto} \PY{n}{tup} \PY{o}{:} \PY{n}{zip}\PY{p}{(}\PY{n}{rhod\PYZus{}cont}\PY{p}{,} \PY{n}{th\PYZus{}cont}\PY{p}{,} \PY{n}{rv\PYZus{}cont}\PY{p}{,} \PY{n}{rc\PYZus{}cont}\PY{p}{,} \PY{n}{rr\PYZus{}cont}\PY{p}{)}\PY{p}{)}
      \PY{p}{\PYZob{}}
        \PY{k}{const} \PY{k+kt}{real\PYZus{}t}
          \PY{o}{\PYZam{}}\PY{n}{rhod} \PY{o}{=} \PY{n}{boost}\PY{o}{:}\PY{o}{:}\PY{n}{get}\PY{o}{\PYZlt{}}\PY{l+m+mi}{0}\PY{o}{\PYZgt{}}\PY{p}{(}\PY{n}{tup}\PY{p}{)}\PY{p}{;}
        \PY{k+kt}{real\PYZus{}t} 
          \PY{o}{\PYZam{}}\PY{n}{th} \PY{o}{=} \PY{n}{boost}\PY{o}{:}\PY{o}{:}\PY{n}{get}\PY{o}{\PYZlt{}}\PY{l+m+mi}{1}\PY{o}{\PYZgt{}}\PY{p}{(}\PY{n}{tup}\PY{p}{)}\PY{p}{,} 
          \PY{o}{\PYZam{}}\PY{n}{rv} \PY{o}{=} \PY{n}{boost}\PY{o}{:}\PY{o}{:}\PY{n}{get}\PY{o}{\PYZlt{}}\PY{l+m+mi}{2}\PY{o}{\PYZgt{}}\PY{p}{(}\PY{n}{tup}\PY{p}{)}\PY{p}{,} 
          \PY{o}{\PYZam{}}\PY{n}{rc} \PY{o}{=} \PY{n}{boost}\PY{o}{:}\PY{o}{:}\PY{n}{get}\PY{o}{\PYZlt{}}\PY{l+m+mi}{3}\PY{o}{\PYZgt{}}\PY{p}{(}\PY{n}{tup}\PY{p}{)}\PY{p}{,} 
          \PY{o}{\PYZam{}}\PY{n}{rr} \PY{o}{=} \PY{n}{boost}\PY{o}{:}\PY{o}{:}\PY{n}{get}\PY{o}{\PYZlt{}}\PY{l+m+mi}{4}\PY{o}{\PYZgt{}}\PY{p}{(}\PY{n}{tup}\PY{p}{)}\PY{p}{;}

	\PY{c+c1}{// double\PYZhy{}checking....}
	\PY{n}{assert}\PY{p}{(}\PY{n}{th} \PY{o}{\PYZgt{}}\PY{o}{=} \PY{l+m+mf}{273.15}\PY{p}{)}\PY{p}{;} \PY{c+c1}{// TODO: that\PYZsq{}s theta, not T!}
	\PY{n}{assert}\PY{p}{(}\PY{n}{rc} \PY{o}{\PYZgt{}}\PY{o}{=} \PY{l+m+mi}{0}\PY{p}{)}\PY{p}{;}
	\PY{n}{assert}\PY{p}{(}\PY{n}{rv} \PY{o}{\PYZgt{}}\PY{o}{=} \PY{l+m+mi}{0}\PY{p}{)}\PY{p}{;}
	\PY{n}{assert}\PY{p}{(}\PY{n}{rr} \PY{o}{\PYZgt{}}\PY{o}{=} \PY{l+m+mi}{0}\PY{p}{)}\PY{p}{;} 

	\PY{n}{F}\PY{p}{.}\PY{n}{init}\PY{p}{(}
	  \PY{n}{rhod} \PY{o}{*} \PY{n}{si}\PY{o}{:}\PY{o}{:}\PY{n}{kilograms} \PY{o}{/} \PY{n}{si}\PY{o}{:}\PY{o}{:}\PY{n}{cubic\PYZus{}metres}\PY{p}{,}
	  \PY{n}{th}   \PY{o}{*} \PY{n}{si}\PY{o}{:}\PY{o}{:}\PY{n}{kelvins}\PY{p}{,}
	  \PY{n}{rv}   \PY{o}{*} \PY{n}{si}\PY{o}{:}\PY{o}{:}\PY{n}{dimensionless}\PY{p}{(}\PY{p}{)}
	\PY{p}{)}\PY{p}{;}

	\PY{k+kt}{real\PYZus{}t} \PY{n}{vapour\PYZus{}excess}\PY{p}{;}
	\PY{k+kt}{real\PYZus{}t} \PY{n}{drr\PYZus{}max} \PY{o}{=} \PY{l+m+mi}{0}\PY{p}{;}
	\PY{k}{if} \PY{p}{(}\PY{n}{F}\PY{p}{.}\PY{n}{rs} \PY{o}{\PYZgt{}} \PY{n}{F}\PY{p}{.}\PY{n}{r} \PY{o}{\PYZam{}}\PY{o}{\PYZam{}} \PY{n}{rr} \PY{o}{\PYZgt{}} \PY{l+m+mi}{0} \PY{o}{\PYZam{}}\PY{o}{\PYZam{}} \PY{n}{opts}\PY{p}{.}\PY{n}{revp}\PY{p}{)} 
        \PY{p}{\PYZob{}}
          \PY{n}{drr\PYZus{}max} \PY{o}{=} \PY{p}{(}\PY{n}{dt} \PY{o}{*} \PY{n}{si}\PY{o}{:}\PY{o}{:}\PY{n}{seconds}\PY{p}{)} \PY{o}{*} \PY{n}{formulae}\PY{o}{:}\PY{o}{:}\PY{n}{evaporation\PYZus{}rate}\PY{p}{(}
            \PY{n}{F}\PY{p}{.}\PY{n}{r}\PY{p}{,} \PY{n}{F}\PY{p}{.}\PY{n}{rs}\PY{p}{,} \PY{n}{rr} \PY{o}{*} \PY{n}{si}\PY{o}{:}\PY{o}{:}\PY{n}{dimensionless}\PY{p}{(}\PY{p}{)}\PY{p}{,} \PY{n}{rhod} \PY{o}{*} \PY{n}{si}\PY{o}{:}\PY{o}{:}\PY{n}{kilograms} \PY{o}{/} \PY{n}{si}\PY{o}{:}\PY{o}{:}\PY{n}{cubic\PYZus{}metres}\PY{p}{,} \PY{n}{F}\PY{p}{.}\PY{n}{p}
	  \PY{p}{)}\PY{p}{;}
        \PY{p}{\PYZcb{}}
	\PY{k+kt}{bool} \PY{n}{incloud}\PY{p}{;}

	\PY{c+c1}{// TODO: rethink and document r\PYZus{}eps!!!}
	\PY{k}{while} \PY{p}{(}
	  \PY{c+c1}{// condensation of cloud water if supersaturated more than a threshold}
	  \PY{p}{(}\PY{n}{vapour\PYZus{}excess} \PY{o}{=} \PY{n}{rv} \PY{o}{\PYZhy{}} \PY{n}{F}\PY{p}{.}\PY{n}{rs}\PY{p}{)} \PY{o}{\PYZgt{}} \PY{n}{opts}\PY{p}{.}\PY{n}{r\PYZus{}eps}
	  \PY{o}{|}\PY{o}{|} 
          \PY{p}{(} 
            \PY{n}{opts}\PY{p}{.}\PY{n}{cevp} \PY{o}{\PYZam{}}\PY{o}{\PYZam{}} \PY{n}{vapour\PYZus{}excess} \PY{o}{\PYZlt{}} \PY{o}{\PYZhy{}}\PY{n}{opts}\PY{p}{.}\PY{n}{r\PYZus{}eps} \PY{o}{\PYZam{}}\PY{o}{\PYZam{}} \PY{p}{(} \PY{c+c1}{// or if subsaturated and }
	      \PY{p}{(}\PY{n}{incloud} \PY{o}{=} \PY{p}{(}\PY{n}{rc} \PY{o}{\PYZgt{}} \PY{l+m+mi}{0}\PY{p}{)}\PY{p}{)}  \PY{c+c1}{// in cloud (then cloud evaporation first)}
	      \PY{o}{|}\PY{o}{|}                    \PY{c+c1}{// or }
              \PY{p}{(}\PY{n}{opts}\PY{p}{.}\PY{n}{revp} \PY{o}{\PYZam{}}\PY{o}{\PYZam{}} \PY{n}{rr} \PY{o}{\PYZgt{}} \PY{l+m+mi}{0} \PY{o}{\PYZam{}}\PY{o}{\PYZam{}} \PY{n}{drr\PYZus{}max} \PY{o}{\PYZgt{}} \PY{l+m+mi}{0}\PY{p}{)} \PY{c+c1}{// in rain shaft (rain evaporation out\PYZhy{}of\PYZhy{}cloud)}
	    \PY{p}{)}
          \PY{p}{)}
	\PY{p}{)}
	\PY{p}{\PYZob{}}
          \PY{c+c1}{// an arbitrary initial guess for drv}
	  \PY{k+kt}{real\PYZus{}t} \PY{n}{drv} \PY{o}{=} \PY{o}{\PYZhy{}} \PY{n}{copysign}\PY{p}{(}\PY{n}{std}\PY{o}{:}\PY{o}{:}\PY{n}{min}\PY{p}{(}\PY{l+m+mf}{.5} \PY{o}{*} \PY{n}{opts}\PY{p}{.}\PY{n}{r\PYZus{}eps}\PY{p}{,} \PY{l+m+mf}{.5} \PY{o}{*} \PY{n}{vapour\PYZus{}excess}\PY{p}{)}\PY{p}{,} \PY{n}{vapour\PYZus{}excess}\PY{p}{)}\PY{p}{;} 
          \PY{c+c1}{// preventing negative mixing ratios if evaporating}
	  \PY{k}{if} \PY{p}{(}\PY{n}{vapour\PYZus{}excess} \PY{o}{\PYZlt{}} \PY{l+m+mi}{0}\PY{p}{)} \PY{n}{drv} \PY{o}{=} 
            \PY{n}{incloud} \PY{o}{?} \PY{n}{std}\PY{o}{:}\PY{o}{:}\PY{n}{min}\PY{p}{(}\PY{n}{rc}\PY{p}{,} \PY{n}{drv}\PY{p}{)} \PY{c+c1}{// limiting by rc}
	            \PY{o}{:} \PY{n}{std}\PY{o}{:}\PY{o}{:}\PY{n}{min}\PY{p}{(}\PY{n}{drr\PYZus{}max}\PY{p}{,} \PY{n}{std}\PY{o}{:}\PY{o}{:}\PY{n}{min}\PY{p}{(}\PY{n}{rr}\PY{p}{,} \PY{n}{drv}\PY{p}{)}\PY{p}{)}\PY{p}{;} \PY{c+c1}{// limiting by rr and drr\PYZus{}max}
	  \PY{n}{assert}\PY{p}{(}\PY{n}{drv} \PY{o}{!}\PY{o}{=} \PY{l+m+mi}{0}\PY{p}{)}\PY{p}{;} \PY{c+c1}{// otherwise it should not pass the while condition!}

	  \PY{c+c1}{// theta is modified by do\PYZus{}step, and hence we cannot pass an expression and we need a temp. var.}
	  \PY{n}{quantity}\PY{o}{\PYZlt{}}\PY{n}{si}\PY{o}{:}\PY{o}{:}\PY{n}{temperature}\PY{p}{,} \PY{k+kt}{real\PYZus{}t}\PY{o}{\PYZgt{}} \PY{n}{tmp} \PY{o}{=} \PY{n}{th} \PY{o}{*} \PY{n}{si}\PY{o}{:}\PY{o}{:}\PY{n}{kelvins}\PY{p}{;}

	  \PY{c+c1}{// integrating the First Law for moist air}
	  \PY{n}{S}\PY{p}{.}\PY{n}{do\PYZus{}step}\PY{p}{(}
	    \PY{n}{boost}\PY{o}{:}\PY{o}{:}\PY{n}{ref}\PY{p}{(}\PY{n}{F}\PY{p}{)}\PY{p}{,}
	    \PY{n}{tmp}\PY{p}{,}
	    \PY{n}{rv}  \PY{o}{*} \PY{n}{si}\PY{o}{:}\PY{o}{:}\PY{n}{dimensionless}\PY{p}{(}\PY{p}{)}\PY{p}{,}
	    \PY{n}{drv} \PY{o}{*} \PY{n}{si}\PY{o}{:}\PY{o}{:}\PY{n}{dimensionless}\PY{p}{(}\PY{p}{)}
	  \PY{p}{)}\PY{p}{;}

	  \PY{c+c1}{// latent heat source/sink due to evaporation/condensation}
	  \PY{n}{th} \PY{o}{=} \PY{n}{tmp} \PY{o}{/} \PY{n}{si}\PY{o}{:}\PY{o}{:}\PY{n}{kelvins}\PY{p}{;}

	  \PY{c+c1}{// updating rv}
	  \PY{n}{rv} \PY{o}{+}\PY{o}{=} \PY{n}{drv}\PY{p}{;}
	  \PY{n}{assert}\PY{p}{(}\PY{n}{rv} \PY{o}{\PYZgt{}}\PY{o}{=} \PY{l+m+mi}{0}\PY{p}{)}\PY{p}{;}
	  
	  \PY{k}{if} \PY{p}{(}\PY{n}{vapour\PYZus{}excess} \PY{o}{\PYZgt{}} \PY{l+m+mi}{0} \PY{o}{|}\PY{o}{|} \PY{n}{incloud}\PY{p}{)}
	  \PY{p}{\PYZob{}}
            \PY{c+c1}{// condensation or evaporation of cloud water}
	    \PY{n}{rc} \PY{o}{\PYZhy{}}\PY{o}{=} \PY{n}{drv}\PY{p}{;}
	    \PY{n}{assert}\PY{p}{(}\PY{n}{rc} \PY{o}{\PYZgt{}}\PY{o}{=} \PY{l+m+mi}{0}\PY{p}{)}\PY{p}{;}
	  \PY{p}{\PYZcb{}}
	  \PY{k}{else} 
	  \PY{p}{\PYZob{}}
            \PY{c+c1}{// evaporation of rain water}
	    \PY{n}{assert}\PY{p}{(}\PY{n}{opts}\PY{p}{.}\PY{n}{revp}\PY{p}{)}\PY{p}{;} \PY{c+c1}{// should be guaranteed by the while() condition above}
	    \PY{n}{rr} \PY{o}{\PYZhy{}}\PY{o}{=} \PY{n}{drv}\PY{p}{;}
	    \PY{n}{assert}\PY{p}{(}\PY{n}{rr} \PY{o}{\PYZgt{}}\PY{o}{=} \PY{l+m+mi}{0}\PY{p}{)}\PY{p}{;}
	    \PY{k}{if} \PY{p}{(}\PY{p}{(}\PY{n}{drr\PYZus{}max} \PY{o}{\PYZhy{}}\PY{o}{=} \PY{n}{drv}\PY{p}{)} \PY{o}{=}\PY{o}{=} \PY{l+m+mi}{0}\PY{p}{)} \PY{k}{break}\PY{p}{;} \PY{c+c1}{// but not more than Kessler allows}
	  \PY{p}{\PYZcb{}}
	\PY{p}{\PYZcb{}}

	\PY{c+c1}{// hopefully true for RK4}
	\PY{n}{assert}\PY{p}{(}\PY{n}{F}\PY{p}{.}\PY{n}{r} \PY{o}{=}\PY{o}{=} \PY{n}{rv}\PY{p}{)}\PY{p}{;}
	\PY{c+c1}{// triple\PYZhy{}checking....}
	\PY{n}{assert}\PY{p}{(}\PY{n}{th} \PY{o}{\PYZgt{}}\PY{o}{=} \PY{l+m+mf}{273.15}\PY{p}{)}\PY{p}{;} \PY{c+c1}{// that is theta, not T ! TODO}
	\PY{n}{assert}\PY{p}{(}\PY{n}{rc} \PY{o}{\PYZgt{}}\PY{o}{=} \PY{l+m+mi}{0}\PY{p}{)}\PY{p}{;}
	\PY{n}{assert}\PY{p}{(}\PY{n}{rv} \PY{o}{\PYZgt{}}\PY{o}{=} \PY{l+m+mi}{0}\PY{p}{)}\PY{p}{;}
	\PY{n}{assert}\PY{p}{(}\PY{n}{rr} \PY{o}{\PYZgt{}}\PY{o}{=} \PY{l+m+mi}{0}\PY{p}{)}\PY{p}{;}
      \PY{p}{\PYZcb{}}
    \PY{p}{\PYZcb{}}
  \PY{p}{\PYZcb{}}
\PY{p}{\PYZcb{}}\PY{p}{;}
\end{Verbatim}
  \vspace{-1.4em}%

  \caption{\label{lst:blk_1m_adj}
    \prog{blk\_1m::adj\_cellwise()} signature
  }
\end{Listing}
\begin{Listing}
  \renewcommand*\FancyVerbStartString{\PY{c+c1}{//\PYZlt{}listing\PYZgt{}}}%
  \renewcommand*\FancyVerbStopString{\PY{c+c1}{//\PYZlt{}/listing\PYZgt{}}}%
  \begin{Verbatim}[commandchars=\\\{\}]
\PY{c+cm}{/** @file}
\PY{c+cm}{  * @copyright University of Warsaw}
\PY{c+cm}{  * @brief Autoconversion and collection righ\PYZhy{}hand side terms using Kessler formulae}
\PY{c+cm}{  * @section LICENSE}
\PY{c+cm}{  * GPLv3+ (see the COPYING file or http://www.gnu.org/licenses/)}
\PY{c+cm}{  */}

\PY{c+cp}{\PYZsh{}}\PY{c+cp}{pragma once}

\PY{c+cp}{\PYZsh{}}\PY{c+cp}{include \PYZlt{}libcloudph++}\PY{c+cp}{/}\PY{c+cp}{blk\PYZus{}1m}\PY{c+cp}{/}\PY{c+cp}{extincl.hpp\PYZgt{}}

\PY{k}{namespace} \PY{n}{libcloudphxx}
\PY{p}{\PYZob{}}
  \PY{k}{namespace} \PY{n}{blk\PYZus{}1m}
  \PY{p}{\PYZob{}}
\PY{c+c1}{//\PYZlt{}listing\PYZgt{}}
    \PY{k}{template} \PY{o}{\PYZlt{}}\PY{k}{typename} \PY{k+kt}{real\PYZus{}t}\PY{p}{,} \PY{k}{class} \PY{n+nc}{cont\PYZus{}t}\PY{o}{\PYZgt{}}
    \PY{k+kt}{void} \PY{n}{rhs\PYZus{}cellwise}\PY{p}{(}
      \PY{k}{const} \PY{k+kt}{opts\PYZus{}t}\PY{o}{\PYZlt{}}\PY{k+kt}{real\PYZus{}t}\PY{o}{\PYZgt{}} \PY{o}{\PYZam{}}\PY{n}{opts}\PY{p}{,}
      \PY{k+kt}{cont\PYZus{}t} \PY{o}{\PYZam{}}\PY{n}{dot\PYZus{}rc\PYZus{}cont}\PY{p}{,} 
      \PY{k+kt}{cont\PYZus{}t} \PY{o}{\PYZam{}}\PY{n}{dot\PYZus{}rr\PYZus{}cont}\PY{p}{,}
      \PY{k}{const} \PY{k+kt}{cont\PYZus{}t} \PY{o}{\PYZam{}}\PY{n}{rc\PYZus{}cont}\PY{p}{,}
      \PY{k}{const} \PY{k+kt}{cont\PYZus{}t} \PY{o}{\PYZam{}}\PY{n}{rr\PYZus{}cont}
    \PY{p}{)}   
\PY{c+c1}{//\PYZlt{}/listing\PYZgt{}}
    \PY{p}{\PYZob{}}
      \PY{k}{for} \PY{p}{(}\PY{k}{auto} \PY{n}{tup} \PY{o}{:} \PY{n}{zip}\PY{p}{(}\PY{n}{dot\PYZus{}rc\PYZus{}cont}\PY{p}{,} \PY{n}{dot\PYZus{}rr\PYZus{}cont}\PY{p}{,} \PY{n}{rc\PYZus{}cont}\PY{p}{,} \PY{n}{rr\PYZus{}cont}\PY{p}{)}\PY{p}{)}
      \PY{p}{\PYZob{}}
        \PY{k+kt}{real\PYZus{}t}
          \PY{n}{tmp} \PY{o}{=} \PY{l+m+mi}{0}\PY{p}{,}
          \PY{o}{\PYZam{}}\PY{n}{dot\PYZus{}rc} \PY{o}{=} \PY{n}{boost}\PY{o}{:}\PY{o}{:}\PY{n}{get}\PY{o}{\PYZlt{}}\PY{l+m+mi}{0}\PY{o}{\PYZgt{}}\PY{p}{(}\PY{n}{tup}\PY{p}{)}\PY{p}{,}
          \PY{o}{\PYZam{}}\PY{n}{dot\PYZus{}rr} \PY{o}{=} \PY{n}{boost}\PY{o}{:}\PY{o}{:}\PY{n}{get}\PY{o}{\PYZlt{}}\PY{l+m+mi}{1}\PY{o}{\PYZgt{}}\PY{p}{(}\PY{n}{tup}\PY{p}{)}\PY{p}{;}
        \PY{k}{const} \PY{k+kt}{real\PYZus{}t}
          \PY{o}{\PYZam{}}\PY{n}{rc}     \PY{o}{=} \PY{n}{boost}\PY{o}{:}\PY{o}{:}\PY{n}{get}\PY{o}{\PYZlt{}}\PY{l+m+mi}{2}\PY{o}{\PYZgt{}}\PY{p}{(}\PY{n}{tup}\PY{p}{)}\PY{p}{,}
          \PY{o}{\PYZam{}}\PY{n}{rr}     \PY{o}{=} \PY{n}{boost}\PY{o}{:}\PY{o}{:}\PY{n}{get}\PY{o}{\PYZlt{}}\PY{l+m+mi}{3}\PY{o}{\PYZgt{}}\PY{p}{(}\PY{n}{tup}\PY{p}{)}\PY{p}{;}

        \PY{c+c1}{// autoconversion}
        \PY{k}{if} \PY{p}{(}\PY{n}{opts}\PY{p}{.}\PY{n}{conv}\PY{p}{)}
        \PY{p}{\PYZob{}}
	  \PY{n}{tmp} \PY{o}{+}\PY{o}{=} \PY{p}{(} 
	    \PY{n}{formulae}\PY{o}{:}\PY{o}{:}\PY{n}{autoconversion\PYZus{}rate}\PY{p}{(}
              \PY{n}{rc}        \PY{o}{*} \PY{n}{si}\PY{o}{:}\PY{o}{:}\PY{n}{dimensionless}\PY{p}{(}\PY{p}{)}\PY{p}{,} 
              \PY{n}{opts}\PY{p}{.}\PY{n}{r\PYZus{}c0} \PY{o}{*} \PY{n}{si}\PY{o}{:}\PY{o}{:}\PY{n}{dimensionless}\PY{p}{(}\PY{p}{)}
            \PY{p}{)} \PY{o}{*} \PY{n}{si}\PY{o}{:}\PY{o}{:}\PY{n}{seconds} \PY{c+c1}{// to make it dimensionless}
	  \PY{p}{)}\PY{p}{;}
        \PY{p}{\PYZcb{}}

        \PY{c+c1}{// collection}
        \PY{k}{if} \PY{p}{(}\PY{n}{opts}\PY{p}{.}\PY{n}{accr}\PY{p}{)}
        \PY{p}{\PYZob{}}
	  \PY{n}{tmp} \PY{o}{+}\PY{o}{=} \PY{p}{(}
	    \PY{n}{formulae}\PY{o}{:}\PY{o}{:}\PY{n}{collection\PYZus{}rate}\PY{p}{(}
              \PY{n}{rc} \PY{o}{*} \PY{n}{si}\PY{o}{:}\PY{o}{:}\PY{n}{dimensionless}\PY{p}{(}\PY{p}{)}\PY{p}{,} 
              \PY{n}{rr} \PY{o}{*} \PY{n}{si}\PY{o}{:}\PY{o}{:}\PY{n}{dimensionless}\PY{p}{(}\PY{p}{)}
            \PY{p}{)} \PY{o}{*} \PY{n}{si}\PY{o}{:}\PY{o}{:}\PY{n}{seconds} \PY{c+c1}{// to make it dimensionless}
	  \PY{p}{)}\PY{p}{;}
        \PY{p}{\PYZcb{}}

	\PY{n}{dot\PYZus{}rr} \PY{o}{+}\PY{o}{=} \PY{n}{tmp}\PY{p}{;}
	\PY{n}{dot\PYZus{}rc} \PY{o}{\PYZhy{}}\PY{o}{=} \PY{n}{tmp}\PY{p}{;}
      \PY{p}{\PYZcb{}}
    \PY{p}{\PYZcb{}}    
  \PY{p}{\PYZcb{}}\PY{p}{;}
\PY{p}{\PYZcb{}}\PY{p}{;}
\end{Verbatim}
  \vspace{-1.4em}%

  \caption{\label{lst:blk_1m_elw}
    \prog{blk\_1m::rhs\_cellwise()} signature
  }
\end{Listing}
\begin{Listing}
  \renewcommand*\FancyVerbStartString{\PY{c+c1}{//\PYZlt{}listing\PYZgt{}}}%
  \renewcommand*\FancyVerbStopString{\PY{c+c1}{//\PYZlt{}/listing\PYZgt{}}}%
  \begin{Verbatim}[commandchars=\\\{\}]
\PY{c+cm}{/** @file}
\PY{c+cm}{  * @copyright University of Warsaw}
\PY{c+cm}{  * @brief Rain sedimentation representation for single\PYZhy{}moment bulk microphysics}
\PY{c+cm}{  *   using forcing terms based on the upstrem advection scheme }
\PY{c+cm}{  * @section LICENSE}
\PY{c+cm}{  * GPLv3+ (see the COPYING file or http://www.gnu.org/licenses/)}
\PY{c+cm}{  */}

\PY{c+cp}{\PYZsh{}}\PY{c+cp}{pragma once}

\PY{c+cp}{\PYZsh{}}\PY{c+cp}{include \PYZlt{}libcloudph++}\PY{c+cp}{/}\PY{c+cp}{blk\PYZus{}1m}\PY{c+cp}{/}\PY{c+cp}{extincl.hpp\PYZgt{}}

\PY{k}{namespace} \PY{n}{libcloudphxx}
\PY{p}{\PYZob{}}
  \PY{k}{namespace} \PY{n}{blk\PYZus{}1m}
  \PY{p}{\PYZob{}}
    \PY{c+c1}{// expects the arguments to be columns with begin() pointing to the lowest level}
    \PY{c+c1}{// returns rain flux out of the domain}
\PY{c+c1}{//\PYZlt{}listing\PYZgt{}}
    \PY{k}{template} \PY{o}{\PYZlt{}}\PY{k}{typename} \PY{k+kt}{real\PYZus{}t}\PY{p}{,} \PY{k}{class} \PY{n+nc}{cont\PYZus{}t}\PY{o}{\PYZgt{}}
    \PY{k+kt}{real\PYZus{}t} \PY{n}{rhs\PYZus{}columnwise}\PY{p}{(}
      \PY{k}{const} \PY{k+kt}{opts\PYZus{}t}\PY{o}{\PYZlt{}}\PY{k+kt}{real\PYZus{}t}\PY{o}{\PYZgt{}} \PY{o}{\PYZam{}}\PY{n}{opts}\PY{p}{,}
      \PY{k+kt}{cont\PYZus{}t} \PY{o}{\PYZam{}}\PY{n}{dot\PYZus{}rr\PYZus{}cont}\PY{p}{,}
      \PY{k}{const} \PY{k+kt}{cont\PYZus{}t} \PY{o}{\PYZam{}}\PY{n}{rhod\PYZus{}cont}\PY{p}{,}   
      \PY{k}{const} \PY{k+kt}{cont\PYZus{}t} \PY{o}{\PYZam{}}\PY{n}{rr\PYZus{}cont}\PY{p}{,}
      \PY{k}{const} \PY{k+kt}{real\PYZus{}t} \PY{o}{\PYZam{}}\PY{n}{dz} 
    \PY{p}{)}   
\PY{c+c1}{//\PYZlt{}/listing\PYZgt{}}
    \PY{p}{\PYZob{}}
      \PY{k}{using} \PY{k+kt}{flux\PYZus{}t} \PY{o}{=} \PY{n}{quantity}\PY{o}{\PYZlt{}}\PY{n}{divide\PYZus{}typeof\PYZus{}helper}\PY{o}{\PYZlt{}}\PY{n}{si}\PY{o}{:}\PY{o}{:}\PY{n}{mass\PYZus{}density}\PY{p}{,} \PY{n}{si}\PY{o}{:}\PY{o}{:}\PY{n}{time}\PY{o}{\PYZgt{}}\PY{o}{:}\PY{o}{:}\PY{n}{type}\PY{p}{,} \PY{k+kt}{real\PYZus{}t}\PY{o}{\PYZgt{}}\PY{p}{;}

      \PY{k}{auto} \PY{n}{dot\PYZus{}rr\PYZus{}unit} \PY{o}{=} \PY{n}{si}\PY{o}{:}\PY{o}{:}\PY{n}{hertz}\PY{p}{;}

      \PY{k}{if} \PY{p}{(}\PY{o}{!}\PY{n}{opts}\PY{p}{.}\PY{n}{sedi}\PY{p}{)} \PY{k}{return} \PY{l+m+mi}{0}\PY{p}{;}

      \PY{c+c1}{// }
      \PY{k+kt}{flux\PYZus{}t} \PY{n}{flux\PYZus{}in} \PY{o}{=} \PY{l+m+mi}{0} \PY{o}{*} \PY{n}{si}\PY{o}{:}\PY{o}{:}\PY{n}{kilograms} \PY{o}{/} \PY{n}{si}\PY{o}{:}\PY{o}{:}\PY{n}{cubic\PYZus{}metres} \PY{o}{/} \PY{n}{si}\PY{o}{:}\PY{o}{:}\PY{n}{seconds}\PY{p}{;}
      \PY{k+kt}{real\PYZus{}t} \PY{o}{*}\PY{n}{dot\PYZus{}rr} \PY{o}{=} \PY{n+nb}{NULL}\PY{p}{;}
      \PY{k}{const} \PY{k+kt}{real\PYZus{}t} \PY{n}{zero} \PY{o}{=} \PY{l+m+mi}{0}\PY{p}{;}
  
      \PY{c+c1}{// this should give zero flux from above the domain top}
      \PY{k}{const} \PY{k+kt}{real\PYZus{}t} \PY{o}{*}\PY{n}{rhod} \PY{o}{=} \PY{o}{\PYZam{}}\PY{o}{*}\PY{p}{(}\PY{o}{\PYZhy{}}\PY{o}{\PYZhy{}}\PY{p}{(}\PY{n}{rhod\PYZus{}cont}\PY{p}{.}\PY{n}{end}\PY{p}{(}\PY{p}{)}\PY{p}{)}\PY{p}{)}\PY{p}{,} \PY{o}{*}\PY{n}{rr} \PY{o}{=} \PY{o}{\PYZam{}}\PY{n}{zero}\PY{p}{;}

      \PY{k}{auto} \PY{n}{iter} \PY{o}{=} \PY{n}{zip}\PY{p}{(}\PY{n}{dot\PYZus{}rr\PYZus{}cont}\PY{p}{,} \PY{n}{rhod\PYZus{}cont}\PY{p}{,} \PY{n}{rr\PYZus{}cont}\PY{p}{)}\PY{p}{;}
      \PY{k}{for} \PY{p}{(}\PY{k}{auto} \PY{n}{tup\PYZus{}ptr} \PY{o}{=} \PY{n}{iter}\PY{p}{.}\PY{n}{end}\PY{p}{(}\PY{p}{)}\PY{p}{;} \PY{n}{tup\PYZus{}ptr} \PY{o}{!}\PY{o}{=} \PY{n}{iter}\PY{p}{.}\PY{n}{begin}\PY{p}{(}\PY{p}{)}\PY{p}{;}\PY{p}{)}
      \PY{p}{\PYZob{}}
        \PY{o}{\PYZhy{}}\PY{o}{\PYZhy{}}\PY{n}{tup\PYZus{}ptr}\PY{p}{;}

        \PY{k}{const} \PY{k+kt}{real\PYZus{}t}
          \PY{o}{*}\PY{n}{rhod\PYZus{}below}  \PY{o}{=} \PY{o}{\PYZam{}}\PY{n}{boost}\PY{o}{:}\PY{o}{:}\PY{n}{get}\PY{o}{\PYZlt{}}\PY{l+m+mi}{1}\PY{o}{\PYZgt{}}\PY{p}{(}\PY{o}{*}\PY{n}{tup\PYZus{}ptr}\PY{p}{)}\PY{p}{,}
          \PY{o}{*}\PY{n}{rr\PYZus{}below}    \PY{o}{=} \PY{o}{\PYZam{}}\PY{n}{boost}\PY{o}{:}\PY{o}{:}\PY{n}{get}\PY{o}{\PYZlt{}}\PY{l+m+mi}{2}\PY{o}{\PYZgt{}}\PY{p}{(}\PY{o}{*}\PY{n}{tup\PYZus{}ptr}\PY{p}{)}\PY{p}{;}

        \PY{k}{if} \PY{p}{(}\PY{n}{dot\PYZus{}rr} \PY{o}{!}\PY{o}{=} \PY{n+nb}{NULL}\PY{p}{)} \PY{c+c1}{// i.e. all but first (top) grid cell}
        \PY{p}{\PYZob{}}
          \PY{c+c1}{// terminal momenta at grid\PYZhy{}cell edge (to assure precip mass conservation)}
	  \PY{k+kt}{flux\PYZus{}t} \PY{n}{flux\PYZus{}out} \PY{o}{=} \PY{o}{\PYZhy{}}\PY{k+kt}{real\PYZus{}t}\PY{p}{(}\PY{l+m+mf}{.5}\PY{p}{)} \PY{o}{*} \PY{p}{(} \PY{c+c1}{// averaging + axis orientation}
	    \PY{p}{(}\PY{o}{*}\PY{n}{rhod\PYZus{}below} \PY{o}{*} \PY{n}{si}\PY{o}{:}\PY{o}{:}\PY{n}{kilograms} \PY{o}{/} \PY{n}{si}\PY{o}{:}\PY{o}{:}\PY{n}{cubic\PYZus{}metres}\PY{p}{)} \PY{o}{*} \PY{n}{formulae}\PY{o}{:}\PY{o}{:}\PY{n}{v\PYZus{}term}\PY{p}{(}
              \PY{o}{*}\PY{n}{rr\PYZus{}below}          \PY{o}{*} \PY{n}{si}\PY{o}{:}\PY{o}{:}\PY{n}{kilograms} \PY{o}{/} \PY{n}{si}\PY{o}{:}\PY{o}{:}\PY{n}{kilograms}\PY{p}{,} 
              \PY{o}{*}\PY{n}{rhod\PYZus{}below}        \PY{o}{*} \PY{n}{si}\PY{o}{:}\PY{o}{:}\PY{n}{kilograms} \PY{o}{/} \PY{n}{si}\PY{o}{:}\PY{o}{:}\PY{n}{cubic\PYZus{}metres}\PY{p}{,} 
              \PY{o}{*}\PY{n}{rhod\PYZus{}cont}\PY{p}{.}\PY{n}{begin}\PY{p}{(}\PY{p}{)} \PY{o}{*} \PY{n}{si}\PY{o}{:}\PY{o}{:}\PY{n}{kilograms} \PY{o}{/} \PY{n}{si}\PY{o}{:}\PY{o}{:}\PY{n}{cubic\PYZus{}metres}
            \PY{p}{)} \PY{o}{+} 
	    \PY{p}{(}\PY{o}{*}\PY{n}{rhod} \PY{o}{*} \PY{n}{si}\PY{o}{:}\PY{o}{:}\PY{n}{kilograms} \PY{o}{/} \PY{n}{si}\PY{o}{:}\PY{o}{:}\PY{n}{cubic\PYZus{}metres}\PY{p}{)} \PY{o}{*} \PY{n}{formulae}\PY{o}{:}\PY{o}{:}\PY{n}{v\PYZus{}term}\PY{p}{(}
              \PY{o}{*}\PY{n}{rr}                \PY{o}{*} \PY{n}{si}\PY{o}{:}\PY{o}{:}\PY{n}{kilograms} \PY{o}{/} \PY{n}{si}\PY{o}{:}\PY{o}{:}\PY{n}{kilograms}\PY{p}{,}    
              \PY{o}{*}\PY{n}{rhod}              \PY{o}{*} \PY{n}{si}\PY{o}{:}\PY{o}{:}\PY{n}{kilograms} \PY{o}{/} \PY{n}{si}\PY{o}{:}\PY{o}{:}\PY{n}{cubic\PYZus{}metres}\PY{p}{,} 
              \PY{o}{*}\PY{n}{rhod\PYZus{}cont}\PY{p}{.}\PY{n}{begin}\PY{p}{(}\PY{p}{)} \PY{o}{*} \PY{n}{si}\PY{o}{:}\PY{o}{:}\PY{n}{kilograms} \PY{o}{/} \PY{n}{si}\PY{o}{:}\PY{o}{:}\PY{n}{cubic\PYZus{}metres}
            \PY{p}{)}
	  \PY{p}{)} \PY{o}{*} \PY{p}{(}\PY{o}{*}\PY{n}{rr} \PY{o}{*} \PY{n}{si}\PY{o}{:}\PY{o}{:}\PY{n}{kilograms} \PY{o}{/} \PY{n}{si}\PY{o}{:}\PY{o}{:}\PY{n}{kilograms}\PY{p}{)} \PY{o}{/} \PY{p}{(}\PY{n}{dz} \PY{o}{*} \PY{n}{si}\PY{o}{:}\PY{o}{:}\PY{n}{metres}\PY{p}{)}\PY{p}{;}

	  \PY{o}{*}\PY{n}{dot\PYZus{}rr} \PY{o}{\PYZhy{}}\PY{o}{=} \PY{p}{(}\PY{n}{flux\PYZus{}in} \PY{o}{\PYZhy{}} \PY{n}{flux\PYZus{}out}\PY{p}{)} \PY{o}{/} \PY{p}{(}\PY{o}{*}\PY{n}{rhod} \PY{o}{*} \PY{n}{si}\PY{o}{:}\PY{o}{:}\PY{n}{kilograms} \PY{o}{/} \PY{n}{si}\PY{o}{:}\PY{o}{:}\PY{n}{cubic\PYZus{}metres}\PY{p}{)} \PY{o}{/} \PY{n}{dot\PYZus{}rr\PYZus{}unit}\PY{p}{;}
          \PY{n}{flux\PYZus{}in} \PY{o}{=} \PY{n}{flux\PYZus{}out}\PY{p}{;} \PY{c+c1}{// inflow = outflow from above}
        \PY{p}{\PYZcb{}}

        \PY{n}{dot\PYZus{}rr} \PY{o}{=} \PY{o}{\PYZam{}}\PY{n}{boost}\PY{o}{:}\PY{o}{:}\PY{n}{get}\PY{o}{\PYZlt{}}\PY{l+m+mi}{0}\PY{o}{\PYZgt{}}\PY{p}{(}\PY{o}{*}\PY{n}{tup\PYZus{}ptr}\PY{p}{)}\PY{p}{;}
        \PY{n}{rhod}   \PY{o}{=} \PY{n}{rhod\PYZus{}below}\PY{p}{;}
        \PY{n}{rr}     \PY{o}{=} \PY{n}{rr\PYZus{}below}\PY{p}{;}
      \PY{p}{\PYZcb{}}

      \PY{c+c1}{// the bottom grid cell (with mid\PYZhy{}cell vterm approximation)}
      \PY{k+kt}{flux\PYZus{}t} \PY{n}{flux\PYZus{}out} \PY{o}{=} \PY{o}{\PYZhy{}} \PY{p}{(}\PY{o}{*}\PY{n}{rhod} \PY{o}{*} \PY{n}{si}\PY{o}{:}\PY{o}{:}\PY{n}{kilograms} \PY{o}{/} \PY{n}{si}\PY{o}{:}\PY{o}{:}\PY{n}{cubic\PYZus{}metres}\PY{p}{)} \PY{o}{*} \PY{n}{formulae}\PY{o}{:}\PY{o}{:}\PY{n}{v\PYZus{}term}\PY{p}{(}
	\PY{o}{*}\PY{n}{rr}                \PY{o}{*} \PY{n}{si}\PY{o}{:}\PY{o}{:}\PY{n}{kilograms} \PY{o}{/} \PY{n}{si}\PY{o}{:}\PY{o}{:}\PY{n}{kilograms}\PY{p}{,}    
	\PY{o}{*}\PY{n}{rhod}              \PY{o}{*} \PY{n}{si}\PY{o}{:}\PY{o}{:}\PY{n}{kilograms} \PY{o}{/} \PY{n}{si}\PY{o}{:}\PY{o}{:}\PY{n}{cubic\PYZus{}metres}\PY{p}{,} 
	\PY{o}{*}\PY{n}{rhod\PYZus{}cont}\PY{p}{.}\PY{n}{begin}\PY{p}{(}\PY{p}{)} \PY{o}{*} \PY{n}{si}\PY{o}{:}\PY{o}{:}\PY{n}{kilograms} \PY{o}{/} \PY{n}{si}\PY{o}{:}\PY{o}{:}\PY{n}{cubic\PYZus{}metres}
      \PY{p}{)} \PY{o}{*} \PY{p}{(}\PY{o}{*}\PY{n}{rr} \PY{o}{*} \PY{n}{si}\PY{o}{:}\PY{o}{:}\PY{n}{kilograms} \PY{o}{/} \PY{n}{si}\PY{o}{:}\PY{o}{:}\PY{n}{kilograms}\PY{p}{)} \PY{o}{/} \PY{p}{(}\PY{n}{dz} \PY{o}{*} \PY{n}{si}\PY{o}{:}\PY{o}{:}\PY{n}{metres}\PY{p}{)}\PY{p}{;}
      \PY{o}{*}\PY{n}{dot\PYZus{}rr} \PY{o}{\PYZhy{}}\PY{o}{=} \PY{p}{(}\PY{n}{flux\PYZus{}in} \PY{o}{\PYZhy{}} \PY{n}{flux\PYZus{}out}\PY{p}{)} \PY{o}{/} \PY{p}{(}\PY{o}{*}\PY{n}{rhod} \PY{o}{*} \PY{n}{si}\PY{o}{:}\PY{o}{:}\PY{n}{kilograms} \PY{o}{/} \PY{n}{si}\PY{o}{:}\PY{o}{:}\PY{n}{cubic\PYZus{}metres}\PY{p}{)} \PY{o}{/} \PY{n}{dot\PYZus{}rr\PYZus{}unit}\PY{p}{;}

      \PY{c+c1}{// outflow from the domain}
      \PY{k}{return} \PY{n+nf}{real\PYZus{}t}\PY{p}{(}\PY{n}{flux\PYZus{}out} \PY{o}{/} \PY{p}{(}\PY{n}{si}\PY{o}{:}\PY{o}{:}\PY{n}{kilograms} \PY{o}{/} \PY{n}{si}\PY{o}{:}\PY{o}{:}\PY{n}{cubic\PYZus{}metres} \PY{o}{/} \PY{n}{si}\PY{o}{:}\PY{o}{:}\PY{n}{seconds}\PY{p}{)}\PY{p}{)}\PY{p}{;}
    \PY{p}{\PYZcb{}}    
  \PY{p}{\PYZcb{}}\PY{p}{;}
\PY{p}{\PYZcb{}}\PY{p}{;}
\end{Verbatim}
  \vspace{-1.4em}%

  \caption{\label{lst:blk_1m_clw}
    \prog{blk\_1m::rhs\_columnwise()} signature
  }
\end{Listing}
The function arguments include references to containers storing $\rho_d$
  (read-only) and $\theta, r_v, r_c, r_r$ (to be adjusted).
The last argument \prog{dt} is the timestep length needed to calculate the
  precipitation evaporation limit (see discussion of eq.~\ref{eq:rho_v_bis}).

Forcings due to autoconversion and accretion are obtained through a call
  to the \prog{blk\_1m::rhs\_cellwise()} function whose signature
  is given in Listing~\ref{lst:blk_1m_elw}.
The function modifies $\dot{r}_c$ and $\dot{r}_r$ by
  adding the computed rhs terms to the values already present in $\dot{r}_c$ and $\dot{r}_r$.
The function needs read-access to values of $\rho_d$, $r_c$ and $r_r$
  passed as the last three arguments.

Representation of sedimentation is included in a separate function 
  \prog{rhs\_columnwise()} (signature in Listing~\ref{lst:blk_1m_clw}) as
  it is applicable only to simulation frameworks for which a notion of a column is valid
  (e.g. not applicable to a parcel framework).
The passed \prog{cont\_t} references are assumed to point
  to containers storing vertical columns of data with the last element
  placed at the top of the domain.
The last argument \prog{dz} is the vertical grid spacing.
The function returns the value of $F_{out}$ (see eq.~\ref{eq:flux_first}) for the
  lowermost grid cell within a column.

\subsubsection{Example calling sequence}\label{sec:blk_1m_callseq}

\begin{figure*}[th!]
  \center
  \input{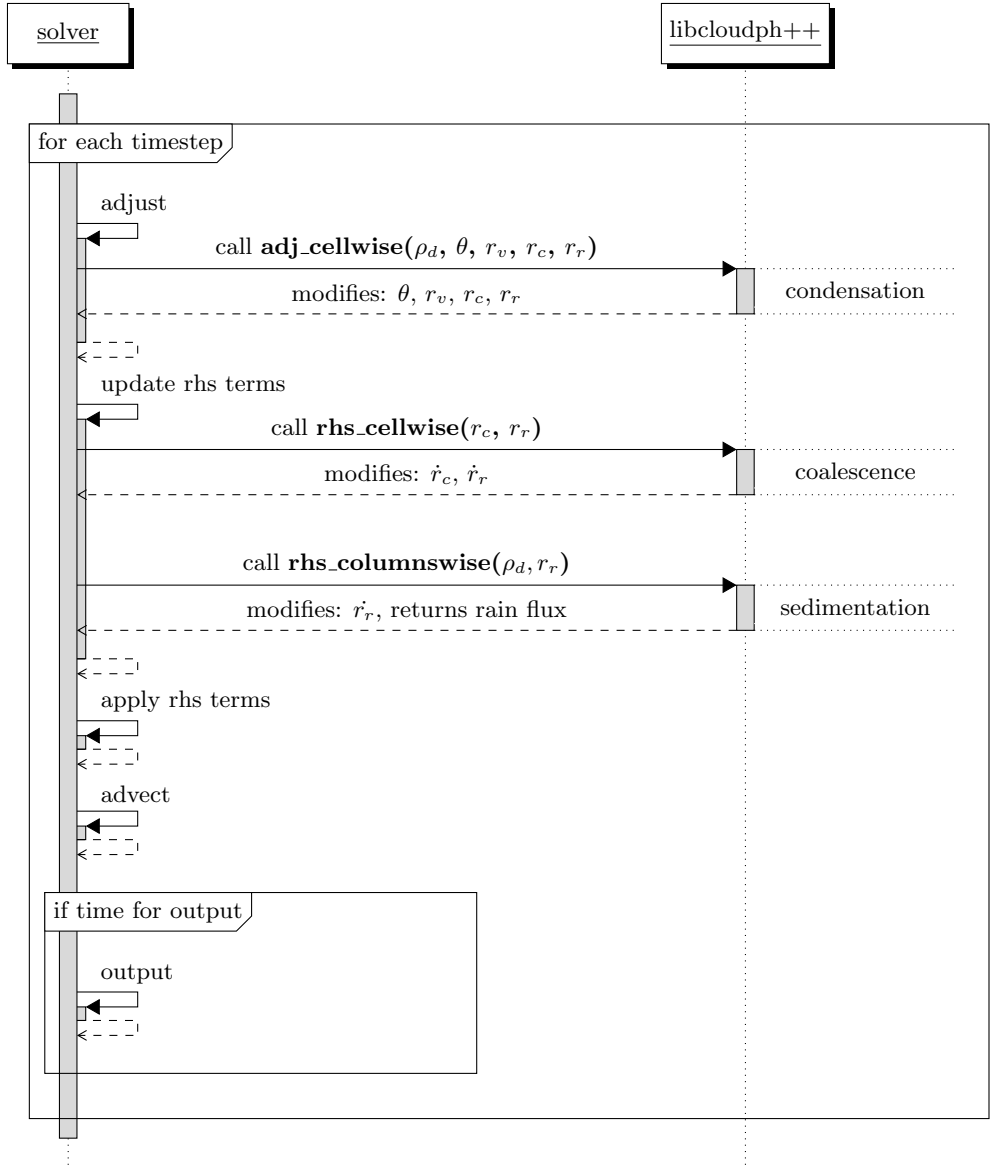}
  \caption{\label{fig:uml_blk_1m}
    Sequence diagram of {\it libcloudph++} API calls for the single-moment bulk scheme
      and a prototype transport equation solver.
    Diagram discussed in section~\ref{sec:blk_1m_callseq}.
    See also caption of Figure~\ref{fig:uml_proto} for description or diagram elements.
  }
\end{figure*}

With the prototype solver concept defined in section~\ref{sec:proto},
  all three functions described above are called once per each timestep.
The diagram in Figure~\ref{fig:uml_blk_1m} depicts the sequence of calls.
As suggested by its name, the \prog{adj\_cellwise()} function (covering
  representation of phase changes) is called within the adjustments step.
Functions \prog{rhs\_cellwise()} and \prog{rhs\_columnwise()} covering
  representation of coalescence and sedimentation, respectively,
  are both called during the rhs-update step.

\subsection{Implementation overview}\label{sec:blk_1m_impl}

The single-moment bulk scheme is implemented as a header-only C++ library.
It requires a C++11-compliant compiler.

Variables, function arguments and return values of physical meaning are
  typed using the Boost.units classes \citep{Schabel_and_Watanabe_2008}.
Consequently, all expressions involving them are subject to dimensional analysis
  at compile time (incurring no runtime overhead).
This reduces the risk of typo-like bugs (e.g. divide instead of multiply by density) 
  but also aids the verification of the model formul\ae.

The integrals in equation~\ref{eq:satadj} defining the saturation adjustment procedure
  are computed using the Boost.Numeric.Odeint library \citep{Ahnert_and_Mulansky_2013}.
The container traversals (e.g.,~iteration over elements of a set of array slices or a set of vectors)
  are performed using the Boost.Iterator library.

\subsection{Example results}\label{sec:blk_1m_exres}

\begin{figure*}[t]
  \center\noindent
  \includegraphics[width=.425\textwidth]{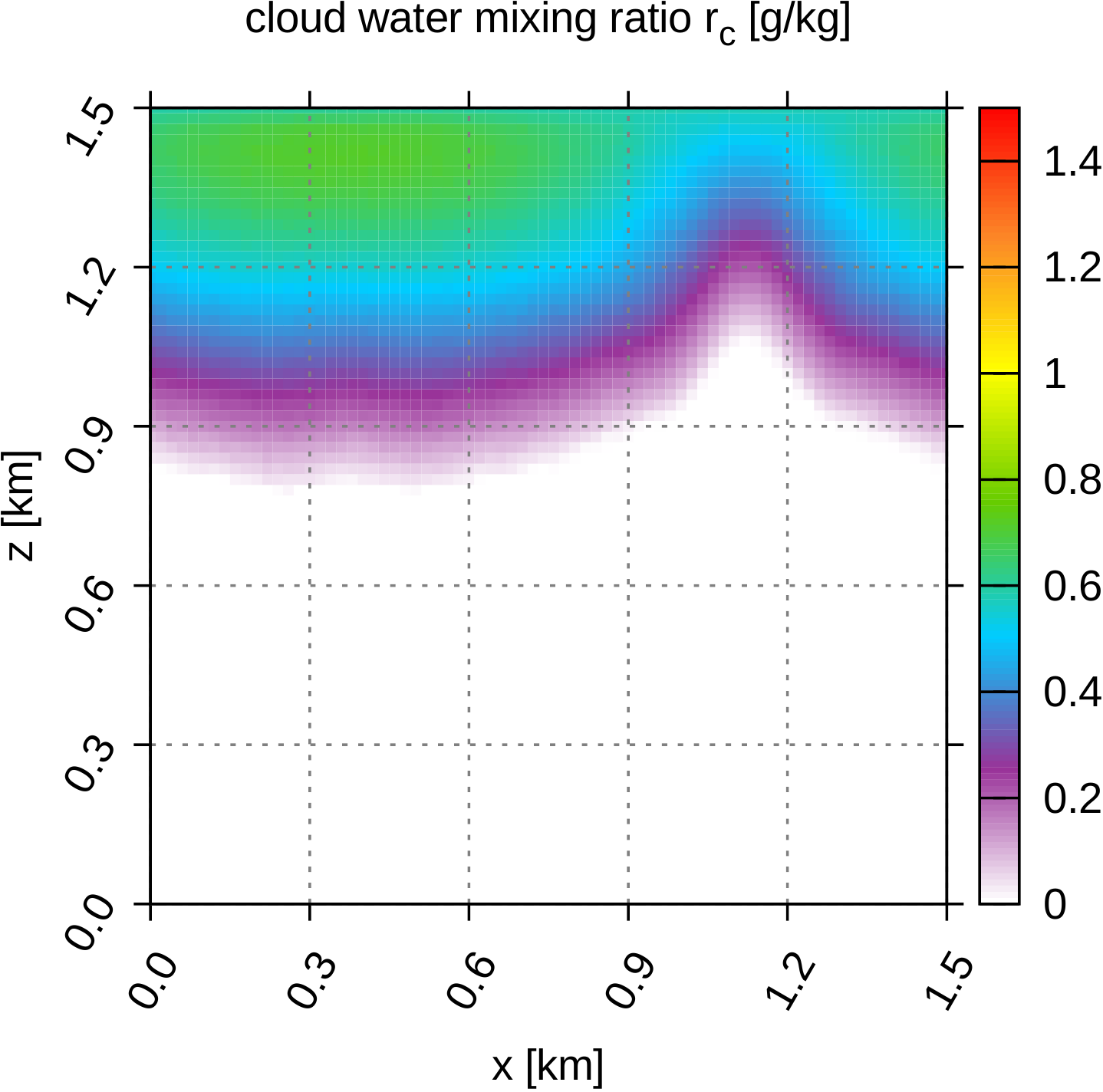}
  ~~~~
  \includegraphics[width=.43\textwidth]{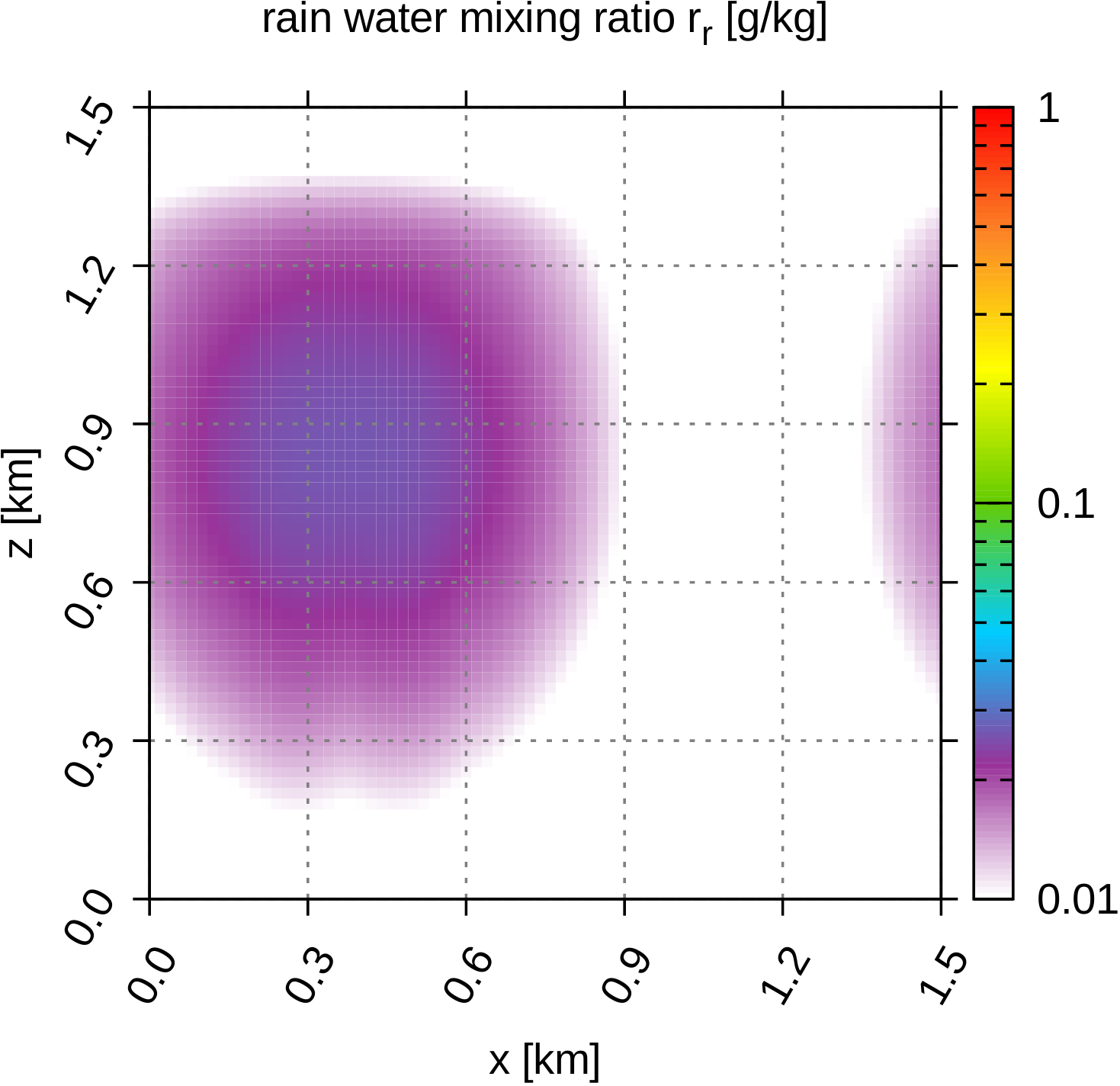}
  \caption{\label{fig:blk_1m_2d}
    Example results from a 2D kinematic simulation using the single-moment bulk scheme.
    All panels depict model state after 30 minutes simulation time (excluding the spin-up period).
    Note~logarithmic colour scale for rain water plots.
    See~section~\ref{sec:blk_1m_exres} for discussion.
  }
\end{figure*}

The simulation framework and setup described in section~\ref{sec:context} and
  implemented using {\it libcloudph++} as described in appendix~\ref{sec:icicle}
  were used to perform an example simulation with the single-moment bulk scheme.
Integration of the transport equations was
  done using the nonoscillatory variant of MPDATA \citep{Smolarkiewicz_2006}.
Figure~\ref{fig:blk_1m_2d} presents a snapshot of cloud and rain water 
  fields after 30~minutes simulation time (excluding the spin-up period).
The cloud deck is located in the upper part of the domain 
  with the cloud water content increasing from the cloud base up to the upper boundary of the domain.
The model has reached a quasi-stationary state and features a drizzle shaft
  that forms in the updraught region in the left-hand side of the domain.
The quasi-stationary state was preceded by a transient rainfall across the entire
  domain in the first minutes of the simulation caused by the initial cloud
  water content exceeding the autoconversion threshold in the upper part 
  of the entire cloud deck.


\FloatBarrier

\section{Double-moment bulk scheme}\label{sec:mm}

A common extension of the single-moment bulk approach is a double-moment bulk scheme.
Similarly to the single-moment approach, the double-moment warm-rain scheme assumes that condensed water 
  is divided into two categories: cloud water and rain water.
In addition to the total mass of water in both categories,
  concentrations of droplets and drops are also predicted.
As a result, the scheme considers two moments of particle size distribution, 
  hence the name.
In the Eulerian framework, four transport equations for cloud droplet concentration
  $n_c$, cloud water mixing ratio $r_c$, rain drop concentration $n_r$ and
  rain water mixing ratio $r_r$ are solved (see table~\ref{tab:vars} for a list of
  model-state variables).
With additional information on the shape of cloud-droplet and rain
  size spectra, the double-moment bulk microphysics scheme is better suited
  than the single-moment scheme for coupling to aerosol and radiative-transfer models 

The double-moment scheme implemented in {\it libcloudph++}
  was introduced by \citet{Morrison_and_Grabowski_2007}.
Their scheme includes, in particular, prediction of the supersaturation.
This makes it well suited for depicting impacts of aerosol on clouds and precipitation.
However, the scheme does not keep track of the changes of aerosol size distribution,
  and hence excludes impacts of clouds and precipitation on aerosol.

\subsection{Formulation}
\subsubsection{Key assumptions}

The model assumes aerosol, cloud and rain spectra shapes (lognormal, gamma and exponential,
  respectively). 
Aerosol is assumed to be well mixed throughout the whole domain and throughout 
  the whole simulation time (uniform concentration per unit mass of dry air).
Cloud water forms only if some of the aerosol particles 
  are activated due to supersaturation.
Activation and subsequent growth by condensation are calculated 
  applying the predicted supersaturation.
As in the single-moment scheme, rain water forms through autoconversion 
  and is further increased by accretion.
Prediction of mean size of cloud droplets
  and rain drops allows to better link the parameterisation of autoconversion 
  and accretion to actual solutions of the collision-coalescence equation.
As in the single-moment scheme, cloud water is assumed to follow the airflow,
  whereas rain water falls relative to the air.
Evaporation of cloud and rain water is included in the formulation
  of phase changes and considers the negligible diffusional growth of rain water.

\subsubsection{Phase changes}

Cloud droplets form from activated aerosol.
The number of activated droplets is derived by
  applying K\"ohler theory to assumed multi-modal lognormal size distribution of aerosols. 
Freshly activated cloud droplets are assumed to have the radius of $1$ $\mu m$;
  for full derivation see \citet[][eqs.~9-13]{Morrison_and_Grabowski_2007} and
  \citet{Khvorostyanov_and_Curry_2006}.
The concentrations of activated droplets are computed separately for each mode
  of the aerosol size distribution and then summed.

The size distribution of aerosols is not resolved by the model.
To take into account the decrease of aerosol concentration due to previous activation, 
  in each timestep the number of available aerosols is approximated
  as the difference between initial 
  aerosol concentration and the concentration of preexisting cloud droplets.
Note that this approximation is valid for weakly precipitating clouds only.
For a strongly raining cloud, the model should include an additional variable,
  the concentration of activated cloud droplets.
It differs from the droplet concentration because of collision-coalescence 
  \citep[see Eqs.~(7) and~(8) in][]{Morrison_and_Grabowski_2008}. 

The changes in cloud and rain water due to condensation and evaporation follow eq.~(8)
  in \citet{Morrison_and_Grabowski_2007} with the phase relaxation times computed
  following eq.~(4)~in~\citet[][]{Morrison_et_al_2005} adapted to fall speed parameterisation
  used in \citet{Morrison_and_Grabowski_2007}.

The decrease in number concentration due to evaporation of cloud droplets and rain drops 
  is computed following the approach of \citet{Khairoutdinov_and_Kogan_2000}.
Cloud droplet concentration is kept constant during evaporation, 
  until all cloud water has to be removed.
Rain drop concentration decreases during evaporation preserving 
  the mean size of rain (drizzle) drops. 

\subsubsection{Coalescence}

Parameterisation of autoconversion and accretion follows 
  the one of \citet{Khairoutdinov_and_Kogan_2000}.
In contrast to the single-moment scheme, the autoconversion
  rate is a continuous function, and the rain onset
  is not controlled by a single threshold. 
Drizzle drops formed due to autoconversion are assumed to have 
  initial radius of $25$ $\mu$m.

\subsubsection{Sedimentation}

Sedimentation is calculated in the same way as in the single-moment scheme 
  (see section~\ref{sec:blk_1m_sedi}), employing upstream advection.
Sedimentation velocities (mass-weighted for the rain density and number-weighted
  for the rain drop concentration) are calculated applying drop terminal
  velocity formulation given in \citet[][Table~2]{Simmel_et_al_2002}.
Sedimentation velocity is multiplied by $\rho_{d0}/\rho_d$ to follow 
  eq.~A4 in \citet{Morrison_et_al_2005}, where $\rho_{d0}$ is the density of dry air
  at standard conditions. 

\begin{Listing}
  \renewcommand*\FancyVerbStartString{\PY{c+c1}{//\PYZlt{}listing\PYZgt{}}}%
  \renewcommand*\FancyVerbStopString{\PY{c+c1}{//\PYZlt{}/listing\PYZgt{}}}%
  \begin{Verbatim}[commandchars=\\\{\}]
\PY{c+cm}{/** @file}
\PY{c+cm}{  * @copyright University of Warsaw}
\PY{c+cm}{  * @section LICENSE}
\PY{c+cm}{  * GPLv3+ (see the COPYING file or http://www.gnu.org/licenses/)}
\PY{c+cm}{  * @brief Definition of a structure holding options for double\PYZhy{}moment bulk microphysics }
\PY{c+cm}{  */}

\PY{c+cp}{\PYZsh{}}\PY{c+cp}{pragma once}

\PY{c+cp}{\PYZsh{}}\PY{c+cp}{include \PYZlt{}vector\PYZgt{}}

\PY{k}{namespace} \PY{n}{libcloudphxx}
\PY{p}{\PYZob{}}
  \PY{k}{namespace} \PY{n}{blk\PYZus{}2m}
  \PY{p}{\PYZob{}}
\PY{c+c1}{//\PYZlt{}listing\PYZgt{}}
    \PY{k}{template}\PY{o}{\PYZlt{}}\PY{k}{typename} \PY{k+kt}{real\PYZus{}t}\PY{o}{\PYZgt{}}
    \PY{k}{struct} \PY{k+kt}{opts\PYZus{}t}
    \PY{p}{\PYZob{}}
      \PY{k+kt}{bool} 
        \PY{n}{acti} \PY{o}{=} \PY{n+nb}{true}\PY{p}{,} \PY{c+c1}{// activation}
        \PY{n}{cond} \PY{o}{=} \PY{n+nb}{true}\PY{p}{,} \PY{c+c1}{// condensation}
        \PY{n}{acnv} \PY{o}{=} \PY{n+nb}{true}\PY{p}{,} \PY{c+c1}{// autoconversion}
        \PY{n}{accr} \PY{o}{=} \PY{n+nb}{true}\PY{p}{,} \PY{c+c1}{// accretion}
        \PY{n}{sedi} \PY{o}{=} \PY{n+nb}{true}\PY{p}{;} \PY{c+c1}{// sedimentation}

      \PY{c+c1}{// RH limit for activation}
      \PY{k+kt}{real\PYZus{}t} \PY{n}{RH\PYZus{}max} \PY{o}{=} \PY{l+m+mi}{44}\PY{p}{;} 
      
      \PY{c+c1}{// aerosol spectrum }
      \PY{k}{struct} \PY{k+kt}{lognormal\PYZus{}mode\PYZus{}t} 
      \PY{p}{\PYZob{}} 
        \PY{k+kt}{real\PYZus{}t}
          \PY{n}{mean\PYZus{}rd}\PY{p}{,}   \PY{c+c1}{// [m]}
          \PY{n}{sdev\PYZus{}rd}\PY{p}{,}   \PY{c+c1}{// [1]}
          \PY{n}{N\PYZus{}stp}\PY{p}{,}     \PY{c+c1}{// [m\PYZhy{}3] @STP}
          \PY{n}{chem\PYZus{}b}\PY{p}{;}    \PY{c+c1}{// [1]}
      \PY{p}{\PYZcb{}}\PY{p}{;}
      \PY{n}{std}\PY{o}{:}\PY{o}{:}\PY{n}{vector}\PY{o}{\PYZlt{}}\PY{k+kt}{lognormal\PYZus{}mode\PYZus{}t}\PY{o}{\PYZgt{}} \PY{n}{dry\PYZus{}distros}\PY{p}{;}
    \PY{p}{\PYZcb{}}\PY{p}{;}
\PY{c+c1}{//\PYZlt{}/listing\PYZgt{}}
  \PY{p}{\PYZcb{}}
 \PY{p}{\PYZcb{}}\PY{p}{;}
\end{Verbatim}
  \vspace{-1.4em}%

  \caption{\label{lst:blk_2m_opt}
    \prog{blk\_2m::opts\_t} definition
  }
\end{Listing}
\vspace{-1em}
\begin{Listing}
  \renewcommand*\FancyVerbStartString{\PY{c+c1}{//\PYZlt{}listing\PYZgt{}}}%
  \renewcommand*\FancyVerbStopString{\PY{c+c1}{//\PYZlt{}/listing\PYZgt{}}}%
  \begin{Verbatim}[commandchars=\\\{\}]
\PY{c+cm}{/** @file}
\PY{c+cm}{  * @copyright University of Warsaw}
\PY{c+cm}{  * @brief Autoconversion and collection righ\PYZhy{}hand side terms using Kessler formulae}
\PY{c+cm}{  * @section LICENSE}
\PY{c+cm}{  * GPLv3+ (see the COPYING file or http://www.gnu.org/licenses/)}
\PY{c+cm}{  */}

\PY{c+cp}{\PYZsh{}}\PY{c+cp}{pragma once}

\PY{c+cp}{\PYZsh{}}\PY{c+cp}{include \PYZlt{}libcloudph++}\PY{c+cp}{/}\PY{c+cp}{blk\PYZus{}2m}\PY{c+cp}{/}\PY{c+cp}{extincl.hpp\PYZgt{}}

\PY{k}{namespace} \PY{n}{libcloudphxx}
\PY{p}{\PYZob{}}
  \PY{k}{namespace} \PY{n}{blk\PYZus{}2m}
  \PY{p}{\PYZob{}}
\PY{c+c1}{//\PYZlt{}listing\PYZgt{}}
    \PY{k}{template} \PY{o}{\PYZlt{}}\PY{k}{typename} \PY{k+kt}{real\PYZus{}t}\PY{p}{,} \PY{k}{class} \PY{n+nc}{cont\PYZus{}t}\PY{o}{\PYZgt{}}
    \PY{k+kt}{void} \PY{n}{rhs\PYZus{}cellwise}\PY{p}{(}
      \PY{k}{const} \PY{k+kt}{opts\PYZus{}t}\PY{o}{\PYZlt{}}\PY{k+kt}{real\PYZus{}t}\PY{o}{\PYZgt{}} \PY{o}{\PYZam{}}\PY{n}{opts}\PY{p}{,}
      \PY{k+kt}{cont\PYZus{}t} \PY{o}{\PYZam{}}\PY{n}{dot\PYZus{}th\PYZus{}cont}\PY{p}{,}
      \PY{k+kt}{cont\PYZus{}t} \PY{o}{\PYZam{}}\PY{n}{dot\PYZus{}rv\PYZus{}cont}\PY{p}{,}
      \PY{k+kt}{cont\PYZus{}t} \PY{o}{\PYZam{}}\PY{n}{dot\PYZus{}rc\PYZus{}cont}\PY{p}{,}
      \PY{k+kt}{cont\PYZus{}t} \PY{o}{\PYZam{}}\PY{n}{dot\PYZus{}nc\PYZus{}cont}\PY{p}{,}
      \PY{k+kt}{cont\PYZus{}t} \PY{o}{\PYZam{}}\PY{n}{dot\PYZus{}rr\PYZus{}cont}\PY{p}{,}
      \PY{k+kt}{cont\PYZus{}t} \PY{o}{\PYZam{}}\PY{n}{dot\PYZus{}nr\PYZus{}cont}\PY{p}{,}
      \PY{k}{const} \PY{k+kt}{cont\PYZus{}t} \PY{o}{\PYZam{}}\PY{n}{rhod\PYZus{}cont}\PY{p}{,}   
      \PY{k}{const} \PY{k+kt}{cont\PYZus{}t} \PY{o}{\PYZam{}}\PY{n}{th\PYZus{}cont}\PY{p}{,}
      \PY{k}{const} \PY{k+kt}{cont\PYZus{}t} \PY{o}{\PYZam{}}\PY{n}{rv\PYZus{}cont}\PY{p}{,}
      \PY{k}{const} \PY{k+kt}{cont\PYZus{}t} \PY{o}{\PYZam{}}\PY{n}{rc\PYZus{}cont}\PY{p}{,}
      \PY{k}{const} \PY{k+kt}{cont\PYZus{}t} \PY{o}{\PYZam{}}\PY{n}{nc\PYZus{}cont}\PY{p}{,}
      \PY{k}{const} \PY{k+kt}{cont\PYZus{}t} \PY{o}{\PYZam{}}\PY{n}{rr\PYZus{}cont}\PY{p}{,}
      \PY{k}{const} \PY{k+kt}{cont\PYZus{}t} \PY{o}{\PYZam{}}\PY{n}{nr\PYZus{}cont}\PY{p}{,}
      \PY{k}{const} \PY{k+kt}{real\PYZus{}t} \PY{o}{\PYZam{}}\PY{n}{dt}
    \PY{p}{)}   
\PY{c+c1}{//\PYZlt{}/listing\PYZgt{}}
    \PY{p}{\PYZob{}}  
      \PY{c+c1}{// sanity checks}
      \PY{n}{assert}\PY{p}{(}\PY{n}{min}\PY{p}{(}\PY{n}{rv\PYZus{}cont}\PY{p}{)} \PY{o}{\PYZgt{}}\PY{o}{=} \PY{l+m+mi}{0}\PY{p}{)}\PY{p}{;}
      \PY{n}{assert}\PY{p}{(}\PY{n}{min}\PY{p}{(}\PY{n}{th\PYZus{}cont}\PY{p}{)} \PY{o}{\PYZgt{}} \PY{l+m+mi}{0}\PY{p}{)}\PY{p}{;}
      \PY{n}{assert}\PY{p}{(}\PY{n}{min}\PY{p}{(}\PY{n}{rc\PYZus{}cont}\PY{p}{)} \PY{o}{\PYZgt{}}\PY{o}{=} \PY{l+m+mi}{0}\PY{p}{)}\PY{p}{;}
      \PY{n}{assert}\PY{p}{(}\PY{n}{min}\PY{p}{(}\PY{n}{rr\PYZus{}cont}\PY{p}{)} \PY{o}{\PYZgt{}}\PY{o}{=} \PY{l+m+mi}{0}\PY{p}{)}\PY{p}{;}
      \PY{n}{assert}\PY{p}{(}\PY{n}{min}\PY{p}{(}\PY{n}{nc\PYZus{}cont}\PY{p}{)} \PY{o}{\PYZgt{}}\PY{o}{=} \PY{l+m+mi}{0}\PY{p}{)}\PY{p}{;}
      \PY{n}{assert}\PY{p}{(}\PY{n}{min}\PY{p}{(}\PY{n}{nr\PYZus{}cont}\PY{p}{)} \PY{o}{\PYZgt{}}\PY{o}{=} \PY{l+m+mi}{0}\PY{p}{)}\PY{p}{;}

      \PY{c+c1}{// TODO: rewrite so thet\PYZsq{}s not needed}
      \PY{n}{assert}\PY{p}{(}\PY{n}{min}\PY{p}{(}\PY{n}{dot\PYZus{}nc\PYZus{}cont}\PY{p}{)} \PY{o}{=}\PY{o}{=} \PY{l+m+mi}{0}\PY{p}{)}\PY{p}{;}
      \PY{c+c1}{//assert(min(dot\PYZus{}nr\PYZus{}cont) == 0);}
      \PY{n}{assert}\PY{p}{(}\PY{n}{min}\PY{p}{(}\PY{n}{dot\PYZus{}rc\PYZus{}cont}\PY{p}{)} \PY{o}{=}\PY{o}{=} \PY{l+m+mi}{0}\PY{p}{)}\PY{p}{;}
      \PY{c+c1}{//assert(min(dot\PYZus{}rr\PYZus{}cont) == 0);}
      \PY{n}{assert}\PY{p}{(}\PY{n}{max}\PY{p}{(}\PY{n}{dot\PYZus{}nc\PYZus{}cont}\PY{p}{)} \PY{o}{=}\PY{o}{=} \PY{l+m+mi}{0}\PY{p}{)}\PY{p}{;}
      \PY{n}{assert}\PY{p}{(}\PY{n}{max}\PY{p}{(}\PY{n}{dot\PYZus{}nr\PYZus{}cont}\PY{p}{)} \PY{o}{=}\PY{o}{=} \PY{l+m+mi}{0}\PY{p}{)}\PY{p}{;}
      \PY{n}{assert}\PY{p}{(}\PY{n}{max}\PY{p}{(}\PY{n}{dot\PYZus{}rc\PYZus{}cont}\PY{p}{)} \PY{o}{=}\PY{o}{=} \PY{l+m+mi}{0}\PY{p}{)}\PY{p}{;}
      \PY{n}{assert}\PY{p}{(}\PY{n}{max}\PY{p}{(}\PY{n}{dot\PYZus{}rr\PYZus{}cont}\PY{p}{)} \PY{o}{=}\PY{o}{=} \PY{l+m+mi}{0}\PY{p}{)}\PY{p}{;}

      \PY{k}{using} \PY{k}{namespace} \PY{n}{formulae}\PY{p}{;}
      \PY{k}{using} \PY{k}{namespace} \PY{n}{common}\PY{o}{:}\PY{o}{:}\PY{n}{moist\PYZus{}air}\PY{p}{;}
      \PY{k}{using} \PY{k}{namespace} \PY{n}{common}\PY{o}{:}\PY{o}{:}\PY{n}{theta\PYZus{}dry}\PY{p}{;}

      \PY{c+c1}{// if something is too small e\PYZhy{}179 it becomes negative \PYZsh{} TODO WHY! \PYZhy{} explain or fix}
      \PY{c+c1}{// so instead of n\PYZus{}l == 0 we have n\PYZus{}l \PYZlt{} eps}
      \PY{c+c1}{// also \PYZhy{}1e\PYZhy{}30 + 1e\PYZhy{}30 is not equal to zero}
      \PY{k}{const} \PY{n}{quantity}\PY{o}{\PYZlt{}}\PY{n}{si}\PY{o}{:}\PY{o}{:}\PY{n}{dimensionless}\PY{p}{,} \PY{k+kt}{real\PYZus{}t}\PY{o}{\PYZgt{}}                                       \PY{n}{eps\PYZus{}d} \PY{o}{=} \PY{k+kt}{real\PYZus{}t}\PY{p}{(}\PY{l+m+mi}{0}\PY{p}{)}\PY{p}{;}
      \PY{k}{const} \PY{n}{quantity}\PY{o}{\PYZlt{}}\PY{n}{divide\PYZus{}typeof\PYZus{}helper}\PY{o}{\PYZlt{}}\PY{n}{si}\PY{o}{:}\PY{o}{:}\PY{n}{dimensionless}\PY{p}{,} \PY{n}{si}\PY{o}{:}\PY{o}{:}\PY{n}{mass}\PY{o}{\PYZgt{}}\PY{o}{:}\PY{o}{:}\PY{n}{type}\PY{p}{,} \PY{k+kt}{real\PYZus{}t}\PY{o}{\PYZgt{}} \PY{n}{eps\PYZus{}n} \PY{o}{=} \PY{k+kt}{real\PYZus{}t}\PY{p}{(}\PY{l+m+mi}{0}\PY{p}{)} \PY{o}{/} \PY{n}{si}\PY{o}{:}\PY{o}{:}\PY{n}{kilograms}\PY{p}{;}

      \PY{c+c1}{//TODO: }
      \PY{c+c1}{//unfortunately can\PYZsq{}t zip through more than 10 arguments }
      \PY{c+c1}{//so instead one loop over all forcings, there will be a few }
      \PY{k}{for} \PY{p}{(}\PY{k}{auto} \PY{n}{tup} \PY{o}{:} \PY{n}{zip}\PY{p}{(}
        \PY{n}{dot\PYZus{}th\PYZus{}cont}\PY{p}{,} 
        \PY{n}{dot\PYZus{}rv\PYZus{}cont}\PY{p}{,} 
        \PY{n}{dot\PYZus{}rc\PYZus{}cont}\PY{p}{,} 
        \PY{n}{dot\PYZus{}nc\PYZus{}cont}\PY{p}{,}
        \PY{n}{rhod\PYZus{}cont}\PY{p}{,} 
        \PY{n}{th\PYZus{}cont}\PY{p}{,}  
        \PY{n}{rv\PYZus{}cont}\PY{p}{,}  
        \PY{n}{rc\PYZus{}cont}\PY{p}{,}  
        \PY{n}{nc\PYZus{}cont}
      \PY{p}{)}\PY{p}{)}
      \PY{p}{\PYZob{}}
        \PY{k+kt}{real\PYZus{}t}
          \PY{o}{\PYZam{}}\PY{n}{dot\PYZus{}th} \PY{o}{=} \PY{n}{boost}\PY{o}{:}\PY{o}{:}\PY{n}{get}\PY{o}{\PYZlt{}}\PY{l+m+mi}{0}\PY{o}{\PYZgt{}}\PY{p}{(}\PY{n}{tup}\PY{p}{)}\PY{p}{,}
          \PY{o}{\PYZam{}}\PY{n}{dot\PYZus{}rv} \PY{o}{=} \PY{n}{boost}\PY{o}{:}\PY{o}{:}\PY{n}{get}\PY{o}{\PYZlt{}}\PY{l+m+mi}{1}\PY{o}{\PYZgt{}}\PY{p}{(}\PY{n}{tup}\PY{p}{)}\PY{p}{,}
          \PY{o}{\PYZam{}}\PY{n}{dot\PYZus{}rc} \PY{o}{=} \PY{n}{boost}\PY{o}{:}\PY{o}{:}\PY{n}{get}\PY{o}{\PYZlt{}}\PY{l+m+mi}{2}\PY{o}{\PYZgt{}}\PY{p}{(}\PY{n}{tup}\PY{p}{)}\PY{p}{,}
          \PY{o}{\PYZam{}}\PY{n}{dot\PYZus{}nc} \PY{o}{=} \PY{n}{boost}\PY{o}{:}\PY{o}{:}\PY{n}{get}\PY{o}{\PYZlt{}}\PY{l+m+mi}{3}\PY{o}{\PYZgt{}}\PY{p}{(}\PY{n}{tup}\PY{p}{)}\PY{p}{;}
        \PY{k}{const} \PY{n}{quantity}\PY{o}{\PYZlt{}}\PY{n}{si}\PY{o}{:}\PY{o}{:}\PY{n}{mass\PYZus{}density}\PY{p}{,}  \PY{k+kt}{real\PYZus{}t}\PY{o}{\PYZgt{}} \PY{o}{\PYZam{}}\PY{n}{rhod}  \PY{o}{=} \PY{n}{boost}\PY{o}{:}\PY{o}{:}\PY{n}{get}\PY{o}{\PYZlt{}}\PY{l+m+mi}{4}\PY{o}{\PYZgt{}}\PY{p}{(}\PY{n}{tup}\PY{p}{)} \PY{o}{*} \PY{n}{si}\PY{o}{:}\PY{o}{:}\PY{n}{kilograms} \PY{o}{/} \PY{n}{si}\PY{o}{:}\PY{o}{:}\PY{n}{cubic\PYZus{}metres}\PY{p}{;}
        \PY{k}{const} \PY{n}{quantity}\PY{o}{\PYZlt{}}\PY{n}{si}\PY{o}{:}\PY{o}{:}\PY{n}{temperature}\PY{p}{,}   \PY{k+kt}{real\PYZus{}t}\PY{o}{\PYZgt{}} \PY{o}{\PYZam{}}\PY{n}{th}    \PY{o}{=} \PY{n}{boost}\PY{o}{:}\PY{o}{:}\PY{n}{get}\PY{o}{\PYZlt{}}\PY{l+m+mi}{5}\PY{o}{\PYZgt{}}\PY{p}{(}\PY{n}{tup}\PY{p}{)} \PY{o}{*} \PY{n}{si}\PY{o}{:}\PY{o}{:}\PY{n}{kelvins}\PY{p}{;}
        \PY{k}{const} \PY{n}{quantity}\PY{o}{\PYZlt{}}\PY{n}{si}\PY{o}{:}\PY{o}{:}\PY{n}{dimensionless}\PY{p}{,} \PY{k+kt}{real\PYZus{}t}\PY{o}{\PYZgt{}} \PY{o}{\PYZam{}}\PY{n}{rv}    \PY{o}{=} \PY{n}{boost}\PY{o}{:}\PY{o}{:}\PY{n}{get}\PY{o}{\PYZlt{}}\PY{l+m+mi}{6}\PY{o}{\PYZgt{}}\PY{p}{(}\PY{n}{tup}\PY{p}{)} \PY{o}{*} \PY{n}{si}\PY{o}{:}\PY{o}{:}\PY{n}{dimensionless}\PY{p}{(}\PY{p}{)}\PY{p}{;}
        \PY{k}{const} \PY{n}{quantity}\PY{o}{\PYZlt{}}\PY{n}{si}\PY{o}{:}\PY{o}{:}\PY{n}{dimensionless}\PY{p}{,} \PY{k+kt}{real\PYZus{}t}\PY{o}{\PYZgt{}} \PY{o}{\PYZam{}}\PY{n}{rc}    \PY{o}{=} \PY{n}{boost}\PY{o}{:}\PY{o}{:}\PY{n}{get}\PY{o}{\PYZlt{}}\PY{l+m+mi}{7}\PY{o}{\PYZgt{}}\PY{p}{(}\PY{n}{tup}\PY{p}{)} \PY{o}{*} \PY{n}{si}\PY{o}{:}\PY{o}{:}\PY{n}{dimensionless}\PY{p}{(}\PY{p}{)}\PY{p}{;}
        \PY{k}{const} \PY{n}{quantity}\PY{o}{\PYZlt{}}
          \PY{n}{divide\PYZus{}typeof\PYZus{}helper}\PY{o}{\PYZlt{}}\PY{n}{si}\PY{o}{:}\PY{o}{:}\PY{n}{dimensionless}\PY{p}{,} \PY{n}{si}\PY{o}{:}\PY{o}{:}\PY{n}{mass}
        \PY{o}{\PYZgt{}}\PY{o}{:}\PY{o}{:}\PY{n}{type}\PY{p}{,}                          \PY{k+kt}{real\PYZus{}t}\PY{o}{\PYZgt{}} \PY{o}{\PYZam{}}\PY{n}{nc}    \PY{o}{=} \PY{n}{boost}\PY{o}{:}\PY{o}{:}\PY{n}{get}\PY{o}{\PYZlt{}}\PY{l+m+mi}{8}\PY{o}{\PYZgt{}}\PY{p}{(}\PY{n}{tup}\PY{p}{)} \PY{o}{/} \PY{n}{si}\PY{o}{:}\PY{o}{:}\PY{n}{kilograms}\PY{p}{;}

        \PY{c+c1}{//helper temperature and pressure}
        \PY{n}{quantity}\PY{o}{\PYZlt{}}\PY{n}{si}\PY{o}{:}\PY{o}{:}\PY{n}{temperature}\PY{p}{,} \PY{k+kt}{real\PYZus{}t}\PY{o}{\PYZgt{}} \PY{n}{T} \PY{o}{=} \PY{n}{common}\PY{o}{:}\PY{o}{:}\PY{n}{theta\PYZus{}dry}\PY{o}{:}\PY{o}{:}\PY{n}{T}\PY{o}{\PYZlt{}}\PY{k+kt}{real\PYZus{}t}\PY{o}{\PYZgt{}}\PY{p}{(}\PY{n}{th}\PY{p}{,} \PY{n}{rhod}\PY{p}{)}\PY{p}{;}
        \PY{n}{quantity}\PY{o}{\PYZlt{}}\PY{n}{si}\PY{o}{:}\PY{o}{:}\PY{n}{pressure}\PY{p}{,} \PY{k+kt}{real\PYZus{}t}\PY{o}{\PYZgt{}}    \PY{n}{p} \PY{o}{=} \PY{n}{common}\PY{o}{:}\PY{o}{:}\PY{n}{theta\PYZus{}dry}\PY{o}{:}\PY{o}{:}\PY{n}{p}\PY{o}{\PYZlt{}}\PY{k+kt}{real\PYZus{}t}\PY{o}{\PYZgt{}}\PY{p}{(}\PY{n}{rhod}\PY{p}{,} \PY{n}{rv}\PY{p}{,} \PY{n}{T}\PY{p}{)}\PY{p}{;}

        \PY{c+c1}{// activation (see Morrison \PYZam{} Grabowski 2007)}
        \PY{k}{if} \PY{p}{(}\PY{n}{opts}\PY{p}{.}\PY{n}{acti}\PY{p}{)}
        \PY{p}{\PYZob{}} \PY{c+c1}{//TODO what if we have some other source terms (that happen somewhere before here), like diffusion?}
          \PY{n}{assert}\PY{p}{(}\PY{n}{dot\PYZus{}rc} \PY{o}{=}\PY{o}{=} \PY{l+m+mi}{0} \PY{o}{\PYZam{}}\PY{o}{\PYZam{}} \PY{l+s}{\PYZdq{}}\PY{l+s}{activation is first}\PY{l+s}{\PYZdq{}}\PY{p}{)}\PY{p}{;}
          \PY{n}{assert}\PY{p}{(}\PY{n}{dot\PYZus{}nc} \PY{o}{=}\PY{o}{=} \PY{l+m+mi}{0} \PY{o}{\PYZam{}}\PY{o}{\PYZam{}} \PY{l+s}{\PYZdq{}}\PY{l+s}{activation is first}\PY{l+s}{\PYZdq{}}\PY{p}{)}\PY{p}{;}
          \PY{n}{assert}\PY{p}{(}\PY{n}{dot\PYZus{}th} \PY{o}{=}\PY{o}{=} \PY{l+m+mi}{0} \PY{o}{\PYZam{}}\PY{o}{\PYZam{}} \PY{l+s}{\PYZdq{}}\PY{l+s}{activation is first}\PY{l+s}{\PYZdq{}}\PY{p}{)}\PY{p}{;}

          \PY{k}{if} \PY{p}{(}\PY{n}{rv} \PY{o}{\PYZgt{}} \PY{n}{common}\PY{o}{:}\PY{o}{:}\PY{n}{const\PYZus{}cp}\PY{o}{:}\PY{o}{:}\PY{n}{r\PYZus{}vs}\PY{o}{\PYZlt{}}\PY{k+kt}{real\PYZus{}t}\PY{o}{\PYZgt{}}\PY{p}{(}\PY{n}{T}\PY{p}{,} \PY{n}{p}\PY{p}{)}\PY{p}{)}
          \PY{p}{\PYZob{}}
            \PY{c+c1}{// summing by looping over lognormal modes}
            \PY{n}{quantity}\PY{o}{\PYZlt{}}\PY{n}{divide\PYZus{}typeof\PYZus{}helper}\PY{o}{\PYZlt{}}\PY{n}{si}\PY{o}{:}\PY{o}{:}\PY{n}{dimensionless}\PY{p}{,} \PY{n}{si}\PY{o}{:}\PY{o}{:}\PY{n}{mass}\PY{o}{\PYZgt{}}\PY{o}{:}\PY{o}{:}\PY{n}{type}\PY{p}{,} \PY{k+kt}{real\PYZus{}t}\PY{o}{\PYZgt{}} \PY{n}{n\PYZus{}ccn} \PY{o}{=} \PY{l+m+mi}{0}\PY{p}{;}
            \PY{k}{for} \PY{p}{(}\PY{k}{const} \PY{k}{auto} \PY{o}{\PYZam{}}\PY{n}{mode} \PY{o}{:} \PY{n}{opts}\PY{p}{.}\PY{n}{dry\PYZus{}distros}\PY{p}{)}
            \PY{p}{\PYZob{}} 
              \PY{n}{n\PYZus{}ccn} \PY{o}{+}\PY{o}{=} \PY{n}{n\PYZus{}c\PYZus{}p}\PY{o}{\PYZlt{}}\PY{k+kt}{real\PYZus{}t}\PY{o}{\PYZgt{}}\PY{p}{(}
                \PY{n}{p}\PY{p}{,} \PY{n}{T}\PY{p}{,} \PY{n}{rv}\PY{p}{,} 
                \PY{n}{mode}\PY{p}{.}\PY{n}{mean\PYZus{}rd} \PY{o}{*} \PY{n}{si}\PY{o}{:}\PY{o}{:}\PY{n}{metres}\PY{p}{,} 
                \PY{n}{mode}\PY{p}{.}\PY{n}{sdev\PYZus{}rd}\PY{p}{,} 
                \PY{n}{mode}\PY{p}{.}\PY{n}{N\PYZus{}stp} \PY{o}{/} \PY{n}{si}\PY{o}{:}\PY{o}{:}\PY{n}{cubic\PYZus{}metres}\PY{p}{,} 
                \PY{n}{mode}\PY{p}{.}\PY{n}{chem\PYZus{}b}\PY{p}{,}
                \PY{n}{opts}\PY{p}{.}\PY{n}{RH\PYZus{}max}
              \PY{p}{)}\PY{p}{;} 
            \PY{p}{\PYZcb{}}

            \PY{n}{quantity}\PY{o}{\PYZlt{}}\PY{n}{divide\PYZus{}typeof\PYZus{}helper}\PY{o}{\PYZlt{}}\PY{n}{si}\PY{o}{:}\PY{o}{:}\PY{n}{frequency}\PY{p}{,} \PY{n}{si}\PY{o}{:}\PY{o}{:}\PY{n}{mass}\PY{o}{\PYZgt{}}\PY{o}{:}\PY{o}{:}\PY{n}{type}\PY{p}{,} \PY{k+kt}{real\PYZus{}t}\PY{o}{\PYZgt{}} \PY{n}{tmp} \PY{o}{=} 
              \PY{n}{activation\PYZus{}rate}\PY{o}{\PYZlt{}}\PY{k+kt}{real\PYZus{}t}\PY{o}{\PYZgt{}}\PY{p}{(}\PY{n}{n\PYZus{}ccn}\PY{p}{,} \PY{n}{nc}\PY{p}{,} \PY{n}{dt} \PY{o}{*} \PY{n}{si}\PY{o}{:}\PY{o}{:}\PY{n}{seconds}\PY{p}{)}\PY{p}{;}

	    \PY{n}{dot\PYZus{}nc} \PY{o}{+}\PY{o}{=} \PY{n}{tmp} \PY{o}{*} \PY{n}{si}\PY{o}{:}\PY{o}{:}\PY{n}{kilograms} \PY{o}{*} \PY{n}{si}\PY{o}{:}\PY{o}{:}\PY{n}{seconds}\PY{p}{;}  
            \PY{n}{dot\PYZus{}rv} \PY{o}{\PYZhy{}}\PY{o}{=} \PY{n}{tmp} \PY{o}{*} \PY{n}{ccnmass}\PY{o}{\PYZlt{}}\PY{k+kt}{real\PYZus{}t}\PY{o}{\PYZgt{}}\PY{p}{(}\PY{p}{)} \PY{o}{*} \PY{n}{si}\PY{o}{:}\PY{o}{:}\PY{n}{seconds}\PY{p}{;}
            \PY{n}{dot\PYZus{}rc} \PY{o}{+}\PY{o}{=} \PY{n}{tmp} \PY{o}{*} \PY{n}{ccnmass}\PY{o}{\PYZlt{}}\PY{k+kt}{real\PYZus{}t}\PY{o}{\PYZgt{}}\PY{p}{(}\PY{p}{)} \PY{o}{*} \PY{n}{si}\PY{o}{:}\PY{o}{:}\PY{n}{seconds}\PY{p}{;}

            \PY{c+c1}{//TODO maybe some common part for all the forcings (for example dot\PYZus{}rho\PYZus{}e)?}
            \PY{n}{dot\PYZus{}th} \PY{o}{\PYZhy{}}\PY{o}{=} \PY{n}{tmp} \PY{o}{*} \PY{n}{ccnmass}\PY{o}{\PYZlt{}}\PY{k+kt}{real\PYZus{}t}\PY{o}{\PYZgt{}}\PY{p}{(}\PY{p}{)} \PY{o}{*} \PY{n}{d\PYZus{}th\PYZus{}d\PYZus{}rv}\PY{o}{\PYZlt{}}\PY{k+kt}{real\PYZus{}t}\PY{o}{\PYZgt{}}\PY{p}{(}\PY{n}{T}\PY{p}{,} \PY{n}{th}\PY{p}{)} \PY{o}{/} \PY{n}{si}\PY{o}{:}\PY{o}{:}\PY{n}{kelvins} \PY{o}{*} \PY{n}{si}\PY{o}{:}\PY{o}{:}\PY{n}{seconds}\PY{p}{;} 
          \PY{p}{\PYZcb{}}

          \PY{n}{assert}\PY{p}{(}\PY{n}{dot\PYZus{}nc} \PY{o}{\PYZgt{}}\PY{o}{=} \PY{l+m+mi}{0} \PY{o}{\PYZam{}}\PY{o}{\PYZam{}} \PY{l+s}{\PYZdq{}}\PY{l+s}{activation can only increase cloud droplet concentration}\PY{l+s}{\PYZdq{}}\PY{p}{)}\PY{p}{;}
          \PY{n}{assert}\PY{p}{(}\PY{n}{dot\PYZus{}rc} \PY{o}{\PYZgt{}}\PY{o}{=} \PY{l+m+mi}{0} \PY{o}{\PYZam{}}\PY{o}{\PYZam{}} \PY{l+s}{\PYZdq{}}\PY{l+s}{activation can only increase cloud water}\PY{l+s}{\PYZdq{}}\PY{p}{)}\PY{p}{;}
          \PY{n}{assert}\PY{p}{(}\PY{n}{dot\PYZus{}th} \PY{o}{\PYZgt{}}\PY{o}{=} \PY{l+m+mi}{0} \PY{o}{\PYZam{}}\PY{o}{\PYZam{}} \PY{l+s}{\PYZdq{}}\PY{l+s}{activation can only increase theta}\PY{l+s}{\PYZdq{}}\PY{p}{)}\PY{p}{;}
         \PY{p}{\PYZcb{}}

        \PY{c+c1}{// condensation/evaporation of cloud water (see Morrison \PYZam{} Grabowski 2007)}
        \PY{k}{if} \PY{p}{(}\PY{n}{opts}\PY{p}{.}\PY{n}{cond}\PY{p}{)}
        \PY{p}{\PYZob{}}                          
          \PY{k}{if} \PY{p}{(}\PY{n}{rc} \PY{o}{\PYZgt{}} \PY{n}{eps\PYZus{}d} \PY{o}{\PYZam{}}\PY{o}{\PYZam{}} \PY{n}{nc} \PY{o}{\PYZgt{}} \PY{n}{eps\PYZus{}n}\PY{p}{)}
          \PY{p}{\PYZob{}}          \PY{c+c1}{//  \PYZca{}\PYZca{}   TODO is it possible?}
            \PY{n}{quantity}\PY{o}{\PYZlt{}}\PY{n}{divide\PYZus{}typeof\PYZus{}helper}\PY{o}{\PYZlt{}}\PY{n}{si}\PY{o}{:}\PY{o}{:}\PY{n}{dimensionless}\PY{p}{,} \PY{n}{si}\PY{o}{:}\PY{o}{:}\PY{n}{time}\PY{o}{\PYZgt{}}\PY{o}{:}\PY{o}{:}\PY{n}{type}\PY{p}{,} \PY{k+kt}{real\PYZus{}t}\PY{o}{\PYZgt{}} \PY{n}{tmp} \PY{o}{=} 
              \PY{n}{cond\PYZus{}evap\PYZus{}rate}\PY{o}{\PYZlt{}}\PY{k+kt}{real\PYZus{}t}\PY{o}{\PYZgt{}}\PY{p}{(}\PY{n}{T}\PY{p}{,} \PY{n}{p}\PY{p}{,} \PY{n}{rv}\PY{p}{,} \PY{n}{tau\PYZus{}relax\PYZus{}c}\PY{p}{(}\PY{n}{T}\PY{p}{,} \PY{n}{p}\PY{p}{,} \PY{n}{r\PYZus{}drop\PYZus{}c}\PY{p}{(}\PY{n}{rc}\PY{p}{,} \PY{n}{nc}\PY{p}{,} \PY{n}{rhod}\PY{p}{)}\PY{p}{,} \PY{n}{rhod} \PY{o}{*} \PY{n}{nc}\PY{p}{)}\PY{p}{)}\PY{p}{;}

            \PY{n}{assert}\PY{p}{(}\PY{n}{r\PYZus{}drop\PYZus{}c}\PY{p}{(}\PY{n}{rc}\PY{p}{,} \PY{n}{nc}\PY{p}{,} \PY{n}{rhod}\PY{p}{)} \PY{o}{\PYZgt{}}\PY{o}{=} \PY{l+m+mi}{0} \PY{o}{*} \PY{n}{si}\PY{o}{:}\PY{o}{:}\PY{n}{metres}  \PY{o}{\PYZam{}}\PY{o}{\PYZam{}} \PY{l+s}{\PYZdq{}}\PY{l+s}{mean droplet radius cannot be \PYZlt{} 0}\PY{l+s}{\PYZdq{}}\PY{p}{)}\PY{p}{;}

            \PY{k}{if} \PY{p}{(}\PY{n}{rc} \PY{o}{+} \PY{p}{(}\PY{p}{(}\PY{n}{dot\PYZus{}rc} \PY{o}{/} \PY{n}{si}\PY{o}{:}\PY{o}{:}\PY{n}{seconds} \PY{o}{+} \PY{n}{tmp}\PY{p}{)} \PY{o}{*} \PY{p}{(}\PY{n}{dt} \PY{o}{*} \PY{n}{si}\PY{o}{:}\PY{o}{:}\PY{n}{seconds}\PY{p}{)}\PY{p}{)}  \PY{o}{\PYZlt{}} \PY{l+m+mi}{0}\PY{p}{)}
            \PY{p}{\PYZob{}}   \PY{c+c1}{// so that we don\PYZsq{}t evaporate more cloud water than there is}
              \PY{n}{tmp}     \PY{o}{=} \PY{o}{\PYZhy{}}\PY{p}{(}\PY{n}{rc} \PY{o}{+} \PY{p}{(}\PY{n}{dt} \PY{o}{*} \PY{n}{dot\PYZus{}rc}\PY{p}{)}\PY{p}{)} \PY{o}{/} \PY{p}{(}\PY{n}{dt} \PY{o}{*} \PY{n}{si}\PY{o}{:}\PY{o}{:}\PY{n}{seconds}\PY{p}{)}\PY{p}{;}  \PY{c+c1}{// evaporate all rc}
              \PY{n}{dot\PYZus{}nc}  \PY{o}{=} \PY{o}{\PYZhy{}}\PY{n}{nc} \PY{o}{/} \PY{n}{dt} \PY{o}{*} \PY{n}{si}\PY{o}{:}\PY{o}{:}\PY{n}{kilograms}\PY{p}{;} \PY{c+c1}{// and all nc}
            \PY{p}{\PYZcb{}}

            \PY{n}{dot\PYZus{}rc} \PY{o}{+}\PY{o}{=} \PY{n}{tmp} \PY{o}{*} \PY{n}{si}\PY{o}{:}\PY{o}{:}\PY{n}{seconds}\PY{p}{;}
            \PY{n}{dot\PYZus{}rv} \PY{o}{\PYZhy{}}\PY{o}{=} \PY{n}{tmp} \PY{o}{*} \PY{n}{si}\PY{o}{:}\PY{o}{:}\PY{n}{seconds}\PY{p}{;}

            \PY{n}{dot\PYZus{}th} \PY{o}{\PYZhy{}}\PY{o}{=} \PY{n}{tmp}  \PY{o}{*} \PY{n}{d\PYZus{}th\PYZus{}d\PYZus{}rv}\PY{o}{\PYZlt{}}\PY{k+kt}{real\PYZus{}t}\PY{o}{\PYZgt{}}\PY{p}{(}\PY{n}{T}\PY{p}{,} \PY{n}{th}\PY{p}{)} \PY{o}{/} \PY{n}{si}\PY{o}{:}\PY{o}{:}\PY{n}{kelvins} \PY{o}{*} \PY{n}{si}\PY{o}{:}\PY{o}{:}\PY{n}{seconds}\PY{p}{;} 
          \PY{p}{\PYZcb{}}

          \PY{n}{assert}\PY{p}{(}\PY{n}{rc} \PY{o}{+} \PY{n}{dot\PYZus{}rc} \PY{o}{*} \PY{n}{dt} \PY{o}{\PYZgt{}}\PY{o}{=} \PY{l+m+mi}{0} \PY{o}{\PYZam{}}\PY{o}{\PYZam{}} \PY{l+s}{\PYZdq{}}\PY{l+s}{condensation/evaporation can\PYZsq{}t make rho\PYZus{}c \PYZlt{} 0}\PY{l+s}{\PYZdq{}}\PY{p}{)}\PY{p}{;}
          \PY{n}{assert}\PY{p}{(}\PY{n}{rv} \PY{o}{+} \PY{n}{dot\PYZus{}rv} \PY{o}{*} \PY{n}{dt} \PY{o}{\PYZgt{}}\PY{o}{=} \PY{l+m+mi}{0} \PY{o}{\PYZam{}}\PY{o}{\PYZam{}} \PY{l+s}{\PYZdq{}}\PY{l+s}{condensation/evaporation can\PYZsq{}t make rho\PYZus{}v \PYZlt{} 0}\PY{l+s}{\PYZdq{}}\PY{p}{)}\PY{p}{;}
          \PY{n}{assert}\PY{p}{(}\PY{n}{th} \PY{o}{/} \PY{n}{si}\PY{o}{:}\PY{o}{:}\PY{n}{kelvin} \PY{o}{+} \PY{n}{dot\PYZus{}th} \PY{o}{*} \PY{n}{dt} \PY{o}{\PYZgt{}}\PY{o}{=} \PY{l+m+mi}{0} \PY{o}{\PYZam{}}\PY{o}{\PYZam{}} \PY{l+s}{\PYZdq{}}\PY{l+s}{condensation/evaporation can\PYZsq{}t make rho\PYZus{}e \PYZlt{} 0}\PY{l+s}{\PYZdq{}}\PY{p}{)}\PY{p}{;}
        \PY{p}{\PYZcb{}}
      \PY{p}{\PYZcb{}}

      \PY{k}{for} \PY{p}{(}\PY{k}{auto} \PY{n}{tup} \PY{o}{:} \PY{n}{zip}\PY{p}{(}
        \PY{n}{dot\PYZus{}rc\PYZus{}cont}\PY{p}{,} 
        \PY{n}{dot\PYZus{}nc\PYZus{}cont}\PY{p}{,} 
        \PY{n}{dot\PYZus{}rr\PYZus{}cont}\PY{p}{,} 
        \PY{n}{dot\PYZus{}nr\PYZus{}cont}\PY{p}{,}
        \PY{n}{rhod\PYZus{}cont}\PY{p}{,}
        \PY{n}{rc\PYZus{}cont}\PY{p}{,}  
        \PY{n}{nc\PYZus{}cont}\PY{p}{,} 
        \PY{n}{rr\PYZus{}cont}
      \PY{p}{)}\PY{p}{)}
      \PY{p}{\PYZob{}}
        \PY{k+kt}{real\PYZus{}t}
          \PY{o}{\PYZam{}}\PY{n}{dot\PYZus{}rc} \PY{o}{=} \PY{n}{boost}\PY{o}{:}\PY{o}{:}\PY{n}{get}\PY{o}{\PYZlt{}}\PY{l+m+mi}{0}\PY{o}{\PYZgt{}}\PY{p}{(}\PY{n}{tup}\PY{p}{)}\PY{p}{,}
          \PY{o}{\PYZam{}}\PY{n}{dot\PYZus{}nc} \PY{o}{=} \PY{n}{boost}\PY{o}{:}\PY{o}{:}\PY{n}{get}\PY{o}{\PYZlt{}}\PY{l+m+mi}{1}\PY{o}{\PYZgt{}}\PY{p}{(}\PY{n}{tup}\PY{p}{)}\PY{p}{,}
          \PY{o}{\PYZam{}}\PY{n}{dot\PYZus{}rr} \PY{o}{=} \PY{n}{boost}\PY{o}{:}\PY{o}{:}\PY{n}{get}\PY{o}{\PYZlt{}}\PY{l+m+mi}{2}\PY{o}{\PYZgt{}}\PY{p}{(}\PY{n}{tup}\PY{p}{)}\PY{p}{,}
          \PY{o}{\PYZam{}}\PY{n}{dot\PYZus{}nr} \PY{o}{=} \PY{n}{boost}\PY{o}{:}\PY{o}{:}\PY{n}{get}\PY{o}{\PYZlt{}}\PY{l+m+mi}{3}\PY{o}{\PYZgt{}}\PY{p}{(}\PY{n}{tup}\PY{p}{)}\PY{p}{;}
        \PY{k}{const} \PY{n}{quantity}\PY{o}{\PYZlt{}}\PY{n}{si}\PY{o}{:}\PY{o}{:}\PY{n}{mass\PYZus{}density}\PY{p}{,} \PY{k+kt}{real\PYZus{}t}\PY{o}{\PYZgt{}}  \PY{o}{\PYZam{}}\PY{n}{rhod} \PY{o}{=} \PY{n}{boost}\PY{o}{:}\PY{o}{:}\PY{n}{get}\PY{o}{\PYZlt{}}\PY{l+m+mi}{4}\PY{o}{\PYZgt{}}\PY{p}{(}\PY{n}{tup}\PY{p}{)} \PY{o}{*} \PY{n}{si}\PY{o}{:}\PY{o}{:}\PY{n}{kilograms} \PY{o}{/} \PY{n}{si}\PY{o}{:}\PY{o}{:}\PY{n}{cubic\PYZus{}metres}\PY{p}{;}
        \PY{k}{const} \PY{n}{quantity}\PY{o}{\PYZlt{}}\PY{n}{si}\PY{o}{:}\PY{o}{:}\PY{n}{dimensionless}\PY{p}{,} \PY{k+kt}{real\PYZus{}t}\PY{o}{\PYZgt{}}   \PY{o}{\PYZam{}}\PY{n}{rc} \PY{o}{=} \PY{n}{boost}\PY{o}{:}\PY{o}{:}\PY{n}{get}\PY{o}{\PYZlt{}}\PY{l+m+mi}{5}\PY{o}{\PYZgt{}}\PY{p}{(}\PY{n}{tup}\PY{p}{)} \PY{o}{*} \PY{n}{si}\PY{o}{:}\PY{o}{:}\PY{n}{dimensionless}\PY{p}{(}\PY{p}{)}\PY{p}{;}
        \PY{k}{const} \PY{n}{quantity}\PY{o}{\PYZlt{}}\PY{n}{divide\PYZus{}typeof\PYZus{}helper}\PY{o}{\PYZlt{}}\PY{n}{si}\PY{o}{:}\PY{o}{:}\PY{n}{dimensionless}\PY{p}{,} \PY{n}{si}\PY{o}{:}\PY{o}{:}\PY{n}{mass}\PY{o}{\PYZgt{}}\PY{o}{:}\PY{o}{:}\PY{n}{type}\PY{p}{,} \PY{k+kt}{real\PYZus{}t}\PY{o}{\PYZgt{}}  
                                                    \PY{o}{\PYZam{}}\PY{n}{nc} \PY{o}{=} \PY{n}{boost}\PY{o}{:}\PY{o}{:}\PY{n}{get}\PY{o}{\PYZlt{}}\PY{l+m+mi}{6}\PY{o}{\PYZgt{}}\PY{p}{(}\PY{n}{tup}\PY{p}{)} \PY{o}{/} \PY{n}{si}\PY{o}{:}\PY{o}{:}\PY{n}{kilograms}\PY{p}{;}
        \PY{k}{const} \PY{n}{quantity}\PY{o}{\PYZlt{}}\PY{n}{si}\PY{o}{:}\PY{o}{:}\PY{n}{dimensionless}\PY{p}{,} \PY{k+kt}{real\PYZus{}t}\PY{o}{\PYZgt{}}   \PY{o}{\PYZam{}}\PY{n}{rr} \PY{o}{=} \PY{n}{boost}\PY{o}{:}\PY{o}{:}\PY{n}{get}\PY{o}{\PYZlt{}}\PY{l+m+mi}{7}\PY{o}{\PYZgt{}}\PY{p}{(}\PY{n}{tup}\PY{p}{)} \PY{o}{*} \PY{n}{si}\PY{o}{:}\PY{o}{:}\PY{n}{dimensionless}\PY{p}{(}\PY{p}{)}\PY{p}{;}
 
\PY{c+c1}{//        if (rc + dot\PYZus{}rc * dt \PYZgt{} 0)}
\PY{c+c1}{//        \PYZob{}}
          \PY{c+c1}{// autoconversion rate (as in Khairoutdinov and Kogan 2000, but see Wood 2005 table 1)}
          \PY{k}{if} \PY{p}{(}\PY{n}{opts}\PY{p}{.}\PY{n}{acnv}\PY{p}{)}
          \PY{p}{\PYZob{}}                                  
           \PY{k}{if} \PY{p}{(}\PY{n}{rc} \PY{o}{\PYZgt{}} \PY{n}{eps\PYZus{}d} \PY{o}{\PYZam{}}\PY{o}{\PYZam{}} \PY{n}{nc} \PY{o}{\PYZgt{}} \PY{n}{eps\PYZus{}n}\PY{p}{)}
            \PY{p}{\PYZob{}}   
              \PY{n}{quantity}\PY{o}{\PYZlt{}}\PY{n}{si}\PY{o}{:}\PY{o}{:}\PY{n}{frequency}\PY{p}{,} \PY{k+kt}{real\PYZus{}t}\PY{o}{\PYZgt{}} \PY{n}{tmp} \PY{o}{=} \PY{n}{autoconv\PYZus{}rate}\PY{p}{(}\PY{n}{rc}\PY{p}{,} \PY{n}{rhod} \PY{o}{*} \PY{n}{nc}\PY{p}{)}\PY{p}{;}

              \PY{c+c1}{// so that autoconversion doesn\PYZsq{}t take more rc than there is}
              \PY{n}{tmp} \PY{o}{=} \PY{n}{std}\PY{o}{:}\PY{o}{:}\PY{n}{min}\PY{p}{(}\PY{n}{tmp}\PY{p}{,} \PY{p}{(}\PY{n}{rc} \PY{o}{+} \PY{n}{dt} \PY{o}{*} \PY{n}{dot\PYZus{}rc}\PY{p}{)} \PY{o}{/} \PY{p}{(}\PY{n}{dt} \PY{o}{*} \PY{n}{si}\PY{o}{:}\PY{o}{:}\PY{n}{seconds}\PY{p}{)}\PY{p}{)}\PY{p}{;}
              \PY{n}{assert}\PY{p}{(}\PY{n}{tmp} \PY{o}{*} \PY{n}{si}\PY{o}{:}\PY{o}{:}\PY{n}{seconds} \PY{o}{\PYZgt{}}\PY{o}{=} \PY{l+m+mi}{0} \PY{o}{\PYZam{}}\PY{o}{\PYZam{}} \PY{l+s}{\PYZdq{}}\PY{l+s}{autoconv rate has to be \PYZgt{}= 0}\PY{l+s}{\PYZdq{}}\PY{p}{)}\PY{p}{;}

              \PY{n}{dot\PYZus{}rc} \PY{o}{\PYZhy{}}\PY{o}{=} \PY{n}{tmp} \PY{o}{*} \PY{n}{si}\PY{o}{:}\PY{o}{:}\PY{n}{seconds}\PY{p}{;}
              \PY{n}{dot\PYZus{}rr} \PY{o}{+}\PY{o}{=} \PY{n}{tmp} \PY{o}{*} \PY{n}{si}\PY{o}{:}\PY{o}{:}\PY{n}{seconds}\PY{p}{;}

              \PY{c+c1}{// sink of N for cloud droplets is combined with the sink due to accretion}
              \PY{c+c1}{// source of N for drizzle assumes that all the drops have the same radius}
              \PY{n}{dot\PYZus{}nr} \PY{o}{+}\PY{o}{=} \PY{n}{tmp} \PY{o}{/} \PY{p}{(}\PY{k+kt}{real\PYZus{}t}\PY{p}{(}\PY{l+m+mi}{4}\PY{p}{)}\PY{o}{/}\PY{l+m+mi}{3} \PY{o}{*} \PY{n}{pi}\PY{o}{\PYZlt{}}\PY{k+kt}{real\PYZus{}t}\PY{o}{\PYZgt{}}\PY{p}{(}\PY{p}{)} \PY{o}{*} \PY{n}{rho\PYZus{}w}\PY{o}{\PYZlt{}}\PY{k+kt}{real\PYZus{}t}\PY{o}{\PYZgt{}}\PY{p}{(}\PY{p}{)} \PY{o}{*} \PY{n}{pow}\PY{o}{\PYZlt{}}\PY{l+m+mi}{3}\PY{o}{\PYZgt{}}\PY{p}{(}\PY{n}{drizzle\PYZus{}radius}\PY{o}{\PYZlt{}}\PY{k+kt}{real\PYZus{}t}\PY{o}{\PYZgt{}}\PY{p}{(}\PY{p}{)}\PY{p}{)}\PY{p}{)}
                \PY{o}{*} \PY{n}{si}\PY{o}{:}\PY{o}{:}\PY{n}{kilograms} \PY{o}{*} \PY{n}{si}\PY{o}{:}\PY{o}{:}\PY{n}{seconds}\PY{p}{;} \PY{c+c1}{// to make it dimensionless}
            \PY{p}{\PYZcb{}}

            \PY{n}{assert}\PY{p}{(}\PY{n}{rc} \PY{o}{+} \PY{n}{dot\PYZus{}rc} \PY{o}{*} \PY{n}{dt} \PY{o}{\PYZgt{}}\PY{o}{=} \PY{l+m+mi}{0} \PY{o}{\PYZam{}}\PY{o}{\PYZam{}} \PY{l+s}{\PYZdq{}}\PY{l+s}{autoconversion can\PYZsq{}t make rho\PYZus{}c negative}\PY{l+s}{\PYZdq{}}\PY{p}{)}\PY{p}{;}
          \PY{p}{\PYZcb{}}

\PY{c+c1}{//          if (rc + dot\PYZus{}rc * dt \PYZgt{} 0)}
\PY{c+c1}{//          \PYZob{}}
            \PY{c+c1}{// accretion rate (as in Khairoutdinov and Kogan 2000, but see Wood 2005 table 1)}
            \PY{k}{if} \PY{p}{(}\PY{n}{opts}\PY{p}{.}\PY{n}{accr}\PY{p}{)}
            \PY{p}{\PYZob{}}              
              \PY{k}{if} \PY{p}{(}\PY{n}{rc} \PY{o}{\PYZgt{}} \PY{n}{eps\PYZus{}d} \PY{o}{\PYZam{}}\PY{o}{\PYZam{}} \PY{n}{nc} \PY{o}{\PYZgt{}} \PY{n}{eps\PYZus{}n} \PY{o}{\PYZam{}}\PY{o}{\PYZam{}} \PY{n}{rr} \PY{o}{\PYZgt{}} \PY{n}{eps\PYZus{}d}\PY{p}{)}  
              \PY{p}{\PYZob{}}                   
                \PY{n}{quantity}\PY{o}{\PYZlt{}}\PY{n}{si}\PY{o}{:}\PY{o}{:}\PY{n}{frequency}\PY{p}{,} \PY{k+kt}{real\PYZus{}t}\PY{o}{\PYZgt{}} \PY{n}{tmp} \PY{o}{=} \PY{n}{accretion\PYZus{}rate}\PY{p}{(}\PY{n}{rc}\PY{p}{,} \PY{n}{rr}\PY{p}{)}\PY{p}{;}
                \PY{c+c1}{// so that accretion doesn\PYZsq{}t take more rho\PYZus{}c than there is}
                \PY{n}{tmp} \PY{o}{=} \PY{n}{std}\PY{o}{:}\PY{o}{:}\PY{n}{min}\PY{p}{(}\PY{n}{tmp}\PY{p}{,} \PY{p}{(}\PY{n}{rc} \PY{o}{+} \PY{n}{dt} \PY{o}{*} \PY{n}{dot\PYZus{}rc}\PY{p}{)} \PY{o}{/} \PY{p}{(}\PY{n}{dt} \PY{o}{*} \PY{n}{si}\PY{o}{:}\PY{o}{:}\PY{n}{seconds}\PY{p}{)}\PY{p}{)}\PY{p}{;}
                \PY{n}{assert}\PY{p}{(}\PY{n}{tmp} \PY{o}{*} \PY{n}{si}\PY{o}{:}\PY{o}{:}\PY{n}{seconds} \PY{o}{\PYZgt{}}\PY{o}{=} \PY{l+m+mi}{0} \PY{o}{\PYZam{}}\PY{o}{\PYZam{}} \PY{l+s}{\PYZdq{}}\PY{l+s}{accretion rate has to be \PYZgt{}= 0}\PY{l+s}{\PYZdq{}}\PY{p}{)}\PY{p}{;}
          
                \PY{n}{dot\PYZus{}rr} \PY{o}{+}\PY{o}{=} \PY{n}{tmp} \PY{o}{*} \PY{n}{si}\PY{o}{:}\PY{o}{:}\PY{n}{seconds}\PY{p}{;}
                \PY{n}{dot\PYZus{}rc} \PY{o}{\PYZhy{}}\PY{o}{=} \PY{n}{tmp} \PY{o}{*} \PY{n}{si}\PY{o}{:}\PY{o}{:}\PY{n}{seconds}\PY{p}{;}
                \PY{c+c1}{// the sink of N for cloud droplets is combined with sink due to autoconversion}
                \PY{c+c1}{// accretion does not change N for drizzle }
              \PY{p}{\PYZcb{}}

              \PY{n}{assert}\PY{p}{(}\PY{n}{rc} \PY{o}{+} \PY{n}{dot\PYZus{}rc} \PY{o}{*} \PY{n}{dt} \PY{o}{\PYZgt{}}\PY{o}{=} \PY{l+m+mi}{0} \PY{o}{\PYZam{}}\PY{o}{\PYZam{}} \PY{l+s}{\PYZdq{}}\PY{l+s}{accretion can\PYZsq{}t make rho\PYZus{}c negative}\PY{l+s}{\PYZdq{}}\PY{p}{)}\PY{p}{;}
            \PY{p}{\PYZcb{}}
\PY{c+c1}{//          \PYZcb{}}

          \PY{c+c1}{// sink of n\PYZus{}c due to autoconversion and accretion (see Khairoutdinov and Kogan 2000 eq 35)}
          \PY{c+c1}{//                                                 (be careful cause \PYZdq{}q\PYZdq{} there actually means mixing ratio, not water content)}
          \PY{c+c1}{// has to be just after autoconv. and accretion so that dot\PYZus{}rho\PYZus{}r is a sum of only those two}
          \PY{k}{if} \PY{p}{(}\PY{n}{opts}\PY{p}{.}\PY{n}{acnv} \PY{o}{|}\PY{o}{|} \PY{n}{opts}\PY{p}{.}\PY{n}{accr}\PY{p}{)}
          \PY{p}{\PYZob{}}
            \PY{k}{if} \PY{p}{(}\PY{n}{nc} \PY{o}{\PYZgt{}} \PY{n}{eps\PYZus{}n} \PY{o}{\PYZam{}}\PY{o}{\PYZam{}} \PY{n}{dot\PYZus{}rr} \PY{o}{\PYZgt{}} \PY{n}{eps\PYZus{}d}\PY{p}{)}  
            \PY{p}{\PYZob{}}                           
              \PY{n}{quantity}\PY{o}{\PYZlt{}}\PY{n}{divide\PYZus{}typeof\PYZus{}helper}\PY{o}{\PYZlt{}}\PY{n}{si}\PY{o}{:}\PY{o}{:}\PY{n}{frequency}\PY{p}{,} \PY{n}{si}\PY{o}{:}\PY{o}{:}\PY{n}{mass}\PY{o}{\PYZgt{}}\PY{o}{:}\PY{o}{:}\PY{n}{type}\PY{p}{,} \PY{k+kt}{real\PYZus{}t}\PY{o}{\PYZgt{}} \PY{n}{tmp} \PY{o}{=}
                \PY{n}{collision\PYZus{}sink\PYZus{}rate}\PY{p}{(}\PY{n}{dot\PYZus{}rr} \PY{o}{/} \PY{n}{si}\PY{o}{:}\PY{o}{:}\PY{n}{seconds}\PY{p}{,} \PY{n}{r\PYZus{}drop\PYZus{}c}\PY{p}{(}\PY{n}{rc}\PY{p}{,} \PY{n}{nc}\PY{p}{,} \PY{n}{rhod}\PY{p}{)}\PY{p}{)}\PY{p}{;}

              \PY{n}{assert}\PY{p}{(}\PY{n}{r\PYZus{}drop\PYZus{}c}\PY{p}{(}\PY{n}{rc}\PY{p}{,} \PY{n}{nc}\PY{p}{,} \PY{n}{rhod}\PY{p}{)} \PY{o}{\PYZgt{}}\PY{o}{=} \PY{l+m+mi}{0} \PY{o}{*} \PY{n}{si}\PY{o}{:}\PY{o}{:}\PY{n}{metres}  \PY{o}{\PYZam{}}\PY{o}{\PYZam{}} \PY{l+s}{\PYZdq{}}\PY{l+s}{mean droplet radius cannot be \PYZlt{} 0}\PY{l+s}{\PYZdq{}}\PY{p}{)}\PY{p}{;}
              \PY{n}{assert}\PY{p}{(}\PY{n}{tmp} \PY{o}{\PYZgt{}}\PY{o}{=} \PY{l+m+mi}{0} \PY{o}{/} \PY{n}{si}\PY{o}{:}\PY{o}{:}\PY{n}{kilograms} \PY{o}{/} \PY{n}{si}\PY{o}{:}\PY{o}{:}\PY{n}{seconds} \PY{o}{\PYZam{}}\PY{o}{\PYZam{}} \PY{l+s}{\PYZdq{}}\PY{l+s}{tmp}\PY{l+s}{\PYZdq{}}\PY{p}{)}\PY{p}{;}
 
              \PY{c+c1}{// so that collisions don\PYZsq{}t take more n\PYZus{}c than there is}
              \PY{n}{tmp} \PY{o}{=} \PY{n}{std}\PY{o}{:}\PY{o}{:}\PY{n}{min}\PY{p}{(}\PY{n}{tmp}\PY{p}{,} \PY{p}{(}\PY{n}{nc} \PY{o}{+} \PY{n}{dt} \PY{o}{*} \PY{n}{dot\PYZus{}nc} \PY{o}{/} \PY{n}{si}\PY{o}{:}\PY{o}{:}\PY{n}{kilograms}\PY{p}{)} \PY{o}{/} \PY{p}{(}\PY{n}{dt} \PY{o}{*} \PY{n}{si}\PY{o}{:}\PY{o}{:}\PY{n}{seconds}\PY{p}{)}\PY{p}{)}\PY{p}{;}
 
              \PY{n}{dot\PYZus{}nc} \PY{o}{\PYZhy{}}\PY{o}{=} \PY{n}{tmp} \PY{o}{*} \PY{n}{si}\PY{o}{:}\PY{o}{:}\PY{n}{kilograms} \PY{o}{*} \PY{n}{si}\PY{o}{:}\PY{o}{:}\PY{n}{seconds}\PY{p}{;}
            \PY{p}{\PYZcb{}}
          
          \PY{n}{assert}\PY{p}{(}\PY{n}{nc} \PY{o}{*} \PY{n}{si}\PY{o}{:}\PY{o}{:}\PY{n}{kilograms} \PY{o}{+} \PY{n}{dot\PYZus{}nc} \PY{o}{*} \PY{n}{dt} \PY{o}{\PYZgt{}}\PY{o}{=} \PY{l+m+mi}{0} \PY{o}{\PYZam{}}\PY{o}{\PYZam{}} \PY{l+s}{\PYZdq{}}\PY{l+s}{collisions can\PYZsq{}t make n\PYZus{}c negative}\PY{l+s}{\PYZdq{}}\PY{p}{)}\PY{p}{;}
          \PY{p}{\PYZcb{}} 
\PY{c+c1}{//        \PYZcb{}}
      \PY{p}{\PYZcb{}}

      \PY{k}{for} \PY{p}{(}\PY{k}{auto} \PY{n}{tup} \PY{o}{:} \PY{n}{zip}\PY{p}{(}
        \PY{n}{dot\PYZus{}th\PYZus{}cont}\PY{p}{,}
        \PY{n}{dot\PYZus{}rv\PYZus{}cont}\PY{p}{,} 
        \PY{n}{dot\PYZus{}rr\PYZus{}cont}\PY{p}{,} 
        \PY{n}{dot\PYZus{}nr\PYZus{}cont}\PY{p}{,}
        \PY{n}{rhod\PYZus{}cont}\PY{p}{,}
        \PY{n}{th\PYZus{}cont}\PY{p}{,} 
        \PY{n}{rv\PYZus{}cont}\PY{p}{,} 
        \PY{n}{rr\PYZus{}cont}\PY{p}{,}
        \PY{n}{nr\PYZus{}cont}
      \PY{p}{)}\PY{p}{)} \PY{p}{\PYZob{}}
        \PY{k+kt}{real\PYZus{}t}
          \PY{o}{\PYZam{}}\PY{n}{dot\PYZus{}th} \PY{o}{=} \PY{n}{boost}\PY{o}{:}\PY{o}{:}\PY{n}{get}\PY{o}{\PYZlt{}}\PY{l+m+mi}{0}\PY{o}{\PYZgt{}}\PY{p}{(}\PY{n}{tup}\PY{p}{)}\PY{p}{,}
          \PY{o}{\PYZam{}}\PY{n}{dot\PYZus{}rv} \PY{o}{=} \PY{n}{boost}\PY{o}{:}\PY{o}{:}\PY{n}{get}\PY{o}{\PYZlt{}}\PY{l+m+mi}{1}\PY{o}{\PYZgt{}}\PY{p}{(}\PY{n}{tup}\PY{p}{)}\PY{p}{,}
          \PY{o}{\PYZam{}}\PY{n}{dot\PYZus{}rr} \PY{o}{=} \PY{n}{boost}\PY{o}{:}\PY{o}{:}\PY{n}{get}\PY{o}{\PYZlt{}}\PY{l+m+mi}{2}\PY{o}{\PYZgt{}}\PY{p}{(}\PY{n}{tup}\PY{p}{)}\PY{p}{,}
          \PY{o}{\PYZam{}}\PY{n}{dot\PYZus{}nr} \PY{o}{=} \PY{n}{boost}\PY{o}{:}\PY{o}{:}\PY{n}{get}\PY{o}{\PYZlt{}}\PY{l+m+mi}{3}\PY{o}{\PYZgt{}}\PY{p}{(}\PY{n}{tup}\PY{p}{)}\PY{p}{;}
        \PY{k}{const} \PY{n}{quantity}\PY{o}{\PYZlt{}}\PY{n}{si}\PY{o}{:}\PY{o}{:}\PY{n}{mass\PYZus{}density}\PY{p}{,}  \PY{k+kt}{real\PYZus{}t}\PY{o}{\PYZgt{}} \PY{o}{\PYZam{}}\PY{n}{rhod}  \PY{o}{=} \PY{n}{boost}\PY{o}{:}\PY{o}{:}\PY{n}{get}\PY{o}{\PYZlt{}}\PY{l+m+mi}{4}\PY{o}{\PYZgt{}}\PY{p}{(}\PY{n}{tup}\PY{p}{)} \PY{o}{*} \PY{n}{si}\PY{o}{:}\PY{o}{:}\PY{n}{kilograms} \PY{o}{/} \PY{n}{si}\PY{o}{:}\PY{o}{:}\PY{n}{cubic\PYZus{}metres}\PY{p}{;}
        \PY{k}{const} \PY{n}{quantity}\PY{o}{\PYZlt{}}\PY{n}{si}\PY{o}{:}\PY{o}{:}\PY{n}{temperature}\PY{p}{,}   \PY{k+kt}{real\PYZus{}t}\PY{o}{\PYZgt{}} \PY{o}{\PYZam{}}\PY{n}{th}    \PY{o}{=} \PY{n}{boost}\PY{o}{:}\PY{o}{:}\PY{n}{get}\PY{o}{\PYZlt{}}\PY{l+m+mi}{5}\PY{o}{\PYZgt{}}\PY{p}{(}\PY{n}{tup}\PY{p}{)} \PY{o}{*} \PY{n}{si}\PY{o}{:}\PY{o}{:}\PY{n}{kelvins}\PY{p}{;}
        \PY{k}{const} \PY{n}{quantity}\PY{o}{\PYZlt{}}\PY{n}{si}\PY{o}{:}\PY{o}{:}\PY{n}{dimensionless}\PY{p}{,} \PY{k+kt}{real\PYZus{}t}\PY{o}{\PYZgt{}} \PY{o}{\PYZam{}}\PY{n}{rv}    \PY{o}{=} \PY{n}{boost}\PY{o}{:}\PY{o}{:}\PY{n}{get}\PY{o}{\PYZlt{}}\PY{l+m+mi}{6}\PY{o}{\PYZgt{}}\PY{p}{(}\PY{n}{tup}\PY{p}{)} \PY{o}{*} \PY{n}{si}\PY{o}{:}\PY{o}{:}\PY{n}{dimensionless}\PY{p}{(}\PY{p}{)}\PY{p}{;}
        \PY{k}{const} \PY{n}{quantity}\PY{o}{\PYZlt{}}\PY{n}{si}\PY{o}{:}\PY{o}{:}\PY{n}{dimensionless}\PY{p}{,} \PY{k+kt}{real\PYZus{}t}\PY{o}{\PYZgt{}} \PY{o}{\PYZam{}}\PY{n}{rr}    \PY{o}{=} \PY{n}{boost}\PY{o}{:}\PY{o}{:}\PY{n}{get}\PY{o}{\PYZlt{}}\PY{l+m+mi}{7}\PY{o}{\PYZgt{}}\PY{p}{(}\PY{n}{tup}\PY{p}{)} \PY{o}{*} \PY{n}{si}\PY{o}{:}\PY{o}{:}\PY{n}{dimensionless}\PY{p}{(}\PY{p}{)}\PY{p}{;}
        \PY{k}{const} \PY{n}{quantity}\PY{o}{\PYZlt{}}\PY{n}{divide\PYZus{}typeof\PYZus{}helper}\PY{o}{\PYZlt{}}\PY{n}{si}\PY{o}{:}\PY{o}{:}\PY{n}{dimensionless}\PY{p}{,} \PY{n}{si}\PY{o}{:}\PY{o}{:}\PY{n}{mass}\PY{o}{\PYZgt{}}\PY{o}{:}\PY{o}{:}\PY{n}{type}\PY{p}{,} \PY{k+kt}{real\PYZus{}t}\PY{o}{\PYZgt{}}  
                                                  \PY{o}{\PYZam{}}\PY{n}{nr}    \PY{o}{=} \PY{n}{boost}\PY{o}{:}\PY{o}{:}\PY{n}{get}\PY{o}{\PYZlt{}}\PY{l+m+mi}{8}\PY{o}{\PYZgt{}}\PY{p}{(}\PY{n}{tup}\PY{p}{)} \PY{o}{/} \PY{n}{si}\PY{o}{:}\PY{o}{:}\PY{n}{kilograms}\PY{p}{;}

        \PY{c+c1}{// helper temperature and pressure (TODO: it is repeated above!)}
        \PY{n}{quantity}\PY{o}{\PYZlt{}}\PY{n}{si}\PY{o}{:}\PY{o}{:}\PY{n}{temperature}\PY{p}{,} \PY{k+kt}{real\PYZus{}t}\PY{o}{\PYZgt{}} \PY{n}{T} \PY{o}{=} \PY{n}{common}\PY{o}{:}\PY{o}{:}\PY{n}{theta\PYZus{}dry}\PY{o}{:}\PY{o}{:}\PY{n}{T}\PY{o}{\PYZlt{}}\PY{k+kt}{real\PYZus{}t}\PY{o}{\PYZgt{}}\PY{p}{(}\PY{n}{th}\PY{p}{,} \PY{n}{rhod}\PY{p}{)}\PY{p}{;}
        \PY{n}{quantity}\PY{o}{\PYZlt{}}\PY{n}{si}\PY{o}{:}\PY{o}{:}\PY{n}{pressure}\PY{p}{,} \PY{k+kt}{real\PYZus{}t}\PY{o}{\PYZgt{}}    \PY{n}{p} \PY{o}{=} \PY{n}{common}\PY{o}{:}\PY{o}{:}\PY{n}{theta\PYZus{}dry}\PY{o}{:}\PY{o}{:}\PY{n}{p}\PY{o}{\PYZlt{}}\PY{k+kt}{real\PYZus{}t}\PY{o}{\PYZgt{}}\PY{p}{(}\PY{n}{rhod}\PY{p}{,} \PY{n}{rv}\PY{p}{,} \PY{n}{T}\PY{p}{)}\PY{p}{;}

        \PY{c+c1}{// evaporation of rain (see Morrison \PYZam{} Grabowski 2007)}
        \PY{k}{if} \PY{p}{(}\PY{n}{opts}\PY{p}{.}\PY{n}{cond}\PY{p}{)}
        \PY{p}{\PYZob{}}
          \PY{k}{if} \PY{p}{(}\PY{n}{rr} \PY{o}{\PYZgt{}} \PY{n}{eps\PYZus{}d} \PY{o}{\PYZam{}}\PY{o}{\PYZam{}} \PY{n}{nr} \PY{o}{\PYZgt{}} \PY{n}{eps\PYZus{}n}\PY{p}{)}
          \PY{p}{\PYZob{}} \PY{c+c1}{// cond/evap for rr}
            \PY{n}{assert}\PY{p}{(}\PY{n}{rr} \PY{o}{+} \PY{n}{dot\PYZus{}rr} \PY{o}{*} \PY{n}{dt} \PY{o}{\PYZgt{}}\PY{o}{=} \PY{l+m+mi}{0} \PY{o}{\PYZam{}}\PY{o}{\PYZam{}} \PY{l+s}{\PYZdq{}}\PY{l+s}{before rain cond\PYZhy{}evap}\PY{l+s}{\PYZdq{}}\PY{p}{)}\PY{p}{;}
            \PY{n}{assert}\PY{p}{(}\PY{n}{rv} \PY{o}{+} \PY{n}{dot\PYZus{}rv} \PY{o}{*} \PY{n}{dt} \PY{o}{\PYZgt{}}\PY{o}{=} \PY{l+m+mi}{0} \PY{o}{\PYZam{}}\PY{o}{\PYZam{}} \PY{l+s}{\PYZdq{}}\PY{l+s}{before rain cond\PYZhy{}evap}\PY{l+s}{\PYZdq{}}\PY{p}{)}\PY{p}{;}
            \PY{n}{assert}\PY{p}{(}\PY{n}{nr} \PY{o}{*} \PY{n}{si}\PY{o}{:}\PY{o}{:}\PY{n}{kilograms} \PY{o}{+} \PY{n}{dot\PYZus{}nr} \PY{o}{*} \PY{n}{dt} \PY{o}{\PYZgt{}}\PY{o}{=} \PY{l+m+mi}{0} \PY{o}{\PYZam{}}\PY{o}{\PYZam{}} \PY{l+s}{\PYZdq{}}\PY{l+s}{before rain cond\PYZhy{}evap}\PY{l+s}{\PYZdq{}}\PY{p}{)}\PY{p}{;}
            \PY{n}{assert}\PY{p}{(}\PY{n}{th} \PY{o}{/} \PY{n}{si}\PY{o}{:}\PY{o}{:}\PY{n}{kelvin} \PY{o}{+} \PY{n}{dot\PYZus{}th} \PY{o}{*} \PY{n}{dt} \PY{o}{\PYZgt{}}\PY{o}{=} \PY{l+m+mi}{0} \PY{o}{\PYZam{}}\PY{o}{\PYZam{}} \PY{l+s}{\PYZdq{}}\PY{l+s}{before rain cond\PYZhy{}evap}\PY{l+s}{\PYZdq{}}\PY{p}{)}\PY{p}{;}

            \PY{n}{quantity}\PY{o}{\PYZlt{}}\PY{n}{si}\PY{o}{:}\PY{o}{:}\PY{n}{frequency}\PY{p}{,} \PY{k+kt}{real\PYZus{}t}\PY{o}{\PYZgt{}} \PY{n}{tmp} \PY{o}{=} 
              \PY{n}{cond\PYZus{}evap\PYZus{}rate}\PY{o}{\PYZlt{}}\PY{k+kt}{real\PYZus{}t}\PY{o}{\PYZgt{}}\PY{p}{(}\PY{n}{T}\PY{p}{,} \PY{n}{p}\PY{p}{,} \PY{n}{rv}\PY{p}{,} \PY{n}{tau\PYZus{}relax\PYZus{}r}\PY{p}{(}\PY{n}{T}\PY{p}{,} \PY{n}{rhod}\PY{p}{,} \PY{n}{rr}\PY{p}{,} \PY{n}{nr}\PY{p}{)}\PY{p}{)}\PY{p}{;}

            \PY{n}{assert}\PY{p}{(}\PY{n}{r\PYZus{}drop\PYZus{}r}\PY{p}{(}\PY{n}{rr}\PY{p}{,} \PY{n}{nr}\PY{p}{)} \PY{o}{\PYZgt{}}\PY{o}{=} \PY{l+m+mi}{0} \PY{o}{*} \PY{n}{si}\PY{o}{:}\PY{o}{:}\PY{n}{metres}  \PY{o}{\PYZam{}}\PY{o}{\PYZam{}} \PY{l+s}{\PYZdq{}}\PY{l+s}{mean drop radius cannot be \PYZlt{} 0}\PY{l+s}{\PYZdq{}}\PY{p}{)}\PY{p}{;}

            \PY{n}{tmp} \PY{o}{=} \PY{n}{std}\PY{o}{:}\PY{o}{:}\PY{n}{min}\PY{p}{(}\PY{n}{tmp}\PY{p}{,} \PY{k+kt}{real\PYZus{}t}\PY{p}{(}\PY{l+m+mi}{0}\PY{p}{)} \PY{o}{/} \PY{n}{si}\PY{o}{:}\PY{o}{:}\PY{n}{seconds}\PY{p}{)}\PY{p}{;}
            \PY{k}{if} \PY{p}{(}\PY{n}{rr} \PY{o}{+} \PY{p}{(}\PY{n}{dot\PYZus{}rr} \PY{o}{/} \PY{n}{si}\PY{o}{:}\PY{o}{:}\PY{n}{seconds} \PY{o}{+} \PY{n}{tmp}\PY{p}{)} \PY{o}{*} \PY{p}{(}\PY{n}{dt} \PY{o}{*} \PY{n}{si}\PY{o}{:}\PY{o}{:}\PY{n}{seconds}\PY{p}{)} \PY{o}{\PYZlt{}} \PY{l+m+mi}{0}\PY{p}{)} \PY{c+c1}{// so that we don\PYZsq{}t evaporate more than we have}
            \PY{p}{\PYZob{}}
              \PY{n}{tmp} \PY{o}{=} \PY{o}{\PYZhy{}} \PY{p}{(}\PY{n}{rr} \PY{o}{+} \PY{n}{dt} \PY{o}{*} \PY{n}{dot\PYZus{}rr}\PY{p}{)} \PY{o}{/} \PY{p}{(}\PY{n}{dt} \PY{o}{*} \PY{n}{si}\PY{o}{:}\PY{o}{:}\PY{n}{seconds}\PY{p}{)}\PY{p}{;} \PY{c+c1}{// evaporate all rho\PYZus{}r}

              \PY{n}{dot\PYZus{}rv} \PY{o}{\PYZhy{}}\PY{o}{=} \PY{n}{tmp} \PY{o}{*} \PY{n}{si}\PY{o}{:}\PY{o}{:}\PY{n}{seconds}\PY{p}{;}
              \PY{n}{dot\PYZus{}rr} \PY{o}{+}\PY{o}{=} \PY{n}{tmp} \PY{o}{*} \PY{n}{si}\PY{o}{:}\PY{o}{:}\PY{n}{seconds}\PY{p}{;}

              \PY{n}{dot\PYZus{}nr}  \PY{o}{=} \PY{o}{\PYZhy{}}\PY{n}{nr} \PY{o}{/} \PY{n}{dt} \PY{o}{*} \PY{n}{si}\PY{o}{:}\PY{o}{:}\PY{n}{kilograms}\PY{p}{;} \PY{c+c1}{// and all n\PYZus{}r}
       
              \PY{n}{dot\PYZus{}th} \PY{o}{+}\PY{o}{=} \PY{o}{\PYZhy{}}\PY{n}{tmp}  \PY{o}{*} \PY{n}{d\PYZus{}th\PYZus{}d\PYZus{}rv}\PY{o}{\PYZlt{}}\PY{k+kt}{real\PYZus{}t}\PY{o}{\PYZgt{}}\PY{p}{(}\PY{n}{T}\PY{p}{,} \PY{n}{th}\PY{p}{)} \PY{o}{/} \PY{n}{si}\PY{o}{:}\PY{o}{:}\PY{n}{kelvins} \PY{o}{*} \PY{n}{si}\PY{o}{:}\PY{o}{:}\PY{n}{seconds}\PY{p}{;} 
            \PY{p}{\PYZcb{}}
            \PY{k}{else}
            \PY{p}{\PYZob{}}
              \PY{n}{dot\PYZus{}rv} \PY{o}{\PYZhy{}}\PY{o}{=} \PY{n}{tmp} \PY{o}{*} \PY{n}{si}\PY{o}{:}\PY{o}{:}\PY{n}{seconds}\PY{p}{;}
              \PY{n}{dot\PYZus{}rr} \PY{o}{+}\PY{o}{=} \PY{n}{tmp} \PY{o}{*} \PY{n}{si}\PY{o}{:}\PY{o}{:}\PY{n}{seconds}\PY{p}{;}
           
              \PY{n}{dot\PYZus{}th} \PY{o}{+}\PY{o}{=} \PY{o}{\PYZhy{}}\PY{n}{tmp} \PY{o}{*} \PY{n}{d\PYZus{}th\PYZus{}d\PYZus{}rv}\PY{o}{\PYZlt{}}\PY{k+kt}{real\PYZus{}t}\PY{o}{\PYZgt{}}\PY{p}{(}\PY{n}{T}\PY{p}{,} \PY{n}{th}\PY{p}{)} \PY{o}{/} \PY{n}{si}\PY{o}{:}\PY{o}{:}\PY{n}{kelvins} \PY{o}{*} \PY{n}{si}\PY{o}{:}\PY{o}{:}\PY{n}{seconds}\PY{p}{;} 

              \PY{c+c1}{// during evaporation n\PYZus{}r is reduced so that a constant mean drizzle/raindrop radius is mantained}
              \PY{k}{if} \PY{p}{(}\PY{n}{tmp} \PY{o}{\PYZlt{}} \PY{l+m+mi}{0} \PY{o}{/} \PY{n}{si}\PY{o}{:}\PY{o}{:}\PY{n}{seconds}\PY{p}{)} 
              \PY{p}{\PYZob{}}
                \PY{n}{quantity}\PY{o}{\PYZlt{}}\PY{n}{divide\PYZus{}typeof\PYZus{}helper}\PY{o}{\PYZlt{}}\PY{n}{si}\PY{o}{:}\PY{o}{:}\PY{n}{frequency}\PY{p}{,} \PY{n}{si}\PY{o}{:}\PY{o}{:}\PY{n}{mass}\PY{o}{\PYZgt{}}\PY{o}{:}\PY{o}{:}\PY{n}{type}\PY{p}{,} \PY{k+kt}{real\PYZus{}t}\PY{o}{\PYZgt{}} \PY{n}{dot\PYZus{}nr\PYZus{}tmp} \PY{o}{=} \PY{n}{tmp} \PY{o}{*} \PY{n}{nr} \PY{o}{/} \PY{n}{rr}\PY{p}{;}

                \PY{k}{if} \PY{p}{(}\PY{n}{nr} \PY{o}{+} \PY{p}{(}\PY{n}{dot\PYZus{}nr} \PY{o}{/} \PY{n}{si}\PY{o}{:}\PY{o}{:}\PY{n}{kilograms} \PY{o}{/} \PY{n}{si}\PY{o}{:}\PY{o}{:}\PY{n}{seconds} \PY{o}{+} \PY{n}{dot\PYZus{}nr\PYZus{}tmp}\PY{p}{)} \PY{o}{*} \PY{p}{(}\PY{n}{dt} \PY{o}{*} \PY{n}{si}\PY{o}{:}\PY{o}{:}\PY{n}{seconds}\PY{p}{)} \PY{o}{\PYZgt{}} \PY{l+m+mi}{0} \PY{o}{/} \PY{n}{si}\PY{o}{:}\PY{o}{:}\PY{n}{kilograms}\PY{p}{)}
                \PY{p}{\PYZob{}}
                  \PY{n}{dot\PYZus{}nr} \PY{o}{+}\PY{o}{=} \PY{n}{dot\PYZus{}nr\PYZus{}tmp} \PY{o}{*} \PY{n}{si}\PY{o}{:}\PY{o}{:}\PY{n}{kilograms} \PY{o}{*} \PY{n}{si}\PY{o}{:}\PY{o}{:}\PY{n}{seconds}\PY{p}{;}
                \PY{p}{\PYZcb{}}
                \PY{c+c1}{// else do nothing}
              \PY{p}{\PYZcb{}}
            \PY{p}{\PYZcb{}}
          \PY{p}{\PYZcb{}}

          \PY{n}{assert}\PY{p}{(}\PY{n}{rr} \PY{o}{+} \PY{n}{dot\PYZus{}rr} \PY{o}{*} \PY{n}{dt} \PY{o}{\PYZgt{}}\PY{o}{=} \PY{l+m+mi}{0} \PY{o}{\PYZam{}}\PY{o}{\PYZam{}} \PY{l+s}{\PYZdq{}}\PY{l+s}{rain condensation/evaporation can\PYZsq{}t make rho\PYZus{}r \PYZlt{} 0}\PY{l+s}{\PYZdq{}}\PY{p}{)}\PY{p}{;}
          \PY{n}{assert}\PY{p}{(}\PY{n}{rv} \PY{o}{+} \PY{n}{dot\PYZus{}rv} \PY{o}{*} \PY{n}{dt} \PY{o}{\PYZgt{}}\PY{o}{=} \PY{l+m+mi}{0} \PY{o}{\PYZam{}}\PY{o}{\PYZam{}} \PY{l+s}{\PYZdq{}}\PY{l+s}{rain condensation/evaporation can\PYZsq{}t make rho\PYZus{}v \PYZlt{} 0}\PY{l+s}{\PYZdq{}}\PY{p}{)}\PY{p}{;}
          \PY{n}{assert}\PY{p}{(}\PY{n}{nr} \PY{o}{*} \PY{n}{si}\PY{o}{:}\PY{o}{:}\PY{n}{kilograms} \PY{o}{+} \PY{n}{dot\PYZus{}nr} \PY{o}{*} \PY{n}{dt} \PY{o}{\PYZgt{}}\PY{o}{=} \PY{l+m+mi}{0} \PY{o}{\PYZam{}}\PY{o}{\PYZam{}} \PY{l+s}{\PYZdq{}}\PY{l+s}{rain condensation/evaporation can\PYZsq{}t make n\PYZus{}r \PYZlt{} 0}\PY{l+s}{\PYZdq{}}\PY{p}{)}\PY{p}{;}
          \PY{n}{assert}\PY{p}{(}\PY{n}{th} \PY{o}{/} \PY{n}{si}\PY{o}{:}\PY{o}{:}\PY{n}{kelvin} \PY{o}{+} \PY{n}{dot\PYZus{}th} \PY{o}{*} \PY{n}{dt} \PY{o}{\PYZgt{}}\PY{o}{=} \PY{l+m+mi}{0} \PY{o}{\PYZam{}}\PY{o}{\PYZam{}} \PY{l+s}{\PYZdq{}}\PY{l+s}{rain condensation/evaporation can\PYZsq{}t make rho\PYZus{}e \PYZlt{} 0}\PY{l+s}{\PYZdq{}}\PY{p}{)}\PY{p}{;}
        \PY{p}{\PYZcb{}}
      \PY{p}{\PYZcb{}}
    \PY{p}{\PYZcb{}}
  \PY{p}{\PYZcb{}}\PY{p}{;}    
\PY{p}{\PYZcb{}}\PY{p}{;}
\end{Verbatim}
  \vspace{-1.4em}%

  \caption{\label{lst:blk_2m_elw}
    \prog{blk\_2m::rhs\_cellwise()} signature
  }
\end{Listing}
\begin{Listing}
  \renewcommand*\FancyVerbStartString{\PY{c+c1}{//\PYZlt{}listing\PYZgt{}}}%
  \renewcommand*\FancyVerbStopString{\PY{c+c1}{//\PYZlt{}/listing\PYZgt{}}}%
  \begin{Verbatim}[commandchars=\\\{\}]
\PY{c+cm}{/** @file}
\PY{c+cm}{  * @copyright University of Warsaw}
\PY{c+cm}{  * @brief Rain sedimentation representation for single\PYZhy{}moment bulk microphysics}
\PY{c+cm}{  *   using forcing terms based on the upstrem advection scheme }
\PY{c+cm}{  * @section LICENSE}
\PY{c+cm}{  * GPLv3+ (see the COPYING file or http://www.gnu.org/licenses/)}
\PY{c+cm}{  */}

\PY{c+cp}{\PYZsh{}}\PY{c+cp}{pragma once}

\PY{c+cp}{\PYZsh{}}\PY{c+cp}{include \PYZlt{}libcloudph++}\PY{c+cp}{/}\PY{c+cp}{blk\PYZus{}2m}\PY{c+cp}{/}\PY{c+cp}{extincl.hpp\PYZgt{} }

\PY{k}{namespace} \PY{n}{libcloudphxx}
\PY{p}{\PYZob{}}
  \PY{k}{namespace} \PY{n}{blk\PYZus{}2m}
  \PY{p}{\PYZob{}}
    \PY{c+c1}{// expects the arguments to be columns with begin() pointing to the lowest level}
    \PY{c+c1}{// returns rain flux out of the domain}
\PY{c+c1}{//\PYZlt{}listing\PYZgt{}}
    \PY{k}{template} \PY{o}{\PYZlt{}}\PY{k}{typename} \PY{k+kt}{real\PYZus{}t}\PY{p}{,} \PY{k}{class} \PY{n+nc}{cont\PYZus{}t}\PY{o}{\PYZgt{}}
    \PY{k+kt}{real\PYZus{}t} \PY{n}{rhs\PYZus{}columnwise}\PY{p}{(}
      \PY{k}{const} \PY{k+kt}{opts\PYZus{}t}\PY{o}{\PYZlt{}}\PY{k+kt}{real\PYZus{}t}\PY{o}{\PYZgt{}} \PY{o}{\PYZam{}}\PY{n}{opts}\PY{p}{,}
      \PY{k+kt}{cont\PYZus{}t} \PY{o}{\PYZam{}}\PY{n}{dot\PYZus{}rr\PYZus{}cont}\PY{p}{,}
      \PY{k+kt}{cont\PYZus{}t} \PY{o}{\PYZam{}}\PY{n}{dot\PYZus{}nr\PYZus{}cont}\PY{p}{,}
      \PY{k}{const} \PY{k+kt}{cont\PYZus{}t} \PY{o}{\PYZam{}}\PY{n}{rhod\PYZus{}cont}\PY{p}{,}
      \PY{k}{const} \PY{k+kt}{cont\PYZus{}t} \PY{o}{\PYZam{}}\PY{n}{rr\PYZus{}cont}\PY{p}{,}
      \PY{k}{const} \PY{k+kt}{cont\PYZus{}t} \PY{o}{\PYZam{}}\PY{n}{nr\PYZus{}cont}\PY{p}{,}
      \PY{k}{const} \PY{k+kt}{real\PYZus{}t} \PY{o}{\PYZam{}}\PY{n}{dt}\PY{p}{,}
      \PY{k}{const} \PY{k+kt}{real\PYZus{}t} \PY{o}{\PYZam{}}\PY{n}{dz}
    \PY{p}{)}   
\PY{c+c1}{//\PYZlt{}/listing\PYZgt{}}
    \PY{p}{\PYZob{}}
      \PY{k}{if} \PY{p}{(}\PY{o}{!}\PY{n}{opts}\PY{p}{.}\PY{n}{sedi}\PY{p}{)} \PY{k}{return} \PY{l+m+mi}{0}\PY{p}{;}

      \PY{k}{using} \PY{n}{flux\PYZus{}rr} \PY{o}{=} \PY{n}{quantity}\PY{o}{\PYZlt{}}\PY{n}{divide\PYZus{}typeof\PYZus{}helper}\PY{o}{\PYZlt{}}\PY{n}{si}\PY{o}{:}\PY{o}{:}\PY{n}{mass\PYZus{}density}\PY{p}{,} \PY{n}{si}\PY{o}{:}\PY{o}{:}\PY{n}{time}\PY{o}{\PYZgt{}}\PY{o}{:}\PY{o}{:}\PY{n}{type}\PY{p}{,} \PY{k+kt}{real\PYZus{}t}\PY{o}{\PYZgt{}}\PY{p}{;}
      \PY{k}{using} \PY{n}{flux\PYZus{}nr} \PY{o}{=} \PY{n}{quantity}\PY{o}{\PYZlt{}}\PY{n}{divide\PYZus{}typeof\PYZus{}helper}\PY{o}{\PYZlt{}}\PY{n}{si}\PY{o}{:}\PY{o}{:}\PY{n}{frequency}\PY{p}{,} \PY{n}{si}\PY{o}{:}\PY{o}{:}\PY{n}{volume}\PY{o}{\PYZgt{}}\PY{o}{:}\PY{o}{:}\PY{n}{type}\PY{p}{,} \PY{k+kt}{real\PYZus{}t}\PY{o}{\PYZgt{}}\PY{p}{;}
     
      \PY{k}{auto} \PY{n}{dot\PYZus{}rr\PYZus{}unit} \PY{o}{=} \PY{n}{si}\PY{o}{:}\PY{o}{:}\PY{n}{hertz}\PY{p}{;}
      \PY{k}{auto} \PY{n}{dot\PYZus{}nr\PYZus{}unit} \PY{o}{=} \PY{n}{si}\PY{o}{:}\PY{o}{:}\PY{n}{hertz} \PY{o}{/} \PY{n}{si}\PY{o}{:}\PY{o}{:}\PY{n}{kilograms}\PY{p}{;}

      \PY{k}{auto} \PY{n}{rflux\PYZus{}unit} \PY{o}{=} \PY{n}{si}\PY{o}{:}\PY{o}{:}\PY{n}{kilograms} \PY{o}{/} \PY{n}{si}\PY{o}{:}\PY{o}{:}\PY{n}{seconds} \PY{o}{/} \PY{n}{si}\PY{o}{:}\PY{o}{:}\PY{n}{cubic\PYZus{}metres}\PY{p}{;}
      \PY{k}{auto} \PY{n}{nflux\PYZus{}unit} \PY{o}{=} \PY{n}{si}\PY{o}{:}\PY{o}{:}\PY{n}{hertz} \PY{o}{/} \PY{n}{si}\PY{o}{:}\PY{o}{:}\PY{n}{cubic\PYZus{}metres}\PY{p}{;}

      \PY{n}{flux\PYZus{}rr} \PY{n}{flux\PYZus{}rr\PYZus{}in} \PY{o}{=} \PY{l+m+mi}{0} \PY{o}{*} \PY{n}{si}\PY{o}{:}\PY{o}{:}\PY{n}{kilograms} \PY{o}{/} \PY{n}{si}\PY{o}{:}\PY{o}{:}\PY{n}{cubic\PYZus{}metres} \PY{o}{/} \PY{n}{si}\PY{o}{:}\PY{o}{:}\PY{n}{seconds}\PY{p}{;}
      \PY{n}{flux\PYZus{}nr} \PY{n}{flux\PYZus{}nr\PYZus{}in} \PY{o}{=} \PY{l+m+mi}{0} \PY{o}{/} \PY{n}{si}\PY{o}{:}\PY{o}{:}\PY{n}{cubic\PYZus{}metres} \PY{o}{/} \PY{n}{si}\PY{o}{:}\PY{o}{:}\PY{n}{seconds}\PY{p}{;}

      \PY{k+kt}{real\PYZus{}t} \PY{o}{*}\PY{n}{dot\PYZus{}rr} \PY{o}{=} \PY{n+nb}{NULL}\PY{p}{;}
      \PY{k+kt}{real\PYZus{}t} \PY{o}{*}\PY{n}{dot\PYZus{}nr} \PY{o}{=} \PY{n+nb}{NULL}\PY{p}{;}
      \PY{k}{const} \PY{k+kt}{real\PYZus{}t} \PY{n}{zero} \PY{o}{=} \PY{l+m+mi}{0}\PY{p}{;}

      \PY{c+c1}{// initial values that should give zero flux from above the domain top}
      \PY{k}{const} \PY{k+kt}{real\PYZus{}t} 
        \PY{o}{*}\PY{n}{rr} \PY{o}{=} \PY{o}{\PYZam{}}\PY{n}{zero}\PY{p}{,} 
        \PY{o}{*}\PY{n}{nr} \PY{o}{=} \PY{o}{\PYZam{}}\PY{n}{zero}\PY{p}{,} 
        \PY{o}{*}\PY{n}{rhod} \PY{o}{=} \PY{o}{\PYZam{}}\PY{o}{*}\PY{p}{(}\PY{o}{\PYZhy{}}\PY{o}{\PYZhy{}}\PY{p}{(}\PY{n}{rhod\PYZus{}cont}\PY{p}{.}\PY{n}{end}\PY{p}{(}\PY{p}{)}\PY{p}{)}\PY{p}{)}\PY{p}{;}

      \PY{k}{auto} \PY{n}{iter} \PY{o}{=} \PY{n}{zip}\PY{p}{(}\PY{n}{rhod\PYZus{}cont}\PY{p}{,} \PY{n}{rr\PYZus{}cont}\PY{p}{,} \PY{n}{nr\PYZus{}cont}\PY{p}{,} \PY{n}{dot\PYZus{}rr\PYZus{}cont}\PY{p}{,} \PY{n}{dot\PYZus{}nr\PYZus{}cont}\PY{p}{)}\PY{p}{;}
      \PY{k}{for} \PY{p}{(}\PY{k}{auto} \PY{n}{tup\PYZus{}ptr} \PY{o}{=} \PY{n}{iter}\PY{p}{.}\PY{n}{end}\PY{p}{(}\PY{p}{)}\PY{p}{;} \PY{n}{tup\PYZus{}ptr} \PY{o}{!}\PY{o}{=} \PY{n}{iter}\PY{p}{.}\PY{n}{begin}\PY{p}{(}\PY{p}{)}\PY{p}{;}\PY{p}{)}
      \PY{p}{\PYZob{}}
        \PY{o}{\PYZhy{}}\PY{o}{\PYZhy{}}\PY{n}{tup\PYZus{}ptr}\PY{p}{;}

        \PY{k}{const} \PY{k+kt}{real\PYZus{}t}
          \PY{o}{*}\PY{n}{rhod\PYZus{}below} \PY{o}{=} \PY{o}{\PYZam{}}\PY{n}{boost}\PY{o}{:}\PY{o}{:}\PY{n}{get}\PY{o}{\PYZlt{}}\PY{l+m+mi}{0}\PY{o}{\PYZgt{}}\PY{p}{(}\PY{o}{*}\PY{n}{tup\PYZus{}ptr}\PY{p}{)}\PY{p}{,}
          \PY{o}{*}\PY{n}{rr\PYZus{}below}   \PY{o}{=} \PY{o}{\PYZam{}}\PY{n}{boost}\PY{o}{:}\PY{o}{:}\PY{n}{get}\PY{o}{\PYZlt{}}\PY{l+m+mi}{1}\PY{o}{\PYZgt{}}\PY{p}{(}\PY{o}{*}\PY{n}{tup\PYZus{}ptr}\PY{p}{)}\PY{p}{,}
          \PY{o}{*}\PY{n}{nr\PYZus{}below}   \PY{o}{=} \PY{o}{\PYZam{}}\PY{n}{boost}\PY{o}{:}\PY{o}{:}\PY{n}{get}\PY{o}{\PYZlt{}}\PY{l+m+mi}{2}\PY{o}{\PYZgt{}}\PY{p}{(}\PY{o}{*}\PY{n}{tup\PYZus{}ptr}\PY{p}{)}\PY{p}{;}

        \PY{k}{if} \PY{p}{(}\PY{n}{dot\PYZus{}rr} \PY{o}{!}\PY{o}{=} \PY{n+nb}{NULL}\PY{p}{)} \PY{c+c1}{// i.e. all but first (top) grid cell}
        \PY{p}{\PYZob{}}
          \PY{c+c1}{// terminal velocities at grid\PYZhy{}cell edge (to assure precip mass conservation)}
          \PY{n}{quantity}\PY{o}{\PYZlt{}}\PY{n}{multiply\PYZus{}typeof\PYZus{}helper}\PY{o}{\PYZlt{}}\PY{n}{si}\PY{o}{:}\PY{o}{:}\PY{n}{velocity}\PY{p}{,} \PY{n}{si}\PY{o}{:}\PY{o}{:}\PY{n}{mass\PYZus{}density}\PY{o}{\PYZgt{}}\PY{o}{:}\PY{o}{:}\PY{n}{type}\PY{p}{,} \PY{k+kt}{real\PYZus{}t}\PY{o}{\PYZgt{}} \PY{n}{tmp\PYZus{}mom\PYZus{}m}  \PY{o}{=} \PY{o}{\PYZhy{}}\PY{k+kt}{real\PYZus{}t}\PY{p}{(}\PY{l+m+mf}{.5}\PY{p}{)} \PY{o}{*} \PY{p}{(} \PY{c+c1}{// averaging + axis orientation}
	    \PY{p}{(}\PY{o}{*}\PY{n}{rhod\PYZus{}below} \PY{o}{*} \PY{n}{si}\PY{o}{:}\PY{o}{:}\PY{n}{kilograms} \PY{o}{/} \PY{n}{si}\PY{o}{:}\PY{o}{:}\PY{n}{cubic\PYZus{}metres}\PY{p}{)} \PY{o}{*} \PY{n}{formulae}\PY{o}{:}\PY{o}{:}\PY{n}{v\PYZus{}term\PYZus{}m}\PY{p}{(}
              \PY{o}{*}\PY{n}{rhod\PYZus{}below} \PY{o}{*} \PY{n}{si}\PY{o}{:}\PY{o}{:}\PY{n}{kilograms} \PY{o}{/} \PY{n}{si}\PY{o}{:}\PY{o}{:}\PY{n}{cubic\PYZus{}metres}\PY{p}{,} 
              \PY{o}{*}\PY{n}{rr\PYZus{}below} \PY{o}{*} \PY{n}{si}\PY{o}{:}\PY{o}{:}\PY{n}{kilograms} \PY{o}{/} \PY{n}{si}\PY{o}{:}\PY{o}{:}\PY{n}{kilograms}\PY{p}{,} 
              \PY{o}{*}\PY{n}{nr\PYZus{}below} \PY{o}{/} \PY{n}{si}\PY{o}{:}\PY{o}{:}\PY{n}{kilograms} 
            \PY{p}{)} \PY{o}{+} 
	    \PY{p}{(}\PY{o}{*}\PY{n}{rhod} \PY{o}{*} \PY{n}{si}\PY{o}{:}\PY{o}{:}\PY{n}{kilograms} \PY{o}{/} \PY{n}{si}\PY{o}{:}\PY{o}{:}\PY{n}{cubic\PYZus{}metres}\PY{p}{)} \PY{o}{*} \PY{n}{formulae}\PY{o}{:}\PY{o}{:}\PY{n}{v\PYZus{}term\PYZus{}m}\PY{p}{(}
              \PY{o}{*}\PY{n}{rhod} \PY{o}{*} \PY{n}{si}\PY{o}{:}\PY{o}{:}\PY{n}{kilograms} \PY{o}{/} \PY{n}{si}\PY{o}{:}\PY{o}{:}\PY{n}{cubic\PYZus{}metres}\PY{p}{,}
              \PY{o}{*}\PY{n}{rr} \PY{o}{*} \PY{n}{si}\PY{o}{:}\PY{o}{:}\PY{n}{kilograms} \PY{o}{/} \PY{n}{si}\PY{o}{:}\PY{o}{:}\PY{n}{kilograms}\PY{p}{,}
              \PY{o}{*}\PY{n}{nr} \PY{o}{/} \PY{n}{si}\PY{o}{:}\PY{o}{:}\PY{n}{kilograms}
            \PY{p}{)}
	  \PY{p}{)}\PY{p}{;} 
 
          \PY{n}{quantity}\PY{o}{\PYZlt{}}\PY{n}{multiply\PYZus{}typeof\PYZus{}helper}\PY{o}{\PYZlt{}}\PY{n}{si}\PY{o}{:}\PY{o}{:}\PY{n}{velocity}\PY{p}{,} \PY{n}{si}\PY{o}{:}\PY{o}{:}\PY{n}{mass\PYZus{}density}\PY{o}{\PYZgt{}}\PY{o}{:}\PY{o}{:}\PY{n}{type}\PY{p}{,} \PY{k+kt}{real\PYZus{}t}\PY{o}{\PYZgt{}} \PY{n}{tmp\PYZus{}mom\PYZus{}n}  \PY{o}{=} \PY{o}{\PYZhy{}}\PY{k+kt}{real\PYZus{}t}\PY{p}{(}\PY{l+m+mf}{.5}\PY{p}{)} \PY{o}{*} \PY{p}{(} \PY{c+c1}{// averaging + axis orientation}
	    \PY{p}{(}\PY{o}{*}\PY{n}{rhod\PYZus{}below} \PY{o}{*} \PY{n}{si}\PY{o}{:}\PY{o}{:}\PY{n}{kilograms} \PY{o}{/} \PY{n}{si}\PY{o}{:}\PY{o}{:}\PY{n}{cubic\PYZus{}metres}\PY{p}{)} \PY{o}{*} \PY{n}{formulae}\PY{o}{:}\PY{o}{:}\PY{n}{v\PYZus{}term\PYZus{}n}\PY{p}{(}
              \PY{o}{*}\PY{n}{rhod\PYZus{}below} \PY{o}{*} \PY{n}{si}\PY{o}{:}\PY{o}{:}\PY{n}{kilograms} \PY{o}{/} \PY{n}{si}\PY{o}{:}\PY{o}{:}\PY{n}{cubic\PYZus{}metres}\PY{p}{,} 
              \PY{o}{*}\PY{n}{rr\PYZus{}below} \PY{o}{*} \PY{n}{si}\PY{o}{:}\PY{o}{:}\PY{n}{kilograms} \PY{o}{/} \PY{n}{si}\PY{o}{:}\PY{o}{:}\PY{n}{kilograms}\PY{p}{,} 
              \PY{o}{*}\PY{n}{nr\PYZus{}below} \PY{o}{/} \PY{n}{si}\PY{o}{:}\PY{o}{:}\PY{n}{kilograms}
            \PY{p}{)} \PY{o}{+} 
	    \PY{p}{(}\PY{o}{*}\PY{n}{rhod} \PY{o}{*} \PY{n}{si}\PY{o}{:}\PY{o}{:}\PY{n}{kilograms} \PY{o}{/} \PY{n}{si}\PY{o}{:}\PY{o}{:}\PY{n}{cubic\PYZus{}metres}\PY{p}{)} \PY{o}{*} \PY{n}{formulae}\PY{o}{:}\PY{o}{:}\PY{n}{v\PYZus{}term\PYZus{}n}\PY{p}{(}
              \PY{o}{*}\PY{n}{rhod} \PY{o}{*} \PY{n}{si}\PY{o}{:}\PY{o}{:}\PY{n}{kilograms} \PY{o}{/} \PY{n}{si}\PY{o}{:}\PY{o}{:}\PY{n}{cubic\PYZus{}metres}\PY{p}{,}
              \PY{o}{*}\PY{n}{rr} \PY{o}{*} \PY{n}{si}\PY{o}{:}\PY{o}{:}\PY{n}{kilograms} \PY{o}{/} \PY{n}{si}\PY{o}{:}\PY{o}{:}\PY{n}{kilograms}\PY{p}{,}
              \PY{o}{*}\PY{n}{nr} \PY{o}{/} \PY{n}{si}\PY{o}{:}\PY{o}{:}\PY{n}{kilograms}
            \PY{p}{)}
	  \PY{p}{)}\PY{p}{;} 
          
          \PY{n}{flux\PYZus{}rr} \PY{n}{flux\PYZus{}rr\PYZus{}out} \PY{o}{=} \PY{n}{tmp\PYZus{}mom\PYZus{}m} \PY{o}{*} \PY{p}{(}\PY{o}{*}\PY{n}{rr} \PY{o}{*} \PY{n}{si}\PY{o}{:}\PY{o}{:}\PY{n}{kilograms} \PY{o}{/} \PY{n}{si}\PY{o}{:}\PY{o}{:}\PY{n}{kilograms}\PY{p}{)} \PY{o}{/} \PY{p}{(}\PY{n}{dz} \PY{o}{*} \PY{n}{si}\PY{o}{:}\PY{o}{:}\PY{n}{metres}\PY{p}{)}\PY{p}{;}
          \PY{n}{flux\PYZus{}rr\PYZus{}out} \PY{o}{=} \PY{o}{\PYZhy{}} \PY{n}{std}\PY{o}{:}\PY{o}{:}\PY{n}{min}\PY{p}{(}\PY{k+kt}{real\PYZus{}t}\PY{p}{(}\PY{o}{\PYZhy{}}\PY{n}{flux\PYZus{}rr\PYZus{}out} \PY{o}{/} \PY{n}{rflux\PYZus{}unit}\PY{p}{)}\PY{p}{,} \PY{o}{*}\PY{n}{rhod} \PY{o}{*} \PY{p}{(}\PY{o}{*}\PY{n}{rr} \PY{o}{+} \PY{n}{dt} \PY{o}{*} \PY{o}{*}\PY{n}{dot\PYZus{}rr}\PY{p}{)} \PY{o}{/} \PY{n}{dt}\PY{p}{)} \PY{o}{*} \PY{n}{rflux\PYZus{}unit}\PY{p}{;}

          \PY{n}{flux\PYZus{}nr} \PY{n}{flux\PYZus{}nr\PYZus{}out} \PY{o}{=} \PY{n}{tmp\PYZus{}mom\PYZus{}n} \PY{o}{*} \PY{p}{(}\PY{o}{*}\PY{n}{nr} \PY{o}{/} \PY{n}{si}\PY{o}{:}\PY{o}{:}\PY{n}{kilograms}\PY{p}{)} \PY{o}{/} \PY{p}{(}\PY{n}{dz} \PY{o}{*} \PY{n}{si}\PY{o}{:}\PY{o}{:}\PY{n}{metres}\PY{p}{)}\PY{p}{;}
          \PY{n}{flux\PYZus{}nr\PYZus{}out} \PY{o}{=} \PY{o}{\PYZhy{}} \PY{n}{std}\PY{o}{:}\PY{o}{:}\PY{n}{min}\PY{p}{(}\PY{k+kt}{real\PYZus{}t}\PY{p}{(}\PY{o}{\PYZhy{}}\PY{n}{flux\PYZus{}nr\PYZus{}out} \PY{o}{/} \PY{n}{nflux\PYZus{}unit}\PY{p}{)}\PY{p}{,} \PY{o}{*}\PY{n}{rhod} \PY{o}{*} \PY{p}{(}\PY{o}{*}\PY{n}{nr} \PY{o}{+} \PY{n}{dt} \PY{o}{*} \PY{o}{*}\PY{n}{dot\PYZus{}nr}\PY{p}{)} \PY{o}{/} \PY{n}{dt}\PY{p}{)} \PY{o}{*} \PY{n}{nflux\PYZus{}unit}\PY{p}{;}

	  \PY{o}{*}\PY{n}{dot\PYZus{}rr} \PY{o}{\PYZhy{}}\PY{o}{=} \PY{p}{(}\PY{n}{flux\PYZus{}rr\PYZus{}in} \PY{o}{\PYZhy{}} \PY{n}{flux\PYZus{}rr\PYZus{}out}\PY{p}{)} \PY{o}{/} \PY{p}{(}\PY{o}{*}\PY{n}{rhod} \PY{o}{*} \PY{n}{si}\PY{o}{:}\PY{o}{:}\PY{n}{kilograms} \PY{o}{/} \PY{n}{si}\PY{o}{:}\PY{o}{:}\PY{n}{cubic\PYZus{}metres}\PY{p}{)} \PY{o}{/} \PY{n}{dot\PYZus{}rr\PYZus{}unit}\PY{p}{;}
          \PY{n}{flux\PYZus{}rr\PYZus{}in} \PY{o}{=} \PY{n}{flux\PYZus{}rr\PYZus{}out}\PY{p}{;} \PY{c+c1}{// inflow = outflow from above}
	  \PY{o}{*}\PY{n}{dot\PYZus{}nr} \PY{o}{\PYZhy{}}\PY{o}{=} \PY{p}{(}\PY{n}{flux\PYZus{}nr\PYZus{}in} \PY{o}{\PYZhy{}} \PY{n}{flux\PYZus{}nr\PYZus{}out}\PY{p}{)} \PY{o}{/} \PY{p}{(}\PY{o}{*}\PY{n}{rhod} \PY{o}{*} \PY{n}{si}\PY{o}{:}\PY{o}{:}\PY{n}{kilograms} \PY{o}{/} \PY{n}{si}\PY{o}{:}\PY{o}{:}\PY{n}{cubic\PYZus{}metres}\PY{p}{)} \PY{o}{/} \PY{n}{dot\PYZus{}nr\PYZus{}unit}\PY{p}{;}
          \PY{n}{flux\PYZus{}nr\PYZus{}in} \PY{o}{=} \PY{n}{flux\PYZus{}nr\PYZus{}out}\PY{p}{;} \PY{c+c1}{// inflow = outflow from above}
        \PY{p}{\PYZcb{}}

        \PY{n}{dot\PYZus{}rr} \PY{o}{=} \PY{o}{\PYZam{}}\PY{n}{boost}\PY{o}{:}\PY{o}{:}\PY{n}{get}\PY{o}{\PYZlt{}}\PY{l+m+mi}{3}\PY{o}{\PYZgt{}}\PY{p}{(}\PY{o}{*}\PY{n}{tup\PYZus{}ptr}\PY{p}{)}\PY{p}{;}
        \PY{n}{dot\PYZus{}nr} \PY{o}{=} \PY{o}{\PYZam{}}\PY{n}{boost}\PY{o}{:}\PY{o}{:}\PY{n}{get}\PY{o}{\PYZlt{}}\PY{l+m+mi}{4}\PY{o}{\PYZgt{}}\PY{p}{(}\PY{o}{*}\PY{n}{tup\PYZus{}ptr}\PY{p}{)}\PY{p}{;}
        \PY{n}{rhod} \PY{o}{=} \PY{n}{rhod\PYZus{}below}\PY{p}{;}
        \PY{n}{rr} \PY{o}{=} \PY{n}{rr\PYZus{}below}\PY{p}{;}
        \PY{n}{nr} \PY{o}{=} \PY{n}{nr\PYZus{}below}\PY{p}{;}
      \PY{p}{\PYZcb{}}

      \PY{c+c1}{// the bottom grid cell (with mid\PYZhy{}cell vterm approximation)}
      \PY{n}{quantity}\PY{o}{\PYZlt{}}\PY{n}{multiply\PYZus{}typeof\PYZus{}helper}\PY{o}{\PYZlt{}}\PY{n}{si}\PY{o}{:}\PY{o}{:}\PY{n}{velocity}\PY{p}{,} \PY{n}{si}\PY{o}{:}\PY{o}{:}\PY{n}{mass\PYZus{}density}\PY{o}{\PYZgt{}}\PY{o}{:}\PY{o}{:}\PY{n}{type}\PY{p}{,} \PY{k+kt}{real\PYZus{}t}\PY{o}{\PYZgt{}} 
        \PY{n}{tmp\PYZus{}mom\PYZus{}m} \PY{o}{=} \PY{o}{\PYZhy{}} \PY{p}{(}\PY{o}{*}\PY{n}{rhod} \PY{o}{*} \PY{n}{si}\PY{o}{:}\PY{o}{:}\PY{n}{kilograms} \PY{o}{/} \PY{n}{si}\PY{o}{:}\PY{o}{:}\PY{n}{cubic\PYZus{}metres}\PY{p}{)} \PY{o}{*} \PY{n}{formulae}\PY{o}{:}\PY{o}{:}\PY{n}{v\PYZus{}term\PYZus{}m}\PY{p}{(}
	  \PY{o}{*}\PY{n}{rhod} \PY{o}{*} \PY{n}{si}\PY{o}{:}\PY{o}{:}\PY{n}{kilograms} \PY{o}{/} \PY{n}{si}\PY{o}{:}\PY{o}{:}\PY{n}{cubic\PYZus{}metres}\PY{p}{,} 
	  \PY{o}{*}\PY{n}{rr} \PY{o}{*} \PY{n}{si}\PY{o}{:}\PY{o}{:}\PY{n}{kilograms} \PY{o}{/} \PY{n}{si}\PY{o}{:}\PY{o}{:}\PY{n}{kilograms}\PY{p}{,}
	  \PY{o}{*}\PY{n}{nr} \PY{o}{/} \PY{n}{si}\PY{o}{:}\PY{o}{:}\PY{n}{kilograms}
	\PY{p}{)}\PY{p}{;} 
      \PY{n}{quantity}\PY{o}{\PYZlt{}}\PY{n}{multiply\PYZus{}typeof\PYZus{}helper}\PY{o}{\PYZlt{}}\PY{n}{si}\PY{o}{:}\PY{o}{:}\PY{n}{velocity}\PY{p}{,} \PY{n}{si}\PY{o}{:}\PY{o}{:}\PY{n}{mass\PYZus{}density}\PY{o}{\PYZgt{}}\PY{o}{:}\PY{o}{:}\PY{n}{type}\PY{p}{,} \PY{k+kt}{real\PYZus{}t}\PY{o}{\PYZgt{}} 
	\PY{n}{tmp\PYZus{}mom\PYZus{}n} \PY{o}{=} \PY{o}{\PYZhy{}} \PY{p}{(}\PY{o}{*}\PY{n}{rhod} \PY{o}{*} \PY{n}{si}\PY{o}{:}\PY{o}{:}\PY{n}{kilograms} \PY{o}{/} \PY{n}{si}\PY{o}{:}\PY{o}{:}\PY{n}{cubic\PYZus{}metres}\PY{p}{)} \PY{o}{*} \PY{n}{formulae}\PY{o}{:}\PY{o}{:}\PY{n}{v\PYZus{}term\PYZus{}n}\PY{p}{(}
	  \PY{o}{*}\PY{n}{rhod} \PY{o}{*} \PY{n}{si}\PY{o}{:}\PY{o}{:}\PY{n}{kilograms} \PY{o}{/} \PY{n}{si}\PY{o}{:}\PY{o}{:}\PY{n}{cubic\PYZus{}metres}\PY{p}{,}
	  \PY{o}{*}\PY{n}{rr} \PY{o}{*} \PY{n}{si}\PY{o}{:}\PY{o}{:}\PY{n}{kilograms} \PY{o}{/} \PY{n}{si}\PY{o}{:}\PY{o}{:}\PY{n}{kilograms}\PY{p}{,}
	  \PY{o}{*}\PY{n}{nr} \PY{o}{/} \PY{n}{si}\PY{o}{:}\PY{o}{:}\PY{n}{kilograms}
	\PY{p}{)}\PY{p}{;} 

      \PY{c+c1}{// outflow from the domain}
      \PY{p}{\PYZob{}}
        \PY{n}{flux\PYZus{}nr} \PY{n}{flux\PYZus{}nr\PYZus{}out} \PY{o}{=} \PY{n}{tmp\PYZus{}mom\PYZus{}n} \PY{o}{*} \PY{p}{(}\PY{o}{*}\PY{n}{nr} \PY{o}{/} \PY{n}{si}\PY{o}{:}\PY{o}{:}\PY{n}{kilograms}\PY{p}{)} \PY{o}{/} \PY{p}{(}\PY{n}{dz} \PY{o}{*} \PY{n}{si}\PY{o}{:}\PY{o}{:}\PY{n}{metres}\PY{p}{)}\PY{p}{;}
        \PY{n}{flux\PYZus{}nr\PYZus{}out} \PY{o}{=} \PY{o}{\PYZhy{}} \PY{n}{std}\PY{o}{:}\PY{o}{:}\PY{n}{min}\PY{p}{(}\PY{k+kt}{real\PYZus{}t}\PY{p}{(}\PY{o}{\PYZhy{}}\PY{n}{flux\PYZus{}nr\PYZus{}out} \PY{o}{/} \PY{n}{nflux\PYZus{}unit}\PY{p}{)}\PY{p}{,} \PY{o}{*}\PY{n}{rhod} \PY{o}{*} \PY{p}{(}\PY{o}{*}\PY{n}{nr} \PY{o}{+} \PY{n}{dt} \PY{o}{*} \PY{o}{*}\PY{n}{dot\PYZus{}nr}\PY{p}{)} \PY{o}{/} \PY{n}{dt}\PY{p}{)} \PY{o}{*} \PY{n}{nflux\PYZus{}unit}\PY{p}{;}
        \PY{o}{*}\PY{n}{dot\PYZus{}nr} \PY{o}{\PYZhy{}}\PY{o}{=} \PY{p}{(}\PY{n}{flux\PYZus{}nr\PYZus{}in} \PY{o}{\PYZhy{}} \PY{n}{flux\PYZus{}nr\PYZus{}out}\PY{p}{)} \PY{o}{/} \PY{p}{(}\PY{o}{*}\PY{n}{rhod} \PY{o}{*} \PY{n}{si}\PY{o}{:}\PY{o}{:}\PY{n}{kilograms} \PY{o}{/} \PY{n}{si}\PY{o}{:}\PY{o}{:}\PY{n}{cubic\PYZus{}metres}\PY{p}{)} \PY{o}{/} \PY{n}{dot\PYZus{}nr\PYZus{}unit}\PY{p}{;}
      \PY{p}{\PYZcb{}}

      \PY{p}{\PYZob{}}
        \PY{n}{flux\PYZus{}rr} \PY{n}{flux\PYZus{}rr\PYZus{}out} \PY{o}{=} \PY{n}{tmp\PYZus{}mom\PYZus{}m} \PY{o}{*} \PY{p}{(}\PY{o}{*}\PY{n}{rr} \PY{o}{*} \PY{n}{si}\PY{o}{:}\PY{o}{:}\PY{n}{kilograms} \PY{o}{/} \PY{n}{si}\PY{o}{:}\PY{o}{:}\PY{n}{kilograms}\PY{p}{)} \PY{o}{/} \PY{p}{(}\PY{n}{dz} \PY{o}{*} \PY{n}{si}\PY{o}{:}\PY{o}{:}\PY{n}{metres}\PY{p}{)}\PY{p}{;}
        \PY{n}{flux\PYZus{}rr\PYZus{}out} \PY{o}{=} \PY{o}{\PYZhy{}} \PY{n}{std}\PY{o}{:}\PY{o}{:}\PY{n}{min}\PY{p}{(}\PY{k+kt}{real\PYZus{}t}\PY{p}{(}\PY{o}{\PYZhy{}}\PY{n}{flux\PYZus{}rr\PYZus{}out} \PY{o}{/} \PY{n}{rflux\PYZus{}unit}\PY{p}{)}\PY{p}{,} \PY{o}{*}\PY{n}{rhod} \PY{o}{*} \PY{p}{(}\PY{o}{*}\PY{n}{rr} \PY{o}{+} \PY{n}{dt} \PY{o}{*} \PY{o}{*}\PY{n}{dot\PYZus{}rr}\PY{p}{)} \PY{o}{/} \PY{n}{dt}\PY{p}{)} \PY{o}{*} \PY{n}{rflux\PYZus{}unit}\PY{p}{;}
        \PY{o}{*}\PY{n}{dot\PYZus{}rr} \PY{o}{\PYZhy{}}\PY{o}{=} \PY{p}{(}\PY{n}{flux\PYZus{}rr\PYZus{}in} \PY{o}{\PYZhy{}} \PY{n}{flux\PYZus{}rr\PYZus{}out}\PY{p}{)} \PY{o}{/} \PY{p}{(}\PY{o}{*}\PY{n}{rhod} \PY{o}{*} \PY{n}{si}\PY{o}{:}\PY{o}{:}\PY{n}{kilograms} \PY{o}{/} \PY{n}{si}\PY{o}{:}\PY{o}{:}\PY{n}{cubic\PYZus{}metres}\PY{p}{)} \PY{o}{/} \PY{n}{dot\PYZus{}rr\PYZus{}unit}\PY{p}{;}

        \PY{k}{return} \PY{n}{flux\PYZus{}rr\PYZus{}out} \PY{o}{/} \PY{p}{(}\PY{n}{si}\PY{o}{:}\PY{o}{:}\PY{n}{kilograms} \PY{o}{/} \PY{n}{si}\PY{o}{:}\PY{o}{:}\PY{n}{cubic\PYZus{}metres} \PY{o}{/} \PY{n}{si}\PY{o}{:}\PY{o}{:}\PY{n}{seconds}\PY{p}{)}\PY{p}{;}
      \PY{p}{\PYZcb{}}
    \PY{p}{\PYZcb{}}    
  \PY{p}{\PYZcb{}}\PY{p}{;}
\PY{p}{\PYZcb{}}\PY{p}{;}
\end{Verbatim}
  \vspace{-1.4em}%

  \caption{\label{lst:blk_2m_clw}
    \prog{blk\_1m::rhs\_columnwise()} signature
  }
\end{Listing}

\begin{figure*}[bh!]
  \center
  \input{dia/blk_2m}
  \caption{\label{fig:uml_blk_2m}
    Sequence diagram of {\it libcloudph++} API calls for the double-moment bulk scheme
      and a prototype transport equation solver.
    Diagram discussed in section~\ref{sec:blk_2m_callseq}.
    See also caption of Figure~\ref{fig:uml_proto} for description or diagram elements.
  }
\end{figure*}
\begin{figure*}[th!]
  \center\noindent
  \includegraphics[width=.425\textwidth]{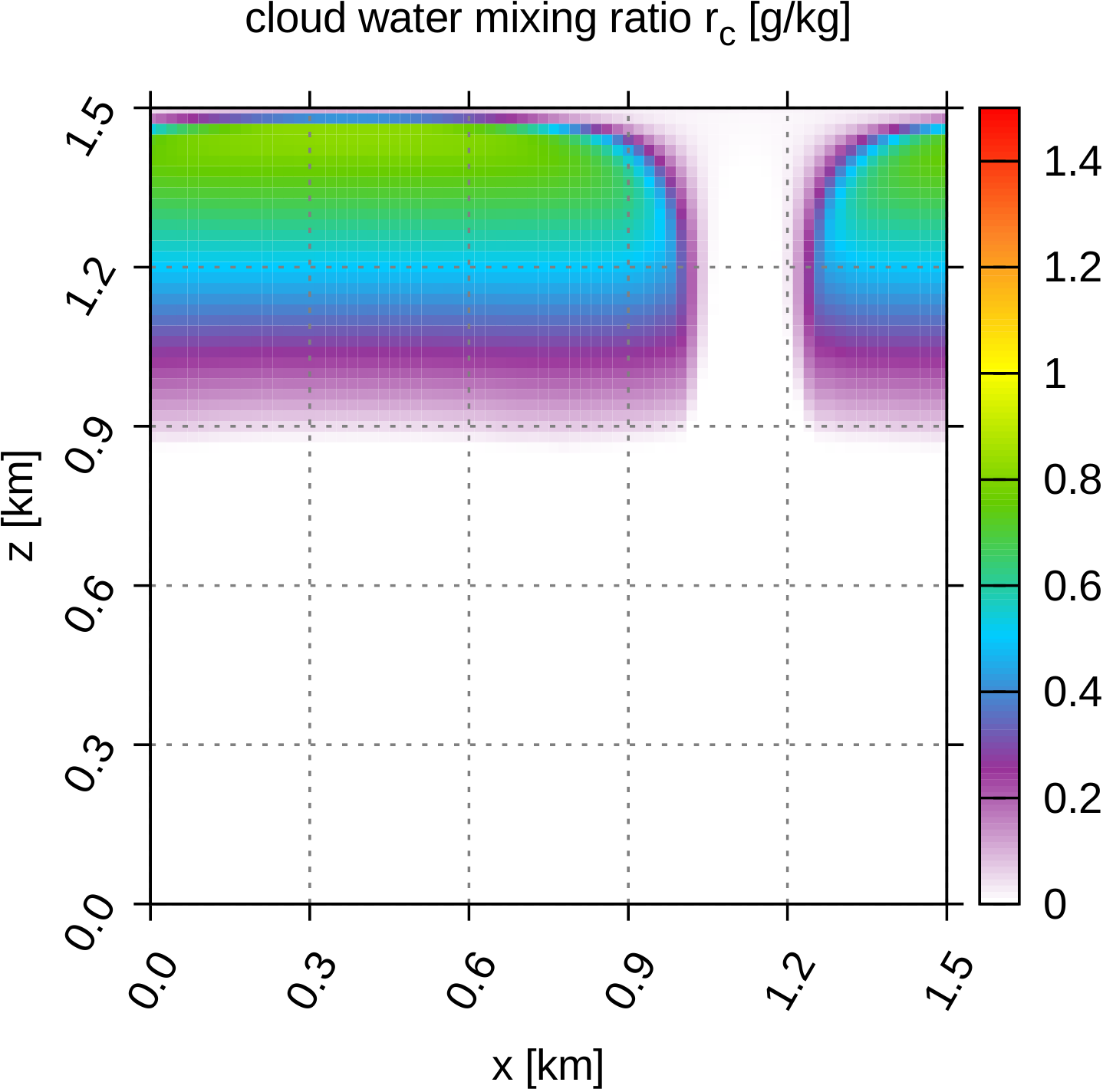}
  ~~~~
  \includegraphics[width=.43\textwidth]{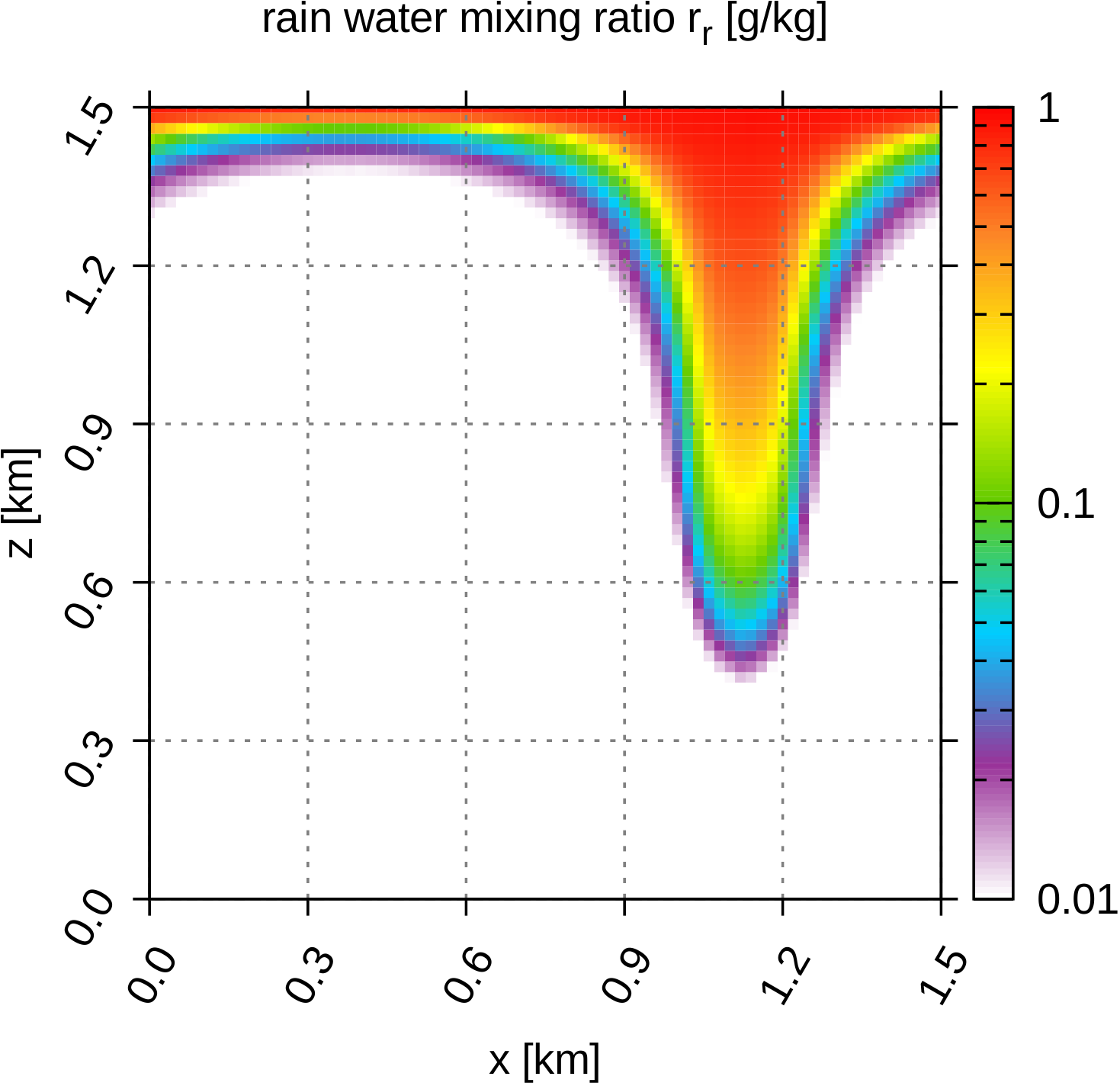}\\
  \vspace{.3em}
  \includegraphics[width=.43\textwidth]{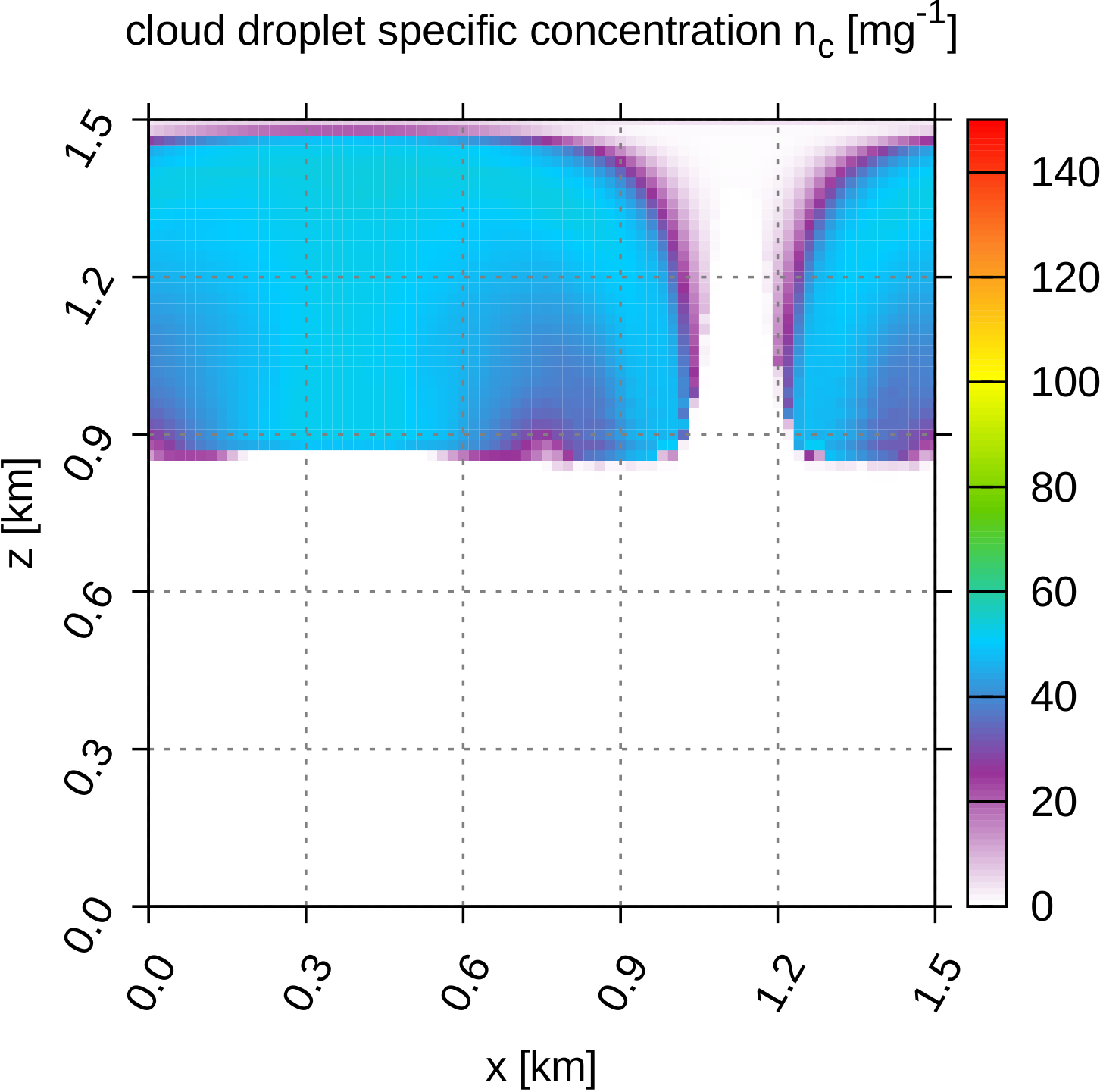}
  ~~~~
  \includegraphics[width=.43\textwidth]{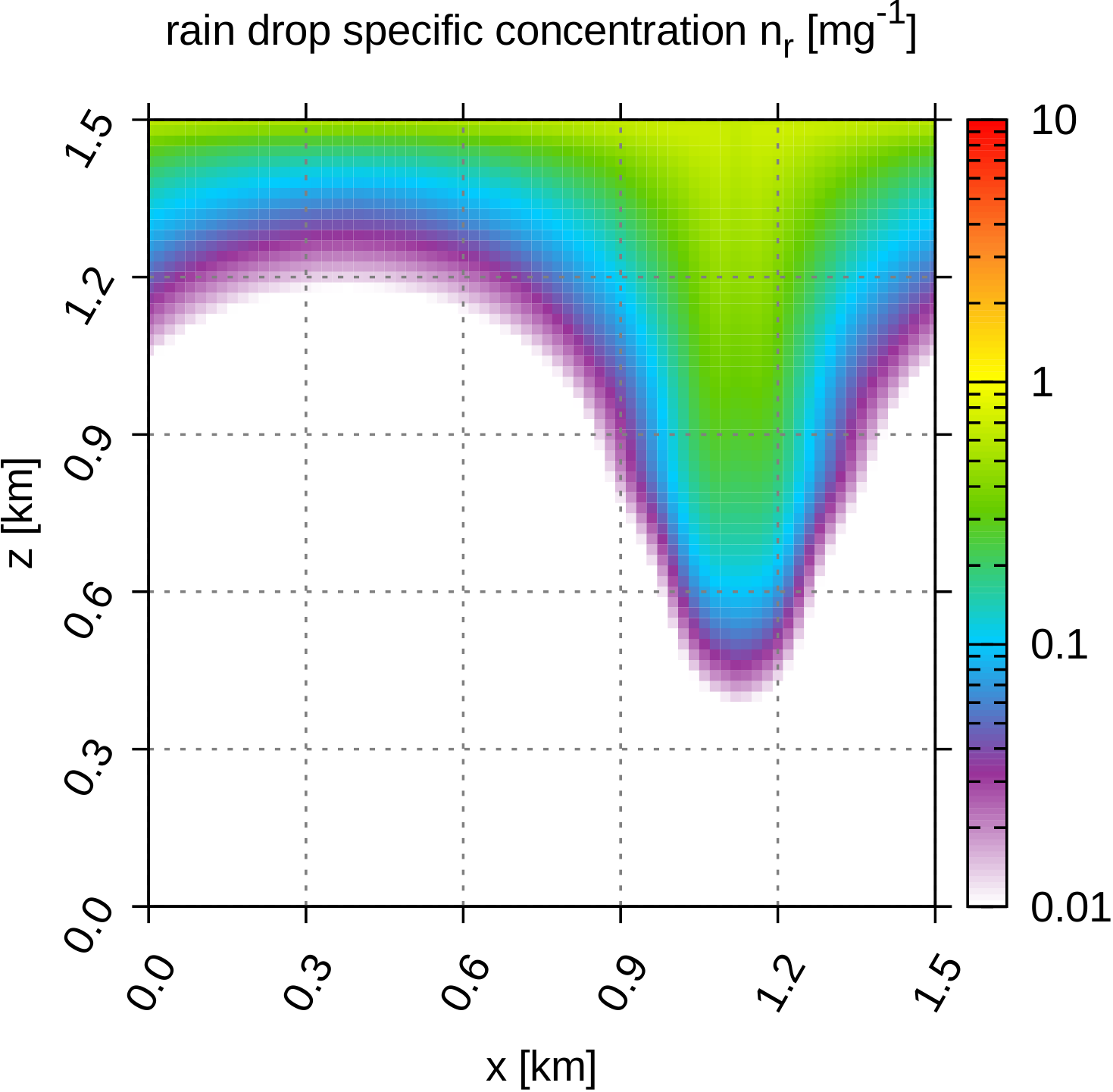}
  \caption{\label{fig:blk_2m_2d}
    Example results from a 2D kinematic simulation using the double-moment bulk scheme.
    All panels depict model state after 30 minutes simulation time (excluding the spin-up period).
    Note logarithmic colour scale for rain water plots.
    See section~\ref{sec:blk_2m_exres} for discussion.
  }
\end{figure*}
\subsection{Programming interface}
\subsubsection{API elements}

The double-moment bulk scheme's API consists of one structure and two functions,
  all defined within the \prog{libcloudphxx::blk\_2m} namespace.
The structure \prog{blk\_2m::opts\_t} holds scheme's options, 
  its definition is provided in Listing~\ref{lst:blk_2m_opt}.
Besides process-toggling Boolean fields, it stores the parameters of 
  the aerosol used in parameterising activation, that is the
  parameters of the lognormal size distribution (see eq.~\ref{eq:lognormal})
  and the parameter $\beta$ defining the solubility of aerosol
  \citep[see][sec.~2.1]{Khvorostyanov_and_Curry_2006}.

All processes are represented as right-hand-side terms in
  the double-moment scheme.
Contributions to the rhs terms due to phase changes and coalescence are
  obtained through a call to \prog{blk\_2m::rhs\_cellwise()} 
  (see Listing~\ref{lst:blk_2m_elw}).
As in the single-moment bulk scheme's API, contribution from sedimentation
  to the rhs terms can be computed by calling
  \prog{blk\_2m::rhs\_columnwise()} (Listing~\ref{lst:blk_2m_clw}).

The meaning of template parameters and function arguments is analogous to
  the single-moment bulk scheme's API (see section~\ref{sec:blk_1m_API}).
The computed values of rhs terms are added (and not assigned) to the 
  arrays passed as arguments.

The cellwise-formulated processes are handled in the following order:
  activation, condensation/evaporation of cloud droplets, autoconversion, accretion, 
  condensation/evaporation of rain.
In principle, there is no guarantee that the summed contributions from all those processes,
  multiplied by the timestep, are smaller than the available water contents
  or droplet concentrations. 
Application of such rhs terms could result in negative values of water contents or concentrations. 
To prevent that, each contribution to the rhs term evaluated within \prog{rhs\_cellwise()} is 
  added to the array $\dot{r}_i$ passed as argument using the following rule: 
\begin{eqnarray}
  \dot{r}_i^{\rm new} \,\, = {\rm min}\!\!\left(
    \!\dot{r}_i^\star, \,
    \frac{r_i + \Delta t \cdot \dot{r}_i^{\rm old}}{\Delta t}
  \right)
\end{eqnarray}
where $\dot{r}_i^{\rm old}$ is the value obtained in evaluation of 
  previously-handled processes, 
   $\dot{r}_i^{\star}$ is the value according to model formul\ae,
  and $\dot{r}_i^{\rm new}$ is the augmented value of rhs term
  that guarantees non-negative values of $r_i$ after its application.
The same rule is applied when evaluating values of outgoing fluxes $F_{out}$ from 
  equation \ref{eq:flux_first} when calculating rhs term within \prog{rhs\_columnwise()}.
The \prog{rhs\_columnwise()} returns the value of the $F_{out}$ flux from the lowermost
  grid cell within a column.

\subsubsection{Example calling sequence}\label{sec:blk_2m_callseq}

A diagram with an example calling sequence for the double-moment
  scheme is presented in figure~\ref{fig:uml_blk_2m}.
The only difference from the single-moment bulk scheme's calling sequence
  presented in section~\ref{sec:blk_1m_callseq} is the lack of an adjustments step.
Condensation is represented using right-hand-side terms and is computed
  together with coalescence in the \prog{blk\_2m::rhs\_cellwise()}.

\subsection{Implementation overview}

The implementation of the double-moment scheme follows closely the implementation 
  of the single-moment scheme (see section~\ref{sec:blk_1m_impl}).
It's a header-only C++ library, using Boost.units classes for dimensional analysis 
  and Boost.Iterator for iterating over sets of array slices.

\subsection{Example results}\label{sec:blk_2m_exres}

Simulations presented in Section~\ref{sec:blk_1m_exres} were repeated with
  the double-moment scheme.
Figure~\ref{fig:blk_2m_2d} presents a snapshots of the cloud and rain water content 
  as well as the cloud and rain drop concentration fields after 30~minutes simulated time
  (excluding the spin-up period).
Because of large differences in the predicted rain, 
  rain water content and drop concentration are plotted using logarithmic colour scale
  in order to keep the same colour range for all three presented schemes.

Similarly to the results from the single-moment scheme presented in Figure~\ref{fig:blk_1m_2d}, 
  cloud water content increases from the cloud base almost up to the upper boundary of the domain.
However, unlike in the case of the single-moment scheme, the cloud deck in Figure~\ref{fig:blk_2m_2d}
  features a ''cloud hole'' above the downdraught region.
The rain forms in the upper part of the updraught and is advected into the downdraught
  region in the right-hand side of the domain.
The double-moment simulation at the thirtieth minute is still to reach the
  quasi-stationary state.
This is because of the differences in the parameterisation of autoconversion that lead
  to different timing of the precipitation onset.

The cloud droplet concentration plot reveals that the model captures the impact of
  the cloud base vertical velocity (and hence the supersaturation) 
  on the concentration of activated cloud droplets.
The highest concentrations are found near the updraught axis, and the lowest near 
  the updraught edges.
The difference in shapes of the rain drop concentration $n_r$ and the rain water
  mixing ratio $r_r$ fields 
  arguably comes from the different fall velocities of $n_r$ and $r_r$.


\FloatBarrier

\section{Particle-based scheme}\label{sec:sdm}

The third scheme available in {\it libcloudph++} 
  differs substantially from the other two bulk schemes.
It does not treat water condensate as continuous medium.
Instead, the scheme employs Lagrangian tracking of particles that represent
  atmospheric aerosol, cloud and drizzle droplets, and rain drops.
However, volumes relevant to atmospheric flows contain far too many
  particles to be individually represented in a numerical model.
Consequently, each ''computational particle'' tracked in the scheme
  represents multiple particles of identical properties 
  (i.e.~spatial coordinates and physicochemical properties).
Such approach was recently applied for modelling precipitating clouds by 
  \citet{Andrejczuk_et_al_2010, Soelch_and_Kaercher_2010, Riechelmann_et_al_2012, Arabas_and_Shima_2013}.
Formulation of the scheme presented here follows the Super Droplet Method of
  \citet{Shima_et_al_2009} to represent collisions and coalescence of particles.

\subsection{Formulation}

The key assumption of the particle-based scheme
  is to assume no distinction between aerosol, 
  cloud, drizzle and rain particles.
All particles tracked by the Lagrangian component of the solver
  are subject to the same set of processes including advection by the flow, gravitational
  sedimentation, diffusional growth, evaporation, and collisional growth.
The Eulerian component of the model is responsible for advecting
  $\theta$ and $r_v$ (see~Appendix~\ref{sec:common}).
Representation of all the processes, as well the method of coupling the Lagrangian and
  Eulerian components of the model is given below.

The Lagrangian component is responsible for tracking the
  computational particles, each having the following attributes:
\begin{itemize}
  \item{multiplicity $N$}
  \item{location (i.e.~spatial coordinates with 0,1,2 or 3 components)}
  \item{wet radius squared $r_w^2$}
  \item{dry radius cubed $r_d^3$}
  \item{hygroscopicity parameter $\kappa$}
\end{itemize}
Multiplicity depicts the number of particles represented by the computational particle.
All particles represented by one computational particle are assumed to be spherical 
  water solution droplets of radius $r_w$.
Following \citet{Shima_et_al_2009} the model is formulated in $r_w^2$ for 
  numerical reasons. 

The amount of solvent is represented with the dry radius $r_d$ 
  (third power is used in the model code as most often $r_d$ 
  serves as a proxy for volume of the solvent).
The hygroscopicity of the solvent is parameterised using the
  single-parameter approach of \citet{Petters_et_al_2007}.

The list of particle attributes can be extended.
For example, parameters describing chemical composition
  of the solution or the electrical charge of a particle can be added.
Adding new particle attributes does not increase the computational expense of 
  the Eulerian component of the solver.
However, extension of the phase space by a new dimension (the added attribute)
  potentially requires using more computational particles to
  achieve sufficient coverage of the phase space. 

\subsubsection{Key assumptions}

Most of the assumptions of the bulk models described in sections \ref{sec:bulk} and \ref{sec:mm}
  are lifted.
All particles are subject to the same set of processes.
It follows that the model represents 
  even dry deposition and collisions between aerosol particles (both being effectively negligible).
The supersaturation in the model domain is resolved taking into account phase change 
  kinetics (i.e.~condensation and evaporation are not treated as instantaneous).
Aerosol may have any initial size distribution and may be internally or externally mixed
  (i.e.~have the same or different solubility for particles of different sizes).
There are no assumptions on the shape of the particle size spectrum.

There are, however, two inherent assumptions in the premise of all particles being spherical and composed
  of water solution.
First, the humidity within the domain and the hygroscopicity of the substance of which aerosol 
  is composed must both be high enough for the solution to be dilute.
For tropospheric conditions and typical complex-composition internally-mixed 
  aerosol this assumption is generally sound 
  \citep{Fernandez-Diaz_et_al_1999, Marcolli_et_al_2004}.
Second, the nonsphericity of large precipitation particles has to be negligible.
It is a valid assumption for drops smaller than 1~mm \citep{Szakall_et_al_2010}.

It is assumed in the present formulation that a particle never breaks up into 
  multiple particles.
It is a reasonable assumption for the evaporation of
  cloud particles into aerosol \citep{Mitra_et_al_1992}.
However, both collision-induced and spontaneous breakup become significant
  (the latter to a much smaller extent) 
  for larger droplets \citep{McFarquhar_2010} 
  and hence the scheme requires an extension in order to allow for
  diagnosing rain spectra for strongly precipitating clouds.

There is not yet any mechanism built into the model to represent aerosol
  sources (other than regeneration of aerosol by evaporation of cloud droplets).

\subsubsection{Advection}
Transport of particles by the flow is computed using the backward Euler scheme:
\begin{eqnarray}\label{eq:adv_implicit}
  x^{[n+1]} = x^{[n]} + \Delta x \cdot C(x^{[n+1]})
\end{eqnarray}
where $C$ is the Courant number field of the 
  Eulerian component of the solver, and $\Delta x$ is the
  grid step (formul\ae~ are given for the x dimension, but
  are applicable to other dimensions as well).
An Arakawa-C staggered grid is used and evaluation of $C(x^{[n+1]})$ is performed 
  using linear approximation (interpolation / extrapolation of the particle velocities
  using fluid velocity values at the grid cell edges):
\begin{eqnarray}\label{eq:adv_c}
  C(x^{[n+1]}) = (1-\omega) \cdot C_{[i-\nicefrac{1}{2}]} + \omega \cdot C_{[i+\nicefrac{1}{2}]} 
\end{eqnarray}
where fractional indices denote Courant numbers on the edges of a grid cell in which
  a given particle is located at time level~$n$.
The Courant number components are defined as velocity components times the ratio of time step and grid step
  in each dimension.
The weight $\omega$ is defined as:
\begin{eqnarray}\label{eq:adv_w}
  \omega = x^{[n+1]} / \Delta x - \lfloor x^{[n]}/\Delta x \rfloor 
\end{eqnarray}
  where $\lfloor x \rfloor$ depicts the largest integer not greater than x.
Substituting (\ref{eq:adv_c}) and (\ref{eq:adv_w}) into (\ref{eq:adv_implicit})
  results in an analytic solution for $x^{[n+1]}$:
\begin{eqnarray}
  x^{[n+1]} = \frac{
    x^{[n]} + \Delta x \left(
      C_{i-\nicefrac{1}{2}} - \lfloor x^{[n]}/\Delta x \rfloor \cdot \Delta C
    \right)
  }{
    1 - \Delta C
  }
\end{eqnarray}
where $\Delta C = C_{i+\nicefrac{1}{2}} - C_{i-\nicefrac{1}{2}}$.

The same procedure is repeated in other spatial dimensions if applicable (i.e.~depending on 
  the dimensionality of the Eulerian component). 
Periodic horizontal boundary conditions are assumed.

\subsubsection{Phase changes}\label{sec:lgrngn_cond}

Representation of condensation and evaporation in the particle-based approach 
  encompasses several phenomena that are often treated individually, namely:
  aerosol humidification, cloud condensation nuclei (CCN) activation and deactivation, 
  cloud droplet growth and evaporation, and finally rain evaporation.
The timescale of some of these processes (notably CCN activation) is much shorter 
  than the characteristic timescale
  of the large-scale air flow solved by the Eulerian component of the solver.
Therefore, representation of condensation and evaporation in the Lagrangian component 
  involves a substepping logic in which the Eulerian component timestep $\Delta t$
  is divided into a number of equal substeps.
For simplicity, this procedure is not depicted explicitly in the following formul\ae.
It is only hinted by labelling subtimestep as $\Delta t'$ and the subtimestep number as $n'$.

Presently, the number of subtimesteps is kept constant throughout the domain and
  throughout the simulation time.
However, the actual constraints for timestep length $\Delta t'$ differ substantially,
  particularly with the distance from cloud base \citep[see figure~2 in][]{Arabas_and_Pawlowska_2011}.
An adaptive timestep choice mechanism is planned for a future release.

Within each substep, the drop growth equation is solved for each computational particle
  with an implicit scheme with respect to
  wet radius but explicit with respect to $r_v$ and $\theta$:
\begin{eqnarray}\label{eq:rw2_implicit}
  r_w^{2^{[n'+1]}} = r_w^{2^{[n']}} + \Delta t \cdot \left.\frac{dr^2_w}{dt}\right|_{r_w^{2^{[n'+1]}}, r_v^{[n']}, \theta^{[n']}}
\end{eqnarray}
Solution to the above equation is sought by employing 
  a predictor-corrector type procedure.
First, the value of the $\nicefrac{dr^2_w}{dt}$ derivative evaluated at $r_w^{2^{[n']}}$
  is used to construct an initial-guess range $a < r^{2^{[n'+1]}}_w < b$ in which roots of 
  equation \ref{eq:rw2_implicit} are to be sought, with:
\begin{eqnarray}
  a = {\rm max}\!\left(
    \!r^2_d,\,\,\,\, 
    r_w^{2^{[n']}}\!+ {\rm min}\!\left(\!2 \cdot \left.\frac{dr^2_w}{dt}\right|_{r_w^{[n']}},\,\,\,\, 0\right)\!\!
  \right) \\
  b =  
    r_w^{2^{[n']}}\!+ {\rm max}\!\left(\!2 \cdot \left.\frac{dr^2_w}{dt}\right|_{r_w^{[n']}},\,\,\,\, 0\right)\!\!
    \,\,\,\,\,
\end{eqnarray}
Second, $r_w^{2^{n'+1}}$ is iteratively searched using the bisection algorithm.
If the initial-guess range choice makes bisection search ill-posed 
  (minimisation function having the same sign at $a$ and $b$),
  the algorithm stops after first iteration returning $(a+b)/2$,
  and reducing the whole procedure to the standard Euler scheme  
  (due to the use of factor $2$ in the definition of~$a$ and~$b$).
It is worth noting, that such treatment of drop growth (i.e.~Lagrangian in radius space,
  the so-called moving sectional or method of lines approach) incurs
  no numerical diffusion.

The growth rate of particles is calculated using the single-equation (so-called Maxwell-Mason)
  approximation to the heat and vapour diffusion process \citep[][rearranged eq.~5.106]{Straka_2009}: 
\begin{eqnarray}\label{eq:maxmason}
  r_w \frac{dr_w}{dt} = \frac{D_{\text{eff}}}{\rho_w} \left(
    \rho_v - \rho_\circ 
  \right)
\end{eqnarray}
where the effective diffusion coefficient is:
\begin{eqnarray}
  D^{-1}_{\text{eff}} = 
    D^{-1}
    +
    K^{-1}
    \frac{\rho_{vs} l_v}{T}                      
    \left( \frac{l_v}{R_v  T} - 1 \right) 
\end{eqnarray}
$\rho_{vs}$ (density of water vapour at saturation with respect to
  plane surface of pure water), $T$ and $\rho_v$ are updated 
  every subtimestep.
The vapour density at drop surface $\rho_\circ$ is modelled as: 
\begin{eqnarray}\label{eq:rho_circ}
  \rho_\circ= \rho_{vs} \,\cdot\, a_w(r_w, r_d) \,\cdot\, {\rm exp}(A/r_w)
\end{eqnarray}
  where water activity $a_w$ and the so-called Kelvin term ${\rm exp}(A/r_w)$
  are evaluated using the $\kappa$-K\"ohler parameterisation 
  of \citet{Petters_et_al_2007}.
See \citet{Arabas_and_Pawlowska_2011} for the formul\ae~for $A$, $l_v$ and $\rho_{vs}$ used.

Vapour and heat diffusion coefficients $D$ and $K$ are evaluated as:
\begin{eqnarray}
  D = D_0 \cdot \beta_M \cdot \frac{{\rm Sh}}{2}\\
  K = K_0 \cdot \beta_T \cdot \frac{{\rm Nu}}{2}
\end{eqnarray}
The Fuchs-Sutugin transition-r\'egime correction factors $\beta_M(r_w, T)$ and $\beta_T(r_w, T, p)$ 
  are used in the form recommended for cloud modelling by \citet[][i.e.~employing mass and heat
  accommodation coefficients of unity]{Laaksonen_et_al_2005}.
The Sherwood number Sh and the Nusselt number Nu (twice the mean ventilation coefficients) are modelled following 
  \citet{Clift_et_al_1978} 
  as advocated by \citet{Smolik_et_al_2001}.

As in the particle-based ice-microphysics model of \citet{Soelch_and_Kaercher_2010}, 
  no interpolation of the Eulerian state variables to particle positions is done
  \citep[in contrast to the approach employed in warm-rain models of][]{Andrejczuk_et_al_2008,Shima_et_al_2009,Riechelmann_et_al_2012}.
It is therefore implicitly assumed, in compliance with the Eulerian solver component logic, 
  that the heat and moisture are homogeneous within a grid cell.
Consequently, the effects of subgrid-scale mixing on the particles follow the so-called
  homogeneous-mixing scenario \citep[see][and references therein]{Jarecka_et_al_2013}.
Furthermore, no effects of vapour field inhomogeneity around particles are taken into
  consideration \citep[see][]{Vaillancourt_et_al_2001, Castellano_and_Avila_2011}. 

Particle terminal velocities used to estimate the Reynolds
  number to evaluate $Sh$ and $Nu$ are calculated
  using the parameterisation of \citet[][see also \ref{sec:lgrngn_sedi} herein]{Khvorostyanov_and_Curry_2002}.

After each substep, the thermodynamic fields $r_v$ and $\theta$ are adjusted to account for water vapour
  content change due to condensation or evaporation on particles within a given
  grid cell and within a given substep by evaluating:
\begin{eqnarray}\label{eq:rhov_update}
  r_v^{[n'+1]} - r_v^{[n']} \nonumber
    \!&\!\!=\!\!& \rho_d^{-1}
    \frac{- 4 \pi \rho_w}{3 \Delta V} \\
    &&
    \!\!\!\!
    \sum\limits_{i\,\in\,\text{grid cell}} 
    \!\!\!\!\!\!
    N_{[i]} \! \left[ r_{w_{[i]}}^{3^{[n'+1]}} - r_{w_{[i]}}^{3^{[n']}} \right]  \\
  \theta^{[n'+1]} - \theta^{[n']} 
    \!&\!\!=\!\!&\! 
    \left( r_v^{[n'+1]} - r_v^{[n']} \right) \! \left.\frac{d \theta}{d r_v}\right|_{r_v^{[n']},\, \theta^{[n']}}\,\,\,\,
\end{eqnarray}
where $\Delta V$ is the grid cell volume, and $\rho_w$ is the density of liquid water.
Noteworthy, such formulation maintains conservation of heat and moisture in the domain
  regardless of the accuracy of integration of the drop growth equation.

Phase change calculations are performed before any other 
  processes, as it is the only process influencing values of $r_v$ and $\theta$
  fields of the Eulerian component.
Consequently, Eulerian component of the solver may continue integration as soon as
  phase-change calculations are completed.
Such asynchronous logic is enabled if performing calculations using a GPU: particle
  advection, sedimentation and collisions are calculated by the Lagrangian component
  of the solver using GPU while the Eulerian component is advecting model state variables
  using CPU(s).

\subsubsection{Coalescence}

The coalescence scheme is an implementation of the Super Droplet Method (SDM)
  described in \citet{Shima_et_al_2009}.
SDM is a Monte-Carlo type algorithm for representing particle collisions.
As it is done for phase changes, coalescence of particles is solved 
  using subtimesteps $\Delta t''$.
In each subtimestep all computational particles within a given grid cell are randomly grouped
  into non-overlapping pairs (i.e.~no~computational particle may belong to more than one pair).
Then the probability of collisions between computational particles $i$ and $j$ in each pair is evaluated as:
\begin{eqnarray}\label{eq:P_ij}
  P_{ij}=max(N_i, N_j) 
   K(r_i, r_j)
  \frac{\Delta t''}{\Delta V} \frac{n (n - 1)}{2 \lfloor n/2 \rfloor}
\end{eqnarray}
where $n$ is the total number of computational particles within the grid cell in a given timestep
  and $K(r_i, r_j)$ is the collection kernel.
In analogy to a target-projectile configuration, scaling the probability of collisions with 
  the larger of the two multiplicities
  ${\rm max}(N_i, N_j)$ (target size) implies that if a collision happens, ${\rm min}(N_i, N_j)$ 
  of particles will collide (number of projectiles).
The last term in equation (\ref{eq:P_ij}) upscales the probability to account for the
  fact that not all ($n(n-1)/2$) possible pairs of computational particles are examined
  but only $\lfloor n/2 \rfloor$ of them.
Evaluation of collision probability for non-overlapping pairs only, instead of 
  for all possible pairs of particles, makes the computational cost of the algorithm
  scale linearly, instead of quadratically, with the number of computational particles 
  (at the cost of increasing the sampling error of the Monte-Carlo scheme). 

The coalescence kernel has the following form if only geometric collisions are
  considered:
\begin{eqnarray}
  K(r_i, r_j) = E(r_i, r_j) \cdot \pi\, (r_i + r_j)^2 \cdot |v_i - v_j|
\end{eqnarray}
where $E(r_i, r_j)$ is the collection efficiency and $v$ is the
  terminal velocity of particles (i.e.~their flow-relative sedimentation velocity).
The collection efficiency differs from unity if hydrodynamic effects \citep[e.g.][]{Vohl_et_al_2007} 
  or van~der~Waals forces \citep{Rogers_and_Davis_1990} are considered.
The whole coalescence kernel may take different form (in particular may be
  nonzero for drops of equal terminal velocity) if turbulence effects are
  taken into account \citep[][and references therein]{Grabowski_and_Wang_2013}.

In each subtimestep the evaluated probability $P_{ij}$ is compared to a random number
  from a uniform distribution over the (0,1) interval.
If the probability is larger than the random number, a collision event is triggered.
During a collision event, all ${\rm min}(N_i, N_j)$ particles collide
  \citep[see Figure~1 and Section~4.1.4 in]{Shima_et_al_2009}.
One of the colliding computational particles (the one with larger multiplicity) 
  retains its multiplicity but
  changes its dry and wet radii to those of the newly formed particles.
The second colliding computational particle (the one with smaller multiplicity)
  retains its dry and wet radii
  but changes its multiplicity to the difference between $N_i$ and $N_j$.

Unlike in the formulation of \citet{Shima_et_al_2009} particles with equal 
  multiplicities collide using the same scheme, leaving one of the particles
  with zero multiplicity.
Particles with zero multiplicity are ``recycled'' at the begining of each timestep.
The recycling procedure first looks for computational particles with highest multiplicites
  and then assignes their properties to the recycled particles halving the multiplicity.

Noteworthy, the collisional growth is represented in a numerical-diffusion-free manner, that is,
  Lagrangian in particle radius space (both dry and wet radius).
This is an advantage over the Eulerian-type schemes based on the Smoluchowski equation 
  which exhibit numerical diffusion \citep[see e.g.][]{Bott_1997}.
Other particle parameters are either summed (i.e.~extensive parameters such as $r_d^3$) 
  or averaged (i.e.~intensive parameters such as $\kappa$). 

Presently, the ''multiple coalescence'' feature of SDM introduced by \citet[][]{Shima_et_al_2009} 
  to robustly cope with an undersampled condition of $P_{ij}>1$ is not implemented.
It is planned, however, to use the values of $P_{ij}$ to control an adaptive timestep logic
  to be introduced in a future release.

\subsubsection{Sedimentation}\label{sec:lgrngn_sedi}

Particle sedimentation velocity is computed using the 
  formula of \citet[][eqs.~2.7, 2.12, 2.13, 3.1]{Khvorostyanov_and_Curry_2002}.
The explicit Euler scheme is used for adjusting particle positions
  (the terminal velocity is effectively constant, taking into account the 
  assumed homogeneity of heat and moisture within a grid cell).

Sedimentation may result in the particles leaving the domain 
  (i.e.~dry deposition or ground-reaching rainfall).
Computational particles that left the domain 
  are flagged with zero multiplicity and hence undergo the same recycling procedure
  as described above for equal-multiplicity collisions.

\subsubsection{Initialisation}\label{sec:lgrngn_init}

One of the key parameters of the particle-based simulation is the number of computational particles used.
As in several recent cloud-studies employing particle-based techniques, the initial particle 
  spatial coordinates are chosen randomly using a uniform distribution.
Consequently, the initial condition has a uniform initial mean density of computational
  particles per cell (assuming all cells have the same volume).
The value of this initial mean density defines the sampling error in particle parameter
  space, particularly in the context of phase changes and coalescence which are both 
  formulated on cellwise basis.
The ranges of values used in recent studies are:
  30--250 \citep[][particles injected throughout simulation]{Soelch_and_Kaercher_2010},
  100--200 \citep[][grid cell size variable in height, particles added throughout simulation]{Andrejczuk_et_al_2010},
  26--186 \citep[][]{Riechelmann_et_al_2012},
  8--512 \citep[][]{Arabas_and_Shima_2013},
  30-260 \citep{Unterstrasser_and_Soelch_2014}.

The dry radii of the computational particles are chosen randomly with a 
  uniform distribution in the logarithm of radius.
The minimal and maximal values of dry radius are chosen automatically
  by evaluating the initial dry-size distribution.
The criterion is that the particle multiplicity (i.e.~the number of particles represented by 
  a computational particle) for both the minimal and the maximal
  radii be greater or equal one.

The initial spectrum shape is arbitrary.
Externally mixed aerosol may be represented using multiple spectra, each 
  characterised by different value of $\kappa$.
The initial particle multiplicities are evaluated treating the input spectra
  as corresponding to the standard atmospheric conditions (STP) and hence the
  concentrations are multiplied
  by the ratio of the dry-air density in a given grid cell to the air density at STP.

Equation~\ref{eq:maxmason} defines the relationships between the dry and the wet spectra
  in the model.
These should, in principle, be fulfilled by the initial condition imposed
  on model state variables.
For cloud-free air, it is possible by assuming an equilibrium defined by putting
  zero on the left-hand side of equation~\ref{eq:maxmason}.
This allows to either diagnose the wet spectrum from the dry one or vice versa.

If the dry size distribution is given as initial condition, bringing all particles to equilibrium
  at a given humidity is done without changing $\theta$ and $r_v$
  to resemble bulk models' initial state.
A small amount of water needed to obtain equilibrium is thus effectively added to the system.

For setups assuming initial presence of cloud water within the model domain,
  the equilibrium condition may be applied only to subsaturated regions within the model
  domain.
The initial wet radius of particles within the supersaturated regions is set to
  its equilibrium value at RH=95\% \citep[following][]{Lebo_and_Seinfeld_2011}.
Subsequent growth is computed within the first few minutes of the simulation.
In order to avoid activation of all available
  aerosol, the drop growth equation~\ref{eq:maxmason} is evaluated limiting the
  value of the supersaturation to 1\% \citep[see also discussion 
  on particle-based simulation initialisation in][sec.~2.2]{Andrejczuk_et_al_2010}.

\subsection{Programming interface}

\subsubsection{API elements}

The particle-based scheme's API differs substantially from bulk schemes' APIs
  as it features object-oriented approach of equipping structures (referred to as classes)
  with functions (referred to as methods).
Furthermore, unlike the bulk schemes' APIs, the particle-based scheme
  is not implemented as a header-only library but requires linking with
  \prog{libcloudphxx\_lgrngn} shared library.
The particle-based scheme's API consists of four structures (classes with all members public), 
  one function and two enumerations, all defined in the \prog{libcloudphxx::lgrngn} namespace.
The often occurring template parameter \prog{real\_t} controls the floating point format.

As in the case of bulk schemes, the options controlling the scheme's 
  course of action in each solver step are stored in a separate 
  structure \prog{lgrngn::opts\_t} 
  whose definition is given in Listing~\ref{lst:lgrngn_opt}.
\begin{Listing}
  \renewcommand*\FancyVerbStartString{\PY{c+c1}{//\PYZlt{}listing\PYZgt{}}}%
  \renewcommand*\FancyVerbStopString{\PY{c+c1}{//\PYZlt{}/listing\PYZgt{}}}%
  \begin{Verbatim}[commandchars=\\\{\}]
\PY{c+cm}{/** @file}
\PY{c+cm}{  * @copyright University of Warsaw}
\PY{c+cm}{  * @brief Definition of a structure holding options for Lagrangian microphysics}
\PY{c+cm}{  * @section LICENSE}
\PY{c+cm}{  * GPLv3+ (see the COPYING file or http://www.gnu.org/licenses/)}
\PY{c+cm}{  */}

\PY{c+cp}{\PYZsh{}}\PY{c+cp}{pragma once}

\PY{c+cp}{\PYZsh{}}\PY{c+cp}{include \PYZlt{}libcloudph++}\PY{c+cp}{/}\PY{c+cp}{lgrngn}\PY{c+cp}{/}\PY{c+cp}{extincl.hpp\PYZgt{}}
\PY{c+cp}{\PYZsh{}}\PY{c+cp}{include \PYZlt{}libcloudph++}\PY{c+cp}{/}\PY{c+cp}{lgrngn}\PY{c+cp}{/}\PY{c+cp}{chem.hpp\PYZgt{}}

\PY{k}{namespace} \PY{n}{libcloudphxx}
\PY{p}{\PYZob{}}
  \PY{k}{namespace} \PY{n}{lgrngn}
  \PY{p}{\PYZob{}}
\PY{c+c1}{//\PYZlt{}listing\PYZgt{}}
    \PY{k}{template}\PY{o}{\PYZlt{}}\PY{k}{typename} \PY{k+kt}{real\PYZus{}t}\PY{o}{\PYZgt{}}
    \PY{k}{struct} \PY{k+kt}{opts\PYZus{}t} 
    \PY{p}{\PYZob{}}
      \PY{c+c1}{// process toggling}
      \PY{k+kt}{bool} \PY{n}{adve}\PY{p}{,} \PY{n}{sedi}\PY{p}{,} \PY{n}{cond}\PY{p}{,} \PY{n}{coal}\PY{p}{;}

      \PY{c+c1}{// RH limit for drop growth}
      \PY{k+kt}{real\PYZus{}t} \PY{n}{RH\PYZus{}max}\PY{p}{;}       

      \PY{c+c1}{// no. of substeps }
      \PY{k+kt}{int} \PY{n}{sstp\PYZus{}cond}\PY{p}{,} \PY{n}{sstp\PYZus{}coal}\PY{p}{;} 
\PY{c+c1}{//\PYZlt{}/listing\PYZgt{}}

      \PY{c+c1}{// chem stuff}
      \PY{k+kt}{bool} \PY{n}{chem}\PY{p}{;}

      \PY{k+kt}{int} \PY{n}{sstp\PYZus{}chem}\PY{p}{;} 

      \PY{n}{std}\PY{o}{:}\PY{o}{:}\PY{n}{vector}\PY{o}{\PYZlt{}}\PY{k+kt}{real\PYZus{}t}\PY{o}{\PYZgt{}} \PY{n}{chem\PYZus{}gas}\PY{p}{;}

      \PY{c+c1}{// ctor with defaults (C++03 compliant) ...}
      \PY{k+kt}{opts\PYZus{}t}\PY{p}{(}\PY{p}{)} \PY{o}{:} 
        \PY{n}{adve}\PY{p}{(}\PY{n+nb}{true}\PY{p}{)}\PY{p}{,} \PY{n}{sedi}\PY{p}{(}\PY{n+nb}{true}\PY{p}{)}\PY{p}{,} \PY{n}{cond}\PY{p}{(}\PY{n+nb}{true}\PY{p}{)}\PY{p}{,} \PY{n}{coal}\PY{p}{(}\PY{n+nb}{true}\PY{p}{)}\PY{p}{,} \PY{n}{chem}\PY{p}{(}\PY{n+nb}{false}\PY{p}{)}\PY{p}{,} 
        \PY{n}{sstp\PYZus{}cond}\PY{p}{(}\PY{l+m+mi}{10}\PY{p}{)}\PY{p}{,} \PY{n}{sstp\PYZus{}coal}\PY{p}{(}\PY{l+m+mi}{10}\PY{p}{)}\PY{p}{,} \PY{n}{sstp\PYZus{}chem}\PY{p}{(}\PY{l+m+mi}{10}\PY{p}{)}\PY{p}{,}
        \PY{n}{RH\PYZus{}max}\PY{p}{(}\PY{l+m+mi}{44}\PY{p}{)}\PY{p}{,} \PY{c+c1}{// :) (anything greater than 1.1 would be enough}
        \PY{n}{chem\PYZus{}gas}\PY{p}{(}\PY{n}{chem\PYZus{}gas\PYZus{}n}\PY{p}{)}
      \PY{p}{\PYZob{}}
        \PY{k}{for}\PY{p}{(}\PY{k+kt}{int} \PY{n}{i}\PY{o}{=}\PY{l+m+mi}{0}\PY{p}{;} \PY{n}{i}\PY{o}{\PYZlt{}}\PY{n}{chem\PYZus{}gas\PYZus{}n}\PY{p}{;} \PY{o}{+}\PY{o}{+}\PY{n}{i}\PY{p}{)}
        \PY{p}{\PYZob{}}
          \PY{n}{chem\PYZus{}gas}\PY{p}{[}\PY{n}{i}\PY{p}{]} \PY{o}{=} \PY{l+m+mi}{0}\PY{p}{;}
        \PY{p}{\PYZcb{}}
      \PY{p}{\PYZcb{}}
    \PY{p}{\PYZcb{}}\PY{p}{;}
  \PY{p}{\PYZcb{}}
\PY{p}{\PYZcb{}}\PY{p}{;}
\end{Verbatim}
  \vspace{-1.4em}%

  \caption{\label{lst:lgrngn_opt}
    \prog{lgrngn::opts\_t} definition
  }
\end{Listing}
The first Boolean fields provide control over process toggling. 
The following fields are the RH limit for evaluating drop growth equation (for the spin-up, see sec.~\ref{sec:setup}), 
  and the numbers of substeps to be taken within one Eulerian timestep
  when calculating condensation and coalescence.
The default values are set in the structure's
  constructor whose definition is not presented in the Listing
  (the C++11 syntax for default parameter values used in \prog{blk\_1m::opts\_t}
  and \prog{blk\_2m::opts\_t} is not used to maintain compatibility with C++03
  required to compile the code for use on a GPU, see discussion in sec.~\ref{sec:lgrngn_impl}).

Several other options of the particle-based scheme not meant to be altered during
  simulation are grouped into a structure named \prog{lgrngn::opts\_init\_t} 
  (Listing \ref{lst:lgrngn_opt_init}).
\begin{Listing}
  \renewcommand*\FancyVerbStartString{\PY{c+c1}{//\PYZlt{}listing\PYZgt{}}}%
  \renewcommand*\FancyVerbStopString{\PY{c+c1}{//\PYZlt{}/listing\PYZgt{}}}%
  \begin{Verbatim}[commandchars=\\\{\}]
\PY{c+cm}{/** @file}
\PY{c+cm}{  * @copyright University of Warsaw}
\PY{c+cm}{  * @brief Definition of a structure holding options for Lagrangian microphysics}
\PY{c+cm}{  * @section LICENSE}
\PY{c+cm}{  * GPLv3+ (see the COPYING file or http://www.gnu.org/licenses/)}
\PY{c+cm}{  */}

\PY{c+cp}{\PYZsh{}}\PY{c+cp}{pragma once}

\PY{c+cp}{\PYZsh{}}\PY{c+cp}{include \PYZlt{}libcloudph++}\PY{c+cp}{/}\PY{c+cp}{lgrngn}\PY{c+cp}{/}\PY{c+cp}{extincl.hpp\PYZgt{}}
\PY{c+cp}{\PYZsh{}}\PY{c+cp}{include \PYZlt{}libcloudph++}\PY{c+cp}{/}\PY{c+cp}{lgrngn}\PY{c+cp}{/}\PY{c+cp}{kernel.hpp\PYZgt{}}
\PY{c+cp}{\PYZsh{}}\PY{c+cp}{include \PYZlt{}libcloudph++}\PY{c+cp}{/}\PY{c+cp}{lgrngn}\PY{c+cp}{/}\PY{c+cp}{chem.hpp\PYZgt{}}

\PY{k}{namespace} \PY{n}{libcloudphxx}
\PY{p}{\PYZob{}}
  \PY{k}{namespace} \PY{n}{lgrngn}
  \PY{p}{\PYZob{}}
    \PY{k}{using} \PY{n}{common}\PY{o}{:}\PY{o}{:}\PY{n}{unary\PYZus{}function}\PY{p}{;}

\PY{c+c1}{//\PYZlt{}listing\PYZgt{}}
    \PY{k}{template}\PY{o}{\PYZlt{}}\PY{k}{typename} \PY{k+kt}{real\PYZus{}t}\PY{o}{\PYZgt{}}
    \PY{k}{struct} \PY{k+kt}{opts\PYZus{}init\PYZus{}t} 
    \PY{p}{\PYZob{}}
      \PY{c+c1}{// initial dry sizes of aerosol}
      \PY{k}{typedef} \PY{n}{boost}\PY{o}{:}\PY{o}{:}\PY{n}{ptr\PYZus{}unordered\PYZus{}map}\PY{o}{\PYZlt{}}
        \PY{k+kt}{real\PYZus{}t}\PY{p}{,}                \PY{c+c1}{// kappa}
        \PY{n}{unary\PYZus{}function}\PY{o}{\PYZlt{}}\PY{k+kt}{real\PYZus{}t}\PY{o}{\PYZgt{}} \PY{c+c1}{// n(ln(rd)) @ STP }
      \PY{o}{\PYZgt{}} \PY{k+kt}{dry\PYZus{}distros\PYZus{}t}\PY{p}{;}
      \PY{k+kt}{dry\PYZus{}distros\PYZus{}t} \PY{n}{dry\PYZus{}distros}\PY{p}{;}

      \PY{c+c1}{// Eulerian component parameters}
      \PY{k+kt}{int} \PY{n}{nx}\PY{p}{,} \PY{n}{ny}\PY{p}{,} \PY{n}{nz}\PY{p}{;}
      \PY{k+kt}{real\PYZus{}t} \PY{n}{dx}\PY{p}{,} \PY{n}{dy}\PY{p}{,} \PY{n}{dz}\PY{p}{,} \PY{n}{dt}\PY{p}{;}

      \PY{c+c1}{// Lagrangian domain extents}
      \PY{k+kt}{real\PYZus{}t} \PY{n}{x0}\PY{p}{,} \PY{n}{y0}\PY{p}{,} \PY{n}{z0}\PY{p}{,} \PY{n}{x1}\PY{p}{,} \PY{n}{y1}\PY{p}{,} \PY{n}{z1}\PY{p}{;}

      \PY{c+c1}{// mean no. of super\PYZhy{}droplets per cell}
      \PY{k+kt}{real\PYZus{}t} \PY{n}{sd\PYZus{}conc\PYZus{}mean}\PY{p}{;} 

      \PY{c+c1}{// coalescence Kernel type}
      \PY{k+kt}{kernel\PYZus{}t} \PY{n}{kernel}\PY{p}{;}
\PY{c+c1}{//\PYZlt{}/listing\PYZgt{}}
      \PY{c+c1}{// chem}
      \PY{k+kt}{real\PYZus{}t} \PY{n}{chem\PYZus{}rho}\PY{p}{;}

      \PY{c+c1}{// ctor with defaults (C++03 compliant) ...}
      \PY{k+kt}{opts\PYZus{}init\PYZus{}t}\PY{p}{(}\PY{p}{)} \PY{o}{:} 
        \PY{n}{nx}\PY{p}{(}\PY{l+m+mi}{0}\PY{p}{)}\PY{p}{,} \PY{n}{ny}\PY{p}{(}\PY{l+m+mi}{0}\PY{p}{)}\PY{p}{,} \PY{n}{nz}\PY{p}{(}\PY{l+m+mi}{0}\PY{p}{)}\PY{p}{,} \PY{c+c1}{// parcel setup, 1m3}
        \PY{n}{dx}\PY{p}{(}\PY{l+m+mi}{1}\PY{p}{)}\PY{p}{,} \PY{n}{dy}\PY{p}{(}\PY{l+m+mi}{1}\PY{p}{)}\PY{p}{,} \PY{n}{dz}\PY{p}{(}\PY{l+m+mi}{1}\PY{p}{)}\PY{p}{,} \PY{c+c1}{// parcel setup, 1m3}
        \PY{n}{x0}\PY{p}{(}\PY{l+m+mi}{0}\PY{p}{)}\PY{p}{,} \PY{n}{y0}\PY{p}{(}\PY{l+m+mi}{0}\PY{p}{)}\PY{p}{,} \PY{n}{z0}\PY{p}{(}\PY{l+m+mi}{0}\PY{p}{)}\PY{p}{,} \PY{c+c1}{// parcel setup, 1m3}
        \PY{n}{x1}\PY{p}{(}\PY{l+m+mi}{1}\PY{p}{)}\PY{p}{,} \PY{n}{y1}\PY{p}{(}\PY{l+m+mi}{1}\PY{p}{)}\PY{p}{,} \PY{n}{z1}\PY{p}{(}\PY{l+m+mi}{1}\PY{p}{)}\PY{p}{,} \PY{c+c1}{// parcel setup, 1m3}
        \PY{n}{sd\PYZus{}conc\PYZus{}mean}\PY{p}{(}\PY{l+m+mi}{64}\PY{p}{)}\PY{p}{,} 
        \PY{n}{kernel}\PY{p}{(}\PY{n}{geometric}\PY{p}{)}\PY{p}{,}
        \PY{n}{dt}\PY{p}{(}\PY{l+m+mf}{1e\PYZhy{}3}\PY{p}{)}\PY{p}{,}
        \PY{n}{chem\PYZus{}rho}\PY{p}{(}\PY{l+m+mf}{1.8e3}\PY{p}{)} \PY{c+c1}{// dry particle density}
      \PY{p}{\PYZob{}}\PY{p}{\PYZcb{}}
    \PY{p}{\PYZcb{}}\PY{p}{;}
  \PY{p}{\PYZcb{}}
\PY{p}{\PYZcb{}}\PY{p}{;}
\end{Verbatim}
  \vspace{-1.4em}%

  \caption{\label{lst:lgrngn_opt_init}
    \prog{lgrngn::opts\_init\_t} definition
  }
\end{Listing}
The initial dry size spectrum of aerosol is represented with a map
  associating values of the solubility parameter $\kappa$ with
  pointers to functors returning concentration of particles at STP
  as a function of logarithm of dry radius.
Subsequent fields specify the geometry of the Eulerian grid and the 
  timestep.
It is assumed that the Eulerian component operates on a 
  rectilinear grid with a constant grid cell spacing,
  although this assumption may easily be lifted in future releases if needed.
The parameters \prog{x0}, \prog{y0}, \prog{z0}, \prog{x1}, \prog{y1}, \prog{z1}
  are intended for defining a subregion of the Eulerian domain to be covered
  with computational particles.
The last two fields provide control of the initial mean concentration of 
  computational particles per grid cell and the type of the coalescence kernel
  (only the geometric one implemented so far, see Listing~\ref{lst:lgrngn_knl}).
\begin{Listing}
  \renewcommand*\FancyVerbStartString{\PY{c+c1}{//\PYZlt{}listing\PYZgt{}}}%
  \renewcommand*\FancyVerbStopString{\PY{c+c1}{//\PYZlt{}/listing\PYZgt{}}}%
  \begin{Verbatim}[commandchars=\\\{\}]
\PY{c+cp}{\PYZsh{}}\PY{c+cp}{pragma once }

\PY{k}{namespace} \PY{n}{libcloudphxx}
\PY{p}{\PYZob{}}
  \PY{k}{namespace} \PY{n}{lgrngn}
  \PY{p}{\PYZob{}}
\PY{c+c1}{//\PYZlt{}listing\PYZgt{}}
    \PY{k}{enum} \PY{k+kt}{kernel\PYZus{}t} \PY{p}{\PYZob{}} \PY{n}{geometric} \PY{p}{\PYZcb{}}\PY{p}{;} 
\PY{c+c1}{//\PYZlt{}/listing\PYZgt{}}
  \PY{p}{\PYZcb{}}\PY{p}{;}
\PY{p}{\PYZcb{}}\PY{p}{;}
\end{Verbatim}
  \vspace{-1.4em}%

  \caption{\label{lst:lgrngn_knl}
    \prog{lgrngn::kernel\_t} definition
  }
\end{Listing}

Computational particle spatial coordinates provide the principal link 
  between the particle-based scheme's Lagrangian and Eulerian components.
Consequently, unlike in the case of bulk schemes which use STL-type iterators to traverse 
  array elements without any information on the array dimensionality or shape,
  here the actual geometry and memory layout of the passed arrays need to be known.
The memory layout of array data is represented in the API using the
  \prog{lgrngn::arrinfo\_t} structure (Listing~\ref{lst:lgrngn_arr}).
\begin{Listing}
  \renewcommand*\FancyVerbStartString{\PY{c+c1}{//\PYZlt{}listing\PYZgt{}}}%
  \renewcommand*\FancyVerbStopString{\PY{c+c1}{//\PYZlt{}/listing\PYZgt{}}}%
  \begin{Verbatim}[commandchars=\\\{\}]
\PY{c+cp}{\PYZsh{}}\PY{c+cp}{pragma once }

\PY{c+cp}{\PYZsh{}}\PY{c+cp}{include \PYZlt{}libcloudph++}\PY{c+cp}{/}\PY{c+cp}{lgrngn}\PY{c+cp}{/}\PY{c+cp}{extincl.hpp\PYZgt{}}

\PY{k}{namespace} \PY{n}{libcloudphxx}
\PY{p}{\PYZob{}}
  \PY{k}{namespace} \PY{n}{lgrngn}
  \PY{p}{\PYZob{}}
    \PY{c+c1}{// helper struct to ease passing n\PYZhy{}dimensional arrays}
\PY{c+c1}{//\PYZlt{}listing\PYZgt{}}
    \PY{k}{template} \PY{o}{\PYZlt{}}\PY{k}{typename} \PY{k+kt}{real\PYZus{}t}\PY{o}{\PYZgt{}}
    \PY{k}{struct} \PY{k+kt}{arrinfo\PYZus{}t}
    \PY{p}{\PYZob{}}
      \PY{c+c1}{// member fields:}
      \PY{k+kt}{real\PYZus{}t} \PY{o}{*} \PY{k}{const} \PY{n}{dataZero}\PY{p}{;}
      \PY{k}{const} \PY{k+kt}{ptrdiff\PYZus{}t} \PY{o}{*}\PY{n}{strides}\PY{p}{;}

      \PY{c+c1}{// methods...}
\PY{c+c1}{//\PYZlt{}/listing\PYZgt{}}

      \PY{c+c1}{// ctors}
      \PY{k+kt}{arrinfo\PYZus{}t}\PY{p}{(}\PY{p}{)}
        \PY{o}{:} \PY{n}{dataZero}\PY{p}{(}\PY{n+nb}{NULL}\PY{p}{)}\PY{p}{,} \PY{n}{strides}\PY{p}{(}\PY{n+nb}{NULL}\PY{p}{)} 
      \PY{p}{\PYZob{}}\PY{p}{\PYZcb{}} 

      \PY{k+kt}{arrinfo\PYZus{}t}\PY{p}{(}\PY{k+kt}{real\PYZus{}t} \PY{o}{*} \PY{k}{const} \PY{n}{dataZero}\PY{p}{,} \PY{k}{const} \PY{k+kt}{ptrdiff\PYZus{}t} \PY{o}{*}\PY{n}{strides}\PY{p}{)} 
        \PY{o}{:} \PY{n}{dataZero}\PY{p}{(}\PY{n}{dataZero}\PY{p}{)}\PY{p}{,} \PY{n}{strides}\PY{p}{(}\PY{n}{strides}\PY{p}{)} 
      \PY{p}{\PYZob{}}\PY{p}{\PYZcb{}} 

      \PY{c+c1}{// methods}
      \PY{k+kt}{bool} \PY{n}{is\PYZus{}null}\PY{p}{(}\PY{p}{)} \PY{k}{const} \PY{p}{\PYZob{}} \PY{k}{return} \PY{n}{dataZero}\PY{o}{=}\PY{o}{=}\PY{n+nb}{NULL} \PY{o}{|}\PY{o}{|} \PY{n}{strides}\PY{o}{=}\PY{o}{=}\PY{n+nb}{NULL}\PY{p}{;} \PY{p}{\PYZcb{}}
    \PY{p}{\PYZcb{}}\PY{p}{;}
  \PY{p}{\PYZcb{}}\PY{p}{;}
\PY{p}{\PYZcb{}}\PY{p}{;}
\end{Verbatim}
  \vspace{-1.4em}%

  \caption{\label{lst:lgrngn_arr}
    \prog{lgrngn::arrinfo\_t} definition
  }
\end{Listing}
The meaning of \prog{dataZero} and \prog{strides} fields match those
  of equally-named methods of the Blitz++ Array class.
Quoting Blitz++ documentation \citep{Veldhuizen_2005}: ,,\prog{dataZero} is a pointer to the element (0,0,...,0), 
  even if such an element does not exist in the array. What's the point of having such a pointer? 
  Say you want to access the element (i,j,k). If you add to the pointer the dot product of (i,j,k) 
  with the stride vector \prog{stride}, you get a pointer to the element (i,j,k).''
Using \prog{arrinfo\_t} as the type for API function arguments makes
  the library compatible with a wide range of array containers, Blitz++
  being just an example.
In addition, no assumptions are made with respect to array index ranges, what allows 
  the library to operate on array slabs (e.g. array segments excluding the so-called halo regions).

The state of the Lagrangian component of the model (notably, the values of particle attributes) is stored 
  in an instance of the \prog{lgrngn::particles\_t} class (see Listing \ref{lst:lgrngn_prt}).
Internally, the Lagrangian calculations are implemented using the Thrust library\footnote{\url{http://thrust.github.io/}}
  which, among other, allows to run the particle-based simulations either on CPU[s] or on a GPU.
The second template parameter of \prog{lgrngn::particles\_t} is the type of the backend to be used by the Thrust library,
  and as of current release it has three possible values: serial, OpenMP or CUDA
  (cf. Listing~\ref{lst:lgrngn_bnd} with definition of the \prog{backend\_t} enumeration).
The OpenMP\footnote{\url{http://openmp.org/}} backend offers multi-threading using multiple CPU cores and/or multiple CPUs.
The CUDA\footnote{\url{http://nvidia.com/}} backend enables the user to perform the computations on a GPU.
The serial backend does single-thread computations on a CPU.
The ''backend-aware'' \prog{particles\_t$<$real\_t, backend$>$} inherits from 
  ''backend-unaware'' \prog{particles\_proto\_t$<$real\_t$>$} (definition not shown)
  what allows to use a single pointer to \prog{particles\_proto\_t} with
  different backends (as used in the return value of \prog{lgrngn::factory()} discussed below).
  
Initialisation, time-stepping and data output is performed by calling
  \prog{particles\_t}'s methods whose signatures are given in Listing \ref{lst:lgrngn_prt}
  and discussed in the following three paragraphs.
\begin{Listing}
  \renewcommand*\FancyVerbStartString{\PY{c+c1}{//\PYZlt{}listing\PYZgt{}}}%
  \renewcommand*\FancyVerbStopString{\PY{c+c1}{//\PYZlt{}/listing\PYZgt{}}}%
  \begin{Verbatim}[commandchars=\\\{\}]
\PY{c+cp}{\PYZsh{}}\PY{c+cp}{pragma once }

\PY{c+cp}{\PYZsh{}}\PY{c+cp}{include \PYZlt{}libcloudph++}\PY{c+cp}{/}\PY{c+cp}{lgrngn}\PY{c+cp}{/}\PY{c+cp}{extincl.hpp\PYZgt{}}

\PY{c+cp}{\PYZsh{}}\PY{c+cp}{include \PYZdq{}opts.hpp\PYZdq{}}
\PY{c+cp}{\PYZsh{}}\PY{c+cp}{include \PYZdq{}opts\PYZus{}init.hpp\PYZdq{}}
\PY{c+cp}{\PYZsh{}}\PY{c+cp}{include \PYZdq{}arrinfo.hpp\PYZdq{}}
\PY{c+cp}{\PYZsh{}}\PY{c+cp}{include \PYZdq{}backend.hpp\PYZdq{}}

\PY{k}{namespace} \PY{n}{libcloudphxx}
\PY{p}{\PYZob{}}
  \PY{k}{namespace} \PY{n}{lgrngn}
  \PY{p}{\PYZob{}}
    \PY{c+c1}{// to allow storing instances for multiple backends in one container/pointer}
    \PY{k}{template} \PY{o}{\PYZlt{}}\PY{k}{typename} \PY{k+kt}{real\PYZus{}t}\PY{o}{\PYZgt{}}
    \PY{k}{struct} \PY{k+kt}{particles\PYZus{}proto\PYZus{}t} 
    \PY{p}{\PYZob{}}
      \PY{c+c1}{// 3D version}
      \PY{k}{virtual} \PY{k+kt}{void} \PY{n}{init}\PY{p}{(}
        \PY{k}{const} \PY{k+kt}{arrinfo\PYZus{}t}\PY{o}{\PYZlt{}}\PY{k+kt}{real\PYZus{}t}\PY{o}{\PYZgt{}} \PY{n}{th}\PY{p}{,}
        \PY{k}{const} \PY{k+kt}{arrinfo\PYZus{}t}\PY{o}{\PYZlt{}}\PY{k+kt}{real\PYZus{}t}\PY{o}{\PYZgt{}} \PY{n}{rv}\PY{p}{,}
        \PY{k}{const} \PY{k+kt}{arrinfo\PYZus{}t}\PY{o}{\PYZlt{}}\PY{k+kt}{real\PYZus{}t}\PY{o}{\PYZgt{}} \PY{n}{rhod}\PY{p}{,} 
        \PY{k}{const} \PY{k+kt}{arrinfo\PYZus{}t}\PY{o}{\PYZlt{}}\PY{k+kt}{real\PYZus{}t}\PY{o}{\PYZgt{}} \PY{n}{rhod\PYZus{}courant\PYZus{}x}\PY{p}{,}
        \PY{k}{const} \PY{k+kt}{arrinfo\PYZus{}t}\PY{o}{\PYZlt{}}\PY{k+kt}{real\PYZus{}t}\PY{o}{\PYZgt{}} \PY{n}{rhod\PYZus{}courant\PYZus{}y}\PY{p}{,}
        \PY{k}{const} \PY{k+kt}{arrinfo\PYZus{}t}\PY{o}{\PYZlt{}}\PY{k+kt}{real\PYZus{}t}\PY{o}{\PYZgt{}} \PY{n}{rhod\PYZus{}courant\PYZus{}z}
      \PY{p}{)} \PY{p}{\PYZob{}} 
        \PY{n}{assert}\PY{p}{(}\PY{n+nb}{false}\PY{p}{)}\PY{p}{;} 
      \PY{p}{\PYZcb{}}  

      \PY{c+c1}{// 2D version}
      \PY{k+kt}{void} \PY{n}{init}\PY{p}{(}
        \PY{k}{const} \PY{k+kt}{arrinfo\PYZus{}t}\PY{o}{\PYZlt{}}\PY{k+kt}{real\PYZus{}t}\PY{o}{\PYZgt{}} \PY{n}{th}\PY{p}{,}
        \PY{k}{const} \PY{k+kt}{arrinfo\PYZus{}t}\PY{o}{\PYZlt{}}\PY{k+kt}{real\PYZus{}t}\PY{o}{\PYZgt{}} \PY{n}{rv}\PY{p}{,}
        \PY{k}{const} \PY{k+kt}{arrinfo\PYZus{}t}\PY{o}{\PYZlt{}}\PY{k+kt}{real\PYZus{}t}\PY{o}{\PYZgt{}} \PY{n}{rhod}\PY{p}{,}
        \PY{k}{const} \PY{k+kt}{arrinfo\PYZus{}t}\PY{o}{\PYZlt{}}\PY{k+kt}{real\PYZus{}t}\PY{o}{\PYZgt{}} \PY{n}{rhod\PYZus{}courant\PYZus{}x}\PY{p}{,}
        \PY{k}{const} \PY{k+kt}{arrinfo\PYZus{}t}\PY{o}{\PYZlt{}}\PY{k+kt}{real\PYZus{}t}\PY{o}{\PYZgt{}} \PY{n}{rhod\PYZus{}courant\PYZus{}z}
      \PY{p}{)} \PY{p}{\PYZob{}} 
        \PY{k}{this}\PY{o}{\PYZhy{}}\PY{o}{\PYZgt{}}\PY{n}{init}\PY{p}{(}\PY{n}{th}\PY{p}{,} \PY{n}{rv}\PY{p}{,} \PY{n}{rhod}\PY{p}{,} \PY{n}{rhod\PYZus{}courant\PYZus{}x}\PY{p}{,} \PY{k+kt}{arrinfo\PYZus{}t}\PY{o}{\PYZlt{}}\PY{k+kt}{real\PYZus{}t}\PY{o}{\PYZgt{}}\PY{p}{(}\PY{p}{)}\PY{p}{,} \PY{n}{rhod\PYZus{}courant\PYZus{}z}\PY{p}{)}\PY{p}{;} 
      \PY{p}{\PYZcb{}}  
 
      \PY{c+c1}{// 1D version}
      \PY{k+kt}{void} \PY{n}{init}\PY{p}{(}
        \PY{k}{const} \PY{k+kt}{arrinfo\PYZus{}t}\PY{o}{\PYZlt{}}\PY{k+kt}{real\PYZus{}t}\PY{o}{\PYZgt{}} \PY{n}{th}\PY{p}{,}
        \PY{k}{const} \PY{k+kt}{arrinfo\PYZus{}t}\PY{o}{\PYZlt{}}\PY{k+kt}{real\PYZus{}t}\PY{o}{\PYZgt{}} \PY{n}{rv}\PY{p}{,}
        \PY{k}{const} \PY{k+kt}{arrinfo\PYZus{}t}\PY{o}{\PYZlt{}}\PY{k+kt}{real\PYZus{}t}\PY{o}{\PYZgt{}} \PY{n}{rhod}\PY{p}{,}
        \PY{k}{const} \PY{k+kt}{arrinfo\PYZus{}t}\PY{o}{\PYZlt{}}\PY{k+kt}{real\PYZus{}t}\PY{o}{\PYZgt{}} \PY{n}{rhod\PYZus{}courant\PYZus{}z}
      \PY{p}{)} \PY{p}{\PYZob{}} 
        \PY{k}{this}\PY{o}{\PYZhy{}}\PY{o}{\PYZgt{}}\PY{n}{init}\PY{p}{(}\PY{n}{th}\PY{p}{,} \PY{n}{rv}\PY{p}{,} \PY{n}{rhod}\PY{p}{,} \PY{k+kt}{arrinfo\PYZus{}t}\PY{o}{\PYZlt{}}\PY{k+kt}{real\PYZus{}t}\PY{o}{\PYZgt{}}\PY{p}{(}\PY{p}{)}\PY{p}{,} \PY{k+kt}{arrinfo\PYZus{}t}\PY{o}{\PYZlt{}}\PY{k+kt}{real\PYZus{}t}\PY{o}{\PYZgt{}}\PY{p}{(}\PY{p}{)}\PY{p}{,} \PY{n}{rhod\PYZus{}courant\PYZus{}z}\PY{p}{)}\PY{p}{;} 
      \PY{p}{\PYZcb{}}  

      \PY{c+c1}{// 0D version}
      \PY{k+kt}{void} \PY{n}{init}\PY{p}{(}
        \PY{k}{const} \PY{k+kt}{arrinfo\PYZus{}t}\PY{o}{\PYZlt{}}\PY{k+kt}{real\PYZus{}t}\PY{o}{\PYZgt{}} \PY{n}{th}\PY{p}{,}
        \PY{k}{const} \PY{k+kt}{arrinfo\PYZus{}t}\PY{o}{\PYZlt{}}\PY{k+kt}{real\PYZus{}t}\PY{o}{\PYZgt{}} \PY{n}{rv}\PY{p}{,}
        \PY{k}{const} \PY{k+kt}{arrinfo\PYZus{}t}\PY{o}{\PYZlt{}}\PY{k+kt}{real\PYZus{}t}\PY{o}{\PYZgt{}} \PY{n}{rhod}
      \PY{p}{)} \PY{p}{\PYZob{}} 
        \PY{k}{this}\PY{o}{\PYZhy{}}\PY{o}{\PYZgt{}}\PY{n}{init}\PY{p}{(}\PY{n}{th}\PY{p}{,} \PY{n}{rv}\PY{p}{,} \PY{n}{rhod}\PY{p}{,} \PY{k+kt}{arrinfo\PYZus{}t}\PY{o}{\PYZlt{}}\PY{k+kt}{real\PYZus{}t}\PY{o}{\PYZgt{}}\PY{p}{(}\PY{p}{)}\PY{p}{,} \PY{k+kt}{arrinfo\PYZus{}t}\PY{o}{\PYZlt{}}\PY{k+kt}{real\PYZus{}t}\PY{o}{\PYZgt{}}\PY{p}{(}\PY{p}{)}\PY{p}{,} \PY{k+kt}{arrinfo\PYZus{}t}\PY{o}{\PYZlt{}}\PY{k+kt}{real\PYZus{}t}\PY{o}{\PYZgt{}}\PY{p}{(}\PY{p}{)}\PY{p}{)}\PY{p}{;} 
      \PY{p}{\PYZcb{}}  

      \PY{c+c1}{// 3D variable density version}
      \PY{k}{virtual} \PY{k+kt}{void} \PY{n}{step\PYZus{}sync}\PY{p}{(}
        \PY{k}{const} \PY{k+kt}{opts\PYZus{}t}\PY{o}{\PYZlt{}}\PY{k+kt}{real\PYZus{}t}\PY{o}{\PYZgt{}} \PY{o}{\PYZam{}}\PY{p}{,}
        \PY{k+kt}{arrinfo\PYZus{}t}\PY{o}{\PYZlt{}}\PY{k+kt}{real\PYZus{}t}\PY{o}{\PYZgt{}} \PY{n}{th}\PY{p}{,}
        \PY{k+kt}{arrinfo\PYZus{}t}\PY{o}{\PYZlt{}}\PY{k+kt}{real\PYZus{}t}\PY{o}{\PYZgt{}} \PY{n}{rv}\PY{p}{,}
        \PY{k}{const} \PY{k+kt}{arrinfo\PYZus{}t}\PY{o}{\PYZlt{}}\PY{k+kt}{real\PYZus{}t}\PY{o}{\PYZgt{}} \PY{n}{rhod\PYZus{}courant\PYZus{}x}\PY{p}{,}
        \PY{k}{const} \PY{k+kt}{arrinfo\PYZus{}t}\PY{o}{\PYZlt{}}\PY{k+kt}{real\PYZus{}t}\PY{o}{\PYZgt{}} \PY{n}{rhod\PYZus{}courant\PYZus{}y}\PY{p}{,}
        \PY{k}{const} \PY{k+kt}{arrinfo\PYZus{}t}\PY{o}{\PYZlt{}}\PY{k+kt}{real\PYZus{}t}\PY{o}{\PYZgt{}} \PY{n}{rhod\PYZus{}courant\PYZus{}z}\PY{p}{,}
        \PY{k}{const} \PY{k+kt}{arrinfo\PYZus{}t}\PY{o}{\PYZlt{}}\PY{k+kt}{real\PYZus{}t}\PY{o}{\PYZgt{}} \PY{n}{rhod}
      \PY{p}{)} \PY{p}{\PYZob{}} 
        \PY{n}{assert}\PY{p}{(}\PY{n+nb}{false}\PY{p}{)}\PY{p}{;} 
      \PY{p}{\PYZcb{}}  

      \PY{c+c1}{// returns accumulated rainfall}
      \PY{k}{virtual} \PY{k+kt}{real\PYZus{}t} \PY{n}{step\PYZus{}async}\PY{p}{(}
        \PY{k}{const} \PY{k+kt}{opts\PYZus{}t}\PY{o}{\PYZlt{}}\PY{k+kt}{real\PYZus{}t}\PY{o}{\PYZgt{}} \PY{o}{\PYZam{}}
      \PY{p}{)} \PY{p}{\PYZob{}} 
        \PY{n}{assert}\PY{p}{(}\PY{n+nb}{false}\PY{p}{)}\PY{p}{;} 
        \PY{k}{return} \PY{l+m+mi}{0}\PY{p}{;}
      \PY{p}{\PYZcb{}}  

      \PY{c+c1}{// 3D constant density version}
      \PY{k+kt}{void} \PY{n}{step\PYZus{}sync}\PY{p}{(}
        \PY{k}{const} \PY{k+kt}{opts\PYZus{}t}\PY{o}{\PYZlt{}}\PY{k+kt}{real\PYZus{}t}\PY{o}{\PYZgt{}} \PY{o}{\PYZam{}}\PY{n}{opts}\PY{p}{,}
        \PY{k+kt}{arrinfo\PYZus{}t}\PY{o}{\PYZlt{}}\PY{k+kt}{real\PYZus{}t}\PY{o}{\PYZgt{}} \PY{n}{th}\PY{p}{,}
        \PY{k+kt}{arrinfo\PYZus{}t}\PY{o}{\PYZlt{}}\PY{k+kt}{real\PYZus{}t}\PY{o}{\PYZgt{}} \PY{n}{rv}\PY{p}{,}
        \PY{k}{const} \PY{k+kt}{arrinfo\PYZus{}t}\PY{o}{\PYZlt{}}\PY{k+kt}{real\PYZus{}t}\PY{o}{\PYZgt{}} \PY{n}{rhod\PYZus{}courant\PYZus{}x}\PY{p}{,}
        \PY{k}{const} \PY{k+kt}{arrinfo\PYZus{}t}\PY{o}{\PYZlt{}}\PY{k+kt}{real\PYZus{}t}\PY{o}{\PYZgt{}} \PY{n}{rhod\PYZus{}courant\PYZus{}y}\PY{p}{,}
        \PY{k}{const} \PY{k+kt}{arrinfo\PYZus{}t}\PY{o}{\PYZlt{}}\PY{k+kt}{real\PYZus{}t}\PY{o}{\PYZgt{}} \PY{n}{rhod\PYZus{}courant\PYZus{}z}
      \PY{p}{)} \PY{p}{\PYZob{}} 
        \PY{k}{this}\PY{o}{\PYZhy{}}\PY{o}{\PYZgt{}}\PY{n}{step\PYZus{}sync}\PY{p}{(}\PY{n}{opts}\PY{p}{,} \PY{n}{th}\PY{p}{,} \PY{n}{rv}\PY{p}{,} \PY{n}{rhod\PYZus{}courant\PYZus{}x}\PY{p}{,} \PY{n}{rhod\PYZus{}courant\PYZus{}y}\PY{p}{,} \PY{n}{rhod\PYZus{}courant\PYZus{}z}\PY{p}{,} \PY{k+kt}{arrinfo\PYZus{}t}\PY{o}{\PYZlt{}}\PY{k+kt}{real\PYZus{}t}\PY{o}{\PYZgt{}}\PY{p}{(}\PY{p}{)}\PY{p}{)}\PY{p}{;} 
      \PY{p}{\PYZcb{}}  

      \PY{c+c1}{// 2D constant density version}
      \PY{k+kt}{void} \PY{n}{step\PYZus{}sync}\PY{p}{(}
        \PY{k}{const} \PY{k+kt}{opts\PYZus{}t}\PY{o}{\PYZlt{}}\PY{k+kt}{real\PYZus{}t}\PY{o}{\PYZgt{}} \PY{o}{\PYZam{}}\PY{n}{opts}\PY{p}{,}
        \PY{k+kt}{arrinfo\PYZus{}t}\PY{o}{\PYZlt{}}\PY{k+kt}{real\PYZus{}t}\PY{o}{\PYZgt{}} \PY{n}{th}\PY{p}{,}
        \PY{k+kt}{arrinfo\PYZus{}t}\PY{o}{\PYZlt{}}\PY{k+kt}{real\PYZus{}t}\PY{o}{\PYZgt{}} \PY{n}{rv}\PY{p}{,}
        \PY{k}{const} \PY{k+kt}{arrinfo\PYZus{}t}\PY{o}{\PYZlt{}}\PY{k+kt}{real\PYZus{}t}\PY{o}{\PYZgt{}} \PY{n}{rhod\PYZus{}courant\PYZus{}x}\PY{p}{,}
        \PY{k}{const} \PY{k+kt}{arrinfo\PYZus{}t}\PY{o}{\PYZlt{}}\PY{k+kt}{real\PYZus{}t}\PY{o}{\PYZgt{}} \PY{n}{rhod\PYZus{}courant\PYZus{}z}
      \PY{p}{)} \PY{p}{\PYZob{}} 
        \PY{k}{this}\PY{o}{\PYZhy{}}\PY{o}{\PYZgt{}}\PY{n}{step\PYZus{}sync}\PY{p}{(}
          \PY{n}{opts}\PY{p}{,}
          \PY{n}{th}\PY{p}{,} 
          \PY{n}{rv}\PY{p}{,} 
          \PY{n}{rhod\PYZus{}courant\PYZus{}x}\PY{p}{,} 
          \PY{k+kt}{arrinfo\PYZus{}t}\PY{o}{\PYZlt{}}\PY{k+kt}{real\PYZus{}t}\PY{o}{\PYZgt{}}\PY{p}{(}\PY{p}{)}\PY{p}{,} 
          \PY{n}{rhod\PYZus{}courant\PYZus{}z}\PY{p}{,} 
          \PY{k+kt}{arrinfo\PYZus{}t}\PY{o}{\PYZlt{}}\PY{k+kt}{real\PYZus{}t}\PY{o}{\PYZgt{}}\PY{p}{(}\PY{p}{)}
        \PY{p}{)}\PY{p}{;} 
      \PY{p}{\PYZcb{}}  

      \PY{c+c1}{// 1D constant density version}
      \PY{k+kt}{void} \PY{n}{step\PYZus{}sync}\PY{p}{(}
        \PY{k}{const} \PY{k+kt}{opts\PYZus{}t}\PY{o}{\PYZlt{}}\PY{k+kt}{real\PYZus{}t}\PY{o}{\PYZgt{}} \PY{o}{\PYZam{}}\PY{n}{opts}\PY{p}{,}
        \PY{k+kt}{arrinfo\PYZus{}t}\PY{o}{\PYZlt{}}\PY{k+kt}{real\PYZus{}t}\PY{o}{\PYZgt{}} \PY{n}{th}\PY{p}{,}
        \PY{k+kt}{arrinfo\PYZus{}t}\PY{o}{\PYZlt{}}\PY{k+kt}{real\PYZus{}t}\PY{o}{\PYZgt{}} \PY{n}{rv}\PY{p}{,}
        \PY{k}{const} \PY{k+kt}{arrinfo\PYZus{}t}\PY{o}{\PYZlt{}}\PY{k+kt}{real\PYZus{}t}\PY{o}{\PYZgt{}} \PY{n}{rhod\PYZus{}courant\PYZus{}z}
      \PY{p}{)} \PY{p}{\PYZob{}} 
        \PY{k}{this}\PY{o}{\PYZhy{}}\PY{o}{\PYZgt{}}\PY{n}{step\PYZus{}sync}\PY{p}{(}
          \PY{n}{opts}\PY{p}{,}
          \PY{n}{th}\PY{p}{,} 
          \PY{n}{rv}\PY{p}{,} 
          \PY{k+kt}{arrinfo\PYZus{}t}\PY{o}{\PYZlt{}}\PY{k+kt}{real\PYZus{}t}\PY{o}{\PYZgt{}}\PY{p}{(}\PY{p}{)}\PY{p}{,}
          \PY{k+kt}{arrinfo\PYZus{}t}\PY{o}{\PYZlt{}}\PY{k+kt}{real\PYZus{}t}\PY{o}{\PYZgt{}}\PY{p}{(}\PY{p}{)}\PY{p}{,} 
          \PY{n}{rhod\PYZus{}courant\PYZus{}z}\PY{p}{,} 
          \PY{k+kt}{arrinfo\PYZus{}t}\PY{o}{\PYZlt{}}\PY{k+kt}{real\PYZus{}t}\PY{o}{\PYZgt{}}\PY{p}{(}\PY{p}{)}\PY{p}{)}\PY{p}{;} 
      \PY{p}{\PYZcb{}}  

      \PY{c+c1}{// 0D constant density version}
      \PY{k+kt}{void} \PY{n}{step\PYZus{}sync}\PY{p}{(}
        \PY{k}{const} \PY{k+kt}{opts\PYZus{}t}\PY{o}{\PYZlt{}}\PY{k+kt}{real\PYZus{}t}\PY{o}{\PYZgt{}} \PY{o}{\PYZam{}}\PY{n}{opts}\PY{p}{,}
        \PY{k+kt}{arrinfo\PYZus{}t}\PY{o}{\PYZlt{}}\PY{k+kt}{real\PYZus{}t}\PY{o}{\PYZgt{}} \PY{n}{th}\PY{p}{,}
        \PY{k+kt}{arrinfo\PYZus{}t}\PY{o}{\PYZlt{}}\PY{k+kt}{real\PYZus{}t}\PY{o}{\PYZgt{}} \PY{n}{rv}
      \PY{p}{)} \PY{p}{\PYZob{}} 
        \PY{k}{this}\PY{o}{\PYZhy{}}\PY{o}{\PYZgt{}}\PY{n}{step\PYZus{}sync}\PY{p}{(}
          \PY{n}{opts}\PY{p}{,}
          \PY{n}{th}\PY{p}{,} 
          \PY{n}{rv}\PY{p}{,} 
          \PY{k+kt}{arrinfo\PYZus{}t}\PY{o}{\PYZlt{}}\PY{k+kt}{real\PYZus{}t}\PY{o}{\PYZgt{}}\PY{p}{(}\PY{p}{)}\PY{p}{,} 
          \PY{k+kt}{arrinfo\PYZus{}t}\PY{o}{\PYZlt{}}\PY{k+kt}{real\PYZus{}t}\PY{o}{\PYZgt{}}\PY{p}{(}\PY{p}{)}\PY{p}{,} 
          \PY{k+kt}{arrinfo\PYZus{}t}\PY{o}{\PYZlt{}}\PY{k+kt}{real\PYZus{}t}\PY{o}{\PYZgt{}}\PY{p}{(}\PY{p}{)}\PY{p}{,} 
          \PY{k+kt}{arrinfo\PYZus{}t}\PY{o}{\PYZlt{}}\PY{k+kt}{real\PYZus{}t}\PY{o}{\PYZgt{}}\PY{p}{(}\PY{p}{)}
        \PY{p}{)}\PY{p}{;} 
      \PY{p}{\PYZcb{}}  

      \PY{c+c1}{// method for accessing super\PYZhy{}droplet statistics}
      \PY{k}{virtual} \PY{k+kt}{void} \PY{n}{diag\PYZus{}sd\PYZus{}conc}\PY{p}{(}\PY{p}{)}                             \PY{p}{\PYZob{}} \PY{n}{assert}\PY{p}{(}\PY{n+nb}{false}\PY{p}{)}\PY{p}{;} \PY{p}{\PYZcb{}}
      \PY{k}{virtual} \PY{k+kt}{void} \PY{n}{diag\PYZus{}dry\PYZus{}rng}\PY{p}{(}\PY{k}{const} \PY{k+kt}{real\PYZus{}t}\PY{o}{\PYZam{}}\PY{p}{,} \PY{k}{const} \PY{k+kt}{real\PYZus{}t}\PY{o}{\PYZam{}}\PY{p}{)} \PY{p}{\PYZob{}} \PY{n}{assert}\PY{p}{(}\PY{n+nb}{false}\PY{p}{)}\PY{p}{;} \PY{p}{\PYZcb{}}
      \PY{k}{virtual} \PY{k+kt}{void} \PY{n}{diag\PYZus{}wet\PYZus{}rng}\PY{p}{(}\PY{k}{const} \PY{k+kt}{real\PYZus{}t}\PY{o}{\PYZam{}}\PY{p}{,} \PY{k}{const} \PY{k+kt}{real\PYZus{}t}\PY{o}{\PYZam{}}\PY{p}{)} \PY{p}{\PYZob{}} \PY{n}{assert}\PY{p}{(}\PY{n+nb}{false}\PY{p}{)}\PY{p}{;} \PY{p}{\PYZcb{}}
      \PY{k}{virtual} \PY{k+kt}{void} \PY{n}{diag\PYZus{}dry\PYZus{}mom}\PY{p}{(}\PY{k}{const} \PY{k+kt}{int}\PY{o}{\PYZam{}}\PY{p}{)}                   \PY{p}{\PYZob{}} \PY{n}{assert}\PY{p}{(}\PY{n+nb}{false}\PY{p}{)}\PY{p}{;} \PY{p}{\PYZcb{}}
      \PY{k}{virtual} \PY{k+kt}{void} \PY{n}{diag\PYZus{}wet\PYZus{}mom}\PY{p}{(}\PY{k}{const} \PY{k+kt}{int}\PY{o}{\PYZam{}}\PY{p}{)}                   \PY{p}{\PYZob{}} \PY{n}{assert}\PY{p}{(}\PY{n+nb}{false}\PY{p}{)}\PY{p}{;} \PY{p}{\PYZcb{}}
      \PY{k}{virtual} \PY{k+kt}{void} \PY{n}{diag\PYZus{}chem}\PY{p}{(}\PY{k}{const} \PY{k}{enum} \PY{k+kt}{chem\PYZus{}species\PYZus{}t}\PY{o}{\PYZam{}}\PY{p}{)}      \PY{p}{\PYZob{}} \PY{n}{assert}\PY{p}{(}\PY{n+nb}{false}\PY{p}{)}\PY{p}{;} \PY{p}{\PYZcb{}}
      \PY{k}{virtual} \PY{k+kt}{real\PYZus{}t} \PY{o}{*}\PY{n}{outbuf}\PY{p}{(}\PY{p}{)}                                \PY{p}{\PYZob{}} \PY{n}{assert}\PY{p}{(}\PY{n+nb}{false}\PY{p}{)}\PY{p}{;} \PY{k}{return} \PY{n+nb}{NULL}\PY{p}{;} \PY{p}{\PYZcb{}}
    \PY{p}{\PYZcb{}}\PY{p}{;}  

    \PY{c+c1}{// prototype of what\PYZsq{}s implemented in the .tpp file}
\PY{c+c1}{//\PYZlt{}listing\PYZgt{}}
    \PY{k}{template} \PY{o}{\PYZlt{}}\PY{k}{typename} \PY{k+kt}{real\PYZus{}t}\PY{p}{,} \PY{k+kt}{backend\PYZus{}t} \PY{n}{backend}\PY{o}{\PYZgt{}}
    \PY{k}{struct} \PY{k+kt}{particles\PYZus{}t}\PY{o}{:} \PY{k+kt}{particles\PYZus{}proto\PYZus{}t}\PY{o}{\PYZlt{}}\PY{k+kt}{real\PYZus{}t}\PY{o}{\PYZgt{}}
    \PY{p}{\PYZob{}}
      \PY{c+c1}{// initialisation }
      \PY{k+kt}{void} \PY{n}{init}\PY{p}{(}
        \PY{k}{const} \PY{k+kt}{arrinfo\PYZus{}t}\PY{o}{\PYZlt{}}\PY{k+kt}{real\PYZus{}t}\PY{o}{\PYZgt{}} \PY{n}{th}\PY{p}{,}
        \PY{k}{const} \PY{k+kt}{arrinfo\PYZus{}t}\PY{o}{\PYZlt{}}\PY{k+kt}{real\PYZus{}t}\PY{o}{\PYZgt{}} \PY{n}{rv}\PY{p}{,}
        \PY{k}{const} \PY{k+kt}{arrinfo\PYZus{}t}\PY{o}{\PYZlt{}}\PY{k+kt}{real\PYZus{}t}\PY{o}{\PYZgt{}} \PY{n}{rhod}\PY{p}{,}
        \PY{k}{const} \PY{k+kt}{arrinfo\PYZus{}t}\PY{o}{\PYZlt{}}\PY{k+kt}{real\PYZus{}t}\PY{o}{\PYZgt{}} \PY{n}{rhod\PYZus{}courant\PYZus{}1}\PY{p}{,}
        \PY{k}{const} \PY{k+kt}{arrinfo\PYZus{}t}\PY{o}{\PYZlt{}}\PY{k+kt}{real\PYZus{}t}\PY{o}{\PYZgt{}} \PY{n}{rhod\PYZus{}courant\PYZus{}2}\PY{p}{,} 
        \PY{k}{const} \PY{k+kt}{arrinfo\PYZus{}t}\PY{o}{\PYZlt{}}\PY{k+kt}{real\PYZus{}t}\PY{o}{\PYZgt{}} \PY{n}{rhod\PYZus{}courant\PYZus{}3}
      \PY{p}{)}\PY{p}{;}

      \PY{c+c1}{// time\PYZhy{}stepping methods}
      \PY{k+kt}{void} \PY{n+nf}{step\PYZus{}sync}\PY{p}{(}
        \PY{k}{const} \PY{k+kt}{opts\PYZus{}t}\PY{o}{\PYZlt{}}\PY{k+kt}{real\PYZus{}t}\PY{o}{\PYZgt{}} \PY{o}{\PYZam{}}\PY{p}{,}
        \PY{k+kt}{arrinfo\PYZus{}t}\PY{o}{\PYZlt{}}\PY{k+kt}{real\PYZus{}t}\PY{o}{\PYZgt{}} \PY{n}{th}\PY{p}{,}
        \PY{k+kt}{arrinfo\PYZus{}t}\PY{o}{\PYZlt{}}\PY{k+kt}{real\PYZus{}t}\PY{o}{\PYZgt{}} \PY{n}{rv}\PY{p}{,}
        \PY{k}{const} \PY{k+kt}{arrinfo\PYZus{}t}\PY{o}{\PYZlt{}}\PY{k+kt}{real\PYZus{}t}\PY{o}{\PYZgt{}} \PY{n}{rhod\PYZus{}courant\PYZus{}1}\PY{p}{,}
        \PY{k}{const} \PY{k+kt}{arrinfo\PYZus{}t}\PY{o}{\PYZlt{}}\PY{k+kt}{real\PYZus{}t}\PY{o}{\PYZgt{}} \PY{n}{rhod\PYZus{}courant\PYZus{}2}\PY{p}{,}
        \PY{k}{const} \PY{k+kt}{arrinfo\PYZus{}t}\PY{o}{\PYZlt{}}\PY{k+kt}{real\PYZus{}t}\PY{o}{\PYZgt{}} \PY{n}{rhod\PYZus{}courant\PYZus{}3}\PY{p}{,}
        \PY{k}{const} \PY{k+kt}{arrinfo\PYZus{}t}\PY{o}{\PYZlt{}}\PY{k+kt}{real\PYZus{}t}\PY{o}{\PYZgt{}} \PY{n}{rhod} 
      \PY{p}{)}\PY{p}{;}
      \PY{k+kt}{real\PYZus{}t} \PY{n+nf}{step\PYZus{}async}\PY{p}{(}
        \PY{k}{const} \PY{k+kt}{opts\PYZus{}t}\PY{o}{\PYZlt{}}\PY{k+kt}{real\PYZus{}t}\PY{o}{\PYZgt{}} \PY{o}{\PYZam{}}
      \PY{p}{)}\PY{p}{;}

      \PY{c+c1}{// diagnostic methods}
      \PY{k+kt}{void} \PY{n+nf}{diag\PYZus{}sd\PYZus{}conc}\PY{p}{(}\PY{p}{)}\PY{p}{;}
      \PY{k+kt}{void} \PY{n+nf}{diag\PYZus{}dry\PYZus{}rng}\PY{p}{(}
        \PY{k}{const} \PY{k+kt}{real\PYZus{}t} \PY{o}{\PYZam{}}\PY{n}{r\PYZus{}mi}\PY{p}{,} \PY{k}{const} \PY{k+kt}{real\PYZus{}t} \PY{o}{\PYZam{}}\PY{n}{r\PYZus{}mx}
      \PY{p}{)}\PY{p}{;}
      \PY{k+kt}{void} \PY{n+nf}{diag\PYZus{}wet\PYZus{}rng}\PY{p}{(}
        \PY{k}{const} \PY{k+kt}{real\PYZus{}t} \PY{o}{\PYZam{}}\PY{n}{r\PYZus{}mi}\PY{p}{,} \PY{k}{const} \PY{k+kt}{real\PYZus{}t} \PY{o}{\PYZam{}}\PY{n}{r\PYZus{}mx}
      \PY{p}{)}\PY{p}{;}
      \PY{k+kt}{void} \PY{n+nf}{diag\PYZus{}dry\PYZus{}mom}\PY{p}{(}\PY{k}{const} \PY{k+kt}{int} \PY{o}{\PYZam{}}\PY{n}{k}\PY{p}{)}\PY{p}{;}
      \PY{k+kt}{void} \PY{n+nf}{diag\PYZus{}wet\PYZus{}mom}\PY{p}{(}\PY{k}{const} \PY{k+kt}{int} \PY{o}{\PYZam{}}\PY{n}{k}\PY{p}{)}\PY{p}{;}
      \PY{k+kt}{real\PYZus{}t} \PY{o}{*}\PY{n+nf}{outbuf}\PY{p}{(}\PY{p}{)}\PY{p}{;}

      \PY{c+c1}{// ...}
\PY{c+c1}{//\PYZlt{}/listing\PYZgt{}}

      \PY{k+kt}{void} \PY{n+nf}{diag\PYZus{}chem}\PY{p}{(}\PY{k}{const} \PY{k}{enum} \PY{k+kt}{chem\PYZus{}species\PYZus{}t}\PY{o}{\PYZam{}}\PY{p}{)}\PY{p}{;}

      \PY{k}{struct} \PY{n}{impl}\PY{p}{;}
      \PY{n}{std}\PY{o}{:}\PY{o}{:}\PY{n}{auto\PYZus{}ptr}\PY{o}{\PYZlt{}}\PY{n}{impl}\PY{o}{\PYZgt{}} \PY{n}{pimpl}\PY{p}{;}

      \PY{c+c1}{// constructor}
      \PY{k+kt}{particles\PYZus{}t}\PY{p}{(}\PY{k}{const} \PY{k+kt}{opts\PYZus{}init\PYZus{}t}\PY{o}{\PYZlt{}}\PY{k+kt}{real\PYZus{}t}\PY{o}{\PYZgt{}} \PY{o}{\PYZam{}}\PY{p}{)}\PY{p}{;}

      \PY{c+c1}{// helper typedef}
      \PY{k}{typedef} \PY{k+kt}{particles\PYZus{}proto\PYZus{}t}\PY{o}{\PYZlt{}}\PY{k+kt}{real\PYZus{}t}\PY{o}{\PYZgt{}} \PY{k+kt}{parent\PYZus{}t}\PY{p}{;}
    \PY{p}{\PYZcb{}}\PY{p}{;}
  \PY{p}{\PYZcb{}}\PY{p}{;}
\PY{p}{\PYZcb{}}\PY{p}{;}
\end{Verbatim}
  \vspace{-1.4em}%

  \caption{\label{lst:lgrngn_prt}
    \prog{lgrngn::particles\_t} definition
  }
\end{Listing}

\begin{Listing}
  \renewcommand*\FancyVerbStartString{\PY{c+c1}{//\PYZlt{}listing\PYZgt{}}}%
  \renewcommand*\FancyVerbStopString{\PY{c+c1}{//\PYZlt{}/listing\PYZgt{}}}%
  \begin{Verbatim}[commandchars=\\\{\}]
\PY{c+cp}{\PYZsh{}}\PY{c+cp}{pragma once }

\PY{k}{namespace} \PY{n}{libcloudphxx}
\PY{p}{\PYZob{}}
  \PY{k}{namespace} \PY{n}{lgrngn}
  \PY{p}{\PYZob{}}
    \PY{c+c1}{// to make inclusion of Thrust not neccesarry here}
\PY{c+c1}{//\PYZlt{}listing\PYZgt{}}
    \PY{k}{enum} \PY{k+kt}{backend\PYZus{}t} \PY{p}{\PYZob{}} \PY{n}{serial}\PY{p}{,} \PY{n}{OpenMP}\PY{p}{,} \PY{n}{CUDA} \PY{p}{\PYZcb{}}\PY{p}{;} 
\PY{c+c1}{//\PYZlt{}/listing\PYZgt{}}
  \PY{p}{\PYZcb{}}\PY{p}{;}
\PY{p}{\PYZcb{}}\PY{p}{;}
\end{Verbatim}
  \vspace{-1.4em}%

  \caption{\label{lst:lgrngn_bnd}
    \prog{lgrngn::backend\_t} definition
  }
\end{Listing}

\begin{Listing}
  \renewcommand*\FancyVerbStartString{\PY{c+c1}{//\PYZlt{}listing\PYZgt{}}}%
  \renewcommand*\FancyVerbStopString{\PY{c+c1}{//\PYZlt{}/listing\PYZgt{}}}%
  \begin{Verbatim}[commandchars=\\\{\}]
\PY{c+cp}{\PYZsh{}}\PY{c+cp}{pragma once }

\PY{c+cp}{\PYZsh{}}\PY{c+cp}{include \PYZdq{}particles.hpp\PYZdq{}}

\PY{k}{namespace} \PY{n}{libcloudphxx}
\PY{p}{\PYZob{}}
  \PY{k}{namespace} \PY{n}{lgrngn}
  \PY{p}{\PYZob{}}
    \PY{c+c1}{// factory will be explicitely instantiated}
\PY{c+c1}{//\PYZlt{}listing\PYZgt{}}
    \PY{k}{template} \PY{o}{\PYZlt{}}\PY{k}{typename} \PY{k+kt}{real\PYZus{}t}\PY{o}{\PYZgt{}}
    \PY{k+kt}{particles\PYZus{}proto\PYZus{}t}\PY{o}{\PYZlt{}}\PY{k+kt}{real\PYZus{}t}\PY{o}{\PYZgt{}} \PY{o}{*}\PY{n}{factory}\PY{p}{(}
      \PY{k}{const} \PY{k+kt}{backend\PYZus{}t}\PY{p}{,} 
      \PY{k}{const} \PY{k+kt}{opts\PYZus{}init\PYZus{}t}\PY{o}{\PYZlt{}}\PY{k+kt}{real\PYZus{}t}\PY{o}{\PYZgt{}} \PY{o}{\PYZam{}}
    \PY{p}{)}\PY{p}{;}
\PY{c+c1}{//\PYZlt{}/listing\PYZgt{}}
  \PY{p}{\PYZcb{}}\PY{p}{;}
\PY{p}{\PYZcb{}}\PY{p}{;}
\end{Verbatim}
  \vspace{-1.4em}%

  \caption{\label{lst:lgrngn_ftr}
    \prog{lgrngn::factory()} signature
  }
\end{Listing}

The \prog{particles\_t::init()} method intended to be called once at the beginning of the simulation
  performs the initialisation steps described in section~\ref{sec:lgrngn_init}.
The first three arguments are mandatory and should point to the $\theta$, $r_v$
  and $\rho_d$ fields of the Eulerian component of the solver.
The next arguments should point to Courant number field multiplied by the dry-air density $\rho_d$.
The number of required arguments pointing to Courant field components depends 
  on the dimensionality of the modelling framework,
  and ranges from zero (parcel framework) up to three (3D simulation).
The Courant number components are expected to be laid out on the Arakwa-C grid, 
  thus for the 2D case \prog{courant\_1}'s shape is \prog{(nx+1)}$\times$\prog{nz} and
  \prog{courant\_2}'s shape is \prog{nx}$\times$\prog{(nz+1)}.

Time-stepping is split into two methods: \prog{particles\_t::step\_sync()} and
  \prog{particles\_t::step\_async()}.
The former covers representation of processes that alter the Eulerian fields 
  (i.e.~phase changes).
The latter covers other processes (transport of particles, sedimentation and coalescence) 
  which may be computed asynchronously, fer example while the Eulerian model calculates advection of the Eulerian fields.
Both methods take a reference to an instance of \prog{lgrngn::opts\_t} 
  as their first argument.
Among subsequent arguments of \prog{step\_sync()}, only the first are mandatory.
They should point to $\theta$ and $r_v$ fields which will be overwritten by the method.
The Courant field components need to be specified only if the Eulerian component
  of the model solves air dynamics (they are omitted in the case of the kinematic framework
  used in examples in this paper).
The last argument pointing to a $\rho_d$ array is also optional and needs to be specified
  only if the Eulerian framework allows the density to vary in time.
The \prog{step\_async()} method returns accumulated rain flux through the bottom of the domain.

The \prog{particles\_t}'s methods prefixed with \prog{diag\_} provide a mechanism
  for obtaining statistical information on the droplet parameters gridded on the Eulerian component mesh.
The \prog{particles\_t::diag\_sd\_conc} method calculates the concentration of
  computational particles per cell.
The \prog{particles\_t::diag\_dry\_mom()} and \prog{particles\_t::diag\_wet\_mom()}
  calculate statistical moments of the dry and wet size spectra respectively.
The $k$-th moment $M$ of the dry (d) or wet (w) spectrum is defined here as:
\begin{eqnarray}
  M^{[k]}_{d,w} = 
    (\rho_d \Delta V)\tsup{-1}
    \!\!\!\!\!\!\!\!\!\!\!
    \sum\limits_{\substack{i\,\in\,\text{grid cell}\\ r_{d,w_{[i]}}\,\in\,[r_{\rm mi}, r_{\rm mx}]}} 
    \!\!\!\!\!\!\!\!\!\!\!
    N_{[i]} \, r_{d,w_{[i]}}^{k} 
\end{eqnarray}
  where the index $i$ traverses all computational particles and $N$ is the particle multiplicity.
The moment number $k$ is chosen through the methods' argument \prog{k}.
The range of radii $[r_{\rm mi}, r_{\rm mx}]$ over which the moments are to be calculated is chosen 
  by calling \prog{diag\_dry\_rng()} or \prog{diag\_wet\_rng()} before
  calls to \prog{diag\_dry\_mom()} and \prog{diag\_wet\_mom()}, respectively.
Calling the \prog{particles\_t::outbuf()} method causes the
  calculated fields to be stored in an output buffer,
  and a pointer to the first element of the buffer to be returned.

The last element of the particle-based scheme's API is the \prog{factory()} function.
It returns a pointer to a newly allocated instance of the \prog{particles\_t} class.
Its arguments are the backend type (see Listing~\ref{lst:lgrngn_bnd}) and
  the scheme's options as specified by the \prog{opts\_init\_t} fields (see
  Listing~\ref{lst:lgrngn_opt_init}).
The purpose of introducing the \prog{lgrngn::factory()} function is twofold.
First, it makes the backend choice a runtime mechanism rather than a compile-time one
  (backend is one of the compile-time template parameters of \prog{particles\_t}).
Second, it does report an error if the library was compiled without CUDA (GPU)
  or OpenMP (multi-threading) backend support.

\subsubsection{Example calling sequence}\label{sec:lgrngn_callseq}

\begin{figure*}
  \center
  \input{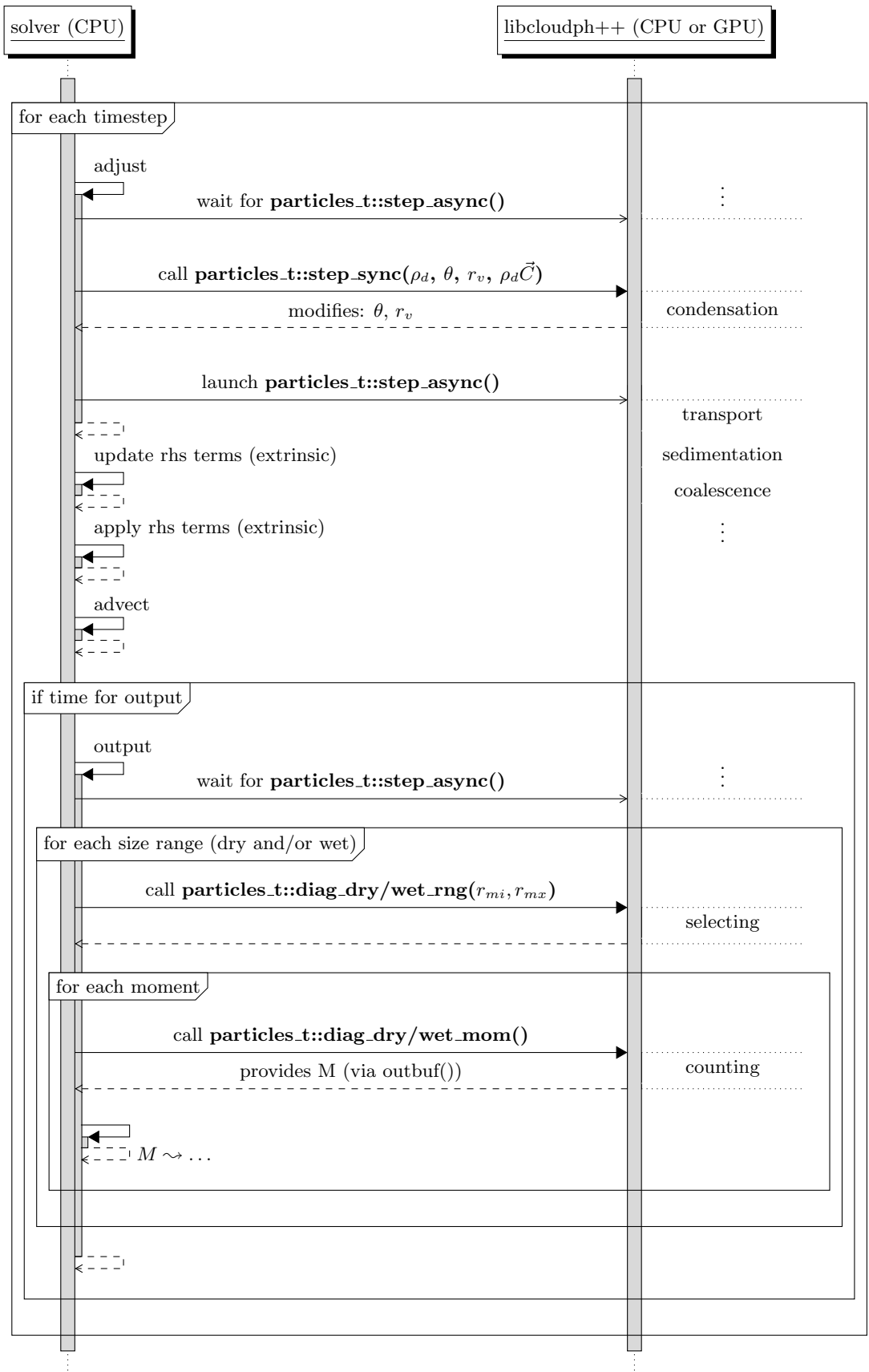}
  \caption{\label{fig:uml_lgrngn}
    Sequence diagram of {\it libcloudph++} API calls for the particle-based scheme
      and a prototype transport equation solver.
    Diagram discussed in section~\ref{sec:lgrngn_callseq}.
    See also caption of Figure~\ref{fig:uml_proto} for description or diagram elements.
  }
\end{figure*}

Figure~\ref{fig:uml_lgrngn} depicts an example calling sequence for the
  particle-based scheme's API.
The API calls are split among the adjustments and output steps of the solver.
The rhs steps are presented in the diagram, but here they refer to
  forcings extrinsic with respect to the cloud microphysics scheme
  (e.g. the relaxation terms in the setup described in Section~\ref{sec:setup}).

In the case of bulk schemes (Figures~\ref{fig:uml_blk_1m} and~\ref{fig:uml_blk_2m})
  both the solver and library flow control was handled by a single thread (or a group of threads performing
  the same operations in case of domain decomposition).
Here, there are two separate threads (or a group of solver threads plus one library thread in case of domain
  decomposition).
The synchronisation between the solver and the library threads is depicted
  in the diagram with ''wait for \ldots'' labels.

In the presented calling sequence the diagnostic methods are only called
  within the output step.
Depending on the modelling framework, such calls may also be needed
  in every timestep, for example to provide data on particle surface
  for a radiative-transfer component, or the data on particle mass for 
  a dynamical component of the solver.
Note that a single call to \prog{diag\_dry/wet\_rng()} may be followed by
  multiple calls to \prog{diag\_dry/wet\_mom()} as depicted by
  nesting the ''for each moment'' loop within the ''for each size range'' loop.

\subsection{Implementation overview}\label{sec:lgrngn_impl}

The Lagrangian component of the model is implemented using the
  Thrust library \citep{Thrust}.
Consequently, all parallelisation logic is hidden behind the
  Thrust API calls.
The parallelisation is obtained by splitting the 
  computational-particle population among several computational units 
  using shared memory.
However, arguably the true power of Thrust is in the possibility to 
  compile the same code for execution on multiple parallel
  architectures including general-purpose GPUs (via CUDA)
  and multi-core CPUs (via OpenMP).
The implemented particle-based scheme is particularly well suited for
  running in a set-up where the Eulerian computations are carried out on a CPU,
  and the Lagrangian computations are delegated to a GPU.
That is due to:
\begin{itemize}
  \item{the low data exchange rate between these two components --
    there is never a need to transfer the state of all computational particles
    to the Eulerian component residing in the main memory, only the aggregated size 
    spectrum parameters defined per each grid box are needed;
  }
  \item{the possibility to perform part of the microphysics computations 
    asynchronously -- simultaneously with other computations carried out on CPU(s) 
    (cf.~Sec.~\ref{sec:lgrngn_cond}).}
\end{itemize}

Since the current release of CUDA compiler does not support C++11, 
  the particle-based scheme is implemented using C++03 constructs only.
Furthermore, the CUDA compiler does not support all C++ constructs used by 
  the Boost.units library.
For this reason, a \prog{fake\_units} drop-in replacement for
  Boost.units was written and is shipped with {\it libcloudph++}.
It causes all quantities in the program to behave as dimensionless.
It is included instead of Boost.units only if compiling the CUDA backend.
Consequently, the particle-based scheme's code is checked for unit correctness
  while compiling other backends.

The asynchronous launch/wait logic is left to be handled by the caller.
In the example program {\it icicle} (see appendix~\ref{sec:icicle}), 
  it is implemented using the C++11's \prog{std::async()} call.

Both in the case of GPU and CPU configurations the
  Mersenne Twister \citep{Matsumoto_and_Nishimura_1998} random number
  generator is used.
If using GPU, the CUDA cuRAND's \prog{MTGP32} is used that offers parallel execution
  with multiple random number streams.
If not using GPU, the C++11 \prog{std::mt19937} is used and 
  the random number generation is done by a single thread only,
  even if using OpenMP.
  
\subsection{Example results}\label{sec:lgrngn_exres}
\begin{figure*}
  \center
  \includegraphics[width=.425\textwidth]{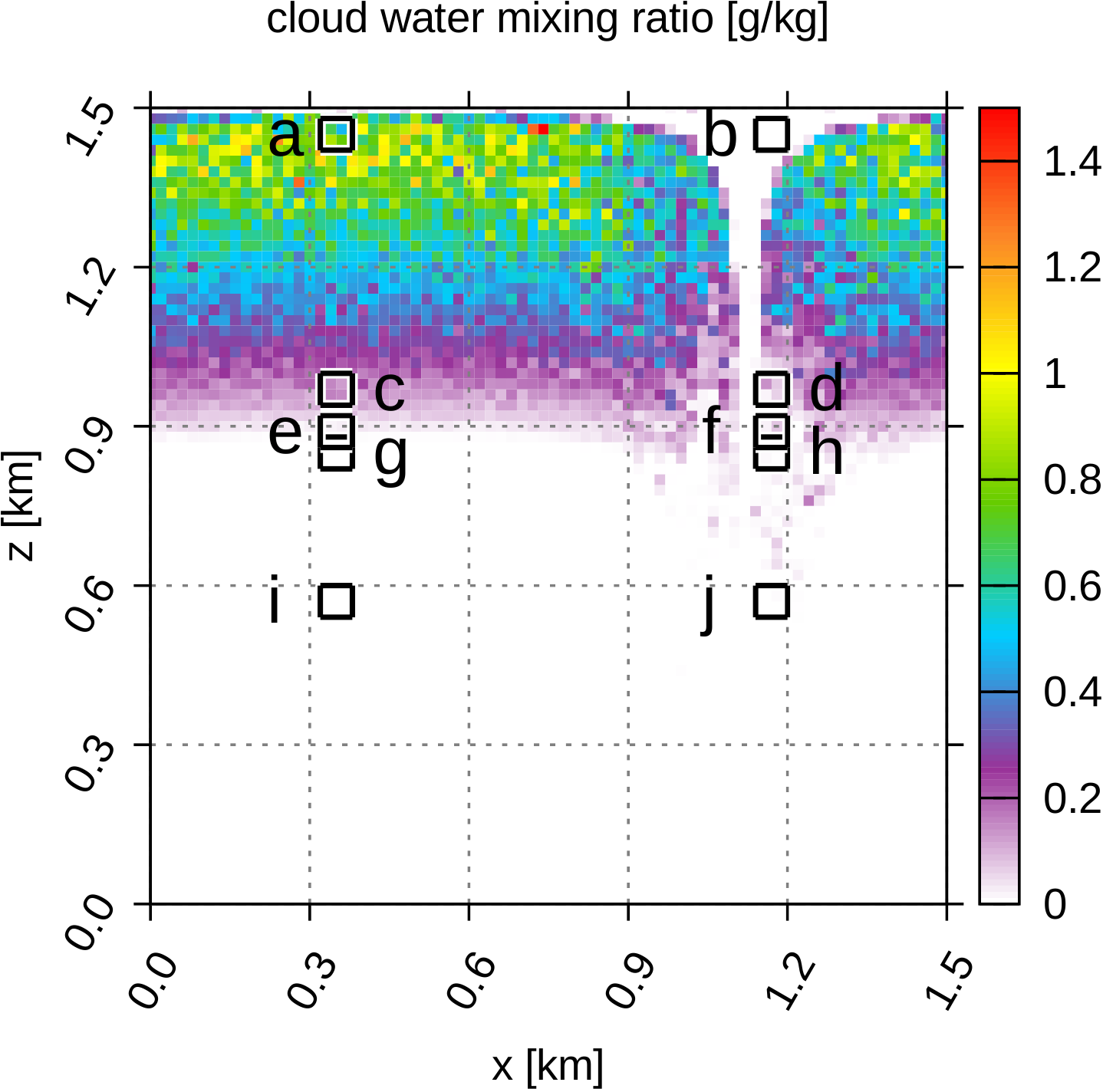}
  ~~~~
  \includegraphics[width=.43\textwidth]{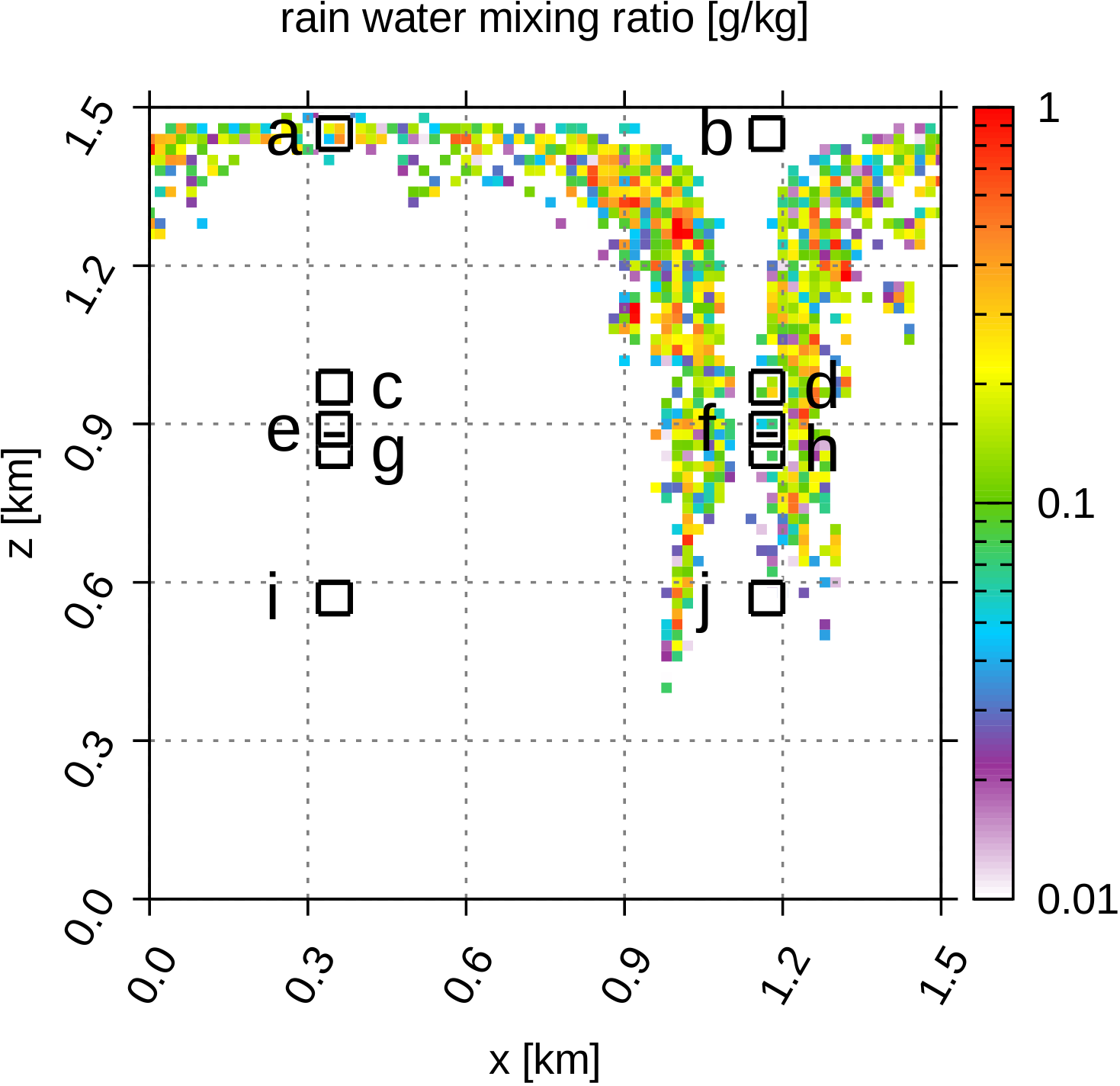}\\
  \vspace{.3em}
  \includegraphics[width=.43\textwidth]{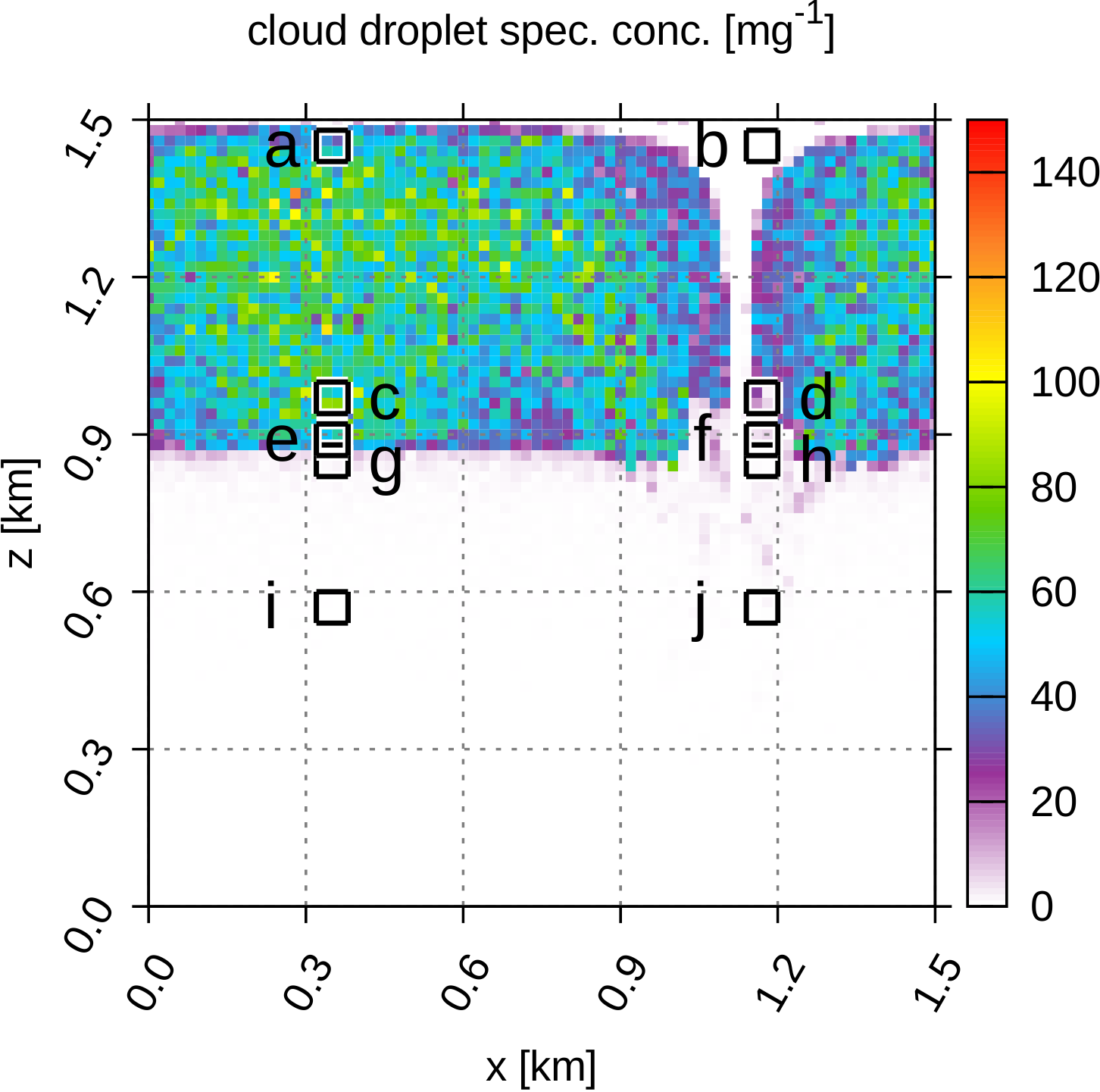}
  ~~~~
  \includegraphics[width=.43\textwidth]{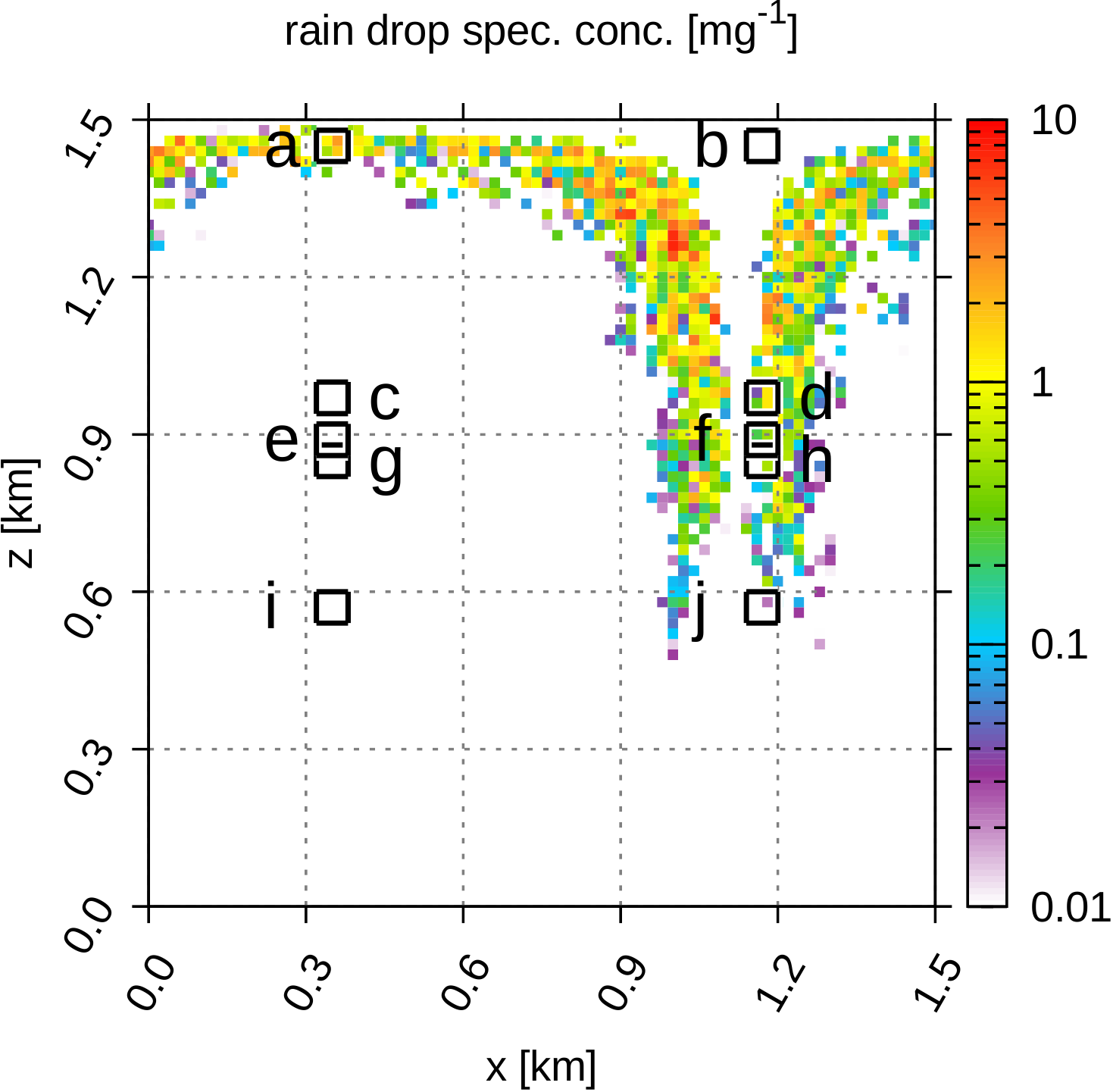}\\
  \vspace{.3em}
  \includegraphics[width=.43\textwidth]{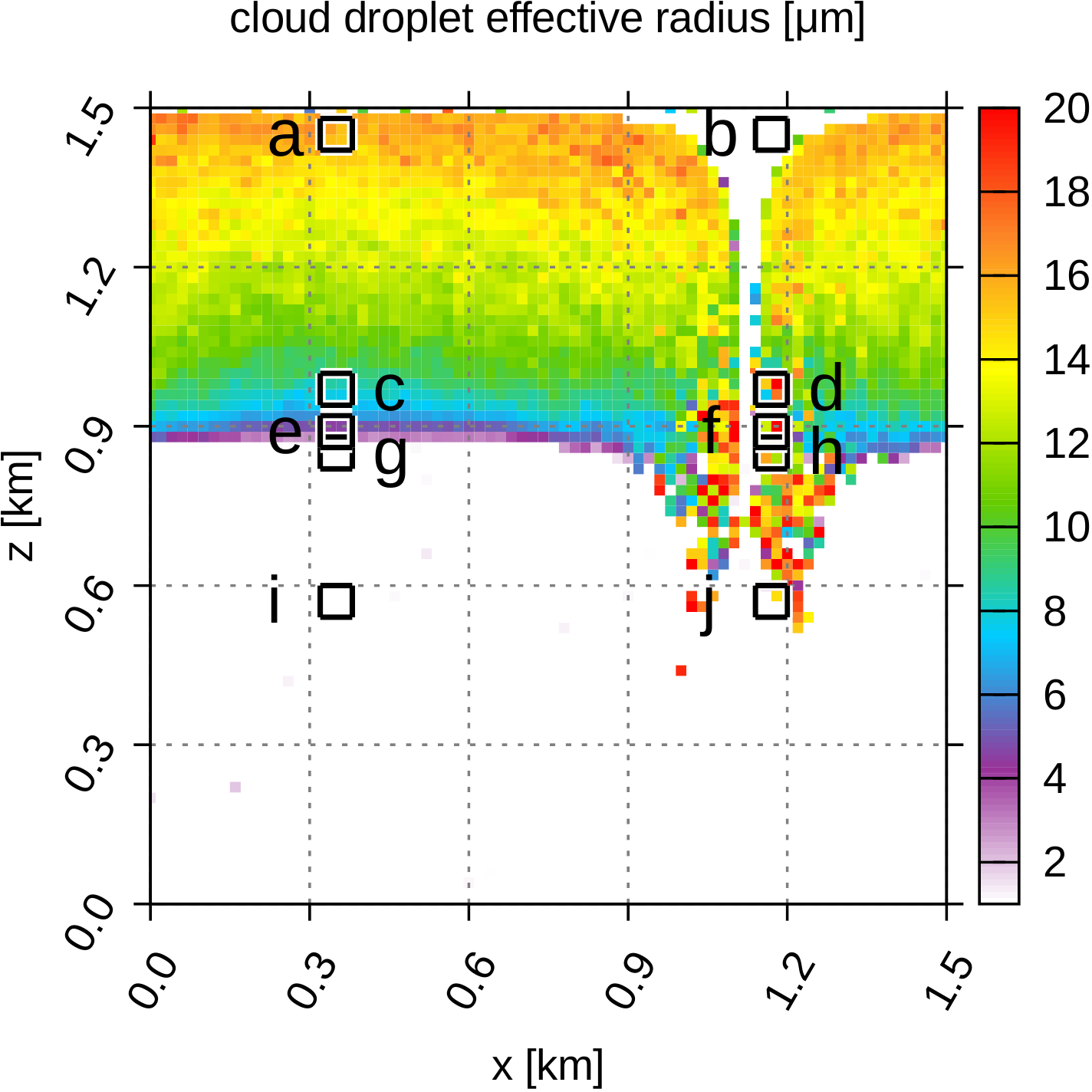}
  ~~~~
  \includegraphics[width=.43\textwidth]{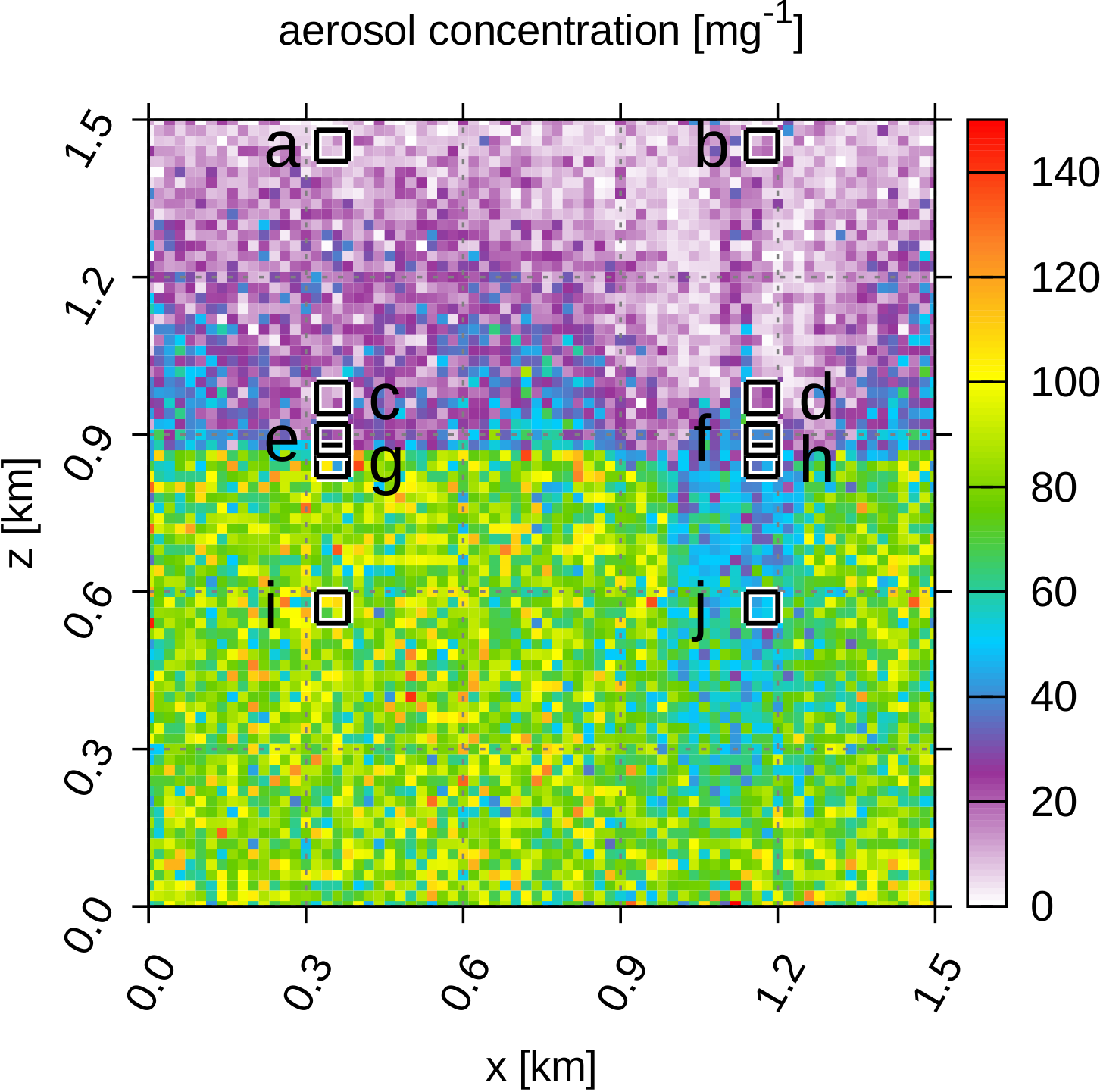}
  \caption{\label{fig:lgrngn_2d}
    Example results from a 2D kinematic simulation using the particle-based scheme.
    All panels depict model state after 30 minutes simulation time (excluding the spin-up period).
    The black overlaid squares mark grid cells for which the dry and wet size spectra are
      shown in Figure~\ref{fig:lgrngn_2d_spec}.
    See section~\ref{sec:lgrngn_exres} for discussion.
  }
\end{figure*}
\begin{figure*}
  \center
  \includegraphics[width=.7\textwidth]{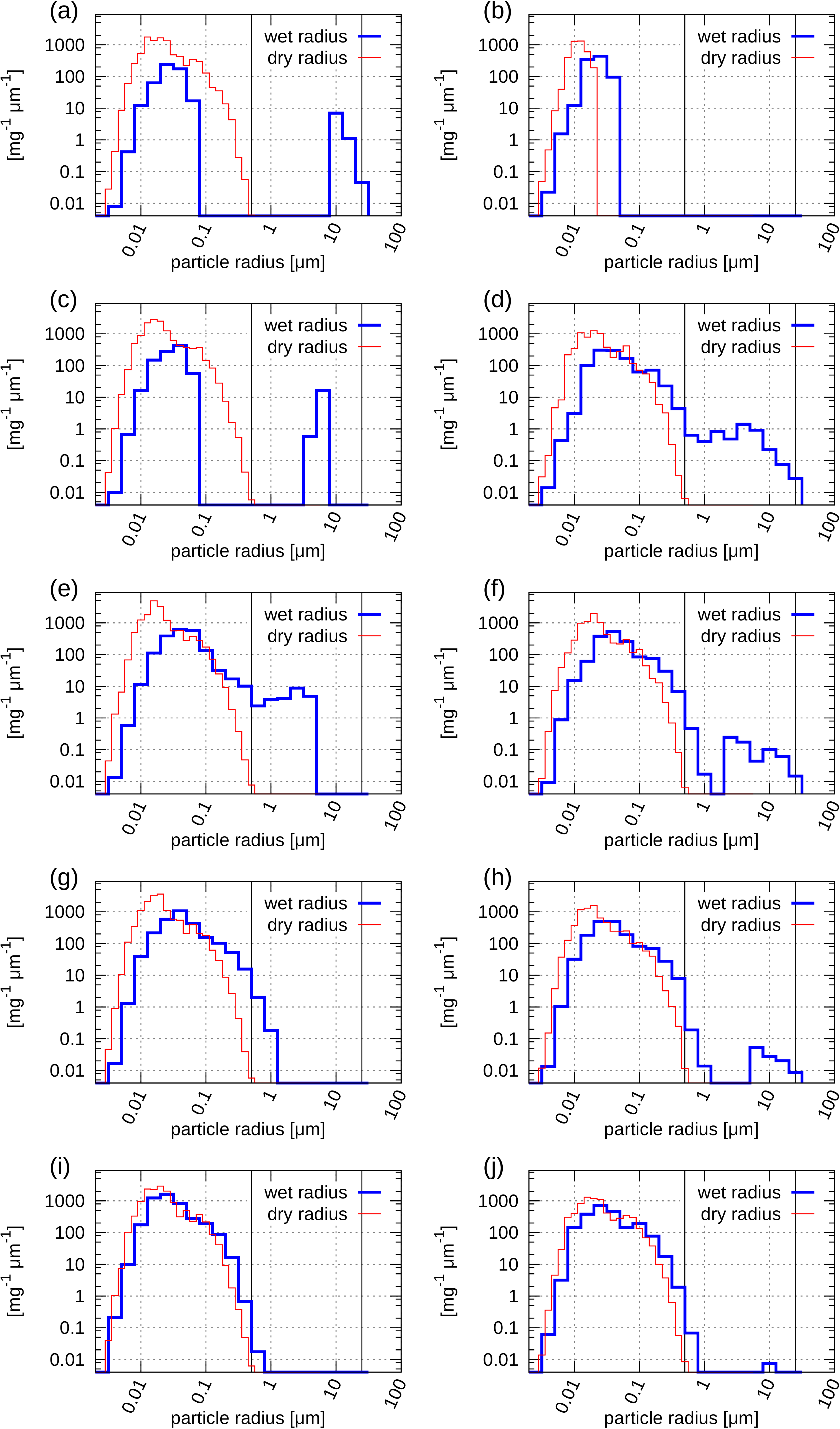}
  \caption{\label{fig:lgrngn_2d_spec}
    Plots of dry and wet size spectra for ten location within the simulation domain.
    The locations and their labels (a--j) are overlaid on plots in Figure~\ref{fig:lgrngn_2d}.
    The vertical bars at 0.5~$\mu$m and 25~$\mu$m indicate the range
      of particle wet radii which is associated with cloud droplets.
    See section~\ref{sec:lgrngn_exres} for discussion.
  }
\end{figure*}

Figures~\ref{fig:lgrngn_2d}~and~\ref{fig:lgrngn_2d_spec}
  present results from an example simulation with the
  particle-based scheme.  
The simulations are analogous to those discussed in Sections~\ref{sec:blk_1m_exres} 
  (single-moment) and \ref{sec:blk_2m_exres} (double-moment).
As before, the plots are for the thirtieth minute of the simulation
  time (again excluding the two-hour-long spin-up period). 
The initial mean concentration of computational particles was set to 64 per
  cell. 
The number of substeps was set to 10 for both condensation
  and coalescence, with coalescence calculated using the geometric
  kernel only.

Figure \ref{fig:lgrngn_2d} depicts gridded aerosol, cloud and rain properties, with
  the gridded data obtained by calculating statistical moments of the
  particle size distribution in each grid cell. 
In addition to quantities corresponding to the bulk model variables 
  $r_c$, $r_r$ (cf.~Figs \ref{fig:blk_1m_2d} and~\ref{fig:blk_2m_2d}) 
  and $n_c$ and $n_r$ (cf.~Fig.~\ref{fig:blk_2m_2d}), Figure~\ref{fig:lgrngn_2d} features
  plots of the effective radius (ratio of the third to the second moment of the
  size spectrum) and the aerosol concentration.  
The distinction between aerosol particles, cloud droplets, and rain drops 
  is made using radius thresholds of 0.5\,$\mu$m and 25\,$\mu$m for 
  aerosol/cloud and cloud/rain boundaries, respectively.
The noise in most panels comes from sampling errors of the
  particle-based scheme, these errors get smaller with increasing number
  of computational particles used (not shown).
The cloud water content and cloud droplet concentration plots both 
  show strong similarities to the results
  of simulation using the double-moment scheme (Fig.~\ref{fig:blk_2m_2d}).
The increase with height of cloud water content, drop concentration 
  approximately constant with height, the maximum droplet concentration 
  near the updraught axis, and the cloud hole are all present in both the
  particle-based and the double-moment simulations.  
The range of values of the rain water content and the rain drop concentration
  predicted by the particle-based model roughly matches those of the
  double-moment scheme, yet the level of agreement is much smaller
  than in the case of cloud water. 
For example, the maximum rain water content in the double-moment 
  simulation is located in the centre of the downdraught, whereas this 
  location features virtually no rain in the particle-based simulation. 
Arguably, this is because of the numerical diffusion of the Eulerian 
  double-moment scheme.
The two schemes agree with respect to the vertical extent of the
  drizzle shaft as it vanishes at about 300 metres above the bottom
  boundary of the domain in both cases.

The plot of the effective radius in Figure~\ref{fig:lgrngn_2d} shows the gradual
  increase of drop sizes from the cloud base up to the top of the cloud.  
The drizzle shaft is the location of the largest particles still classified 
  as cloud droplets.  
The effective radius plot features the smoothest gradients among all 
  presented plots, particularly across the cloud.  
This is likely due to the fact that unlike other plotted quantities, 
  the effective radius is an intensive parameter and hence is not 
  proportional to the drop concentration which inherits random 
  fluctuations of the initial aerosol concentrations.
The aerosol concentration demonstrates anticipated presence of the
  interstitial aerosol within the cloud.  
The regions of largest rain water content correspond to regions of 
  lowered aerosol concentrations, both within and below the cloud.  
This likely demonstrates the effect of scavenging of aerosol particles 
  by the drizzle drops, most likely overpredicted by the geometric 
  collision kernel applied in the simulation.

The ten black squares overlaid on each plot in Figure~\ref{fig:lgrngn_2d} show locations
  of the regions for which the wet and dry particle size spectra are
  plotted in Figure~\ref{fig:lgrngn_2d_spec}.  
The ten locations are composed of a 3$\times$3 grid cells each, 
  and the spectra plotted in the ten panels of Figure~\ref{fig:lgrngn_2d_spec}
  are all averages over the 9 cells.  
The dry spectra are composed of 40~bins in an isologarithmic layout 
  from 1\,nm to 10\,$\mu$m.  
The wet spectra are composed of 25 bins extending the above range up to 100 $\mu$m.  
  Each square in the Figure \ref{fig:lgrngn_2d} and its corresponding panel in
  Figure~\ref{fig:lgrngn_2d_spec} are labelled with a letter (a to j).  
All panels in Figure~\ref{fig:lgrngn_2d_spec} contain two vertical lines at 0.5\,$\mu$m 
  and 25\,$\mu$m that depict the threshold values of particle wet radius 
  used to differentiate between aerosol, cloud droplets and rain drops.

To match the pathway of cloud evolution, we shall discuss the
  panels in Figure~\ref{fig:lgrngn_2d_spec} counterclockwise, 
  starting from panel (i) which presents data on
  the aerosol size spectrum in the updraught below cloud base.
There, the wet spectrum plotted with thick blue line is slightly
  shifted towards larger sizes than the dry spectrum plotted with thin
  red line.
This shift corresponds to humidification of the hygroscopic aerosol.
Panels (g) and (e) show how the wet spectrum evolves while the updraught
  lifts the particles across the cloud base causing the largest aerosol
  to be activated and to form cloud droplets.
Panel (c) shows a distinctly bimodal wet spectrum with an unactivated
  aerosol mode to the left and the cloud droplet mode just below 10\,$\mu$m.
Panel (a) depicting the near-cloud-top conditions shows that some of the
  cloud droplets had already grown pass the 25\,$\mu$m threshold, likely
  through collisional growth.
Such drops have significant fall velocities what causes the air in the
  upper part of the domain to become void of largest aerosol what is
  evident from the shape of the dry spectrum in panel (b) depicting
  conditions above the downdraught.
Panel (d) and panel (c) show size spectra at the same altitude of about
  100 m above cloud base.
Their comparison reveals that the spectrum of cloud droplets in the
  downdraught (panel d, edge of the cloud hole) is much
  wider than near the updraught axis (panel c).
Finally, panels (f), (h) and (j) show gradual evaporation of drizzle
  and cloud droplets back to aerosol-sized particles.

\section{Performance evaluation}\label{sec:perf}

\begin{figure*}[th!]
  \center
  \pgfimage[width=\textwidth]{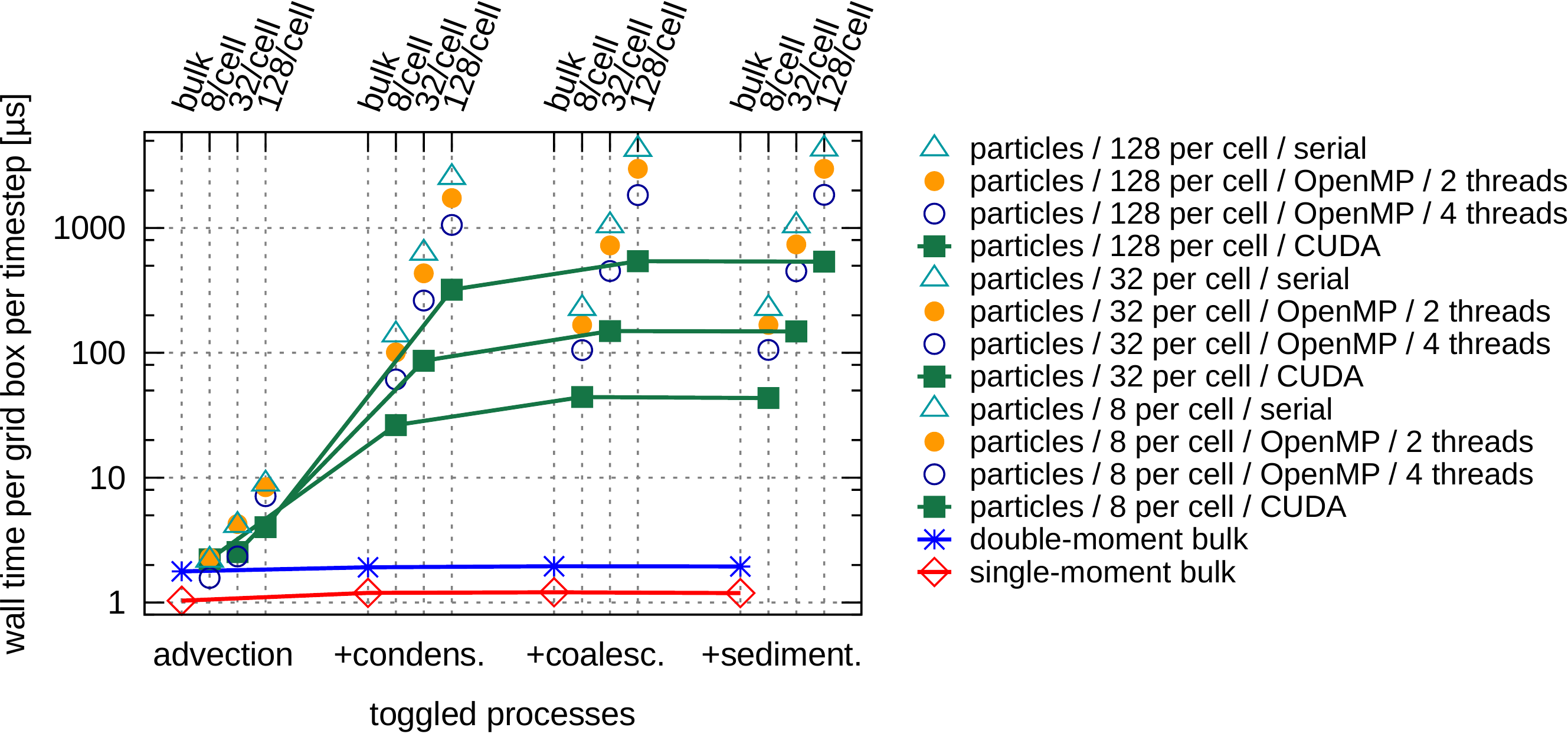}\\ 
  \caption{\label{fig:timing}
    Computational cost of the three microphysics schemes expressed as
      wall-clock time per timestep per grid box.
    Values measured for different settings of process-toggling options
      shown (bottom horizontal axis).
    Results obtained with the particle-based scheme are grouped by
      the number of computational particles used (upper helper horizontal axes).
    See section~\ref{sec:perf} for discussion.
  }
\end{figure*}

Computational cost of a microphysics scheme is one of the key factors
  determining its practical applicability.
Here, we present a basic analysis of the computational cost of the 
  three schemes presented in this paper.
The analysis is based on timing of simulations carried out
  with the kinematic framework and the simulation set-up 
  described in section~\ref{sec:framework} using the {\it icicle}
  tool described in appendix~\ref{sec:icicle}.
In order to depict the contributions of individual elements of the
  schemes, all simulations were repeated with four sets of process-toggling
  options:
\begin{itemize}
  \item{advection only,}
  \item{advection and phase changes,}
  \item{advection, phase changes and coalescence,}
  \item{all above plus sedimentation.}
\end{itemize}
For the particle-based scheme, the advection-only runs include transport
  of particles and the Eulerian fields (moisture and heat).

Simulations were performed 
  with a 6-core AMD Phenom~II CPU and a 96-core nVidia Quadro~600 GPU
  (an example 2010 prosumer desktop computer).
The CPU code was compiled using GCC 4.8 with -Ofast, -march=native and -DNDEBUG options enabled.
The GPU code was compiled with nvcc 5.5 with -arch=sm\_20 and -DNDEBUG options enabled.
No data output was performed.

In order to eliminate from the reported values the time spent on simulation startup,
  all simulations were repeated twice, performing a few timesteps in the first
  run and a dozen timesteps in the second run.
The long and short run times are then subtracted and the result is normalised
  by the difference in number of timesteps.

In order to reduce the chance of an influence of other processes on the
  wall-clock timing, all simulations were additionally thrice-repeated and
  the shortest measured time is reported.

The particle-based simulations were performed with three different mean densities
  of computation particles, 8, 32 and 128 per grid cell,
  and with four ''backend'' settings:
\begin{itemize}
  \item{serial backend,}
  \item{OpenMP backend using 2 threads,}
  \item{OpenMP backend using 4 threads,}
  \item{CUDA backend using the GPU.}
\end{itemize}
The test was completed for single-precision arithmetics.
The GPU used offered about three 
  times higher performance at single precision.
Higher-performance GPU hardware typically applied in computing centres
  is expected to deliver similar performance for double precision.
Execution times for CPU-only calculations hardly change when switching 
  from double to single precision.

Figure~\ref{fig:timing} presents measured wall-clock times 
  for the four sets of processes (bottom x-axis labels) and for all three schemes (different colours and symbols).
The figure reveals the significant spread of times needed to
  compute a single timestep -- spanning over three orders of magnitude.
For simulations with all processes turned on, it takes the double-moment scheme
  roughly twice longer than the single-moment scheme to advance the solution 
  by one timestep.
The particle-based scheme may be anything from about ten- to over hundred-times more costly than 
  the double-moment bulk scheme depending on its settings.

Figure~\ref{fig:timing} also shows how the execution time of the particle-based scheme 
  depends on the backend choice and on the number of computational particles used.
The execution time is also dependent on the number of subtimesteps used for phase changes
  and coalescence (not shown, the default of 10 subtimesteps per one advective step was used here).
It is also evident in Figure~\ref{fig:timing} that computations of
  phase changes for particle-based simulations take most of the simulation time.
The code responsible for the iterative implicit solution of the drop-growth 
  equation is thus the first candidate for optimisation (e.g.,~through employment
  of a faster-converging root-finding algorithm).

Arguably, the most striking feature depicted in Figure~\ref{fig:timing} is the order-of-magnitude 
  speedup between serial execution times for CPU and the GPU execution times.
Even compared to the four-thread OpenMP runs, the GPU backend offers a threefold speedup.
It is worth reiterating here the two reasons why the particle-based scheme is 
  particularly well-suited for GPUs.
First, the large body of data defining the state of all particles never leaves the GPU memory
  (and the GPU-CPU transfer bandwidth is often a major issue for the performance of GPU codes).
Here, all data that are transferred from the GPU are first gridded onto the Eulerian mesh
  before being sent from GPU to the main memory.
Second, a significant part of the computations (i.e.~everything but phase changes) may be computed
  asynchronously, leaving all but one CPU available for other tasks of the solver
  (one thread is busy controlling the GPU).

Finally, Figure~\ref{fig:timing} also depicts the linear scaling of the computational 
  cost of the particle-based method with the number of computational particles (cf. section~\ref{sec:sdm}).
Regardless of the backend choice, increasing the mean number of particles per cell from 
  8 to 32 to 128 gives a linear increase of wall-time as seen in the logarithmic scale of the plot.

The library is still at its initial stage of development, and 
  ongoing work on its code is expected to result in shorter execution
  times.

\section{Summary and outlook}\label{sec:concl}  

The main goal behind the ongoing development of {\it libcloudph++}, as stated in Section~\ref{sec:intro},
  has been to develop and to offer the community a set of reusable software components of applicability in
  modern cloud modelling.
Incorporation of the double-moment bulk and the particle-based schemes suitable for studies on 
  the interactions between clouds and aerosol makes the library applicable
  for research on the widely discussed indirect effects of aerosol on climate.

The implementation of the library was carried out having 
  maintainability and auditability as priorities. 
This is reflected in:
\begin{description}
  \item[{\bf (i)}]{the choice of C++ with its concise and modularity-encouraging syntax\footnote{%
    As of current release, {\it libcloudph++} consists of ca.~70 files with a total of ca.~5500 lines of code (LOC)
      of which ca.~1000 LOC are common to all schemes, and ca.~500, 1000 and 4500 LOC are pertaining to
      the single-moment, double-moment and particle-based schemes, respectively.
  };}
  \item[{\bf (ii)}]{the separation of code elements related to the schemes' formulation (formul\ae) 
    from other elements of the library (API, numerics);}
  \item[{\bf (iii)}]{the adoption of compile-time dimensional analysis for all physically-meaningful expressions in the code;}
  \item[{\bf (iv)}]{the delegation of substantial part of the library implementation to external libraries (including the dimensional analysis, algorithm parallelisation and GPU hardware handling);}
  \item[{\bf (v)}]{the hosting of library development and handling of code dissemination through a public code repository.}
\end{description}
All above, supported by the choice of the GNU General Public License,
  underpins our goal of offering reusable code.\\
  
Development plans for the upcoming releases of {\it libcloudph++} include:
\begin{description}
  \item[{\bf (i)}]{enriching the set of usage examples;}
  \item[{\bf (ii)}]{widening the choice of available schemes, also by integration of third-party codes;} 
  \item[{\bf (iii)}]{extending the spectrum of processes in the particle-based scheme (notably, by covering basic aqueous chemistry, droplet breakup and implementing more realistic collision kernels).}
\end{description}
Python bindings to {\em libcloudph++} offering analogous functionality as
  the original C++ interface are already included in the library code.
Their description along with reports on further developments will be 
  reported in forthcoming communications.

\appendix

\section{Common concepts and nomenclature}\label{sec:common}

This section presents some key elements of a mostly standard approach to 
  analytic description of motion of moist air, particularly in context
  of modelling warm-rain processes.
It is given for the sake of completeness of the formulation and to
  ease referencing particular equations from within the text and the
  source code.

\paragraph{Governing equations}
~\\

There are three key types of matter considered in the model formulation and their
  densities $\rho_i$ and mass mixing ratios $r_i$ are defined as follows:
\begin{eqnarray}
  \notag \rho_d &             &\text{~~dry air}\\
         \rho_v &= r_v \rho_d &\text{~~water vapour}\\
         \rho_l &= r_l \rho_d &\text{~~liquid water}
\end{eqnarray}
The governing equations are 
  the continuity equation for dry air, 
  a conservation law for water vapour,
  and the thermodynamic equation \citep[see e.g.][Sec.~1.6]{Vallis_2006}:
\begin{eqnarray}
  \label{eq:cont_d} \partial_t \rho_d + \nabla \cdot (\vec{u} \rho_d) &=& 0\\
  \label{eq:cont_v} \frac{Dr_v}{Dt}  &=& \dot{r}_v\\
  \label{eq:entrop} \frac{Ds}{Dt} &=& \frac{\dot{q}}{T} 
\end{eqnarray}
where $s$ and $\dot{q}$ represent entropy and heat sources, respectively
  (both defined per unit mass of dry air).
The dot notation 
  is used to distinguish variations due to transport 
  and due to thermodynamic processes.

It is assumed already in (\ref{eq:cont_d}) that the presence of moisture
  and its transformations through phase changes do not influence the
  density of dry air.
Dry-air flow is assumed to act as a carrier flow for trace constituents.
This assumption is corroborated by the fact that in the Earth's atmosphere
  $1 \gg r_v > r_l$.

\paragraph{System of transport equations}
~\\
~\\
Equations (\ref{eq:cont_v}) and~(\ref{eq:entrop}) may be conveniently expressed
  as a pair of transport equations of a similar form to (\ref{eq:cont_d}).

A continuity equation for water vapour density $\rho_v$
  is obtained by summing $(\ref{eq:cont_v}) \cdot \rho_d + r_v \cdot (\ref{eq:cont_d})$:
\begin{eqnarray}\label{eq:cont_rv}
  \partial_t (\rho_d r_v) + \nabla \cdot (\vec{u} \rho_d r_v) = \rho_d \dot{r}_v
\end{eqnarray}

Combining equation (\ref{eq:entrop}) with the definition of
  potential temperature $\theta^\star$:
\begin{eqnarray}
  ds = c_p^\star d({\rm ln}\theta^\star)
\end{eqnarray}
gives:
\begin{eqnarray}\label{eq:thermo}
  c_{p}^\star \frac{D\theta^\star}{Dt} = \frac{\theta^\star}{T} \dot{q}
\end{eqnarray}
At this point no assumption is made on the exact form of $\theta^\star$
  or $c_p^\star$.
Summing $(\ref{eq:cont_d})\cdot \theta^\star c_p^\star + (\ref{eq:thermo})\cdot \rho_d$ and $\rho_d \theta^\star \cdot \frac{D}{Dt} c_p^\star = \rho_d \theta^\star \dot{c}_p^\star$
  results in a continuity equation for $\rho_d c_p^\star \theta^\star$ (akin to energy density):
\begin{eqnarray}\label{eq:cont_th}
  \partial_t (\rho_d c_p^\star \theta^\star) + \nabla \cdot (\vec{u} \rho_d c_p^\star \theta^\star) 
    = \rho_d \theta^\star \left[\dot{c}_p^\star + \dot{q}/T \right]
\end{eqnarray}
Resultant equations (\ref{eq:cont_rv}) and (\ref{eq:cont_th}) share the form of a generalised 
  transport equation \citep[see][sec.~4.1]{Smolarkiewicz_2006}:
\begin{eqnarray}
  \partial_t (\rho_d \phi) + \nabla \cdot (\rho_d \vec{u} \phi) = \rho_d \dot{\phi}
\end{eqnarray}
representing transport of a quantity $\phi$ (equal to $r_v$ or $c_p^\star \theta^\star$)
  by a dry-air carrier flow.

\paragraph{Dry air potential temperature}
~\\

The way the potential temperature was defined in the preceding section
  gives a degree of freedom in the choice of $\theta^\star$ and $\dot{q}$.
For moist air containing suspended water aerosol, assuming thermodynamic equilibrium 
  and neglecting the expansion work of liquid water, $ds$ may be expressed as 
  \citep[eq.~6.10-6.11 in][]{Curry_and_Webster_1999}:
\begin{eqnarray}\label{eq:ds}
  d s &=& \overbrace{
    c_{pd}d({\rm ln} T) - R_d d ({\rm ln} p_d)
  }^{
    c_{pd} d (ln \theta)
  }\\ \notag
  &+&
  \underbrace{
    \left[l_v dr_v + \left(r_v c_{pv} + r_l c_l + r_v l_v / T\right) dT \right]
  }_{
    -dq
  } / T
\end{eqnarray}
where $p_d=\rho_d R_d T$ is the partial pressure of dry air, and the potential temperature $\theta$ is defined here as:
\begin{eqnarray}
  \theta = T \left(\nicefrac{p_{1000}}{p_d}\right)^{\nicefrac{R_d}{c_{pd}}}
\end{eqnarray}
\citep[$p_{1000}=1000 hPa$, note that the definition features the dry air pressure 
as opposed to the total pressure, see e.g.][]{Bryan_2008, Duarte_et_al_2014}.

Substituting $c_{p}^\star = c_{pd} = \text{const}$ and $\theta^\star = \theta$
  into equation (\ref{eq:cont_th})
  and employing the form of $\dot{q}$ hinted with $-dq$ in equation (\ref{eq:ds})
  gives:
\begin{eqnarray}\label{eq:rhod_th}
  \partial_t (\rho_d \theta) \!&\!+\!&\! \nabla \cdot (\vec{u} \rho_d \theta) =\\
    \notag
    \!&\!=\!&\! \frac{-\rho_d \theta}{c_{pd}T} \left[ 
    l_v \dot{r_v} 
    + \cancel{\dot{T} \left( r_v c_{pv} + r_l c_l + \frac{r_v l_v}{T} \right)}
  \right]
\end{eqnarray}
Neglecting of all but the $l_v \dot{r}$ terms on the right-hand side results in an approximation 
  akin to the one employed in \citet{Grabowski_and_Smolarkiewicz_1996} and used herein as well.

Another common choice of $\theta^\star$ and $\dot{q}$ is obtained by putting $\theta^\star=\theta \cdot {\rm exp}\left(\nicefrac{-r_v l_v}{c_{pd} T}\right)$,
  what results in the $l_v dr_v$ term becoming a part of $c_{pd} d(ln\theta^\star)$ instead of $-dq$
  in equation~\ref{eq:ds} \citep[see e.g.][sect.~3]{Grabowski_and_Smolarkiewicz_1990}.

\paragraph{Diagnosing $T$ and $p$ from state variables} 
~\\

The principal role of any cloud-microphysics scheme is to close the equation
  system defined by (\ref{eq:cont_rv}) and (\ref{eq:rhod_th}) with a definition of $\dot{r}_v$
  linked with a representation of liquid water within the model domain.
This requires representation of various thermodynamic processes that depend on temperature and
  pressure which are diagnosed from the model state variables 
  (i.e.~the~quantities for which the transport equations are solved).
With the approach outlined above, the model state variables are:
\begin{description}
  \item[$r_v$]{water vapour mixing ratio}
  \item[$\theta$]{potential temperature}
\end{description}
Assuming $\rho_d$ is known (solved by a dynamical core of a model),
  temperature and pressure may be diagnosed from $r_v$ and $\theta$ with:
\begin{eqnarray}
  T      &=& \left[\theta \left(\frac{\rho_d R_d}{p_{1000}}\right)^{\frac{R_d}{c_{pd}}}\right]^{c_{pd}/(c_{pd} - R_d)}\\
  p      &=& \rho_d \left( R_d + r_v R_v \right) T
\end{eqnarray}

\section{List of symbols}\label{sec:symbols}
A list of symbols is provided in Table~\ref{tab:symbols}.

\begin{table*}
  \caption{\label{tab:symbols}List of symbols}
  \center
  \footnotesize
  \begin{tabular}{lllll}
    Symbol                                                        & SI unit                       & Description                                   \\ \hline
    $A=2 \sigma_w / (R_v T \rho_w)$                               & [m]                           & Kelvin term exponent parameter \\
    $\beta_M$, $\beta_T$                                          & [1]                           & transition-r\'egime correction factors \\
    $\Delta t$, $\Delta x$, $\Delta z$, $\Delta V$                & [s] or [m] or [m\tsup{3}]     & timestep, grid cell dimensions and volume \\
    $\theta_l$                                                    & [K]                           & liquid water potential temperature (cf. Sect.~\ref{sec:setup})\\
    $\theta$                                                      & [K]                           & potential temperature                         \\ 
    $\kappa$                                                      & [1]                           & hygroscopicity parameter                      \\
    $\rho_i$                                                      & depends on i                  & any state variable (density)                  \\
    $\rho_d$, $\rho_v$                                            & [kg\,m\tsup{-3}]              & densities of dry air and vapour vapour        \\
    $\rho_c$, $\rho_r$                                            & [kg\,m\tsup{-3}]              & cloud and rain water densities/content        \\ 
    $\rho_w = 1000$                                               & [kg\,m\tsup{-3}]              & density of liquid water                       \\
    $\rho_{vs}$                                                   & [kg\,m\tsup{-3}]              & saturation vapour density                     \\                                          
    $\rho_\circ$                                                  & [kg\,m\tsup{-3}]              & vapour density at drop surface \\
    $\dot{\rho}_i$, $\dot{\rho}_c$, $\dot{\rho}_r$                & depends on i                  & rhs terms (any, cloud water, rain water)      \\
    $\sigma_m$                                                    & [1]                           & geometric standard deviation (lognormal spectrum)  \\   
    $\sigma_w = 0.072$                                            & [J\,m\tsup{-2}]               & surface tension coefficient of water \\
    $\tau$, $\tau_{\rm rlx}$                                      & [s]                           & relaxation time scale (cf. Sect.~\ref{sec:setup}) \\
    $\phi_i$                                                      & depends on i                  & any advected specific quantity (e.g. mixing ratio) \\
    $\psi$                                                        & [kg\,m\tsup{-1}\,s\tsup{-1}]  & streamfunction                                \\
    $a_w=(r_w^3 - r_d^3)/(r_w^3 - r_d^3 \cdot (1 - \kappa))$      & [1]                           & water activity                                \\
    $a$, $b$                                                      & [m\tsup{2}]                   & initial interval for bisection algorithm \\
    $c_{pd} = 1005$, $c_{pv}=1850$, $c_l=4218$                    & [J\,kg\tsup{-1}\,K\tsup{-1}]  & specific heat at const. pressure (dry air, vapour \& liquid water) \\
    $C$                                                           & [1]                           & Courant number \\
    $d_m$, $r_m$                                                  & [m]                           & mode diameter and radius (lognormal spectrum) \\
    $D$, $D_{\rm eff}$, $D_0$                                     & [m\tsup{2}\,s\tsup{-1}]       & diffusion coefficients for water vapour in air \\
    $E_r$                                                         & [kg\,m\tsup{-3}\,s\tsup{-1}]  & evaporation rate of rain (single-moment scheme) \\
    $E(r_i, r_j)$                                                 & [1]                           & collection efficiency \\
    $F_{\rm in}$, $F_{\rm out}$                                   & [kg\,m\tsup{-3}\,s\tsup{-1}]  & fluxes of $\rho_r$ through the grid cell edges \\
    $K$, $K_0$                                                    & [J\,m\tsup{-1}\,s\tsup{-1}\,K\tsup{-1}] & thermal conductivities of air \\
    $K(r_i, r_j)$                                                 & [m\tsup{3}\,s\tsup{-1}]       & collection kernel \\
    $l_{v0} = 2.5\times10^6$                                      & [J\,kg\tsup{-1}]              & latent heat of evaporation at the triple point                  \\
    $l_v(T) = l_{v0} + (c_{pv} - c_{l}) \cdot (T - T_0)$          & [J\,kg\tsup{-1}]              & latent heat of evaporation at a given temperature \\
    $M^{[k]}$                                                     & [m\tsup{-3+k}]                & $k$-th moment of size spectrum \\
    $n$                                                           & [1]                           & total number of computational particles \\
    $n_c$, $n_r$                                                  & [m\tsup{-3}]                  & cloud droplet and rain drop concentrations \\
    $N$                                                           & [1]                           & multiplicity (attribute of computational particle) \\
    $N_m$                                                         & [m\tsup{-3}]                  & particle concentration (lognormal spectrum)   \\
    $p$, $p_d$                                                    & [Pa]                          & pressure, dry air partial pressure \\
    $P_{ij}$                                                      & [1]                           & probability of collisions \\
    $Q$, $q$                                                      & [J\,m\tsup{-3}], [J\,kg\tsup{-1}] & heat per unit volume and mass \\
    $r_d$, $r_w$                                                  & [m]                           & particle dry and wet radii                    \\
    $r_{c0}$                                                      & [kg kg\tsup{-1}]              & autoconversion threshold (mixing ratio)       \\
    $r_v$, $r_l$, $r_t=r_v+r_l$                                   & [kg kg\tsup{-1}]              & mixing ratios (vapour, liquid, total)         \\
    $R_v$, $R_d$                                                  & [J\,K\tsup{-1}\,kg\tsup{-1}]  & gas constants for water vapour and dry air \\
    $S$, $s$                                                      & [J\,K\tsup{-1}\,m\tsup{-3}], [J\,K\tsup{-1}\,kg\tsup{-1}]  & entropy per unit volume and mass \\
    $T$                                                           & [K]                           & temperature \\
    $\vec{u} = (u,v)$                                             & [m\,s\tsup{-1}]               & velocity field                                \\
    $v_t$, $v_i$, $v_j$                                           & [m\,s\tsup{-1}]               & terminal velocity                             \\
    $w_{\rm max}$                                                 & [m\,s\tsup{-1}]               & maximum velocity (cf. amplitude of $\psi$)    \\
    $w$                                                           & [1]                           & averaging weight in particle advection scheme \\
    $x$, $z$                                                      & [m]                           & spatial coordinate                            \\
    $X$, $Z$                                                      & [m]                           & domain extent
  \end{tabular}
\end{table*}

\section{Example program ``icicle''}\label{sec:icicle}

\begin{figure*}[b!]
  \center
  \begin{tikzpicture}
    \tikzset{sibling distance=8pt}
    \tikzset{level distance=50pt}
    \tikzset{edge from parent/.append style={very thick}}
    \tikzset{every tree node/.style={align=center,anchor=north}} 
    \Tree [.\fbox{icicle}
	      [.\fbox{libcloudph++}
		  [.Thrust 
                      [.CUDA\\or~OpenMP ]
                  ]   
                  [.Boost\\(Units,\\odeint,\\\ldots) ]
	      ]
              [.\fbox{libmpdata++}
                  [.Blitz++ ]
                  [.HDF5 ]
                  [.OpenMP\\(or~Boost.Thread) ]
                  [.Boost\\(\ldots) ]
              ]
              [.Boost\\(program\_options,\\Spirit,\\\ldots) ]
	  ]
  \end{tikzpicture}

  \caption{\label{fig:depend}
    A tree of libcloudph++'s and icicle's major dependencies.
    In addition to these libraries, several components require C++11 compiler and CMake at build time.  
  }
\end{figure*}
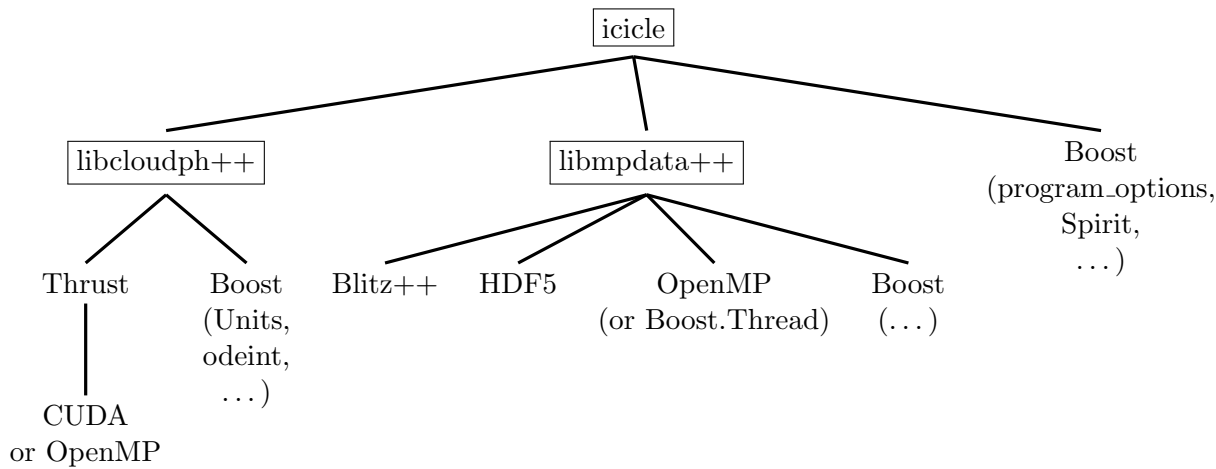

The example simulations discussed in the text were performed with {\it icicle} --
  an implementation of all elements of the example modelling context presented
  in Section~\ref{sec:context}, that is: transport equation solver, 2D kinematic
  framework and simulation set-up.

\paragraph{Dependencies}
~\\

The code of {\it icicle} depend on several libraries: {\it libcloudph++}'s sister project 
  {\em libmpdata++} \citep{Jaruga_et_al_2014}, {\it libcloudph++} itself 
  and several components of the Boost\footnote{\url{http://boost.org/}} collection.
The {\it libmpdata++} components solve the transport equations for the
  Eulerian fields using the MPDATA algorithm \citep{Smolarkiewicz_2006}
  and provide data output facility using the HDF5 library\footnote{\url{http://hdfgroup.org/}}.
Figure~\ref{fig:depend} presents dependency tree of {\it icicle}.
Source code of {\it icicle}, {\it libmpdata++} and {\it libcloudph++} is 
  available for download at \url{http://foss.igf.fuw.edu.pl/}. 
All other {\it icicle} dependencies are available, for instance,
  as Debian\footnote{\url{http://debian.org/}} packages.
All {\it icicle} dependencies are free (gratis) software, and 
  all but CUDA (which is an optional dependency) are 
  additionally libre -- open sourced, and released under freedom-ensuring licenses.

\paragraph{Compilation}
~\\

Build automation for {\it icicle}, {\it libmpdata++} and {\it libcloudph++}
  is handled in a standard way using CMake\footnote{\url{http://cmake.org/}}.
In all three cases, a possible command sequence will resemble:
\begin{verbatim}
$ mkdir build
$ cd build
$ cmake .. -DCMAKE_BUILD_TYPE=Release
$ make 
$ make test 
# make install
\end{verbatim}
  

\paragraph{Usage}
~\\

Control over simulation options of {\it icicle} is available via 
  command-line parameters.
Most of the options correspond to the fields of the \prog{opts\_t} structures
  of the three microphysics schemes discussed in the paper.
A list of general options may be obtained by calling
\begin{verbatim}
$ icicle --help
\end{verbatim}
  and includes, in particular, the \prog{-\,-micro} option that selects the
  microphysics scheme.
Options specific to each of the three available schemes are listed as in the following example:
\begin{verbatim}
$ icicle --micro=lgrngn --help
\end{verbatim}
For the particle-based scheme, the options include such settings as the
  backend type (serial, OpenMP or CUDA) and the size ranges for 
  which to output the moments of the particle size distribution.

Simulations may be stopped at any time by sending the process a SIGTERM
  or SIGINT signal (e.g., using the \prog{kill} utility or with \prog{Ctrl+C}).
It causes the solver to continue integration up to the end of the current timestep,
  close the output file and exit.
After executing the simulation, its progress may be monitored for example
  with \prog{top -H} as the process threads' names are continuously 
  updated with the percentage of work completed.

\relscale{.95}

\begin{acknowledgements}
SA thanks Shin-ichiro Shima (University of Hyogo, Japan) for
  introducing to particle-based simulations.
We thank Dorota Jarecka (University of Warsaw) and 
  Graham Feingold (NOAA) for insightful discussions
  and comments to the initial version of the manuscript.
Development of libcloudph++, libmpdata++ and icicle have been supported by the Polish National
  Science Centre (projects PRELUDIUM 2011/01/N/ST10/01483 and HARMONIA 2012/06/M/ST10/00434).
Additional support was provided by the European Union 7 FP ACTRIS 
  (Aerosol, Clouds, and Trace gases Research InfraStructure network) No. 262254.
WWG's institution NCAR is operated by the University Corporation for Atmospheric Research 
  under sponsorship of the US National Science Foundation.
The authors express their appreciation of the work of the developers of the free/libre/open-source
  software which served as a basis for implementation of the presented library
  (see section \ref{sec:icicle} for a list).
We would like to express our admiration to the way the Clang\footnote{\url{http://clang.llvm.org/}}
  C++ compiler improved the comfort of development and debugging of heavily-templated 
  code based on libraries such as Boost.units and Blitz++.
All figures were generated using gnuplot\footnote{\url{http://gnuplot.info/}}. 
\end{acknowledgements}

\bibliography{paper}

\begin{thebibliography}{65}
\providecommand{\natexlab}[1]{#1}
\providecommand{\url}[1]{{\tt #1}}
\providecommand{\urlprefix}{URL }
\expandafter\ifx\csname urlstyle\endcsname\relax
  \providecommand{\doi}[1]{doi:\discretionary{}{}{}#1}\else
  \providecommand{\doi}{doi:\discretionary{}{}{}\begingroup
  \urlstyle{rm}\Url}\fi

\bibitem[{Ahnert and Mulansky(2013)}]{Ahnert_and_Mulansky_2013}
Ahnert, K. and Mulansky, M.: Boost.Numeric.Odeint: Solving ordinary
  differential equations, in: Boost Library Documentation, (available at
  \url{http://www.boost.org/doc/libs/}), 2013.

\bibitem[{Allen et~al.(2011)Allen, Coe, Clarke, Bretherton, Wood, Abel,
  Barrett, Brown, George, Freitag, McNaughton, Howell, Shank, Kapustin,
  Brekhovskikh, Kleinman, Lee, Springston, Toniazzo, Krejci, Fochesatto, Shaw,
  Krecl, Brooks, McMeeking, Bower, Williams, Crosier, Crawford, Connolly,
  Allan, Covert, Bandy, Russell, Trembath, Bart, McQuaid, Wang, and
  Chand}]{Allen_et_al_2011}
Allen, G., Coe, H., Clarke, A., Bretherton, C., Wood, R., Abel, S.~J., Barrett,
  P., Brown, P., George, R., Freitag, S., McNaughton, C., Howell, S., Shank,
  L., Kapustin, V., Brekhovskikh, V., Kleinman, L., Lee, Y.-N., Springston, S.,
  Toniazzo, T., Krejci, R., Fochesatto, J., Shaw, G., Krecl, P., Brooks, B.,
  McMeeking, G., Bower, K.~N., Williams, P.~I., Crosier, J., Crawford, I.,
  Connolly, P., Allan, J.~D., Covert, D., Bandy, A.~R., Russell, L.~M.,
  Trembath, J., Bart, M., McQuaid, J.~B., Wang, J., and Chand, D.: South East
  Pacific atmospheric composition and variability sampled along
  20\textsuperscript{$\circ$}S during VOCALS-REx, Atmos. Chem. Phys., 11,
  5237--5262, 2011.

\bibitem[{Andrejczuk et~al.(2008)Andrejczuk, Reisner, Henson, Dubey, and
  Jeffery}]{Andrejczuk_et_al_2008}
Andrejczuk, M., Reisner, J., Henson, B., Dubey, M., and Jeffery, C.: The
  potential impacts of pollution on a nondrizzling stratus deck: Does aerosol
  number matter more than type?, J. Geoph. Res., 113, D19\,204,
  \doi{10.1029/2007JD009445}, 2008.

\bibitem[{Andrejczuk et~al.(2010)Andrejczuk, Grabowski, Reisner, and
  Gadian}]{Andrejczuk_et_al_2010}
Andrejczuk, M., Grabowski, W., Reisner, J., and Gadian, A.: Cloud-aerosol
  interactions for boundary layer stratocumulus in the Lagrangian Cloud Model,
  J. Geoph. Res., 115, D22\,214, 2010.

\bibitem[{Arabas and Pawlowska(2011)}]{Arabas_and_Pawlowska_2011}
Arabas, S. and Pawlowska, H.: Adaptive method of lines for multi-component
  aerosol condensational growth and CCN activation, Geosci. Model Dev., 4,
  15--31, 2011.

\bibitem[{Arabas and Shima(2013)}]{Arabas_and_Shima_2013}
Arabas, S. and Shima, S.: Large Eddy Simulations of Trade-Wind Cumuli using
  Particle-Based Microphysics with Monte-Carlo Coalescence, J. Atmos. Sci., 70,
  2768–--2777, \doi{10.1175/JAS-D-12-0295.1}, 2013.

\bibitem[{Bott(1997)}]{Bott_1997}
Bott, A.: A Flux Method for the Numerical Solution of the Stochastic Collection
  Equation, J. Atmos. Sci., 55, 2284--2293, 1997.

\bibitem[{Brokken(2013)}]{Brokken_2013}
Brokken, F.: C++ Annotations, Center of Information Technology, University of
  Groningen, \urlprefix\url{http://cppannotations.sf.net/}, 2013.

\bibitem[{Bryan(2008)}]{Bryan_2008}
Bryan, G.: On the Computation of Pseudoadiabatic Entropy and Equivalent
  Potential Temperature, Mon. Wea. Rev., 136, 5239--5245,
  \doi{0.1175/2008MWR2593.1}, 2008.

\bibitem[{Castellano and \'{A}vila(2011)}]{Castellano_and_Avila_2011}
Castellano, N.~E. and \'{A}vila, E.~E.: Vapour density field of a population of
  cloud droplets, J. Atmos. Solar-Terr. Phys., 73, 2423--2428,
  \doi{10.1016/j.jastp.2011.08.013}, 2011.

\bibitem[{Clift et~al.(1978)Clift, Grace, and Weber}]{Clift_et_al_1978}
Clift, R., Grace, J., and Weber, M.: Bubbles, Drops, and Particles, Academic
  Press, New York, (reprinted by Dover Publications, 2005), 1978.

\bibitem[{Crowe et~al.(2012)Crowe, Schwarzkopf, Sommerfeld, and
  Tsuji}]{Crowe_et_al_2012}
Crowe, C., Schwarzkopf, J., Sommerfeld, M., and Tsuji, Y.: Multiphase flows
  with droplets and particles, CRC Press, Boca Raton, FL, USA, second edn.,
  2012.

\bibitem[{Curry and Webster(1999)}]{Curry_and_Webster_1999}
Curry, J. and Webster, P.: Thermodynamics of Atmospheres and Oceans, Academic
  Press, 1999.

\bibitem[{Duarte et~al.(2014)Duarte, Almgren, Balakrishnan, Bell, and
  Romps}]{Duarte_et_al_2014}
Duarte, M., Almgren, A., Balakrishnan, K., Bell, J., and Romps, D.: A Numerical
  Study of Methods for Moist Atmospheric Flows: Compressible Equations, Mon.
  Wea. Rev., \doi{10.1175/MWR-D-13-00368.1}, 2014.

\bibitem[{Easterbrook and Johns(2009)}]{Easterbrook_and_Johns_2009}
Easterbrook, S.~M. and Johns, T.~C.: Engineering the software for understanding
  climate change, Comp. Sci. Eng., 11, 65--74, 2009.

\bibitem[{Fern\'{a}ndez-D\'{i}az et~al.(1999)Fern\'{a}ndez-D\'{i}az, Bra\~{n}a,
  Garc\'{i}a, Mu\~{n}iz, and Nieto}]{Fernandez-Diaz_et_al_1999}
Fern\'{a}ndez-D\'{i}az, J.~M., Bra\~{n}a, M. A.~R., Garc\'{i}a, B.~A.,
  Mu\~{n}iz, C. G.-P., and Nieto, P. J.~G.: The Goodness of the Internally
  Mixed Aerosol Assumption Under Condensation-Evaporation, Aerosol Sci. Tech.,
  31, 17--23, \doi{10.1080/027868299304327}, 1999.

\bibitem[{Golaz et~al.(2002)Golaz, Larson, and Cotton}]{Golaz_et_al_2002}
Golaz, J.-C., Larson, V., and Cotton, W.: A PDF-Based Model for Boundary Layer
  Clouds. Part {I}: Method and Model Description, J. Atmos. Sci., 59,
  3540--3551, 2002.

\bibitem[{Grabowski and Smolarkiewicz(1990)}]{Grabowski_and_Smolarkiewicz_1990}
Grabowski, W. and Smolarkiewicz, P.: Monotone finite-difference approximations
  to the advection-condensation problem, Mon. Wea. Rev., 118, 2082--2097, 1990.

\bibitem[{Grabowski and Smolarkiewicz(1996)}]{Grabowski_and_Smolarkiewicz_1996}
Grabowski, W. and Smolarkiewicz, P.: Two-time-level semi-lagrangian modeling of
  precipitating clouds, Mon. Weather Rev., pp. 487--497,
  \doi{10.1175/1520-0493(1996)124$<$0487:TTLSLM$>$2.0.CO;2}, 1996.

\bibitem[{Grabowski and Smolarkiewicz(2002)}]{Grabowski_and_Smolarkiewicz_2002}
Grabowski, W. and Smolarkiewicz, P.: A Multiscale Anelastic Model for
  Meteorological Research, Mon. Wea. Rev., 130, 939--956,
  \doi{10.1175/1520-0493(1983)111<0479:ASPDAS>2.0.CO;2}, 2002.

\bibitem[{Grabowski and Wang(2013)}]{Grabowski_and_Wang_2013}
Grabowski, W.~W. and Wang, L.-P.: Growth of Cloud Droplets in a Turbulent
  Environment, Annu. Rev. Fluid Mech., 45, 293--324,
  \doi{10.1146/annurev-fluid-011212-140750}, 2013.

\bibitem[{Hoberock and Bell(2010)}]{Thrust}
Hoberock, J. and Bell, N.: Thrust: a parallel template library,
  \urlprefix\url{http://thrust.github.io/}, 2010.

\bibitem[{Ince et~al.(2012)Ince, Hatton, and Graham-Cumming}]{Ince_et_al_2012}
Ince, D., Hatton, L., and Graham-Cumming, J.: The case for open computer
  programs, Nature, pp. 485--488, \doi{10.1038/nature10836}, 2012.

\bibitem[{Jarecka et~al.(2013)Jarecka, Grabowski, Morrison, and
  Pawlowska}]{Jarecka_et_al_2013}
Jarecka, D., Grabowski, W., Morrison, H., and Pawlowska, H.: Homogeneity of the
  Subgrid-Scale Turbulent Mixing in Large-Eddy Simulation of Shallow
  Convection, J. Atmos. Sci., pp. 2751--–2767, \doi{10.1175/JAS-D-13-042.1},
  2013.

\bibitem[{Jaruga et~al.(2014)Jaruga, Arabas, Jarecka, Pawlowska, Smolarkiewicz,
  and Waruszewski}]{Jaruga_et_al_2014}
Jaruga, A., Arabas, S., Jarecka, D., Pawlowska, H., Smolarkiewicz, P., and
  Waruszewski, M.: libmpdata++ 0.1: a library of parallel MPDATA solvers for
  systems of generalised transport equations, ArXiv e-prints,
  \urlprefix\url{http://arxiv.org/abs/1407.1309}, 2014.

\bibitem[{Kessler(1995)}]{Kessler_1995}
Kessler, E.: On the continuity and distribution of water substance in
  atmospheric circulations, Atmos. Res., 38, 109--145,
  \doi{10.1016/0169-8095(94)00090-Z}, 1995.

\bibitem[{Khairoutdinov and Kogan(2000)}]{Khairoutdinov_and_Kogan_2000}
Khairoutdinov, M. and Kogan, Y.: A new cloud physics parameterization in a
  large-eddy simulation model of marine stratocumulus, Mon.Wea.Rev., 128,
  229--243, 2000.

\bibitem[{Khvorostyanov and Curry(2002)}]{Khvorostyanov_and_Curry_2002}
Khvorostyanov, V. and Curry, J.: Terminal Velocities of Droplets and Crystals:
  Power Laws with Continuous Parameters over the Size Spectrum, J. Atmos. Sci.,
  59, 1872--1884, 2002.

\bibitem[{Khvorostyanov and Curry(2006)}]{Khvorostyanov_and_Curry_2006}
Khvorostyanov, V. and Curry, J.: Aerosol size spectra and CCN activity spectra:
  Reconciling the lognormal, algebraic, and power laws, J.Geophys. Res., 111,
  \doi{10.1029/2005JD006532}, 2006.

\bibitem[{Laaksonen et~al.(2005)Laaksonen, Vesala, Kulmala, Winkler, and
  Wagner}]{Laaksonen_et_al_2005}
Laaksonen, A., Vesala, T., Kulmala, M., Winkler, P., and Wagner, P.: Commentary
  on cloud modelling and the mass accommodation coefficient of water, Atmos.
  Chem. Phys., 5, 461--464, \doi{10.5194/acp-5-461-2005}, 2005.

\bibitem[{Lebo and Seinfeld(2011)}]{Lebo_and_Seinfeld_2011}
Lebo, Z. and Seinfeld, J.: Continuous spectral aerosol-droplet microphysics
  model, Atmos. Chem. Phys., 11, 12\,297--12\,316,
  \doi{10.5194/acp-11-12297-2011}, 2011.

\bibitem[{Marcolli et~al.(2004)Marcolli, Luo, Peter, and
  Wienhold}]{Marcolli_et_al_2004}
Marcolli, C., Luo, B.~P., Peter, T., and Wienhold, F.~G.: Internal mixing of
  the organic aerosol by gas phase diffusion of semivolatile organic compounds,
  Atmos. Chem. and Phys., 4, 2593--2599, 2004.

\bibitem[{Matsumoto and Nishimura(1998)}]{Matsumoto_and_Nishimura_1998}
Matsumoto, M. and Nishimura, T.: Mersenne twister: a 623-dimensionally
  equidistributed uniform pseudo-random number generator, ACM Trans. Model.
  Comput. Simul., 8, 3--30, \doi{10.1145/272991.272995}, 1998.

\bibitem[{Mayer and Kylling(2005)}]{Mayer_and_Kylling_2005}
Mayer, B. and Kylling, A.: Technical note: The {libRadtran} software package
  for radiative transfer calculations - description and examples of use, Atmos.
  Chem. Phys., 5, 1855--1877, \doi{10.5194/acp-5-1855-2005}, 2005.

\bibitem[{McFarquhar(2010)}]{McFarquhar_2010}
McFarquhar, G.: Raindrop size distribution and evolution, pp. 49--59, AGU,
  \doi{10.1029/2010GM000971}, 2010.

\bibitem[{Mitra et~al.(1992)Mitra, Brinkmann, and
  Pruppacher}]{Mitra_et_al_1992}
Mitra, S., Brinkmann, J., and Pruppacher, H.: A wind tunnel study on the
  drop-to-particle conversion, J. Aerosol Sci., 23, 245--256,
  \doi{10.1016/0021-8502(92)90326-Q}, 1992.

\bibitem[{Morin et~al.(2012)Morin, Urban, Adams, Foster, Sali, Baker, and
  Sliz}]{Morin_et_al_2012}
Morin, A., Urban, J., Adams, P., Foster, I., Sali, A., Baker, D., and Sliz, P.:
  Shining Light into Black Boxes, Science, 336, 159--160,
  \doi{10.1126/science.1218263}, 2012.

\bibitem[{Morrison and Grabowski(2007)}]{Morrison_and_Grabowski_2007}
Morrison, H. and Grabowski, W.: Comparison of Bulk and Bin Warm-Rain
  Microphysics Models Using a Kinematic Framework, J. Atmos. Sci., 64,
  2839--2861, \doi{10.1175/JAS3980}, 2007.

\bibitem[{Morrison and Grabowski(2008)}]{Morrison_and_Grabowski_2008}
Morrison, H. and Grabowski, W.: Modeling supersaturation and subgrid-scale
  mixing with two-moment bulk warm microphysics, J. Atmos. Sci., 65, 792--812,
  2008.

\bibitem[{Morrison et~al.(2005)Morrison, Curry, and
  Khvorostyanov}]{Morrison_et_al_2005}
Morrison, H., Curry, J., and Khvorostyanov, V.: A new double-moment
  microphysics parameterization for application in cloud and climate models.
  Part I: Description., J. Atmos. Sci., 62, 1665--1677, 2005.

\bibitem[{Muhlbauer et~al.(2013)Muhlbauer, Grabowski, Malinowski, Ackerman,
  Bryan, Lebo, Milbrandt, Morrison, Ovchinnikov, Tessendorf, Thériault, and
  Thompson}]{Muhlbauer_et_al_2013}
Muhlbauer, A., Grabowski, W.~W., Malinowski, S.~P., Ackerman, T.~P., Bryan,
  G.~H., Lebo, Z.~J., Milbrandt, J.~A., Morrison, H., Ovchinnikov, M.,
  Tessendorf, S., Thériault, J.~M., and Thompson, G.: Reexamination of the
  State-of-the-art of Cloud Modeling Shows Real Improvements, Bull. Amer.
  Meteor. Soc., 94, ES45--–ES48, \doi{10.1175/BAMS-D-12-00188}, 2013.

\bibitem[{Ogura and Takahashi(1971)}]{Ogura_and_Takahashi_1971}
Ogura, Y. and Takahashi, T.: Numerical simulation of the life cycle of a
  thunderstorm cell, Mon. Wea. Rev., 99, 895--911,
  \doi{10.1175/1520-0493(1971)099<0895:NSOTLC>2.3.CO;2}, 1971.

\bibitem[{Pennell and Reichler(2010)}]{Pennel_and_Reichler_2010}
Pennell, C. and Reichler, T.: On the Effective Number of Climate Models, J.
  Climate, 24, 2358--2367, \doi{10.1175/2010JCLI3814.1}, 2010.

\bibitem[{Petters and Kreidenweis(2007)}]{Petters_et_al_2007}
Petters, M. and Kreidenweis, S.: A single parameter representation of
  hygroscopic growth and cloud condensation nucleus activity, Atmos. Chem.
  Phys., 7, 1961--1971, \doi{10.5194/acp-8-6273-2008}, 2007.

\bibitem[{Rasinski et~al.(2011)Rasinski, Pawlowska, and
  Grabowski}]{Rasinski_et_al_2011}
Rasinski, P., Pawlowska, H., and Grabowski, W.: Observations and kinematic
  modeling of drizzling marine stratocumulus, Atmos. Res., 102, 120--135,
  \doi{10.1016/j.atmosres.2011.06.020}, 2011.

\bibitem[{Riechelmann et~al.(2012)Riechelmann, Noh, and
  Raasch}]{Riechelmann_et_al_2012}
Riechelmann, T., Noh, Y., and Raasch, S.: A new method for large-eddy
  simulations of clouds with Lagrangian droplets including the effects of
  turbulent collision, New J. Phys, 14, 065\,008,
  \doi{10.1088/1367-2630/14/6/065008}, 2012.

\bibitem[{Rogers and Davis(1990)}]{Rogers_and_Davis_1990}
Rogers, J. and Davis, R.: The Effects of van der Waals Attractions on Cloud
  Droplet Growth by Coalescence, J. Atmos. Sci., 47, 1075--1080, 1990.

\bibitem[{Schabel and Watanabe(2008)}]{Schabel_and_Watanabe_2008}
Schabel, M. and Watanabe, S.: Boost.Units: Zero-overhead dimensional analysis
  and unit/quantity manipulation and conversion, in: Boost Library
  Documentation, (available at \url{http://www.boost.org/doc/libs/}), 2008.

\bibitem[{Shima et~al.(2009)Shima, Kusano, Kawano, Sugiyama, and
  Kawahara}]{Shima_et_al_2009}
Shima, S., Kusano, K., Kawano, A., Sugiyama, T., and Kawahara, S.: The
  Super-Droplet Method for the numerical simulation of clouds and
  precipitation: A particle-based and probabilistic microphysics model coupled
  with a non-hydrostatic model, Quart. J. Roy. Meteor. Soc., 135, 1307--1320,
  \doi{10.1002/qj.441}, 2009.

\bibitem[{Simmel et~al.(2002)Simmel, Trautmann, and
  Tetzlaff}]{Simmel_et_al_2002}
Simmel, M., Trautmann, T., and Tetzlaff, G.: Numerical solution of the
  stochastic collection equation—comparison of the Linear Discrete Method
  with other methods, Atmos. Res., 61, 135--148, 2002.

\bibitem[{Slawinska et~al.(2009)Slawinska, Grabowski, and
  Morrison}]{Slawinska_et_al_2009}
Slawinska, J., Grabowski, W.~W., and Morrison, H.: The impact of atmospheric
  aerosols on precipitation from deep organized convection: A prescribed-flow
  model study using double-moment bulk microphysics, Quart. J. Roy. Meteor.
  Soc., 135, 1906--1913, \doi{10.1002/qj.450}, 2009.

\bibitem[{Smolarkiewicz(2006)}]{Smolarkiewicz_2006}
Smolarkiewicz, P.: Multidimensional positive definite advection transport
  algorithm: An overview, Int. J. Numer. Meth. Fluids, 50, 1123--1144,
  \doi{10.1002/fld.1071}, 2006.

\bibitem[{Smol\'{i}k et~al.(2001)Smol\'{i}k, D\v{z}umbov\'{a}, Schwarz, and
  Kulmala}]{Smolik_et_al_2001}
Smol\'{i}k, J., D\v{z}umbov\'{a}, L., Schwarz, J., and Kulmala, M.: Evaporation
  of ventilated water droplet: connection between heat and mass transfer,
  Journal of Aerosol Science, 32, 739--748,
  \doi{10.1016/S0021-8502(00)00118-X}, 2001.

\bibitem[{S\"olch and K\"archer(2010)}]{Soelch_and_Kaercher_2010}
S\"olch, I. and K\"archer, B.: A large-eddy model for cirrus clouds with
  explicit aerosol and ice microphysics and Lagrangian ice particle tracking,
  Quart. J. Roy. Meteor. Soc., 136, 2074--2093, \doi{10.1002/qj.689}, 2010.

\bibitem[{Stevens and Feingold(2009)}]{Stevens_and_Feingold_2009}
Stevens, B. and Feingold, G.: Untangling aerosol effects on clouds and
  precipitation in a buffered system, Nature, 461, 607--613,
  \doi{10.1038/nature08281}, 2009.

\bibitem[{Straka(2009)}]{Straka_2009}
Straka, J.: Cloud and Precipitation Microphysics: Principles and
  Parameterizations, Cambridge University Press, 2009.

\bibitem[{Szak\'{a}ll et~al.(2010)Szak\'{a}ll, Mitra, Diehl, and
  Borrmann}]{Szakall_et_al_2010}
Szak\'{a}ll, M., Mitra, S.~K., Diehl, K., and Borrmann, S.: Shapes and
  oscillations of falling raindrops — A review, Atmos. Res., 97, 416--425,
  \doi{10.1016/j.atmosres.2010.03.024}, 2010.

\bibitem[{Szumowski et~al.(1998)Szumowski, Grabowski, and
  Ochs~{III}}]{Szumowski_et_al_1998}
Szumowski, M., Grabowski, W., and Ochs~{III}, H.: Simple two-dimensional
  kinematic framework designed to test warm rain microphysical models, Atmos.
  Res., 45, 299--326, \doi{10.1016/S0169-8095(97)00082-3}, 1998.

\bibitem[{Unterstrasser and S\"olch(2014)}]{Unterstrasser_and_Soelch_2014}
Unterstrasser, S. and S\"olch, I.: Optimisation of the simulation particle
  number in a Lagrangian ice microphysical model, Geoscientific Model
  Development, 7, 695--709, \doi{10.5194/gmd-7-695-2014}, 2014.

\bibitem[{Vaillancourt et~al.(2001)Vaillancourt, Yau, and
  Grabowski}]{Vaillancourt_et_al_2001}
Vaillancourt, P., Yau, M., and Grabowski, W.: Microscopic Approach to Cloud
  Droplet Growth by Condensation. Part I: Model Description and Results without
  Turbulence, J. Atmos. Sci., 58, 1945--–1964,
  \doi{10.1175/1520-0469(2001)058<1945:MATCDG>2.0.CO;2}, 2001.

\bibitem[{Vallis(2006)}]{Vallis_2006}
Vallis, G.: Atmospheric and Oceanic Fluid Dynamics: Fundamentals and
  Large-Scale Circulation, Cambridge University Press, Cambridge, 2006.

\bibitem[{Veldhuizen(2005)}]{Veldhuizen_2005}
Veldhuizen, T.: Blitz++ User's Guide: A C++ class library for scientific
  computing, \urlprefix\url{http://blitz.sf.net/resources/blitz-0.9.pdf},
  version 0.9, 2005.

\bibitem[{Vohl et~al.(2007)Vohl, Mitra, Wurzler, Diehl, and
  Pruppacher}]{Vohl_et_al_2007}
Vohl, O., Mitra, S., Wurzler, S., Diehl, K., and Pruppacher, H.: Collision
  efficiencies empirically determined from laboratory investigations of
  collisional growth of small raindrops in a laminar flow field, Atmos. Res.,
  85, 120--125, 2007.

\bibitem[{Wilson et~al.(2014)Wilson, Aruliah, Titus~Brown, Chue~Hong, Davis,
  Guy, Haddock, Huff, Mitchell, Plumbley, Waugh, White, and
  Wilson}]{Wilson_et_al_2014}
Wilson, G., Aruliah, D.~A., Titus~Brown, C., Chue~Hong, N.~P., Davis, M., Guy,
  R.~T., Haddock, S.~H.~D., Huff, K., Mitchell, I.~M., Plumbley, M., Waugh, B.,
  White, E.~P., and Wilson, P.: Best Practices for Scientific Computing, PLoS
  Biol, 12, e1001\,745, \doi{10.1371/journal.pbio.1001745}, 2014.

\bibitem[{Wood(2005)}]{Wood_2005}
Wood, R.: Drizzle in Stratiform Boundary Layer Clouds. Part II: Microphysical
  Aspects, J. Atmos. Sci., 62, 3034--3050, \doi{10.1175/JAS3530.1}, 2005.

\end{thebibliography}

\end{document}